\documentclass[12pt]{article}
 
\usepackage{graphicx}
\usepackage{amsfonts}
\usepackage[usenames]{color}
\usepackage{amsmath}

\def\Z{\mathbb{Z}}
\def\R{\mathbb{R}}

\def\diag{\mathop{\rm diag}\nolimits}
\def\tr{\mathop{\rm tr}}

\def\SO{\mathop{\rm SO}}

\def\SU{\mathop{\rm SU}}
\def\U{\mathop{\rm U}}

\def\simgt{\mathrel{\lower2.5pt\vbox{\lineskip=0pt\baselineskip=0pt
           \hbox{$>$}\hbox{$\sim$}}}}
\def\simlt{\mathrel{\lower2.5pt\vbox{\lineskip=0pt\baselineskip=0pt
           \hbox{$<$}\hbox{$\sim$}}}}

\newcommand{\bea}{\begin{eqnarray}}
\newcommand{\eea}{\end{eqnarray}}
\newcommand{\nn}{\nonumber}

\newcommand{\vev}[1]{ \left\langle {#1} \right\rangle }
\newcommand{\bra}[1]{ \langle {#1} | }
\newcommand{\ket}[1]{ | {#1} \rangle }

\newcommand{\EV}{ {\rm eV} }

\newcommand{\comment}[1]{}

\begin{document}
\baselineskip 0.6cm


\begin{titlepage}
 
\begin{flushright}
UCB-PTH-07/12 \\
LBNL-62798 \\
CALT-68-2654 \\
UT-07-19 
\end{flushright}

\vskip 1cm
\begin{center}
{\large \bf Statistical Understanding of Quark and Lepton Masses\\ 
in Gaussian Landscapes} \\

\vskip 1.2cm
Lawrence J. Hall,$^a$ Michael P. Salem$^b$ and Taizan Watari$^c$

\vskip 0.4cm
${}^a$ {\it Department of Physics and Lawrence Berkeley National 
Laboratory,\\
University of California, Berkeley, CA 94720, USA} \\

${}^b$ {\it California Institute of Technology, Pasadena, CA 91125, 
USA} \\

${}^c$ {\it Department of Physics, the University of Tokyo, Tokyo, 
113-0033, Japan} \\

\vskip 1.5cm
\abstract{The fundamental theory of nature may allow a large landscape of 
vacua.  Even if the theory contains a unified gauge symmetry, the 22 flavor 
parameters of the Standard Model, including neutrino masses, may be largely 
determined by the statistics of this landscape, and not by any symmetry.   
Then the measured values of the flavor parameters do not lead to any 
fundamental symmetries, but are statistical accidents; their precise values 
do not provide any insights into the fundamental theory, rather the overall 
pattern of flavor reflects the underlying landscape.  We investigate 
whether random selection from the statistics of a simple landscape can 
explain the broad patterns of quark, charged lepton, and neutrino masses 
and mixings.   We propose Gaussian landscapes as simplified models of 
landscapes where Yukawa couplings result from overlap integrals of 
zero-mode wavefunctions in higher-dimensional supersymmetric gauge theories. 
In terms of just five free parameters, such landscapes can account for 
all gross features of flavor, including: the hierarchy of quark and charged 
lepton masses; small quark mixing angles in the basis with quarks arranged 
according to mass, with 13 mixing less than 12 and 23 mixing; very light 
Majorana neutrino masses, with the solar to atmospheric neutrino mass ratio 
consistent with data; distributions for leptonic mixings $\sin 2\theta_{12}$ 
and $\sin 2\theta_{23}$ that are peaked at large values, while the 
distribution for $\sin 2\theta_{13}$ is peaked at low values; and order 
unity CP violating phases in both the quark and lepton sectors.  While the 
statistical distributions for flavor parameters are broad, the distributions 
are robust to changes in the geometry of the extra dimensions.  
Constraining the distributions by loose cuts about observed values leads to 
narrower distributions for neutrino measurements of $\theta_{13}$, CP 
violation, and neutrinoless double beta decay.}

\end{center}
\end{titlepage}

\tableofcontents

\vspace{22pt}

\begin{table}[h!]
\begin{center}
\begin{tabular}{ || l | l || l | l || }
\hline
$\lambda_u$ & $(3.0\pm 1.0)\times 10^{-6}$\,\, &
$\log_{10}\lambda_u$ & $-5.53^{+0.13}_{-0.17}$ 
\phantom{$\Big[\Big]$\!\!\!\!\!}\\ 
\hline
$\lambda_c$ & $(1.4\pm 0.1)\times 10^{-3}$\,\, &
$\log_{10}\lambda_c$ & $-2.87\pm0.03$
\phantom{$\Big[\Big]$\!\!\!\!\!}\\
\hline
$\lambda_t$ & $(4.9\pm 0.3)\times 10^{-1}$\,\, &
$\log_{10}\lambda_t$ & $-0.31\pm0.02$
\phantom{$\Big[\Big]$\!\!\!\!\!}\\
\hline
$\lambda_d$ & $(6.7\pm 2.7)\times 10^{-6}$\,\, &
$\log_{10}\lambda_d$ & $-5.18^{+0.15}_{-0.22}$
\phantom{$\Big[\Big]$\!\!\!\!\!}\\
\hline
$\lambda_s$ & $(1.3\pm 0.4)\times 10^{-4}$\,\, &
$\log_{10}\lambda_s$ & $-3.90^{+0.10}_{-0.13}$
\phantom{$\Big[\Big]$\!\!\!\!\!}\\
\hline
$\lambda_b$ & $(5.7\pm 0.1)\times 10^{-3}$\,\, &
$\log_{10}\lambda_b$ & $-2.24\pm0.01$
\phantom{$\Big[\Big]$\!\!\!\!\!}\\
\hline
$\frac{1}{\pi}\,\theta^{\rm CKM}_{12}$ & $(7.31\pm 0.03)\times10^{-2}$\, &
$\log_{10}\left(\frac{2}{\pi}\,\theta^{\rm CKM}_{12}\right)$ &
$-0.835 \pm 0.002$
\phantom{$\Big[\Big]$\!\!\!\!\!}\\ 
\hline
$\frac{1}{\pi}\,\theta^{\rm CKM}_{23}$ & 
$(1.344^{+0.003}_{-0.025})\times10^{-2}$\, &
$\log_{10}\left(\frac{2}{\pi}\,\theta^{\rm CKM}_{23}\right)$ &
$-1.571^{+0.001}_{-0.008}$
\phantom{$\Big[\Big]$\!\!\!\!\!}\\ 
\hline
$\sin\theta^{\rm CKM}_{13}$ & $(4.01\pm 0.09)\times10^{-3}$\, &
$\log_{10}\left(\sin\theta^{\rm CKM}_{13}\right)$ &
$-2.397\pm0.010$ 
\phantom{$\Big[\Big]$\!\!\!\!\!}\\ 
\hline
$\frac{1}{\pi} \delta^{\rm CKM}$ & $0.29$--$0.34$ & 
 & 
\phantom{$\Big[\Big]$\!\!\!\!\!}\\ 
\hline
$\lambda_e$ & $2.99\times10^{-6}$ &
$\log_{10}\lambda_e$ & $-5.52$
\phantom{$\Big[\Big]$\!\!\!\!\!}\\ 
\hline
$\lambda_\mu$ & $6.31\times10^{-4}$ &
$\log_{10}\lambda_\mu$ & $-3.20$
\phantom{$\Big[\Big]$\!\!\!\!\!}\\ 
\hline
$\lambda_\tau$ & $1.07\times10^{-2}$ &
$\log_{10}\lambda_\tau$ & $-1.97$
\phantom{$\Big[\Big]$\!\!\!\!\!}\\ 
\hline
$\frac{1}{\pi}\,\theta^{\rm PMNS}_{12}$ &  $0.19 \pm 0.01$ & 
$\sin (2\theta_{12}^{\rm PMNS})$ & 
$0.93^{+0.01}_{-0.02}$ (90\% CL)
\phantom{$\Big[\Big]$\!\!\!\!\!}\\ 
\hline
$\frac{1}{\pi}\,\theta^{\rm PMNS}_{23}$ & $0.20$--$0.30$ &
$\sin (2\theta_{23}^{\rm PMNS})$ & 
$0.96$--$1.0$ (90\% CL)
\phantom{$\Big[\Big]$\!\!\!\!\!}\\ 
\hline
$\sin\theta^{\rm PMNS}_{13}$ & $< 0.18$ (95\% CL) &
$\log_{10}\left(\sin\theta^{\rm PMNS}_{13}\right)$ & 
$ < -0.74$
\phantom{$\Big[\Big]$\!\!\!\!\!}\\ 
\hline
$\Delta m^2_{\rm atm}$ & $(1.9\mbox{--}3.0) \times 10^{-3}$ eV$^2$ & 
$\sqrt{\Delta m^2_{\rm atm}}$ & $(4.4\mbox{--}5.5) \times 10^{-2}$ eV  
\phantom{$\Big[\Big]$\!\!\!\!\!}\\
\hline
$\Delta m^2_{\odot}$ & $(8.0^{+0.4}_{-0.3}) \times 10^{-5}$ eV$^2$ &
$\sqrt{\Delta m^2_{\odot}}$ & $(8.9\pm 0.3) \times 10^{-3}$ eV 
\phantom{$\Big[\Big]$\!\!\!\!\!}\\
\hline
$\sqrt{\Delta m^2_{\odot} / \Delta m^2_{\rm atm}}$ & $0.16$--$0.20$ & & 
\phantom{$\Big[\Big]$\!\!\!\!\!}\\
\hline
\end{tabular}
\end{center}
\caption{\label{table1} The 17 measured flavor parameters and a limit on 
a mixing angle in the lepton sector.  All data comes from the pdgLive 
feature from the Particle Data Group~\cite{pdgLive}. 
For comparison to distributions provided throughout this paper, 
the quark and charged lepton Yukawa eigenvalues and the CKM matrix 
elements have been run up to the (reduced) Planck scale 
$M_P=2.4\times10^{18}$ GeV, assuming no physics beyond the Standard Model 
enters up to this scale.  On the other hand, the RG scaling effects are 
not taken into account in the neutrino sector.}
\end{table}

\newpage


\section{Introduction}

The Standard Model\footnote{We everywhere refer to the ``Standard Model'' 
as the theory including the dimension-five operators of (\ref{eq:Lflav}).} 
describes all laboratory data with 28 free parameters.  Of these, 22 
arise from the Yukawa matrices $\lambda^{u,d,e}$ and the coupling matrix 
$C$ that appear in the flavor interactions
\begin{equation}
{\mathcal L}_{flav} = \lambda^u_{ij} \bar{u}_i q_j \, h +
  \lambda^d_{kj} \bar{d}_k q_j \, h^* + 
  \lambda^e_{a i} \bar{e}_a l_i \, h^* + \frac{C_{ij}}{M} l_i l_j hh \,, 
\label{eq:Lflav}
\end{equation}
where $q,l$ ($\bar{u}, \bar{d}, \bar{e}$) are the left (right) handed
quark and lepton fields.  Of these 22 flavor parameters, 17 are measured, 
with varying levels of accuracy.  The remaining five parameters are all 
in the neutrino sector and two of these have upper limits.  Given several 
decades of continued progress on improving the accuracy of the 
experimental measurements, the most striking fact is that there is 
nothing approaching a standard theory of the origin of these parameters.  
Not only are we ignorant about the overall picture of flavor, we do not 
have a convincing explanation for the value of any of the 17 measured 
parameters.

A complete theory of flavor would provide answers to three very different 
questions:
\begin{itemize}
\item What is the origin of the fermion quantum numbers and why are there 
three generations?
\item What determines the qualitative pattern of the quark and lepton 
mass matrices? For example, why do the charged fermion masses 
and mixings have a hierarchical pattern, while in the neutrino sector 
there are large mixing angles? 
\item What determines the precise values of the 22 flavor parameters?
\end{itemize}
The couplings $\lambda^{u,d,e}$ and $C$ are symmetry breaking parameters 
of the flavor symmetry group $U(3)^5$, where one $U(3)$ factor acts on 
each of $q,l, \bar{u}, \bar{d}, \bar{e}$.  The dominant approach to 
constructing theories of flavor is to use symmetries to reduce the number 
of free parameters, $n$.  An underlying flavor group $G_f \subset U(3)^5$ 
and a more unified gauge symmetry both limit $n$, leading to precise 
predictions if $n < n_{obs}$, the number of observables.  A hierarchy of 
symmetry breaking scales can lead to small dimensionless parameters that 
explain qualitative features of the mass matrices \cite{fn}, and there 
are many realizations with $G_f$ both Abelian and non-Abelian. Still, it 
is striking that the progress along these lines is limited, even as more 
precise data have become available.  Perhaps this is a sign that a 
completely new approach is needed.

The cosmological dark energy~\cite{DE} apparently has little to do with 
flavor.  However, it may be the first evidence for a huge landscape of 
vacua, with the observed value of the cosmological constant resulting from 
environmental selection for large scale structure~\cite{CC}.  If we take 
string theory to be a theory of a landscape rather than a theory of a 
single vacuum, then what are the implications for flavor?  The enormous 
number of vacua, required for a sufficiently fine scan of the cosmological 
constant, results from the large number of ways background fluxes can be 
arranged on non-trivial compact manifolds of extra spatial 
dimensions~\cite{bp}. The cosmological constant, however, is not the only
parameter that scans over such a landscape. The Yukawa couplings may also 
vary from one universe to another. 

In any landscape of vacua, the Yukawa couplings have a relevant dependence 
on $n = n_S+n_F$ parameters, where $n_S$ of the parameters scan and $n_F$ 
of them are fixed.  If $n < n_{obs}$ then data can determine the subset of 
vacua in which we happen to live and $n_{obs}-n$ predictions can be made.  
A certain simple model with a single extra dimension has $n_{obs}-n=1$, 
giving a single prediction~\cite{MS}.  If $n > n_{obs}$, however, then no 
such precise predictions are possible.  Some flavor parameters---such as 
the electron, up and down masses---may be determined by environmental 
effects~\cite{Hogan}.  The top mass may also be determined along this 
line~\cite{FHW}, while selecting for leptogenesis may be a key to 
understanding some parameters in the lepton sector.  However, most of the 
flavor parameters do not seem to be strongly selected.  Therefore it could 
be that most flavor parameters have values that simply reflect some 
underlying probability distribution over an enormous number of solutions 
to the fundamental theory.  The precise value of any such observable is 
accidental and not fundamental, since any other nearby value would be just 
as probable.  Although it may be unappealing that there is nothing 
fundamental or beautiful relating the flavor parameters in our universe, 
so far this possibility cannot be dismissed, especially given that no such 
relations have been found.

The large mixing angles observed in atmospheric and solar neutrino 
oscillations inspired the idea that the relevant Yukawa couplings 
of the neutrino sector are governed by randomness rather than by flavor 
symmetries~\cite{HMW}.  By introducing a simple probability distribution 
for the elements in the Dirac and Majorana neutrino mass matrices, 
probability distributions for the neutrino observables were generated 
and found to agree well with data.  However, a complete landscape approach 
to the theory of flavor must explain many other things;  there are rich 
structures among the observed flavor parameters.  Reference~\cite{DDR} 
took a bold step in this direction, studying masses and mixings in both 
the quark and lepton sectors by introducing a simple probability 
distribution for the Yukawa couplings.  We begin in section~\ref{sec:indep} 
by analyzing the toy landscape introduced in~\cite{DDR}. While the main 
purpose of this section is to clarify how hierarchy among quark masses is 
generated in this model and to point out its limitations, we also introduce 
an approximate analytical study of the probability distribution of flavor 
observables.  This turns out to be a useful warm-up exercise for the 
subsequent sections.  

Ideally, a statistical approach to the theory of flavor would be both 
phenomenologically successful and theoretically well-motivated.  However 
to determine the probability distributions for flavor parameters from a 
purely top-down calculation is not an easy task; meanwhile the number
of observables is limited, and such low-energy information does not
constrain the underlying probability distributions in a purely bottom-up
approach.  We consider as a practical strategy to use string theory as a
guide in deducing the form of the probability distributions, while using
known experimental constraints to try to find a phenomenologically 
successful model within this restricted set of possibilities.
In a given compactification of string theory, the Yukawa couplings are 
determined by an overlap integration over the extra spatial
dimensions. Scanning moduli parameters of the gauge field configuration on 
the extra dimensions corresponds to scanning zero-mode wavefunctions, and 
hence this corresponds to scanning Yukawa couplings.  If the wavefunctions 
of the quarks, leptons, and the Higgs are peaked at different locations, 
then the overlap integral can lead to small Yukawa matrix elements. Hence 
localization in extra dimensions is an alternative to symmetries for 
generating fermion mass hierarchies~\cite{Arkani-Hamed:1999dc}. 

In sections~\ref{sec:toy1} through~\ref{sec:Lepton} we introduce simple 
``Gaussian landscapes'' and study the resulting distributions for quark 
and lepton masses and mixings.\footnote{A brief summary of some of the 
major results of this work can be found in~\cite{HSW}.}  These landscapes 
have features expected from certain string landscapes, for which they can 
be viewed as simplified or toy models.  The key feature of a Gaussian 
landscape is that all quark, lepton and Higgs fields have zero-mode 
wavefunctions with Gaussian profiles in the extra dimensions, and that the 
centers of these profiles all scan independently with flat probability 
distributions over the volume of the extra dimensions.   For simplicity, 
the number of free parameters used to describe the geometry of the extra 
dimensions and the widths of the Gaussian profiles is kept to a minimum.  
We find the observed quark and lepton masses and mixings can be typical in 
such simple landscapes.

In section~\ref{sec:toy1} we introduce a Gaussian landscape for quarks on 
a circle $S^1$, with all Gaussian profiles having the same width.  
Numerical probability distributions for the nine CP-conserving flavor 
observables are provided, and a qualitative semi-analytic description of 
these distributions is derived.  The results are compared with those that 
result from introducing approximate flavor symmetries, with some 
similarities and some differences emerging.  Finally, the effects of 
possible environmental selection on the top quark mass is studied.  The 
large number of flavor parameters in the Standard Model allows for a 
reasonably significant evaluation of goodness-of-fit between a Gaussian 
landscape and the observed flavor structure.  This is described in 
section~\ref{sec:statistics}, with the $S^1$ Gaussian landscape used for 
illustration.  The effects on the quark sector from adding more dimensions 
to the Gaussian landscape are examined in section~\ref{sec:Geometry}, 
together with a preliminary study of the effects of geometry.    

In section~\ref{sec:Lepton} the Gaussian landscape is extended to 
include the lepton sector.  Motivated by the expectation of supersymmetry 
in the higher dimensional theory, Yukawa couplings for both Dirac and 
Majorana neutrino masses are introduced in the Gaussian landscape, with
light neutrinos resulting from the usual seesaw mechanism in four 
dimensions.  Larger mixing angles in the lepton sector might arise from 
assigning the appropriate fermions larger Gaussian widths, but these are 
found to largely cancel between charged and neutral sectors.  However, this
cancellation is prevented by introducing CP-violating phases in the 
Gaussian profiles.  This suggests a connection in Gaussian landscapes 
between large CP violation and large leptonic mixing angles.  In addition 
to providing numerical distributions for the observed lepton flavor 
parameters, probability distributions are predicted for the leptonic 
mixing angle $\theta_{13}$, the CP phase in neutrino oscillations, 
and the Majorana mass relevant for neutrinoless double beta decay, 
illustrating how these toy models for the string landscape can connect to 
future experimental data.  Finally, the extent to which these Gaussian 
landscapes arise from unified supersymmetric field theories in higher 
dimensions is examined in section~\ref{sec:SYM}. Such theories have huge 
symmetries---ultimately all entries of the Yukawa matrices arise 
from a unified supersymmetric gauge coupling---so that the complicated 
pattern of observed masses and mixings can arise from a very simple 
mechanism: the scanning of the centers of the Gaussian profiles.

\section{Prelude:  Hierarchy without Flavor Symmetry}
\label{sec:indep}

In previous work it has been suggested that the components of the Yukawa 
matrices $\lambda^{u,d,e}$ and the coupling matrix $C$ are selected 
randomly and independently of each other~\cite{HMW,DDR}.  For example, 
in Neutrino Anarchy~\cite{HMW} one finds that the large mixing angles 
underlying neutrino oscillation are typical of the lepton interactions 
that arise when each element of the matrices $\lambda^e$ and $C$ is 
independently selected from the simple probability distributions
\begin{equation}
\frac{dP(\lambda^e)}{d\lambda^e}={\rm const.} \,, \qquad
\frac{dP(C)}{dC}={\rm const.},
\end{equation}
or from distributions such as 
\begin{equation}
\frac{dP(\lambda^e)}{d\ln\lambda^e}={\rm const.} \quad {\rm for~}
\lambda_{\rm min}^e < \lambda^e < \lambda_{\rm max}^e \,, \qquad
\frac{dP(C)}{d\ln C}={\rm const.} \quad {\rm for~}
C_{\rm min} < C < C_{\rm max} \,,
\end{equation}
where $\lambda^e_{\rm min,max}$ and $C_{\rm min,max}$ are of order unity
(for more details see \cite{HMW,HM}).  Such an absolute anarchy of 
lepton couplings tends to result in comparable mass eigenvalues.  On the 
other hand, \cite{DDR} introduced a power-law probability distribution 
for the Yukawa matrix elements, 
$dP(\lambda)/d\lambda\propto\lambda^{-\delta}$ for 
$\lambda_{\rm min}<\lambda<\lambda_{\rm max}$ and 
$dP(\lambda)/d\lambda=0$ otherwise.  By assuming 
$\lambda_{\rm min}\ll\lambda_{\rm max}$ and choosing $\delta$ 
appropriately, quark Yukawa matrices with each matrix element following 
such a distribution can roughly accommodate the hierarchical pattern of 
quark mass eigenvalues~\cite{DDR}.  According to~\cite{DDR}, 
$\delta=1.16$ provides the best fit to the quark sector.

In this paper we propose a significant modification to these ideas.  
However as a prelude to discussing our proposal it is worthwhile to 
first study the model of~\cite{DDR} in greater detail.  To simplify the 
analysis we specialize to the particular case $\delta=1$, such that 
\begin{equation}
\frac{dP(\lambda)}{d\log_{10}\lambda} = \left\{ \begin{array}{ll}
0  & {\rm for~} \lambda > \lambda_{\rm max} \\
1/\log_{10}(\lambda_{\rm max}/\lambda_{\rm min}) \qquad & 
{\rm for~} \lambda_{\rm min} < \lambda < \lambda_{\rm max} \\
0 &  {\rm for~} \lambda < \lambda_{\rm min} 
\end{array}
\right.\,.
\label{eq:nocorr-distr}
\end{equation}
Henceforth we refer to this distribution as a scale-invariant 
distribution.  In this section we also restrict attention to the quark 
sector.  Note that CP-violating phases are not introduced in this 
landscape because all of the matrix elements are real-valued.  A pair 
of $3\times3$ Yukawa matrices $\lambda^u$ and $\lambda^d$ is generated 
by choosing each of the 18 matrix elements randomly according to the 
distribution (\ref{eq:nocorr-distr}). The Yukawa matrices are then 
diagonalized using separate left- and right-handed unitary 
transformations, and the quark masses and mixings are calculated from 
the eigenvalues of the diagonalized Yukawa matrices and the resulting 
CKM matrix.  The above process is then repeated to generate an ensemble 
of these nine observables.

\subsection{The Distribution of Mass Eigenvalues} 

\begin{figure}[t]
\begin{center}
\begin{tabular}{ccc}
\includegraphics[width=0.3\linewidth]{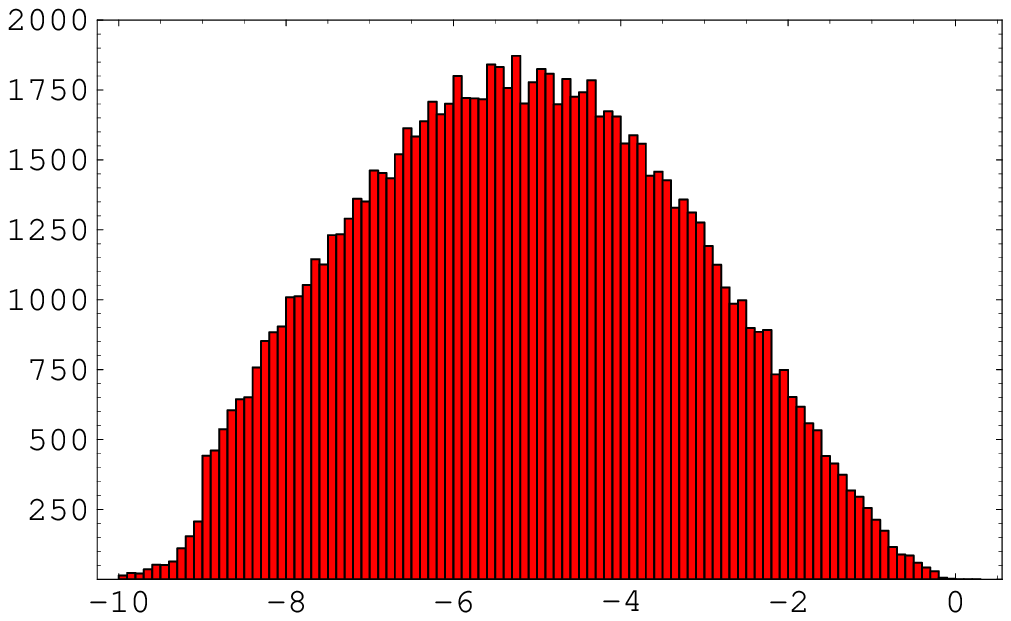} &
\includegraphics[width=0.3\linewidth]{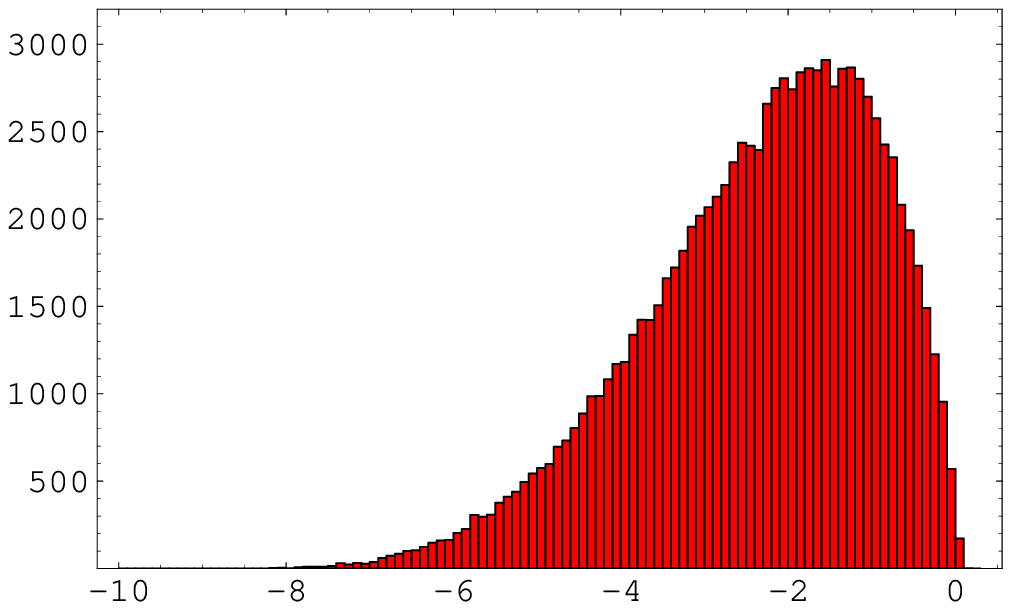} &
\includegraphics[width=0.3\linewidth]{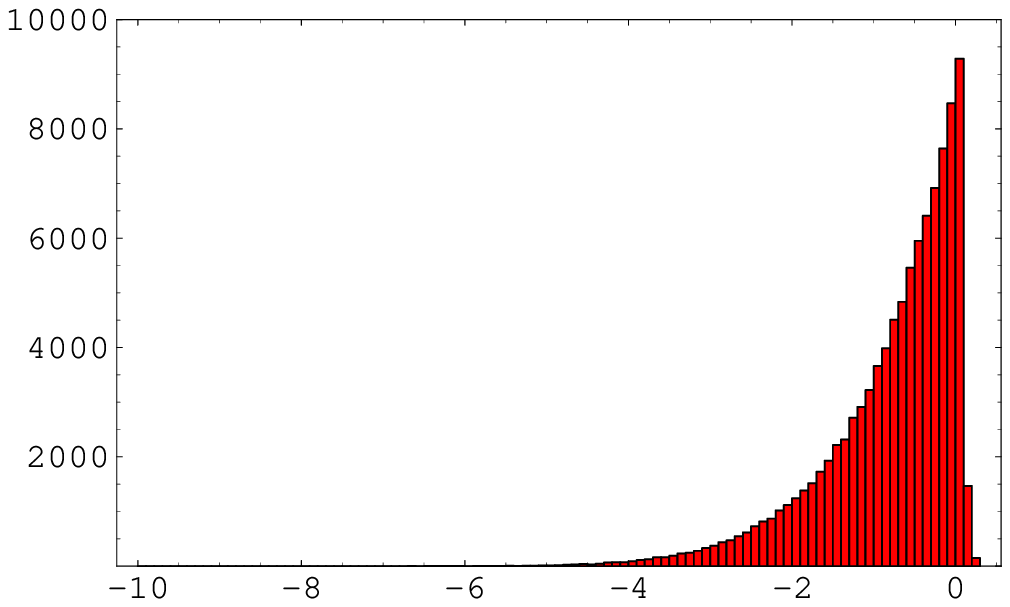}
\end{tabular}
\caption{\label{fig:Donoghue-egval} Distributions of the three 
eigenvalues of Yukawa matrices whose each element follows the 
distribution (\ref{eq:nocorr-distr}). From left to right the three
panels correspond to the smallest, middle and largest eigenvalues.  The 
sum of all three distributions reproduces Figure 9a of \cite{DDR}. We 
used $\log_{10}\lambda_{\rm min}=-9$ and 
$\log_{10}\lambda_{\rm max}=0.1$ for this simulation.}
\end{center}
\end{figure}

We first study the distributions of the quark masses.  Since the up- 
and down-type Yukawa matrices are generated independently of each other 
and in exactly the same way, the distributions of masses for these 
sectors are exactly the same.  Thus we only need to study one of the two 
sectors.  Results of a numerical study are shown in 
Figure~\ref{fig:Donoghue-egval}, where we have chosen 
$\lambda_{\rm min}/\lambda_{\rm max} = 10^{-9.1}$.
With some approximations we can understand the shapes of these 
distributions.  Let $\lambda_3'$ denote the largest element of the 
$3\times3$ matrix $\lambda$.  Meanwhile, the largest element of the 
$2\times2$ sub-matrix of $\lambda$ that excludes $\lambda'_3$ is denoted 
$\lambda'_2$.  For example, in the matrix 
\begin{equation}
\lambda =\left( \begin{array}{c|cc}
	 \lambda_{23} & \lambda'_2 & \lambda_{21} \\
         \lambda_{13} & \lambda_{12} & \lambda'_1 \\
\hline  
         \lambda'_3 & \lambda_{32} & \lambda_{31}
         \end{array}\right), 
\label{eq:3by3}
\end{equation}
$\lambda'_3$ is the largest among the nine entries and $\lambda'_2$ the 
largest in the upper right $2\times2$ sub-matrix.  Given this 
characterization, the probability distribution for the variables 
$\lambda'_{1,2,3}$ and $\lambda_{ij}$ is 
\begin{eqnarray}
dP(x'_{1,2,3}, x_{ij}) \!&=&\! 36\,
\Theta(x'_3 - x'_2)\Theta(x'_3 - x_{32}) \Theta(x'_3 - x_{23}) 
\Theta(x'_3 - x_{31}) \Theta(x'_3 - x_{13}) \Theta(x'_2 - x'_1) \nn\\
&&\! \times \Theta(x'_2 - x_{21})\Theta(x'_2 - x_{12})\,
dx'_1 dx'_2 dx'_3 dx_{12} dx_{21} dx_{13} dx_{31} dx_{23} dx_{32} \,,
\label{eq:distr-all9}
\end{eqnarray}
where $\Theta$ is the step function.  The factor of 36 comes from the 
nine possible locations for $\lambda_3'$ times the four possible 
locations for $\lambda_2'$.  In addition we have introduced the notation
\begin{equation}
x'_i \equiv \frac{\ln (\lambda'_i/\lambda_{\rm min})}
           {\ln (\lambda_{\rm max}/\lambda_{\rm min})}\,, \qquad 
x_{ij} \equiv \frac{\ln (\lambda_{ij}/\lambda_{\rm min})}
           {\ln (\lambda_{\rm max}/\lambda_{\rm min})}\,.
\end{equation}

\begin{figure}[t]
\begin{center}
\begin{tabular}{ccc}
\includegraphics[width=0.3\linewidth]{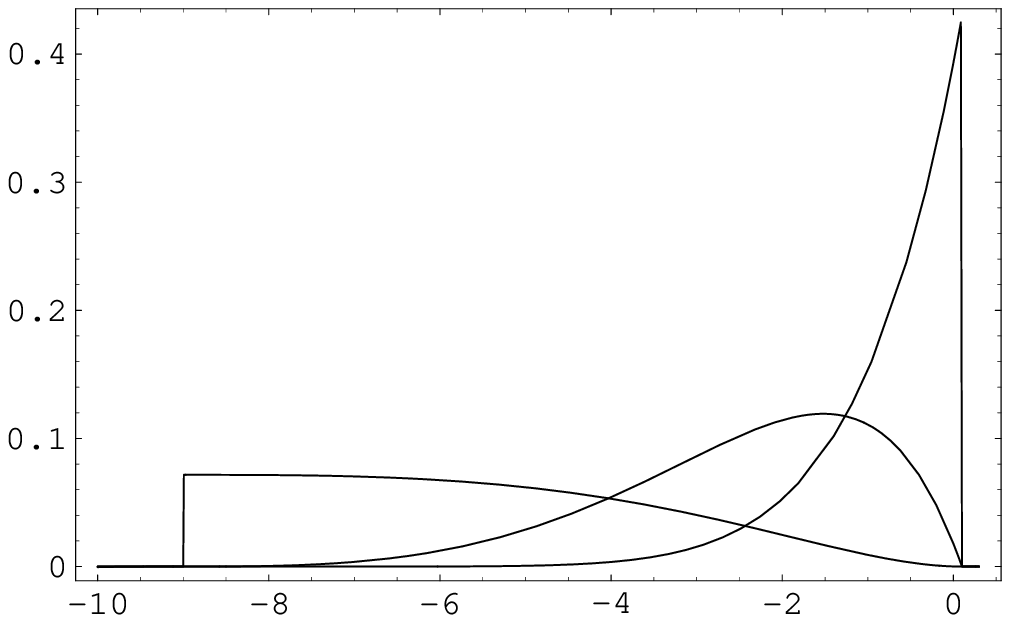} &
\includegraphics[width=0.3\linewidth]{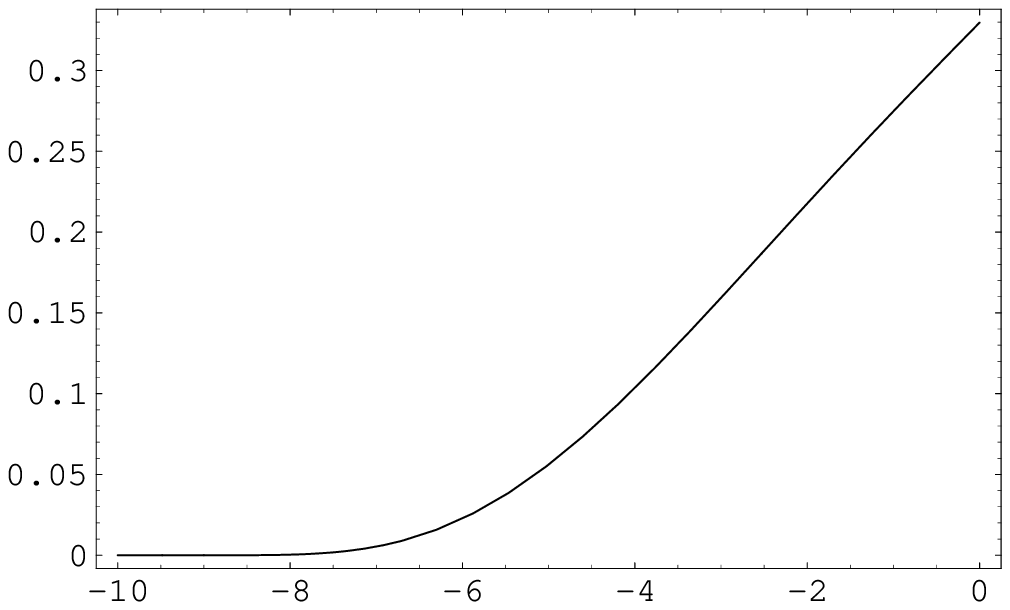} &
\includegraphics[width=0.3\linewidth]{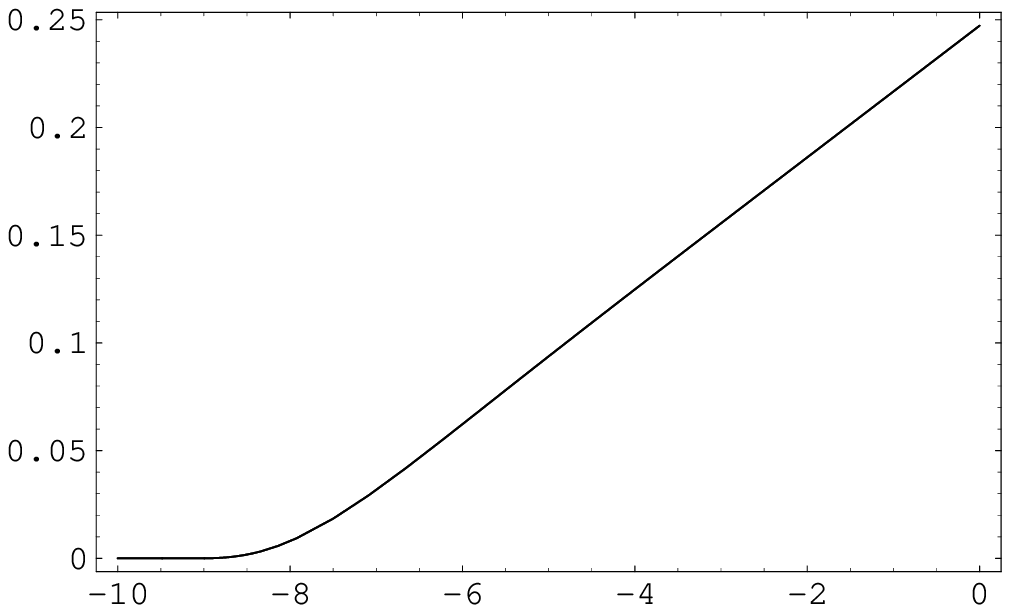} \\
$\log_{10} \lambda_{1,2,3}$  & $\log_{10}\sin\theta_{12}$ &
$\log_{10}\sin\theta_{23,13}$
\end{tabular}
\caption{\label{fig:nocorr-egval-analytic} The approximate distributions 
of the three eigenvalues given in (\ref{eq:distr-x3}--\ref{eq:distr-x1}), 
and those of mixing angles given in (\ref{eq:dnht12}, \ref{eq:dnht23}).  
We have used the same $\lambda_{\rm max}$ and $\lambda_{\rm min}$ as are 
used in Figure~\ref{fig:Donoghue-egval}.}
\end{center}
\end{figure}

The largest mass eigenvalue of (\ref{eq:3by3}) is approximately 
$\lambda'_3$; this approximation is poor if one of $\lambda_{32}$, 
$\lambda_{31}$, $\lambda_{23}$ and $\lambda_{13}$ is almost as large 
$\lambda'_3$, but this is unlikely if 
$\ln (\lambda_{\rm max}/\lambda_{\rm min})$ is large.  We call this 
largest eigenvalue $\lambda_3$ and define $x_3$ analogously to $x_3'$.  
The probability distribution of $x_3$ is therefore approximated by 
integrating out from (\ref{eq:distr-all9}) all the variables except 
$x'_3$:
\begin{equation}
dP(x_3) \simeq 9 x_3^8 \, dx_3\,.
\label{eq:distr-x3}
\end{equation}
Meanwhile, we approximate the middle eigenvalue $\lambda_2$ by 
$\lambda_2'$ and the smallest eigenvalue $\lambda_1$ by $\lambda'_1$.
This approximation is poor when the seesaw contributions 
$(\lambda_{i3}\lambda_{3j})/\lambda'_3$ and 
$(\lambda_{12}\lambda_{21})/\lambda'_2$ are larger than $\lambda'_2$ 
and $\lambda'_1$, respectively. Thus we do not expect this approximation
to be reliable for small values of $x_2$ and $x_1$.  Nevertheless, 
integrating out all of other variables we find 
\begin{eqnarray}
dP(x_2) & = &  \frac{36}{5} x_2^3 (1-x_2^5) \, d x_2,
 \label{eq:distr-x2} \\
dP(x_1) & = & 
\frac{36}{5}\left(\frac{1-x_1^3}{3} - \frac{1-x_1^8}{8}\right) \, dx_1 \,.  
 \label{eq:distr-x1}
\end{eqnarray} 
The probability distributions (\ref{eq:distr-x3}--\ref{eq:distr-x1}) are 
shown in Figure~\ref{fig:nocorr-egval-analytic}.
Remarkably, they capture the gross features of the numerical results 
in Figure~\ref{fig:Donoghue-egval}. Therefore we use these distributions 
to examine the qualitative aspects of the mass distributions that follow 
from the landscape (\ref{eq:nocorr-distr}).

The average $\vev{x_i}$ and the standard deviation $\sigma_i$ of 
the three Yukawa eigenvalues (both on a logarithmic scale) can be 
calculated from the distributions (\ref{eq:distr-x3}--\ref{eq:distr-x1}): 
\begin{eqnarray}
 \vev{x_3} = 0.90, & \sigma_3 = 0.09, & x_3 \sim [0.81 - 0.99]\,, \\
 \vev{x_2} = 0.72, & \sigma_2 = 0.16, & x_2 \sim [0.56 - 0.88]\,, \\
 \vev{x_1} = 0.36, & \sigma_1 = 0.22, & x_1 \sim [0.14 - 0.58]\,.
\end{eqnarray}
The three eigenvalues are on average well-separated and they overlap 
with neighboring eigenvalues only slightly at one standard deviation. 
Even this slight overlap between the distributions is misleading.  
Recall that by definition for any particular set of Yukawa matrices we 
have $x'_3>x'_2>x'_1$.  Thus the combined distribution for the eigenvalues 
is not a naive product of (\ref{eq:distr-x3}--\ref{eq:distr-x1}) but is 
given by integrating the other six variables out of (\ref{eq:distr-all9}).  
This gives
\begin{equation}
dP(x_1,x_2,x_3) \simeq 36\,x_3^4 x_2^2\,  
\Theta(x_3 - x_2)\Theta(x_2-x_1)\, dx_1 dx_2 dx_3 \,.  
\label{eq:distr-all3}
\end{equation}
Thus it happens that in the subset of cases where $x_3$ is small, 
the distribution of $x_2$ is pushed to even smaller values.  Hence the 
three eigenvalues tend to be well-separated even in the logarithmic 
scale, and only rarely are adjacent eigenvalues comparable.  Note that 
none of this depends on the choice of $\lambda_{\rm min}$ and 
$\lambda_{\rm max}$.  In short, hierarchical structure (Yukawa 
eigenvalues well-separated in logarithmic scale) is generated 
statistically in a landscape where each matrix element independently 
follows the scale-invariant distribution (\ref{eq:nocorr-distr}).
Whether the hierarchy is large or small is determined by whether 
$\log_{10} (\lambda_{\rm max}/\lambda_{\rm min})$ is large or small.

\subsection{Pairing Structure in Electroweak Interactions} 

\begin{figure}[t]
\begin{center}
\begin{tabular}{ccc}
\includegraphics[width=0.3\linewidth]{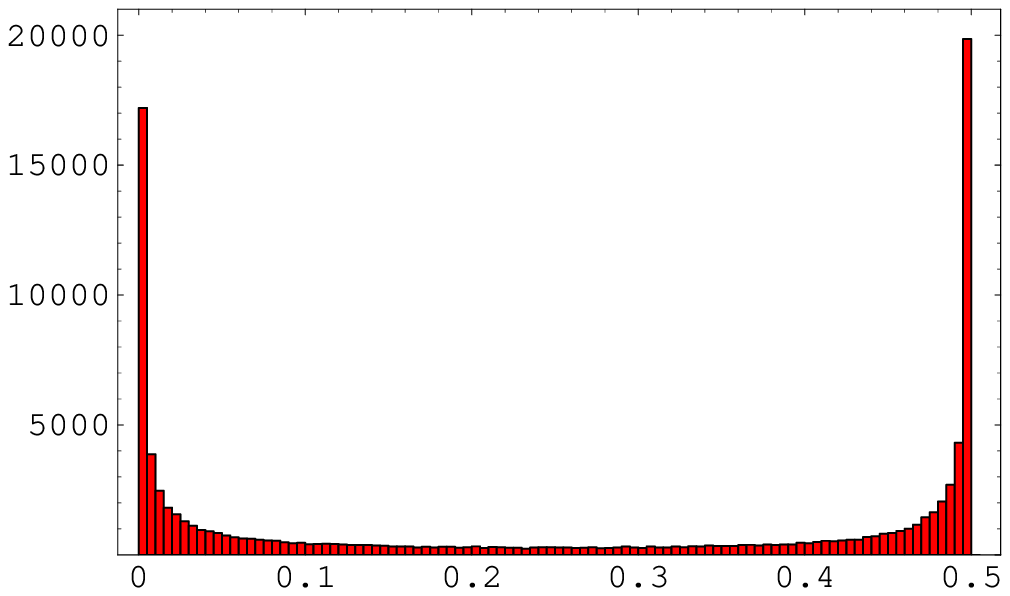} &
\includegraphics[width=0.3\linewidth]{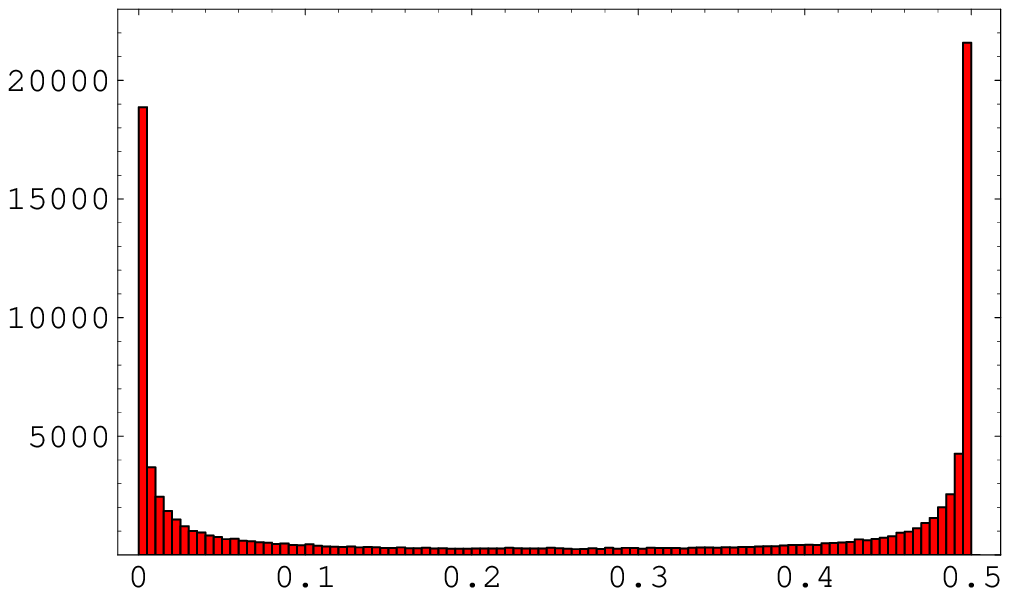} &
\includegraphics[width=0.3\linewidth]{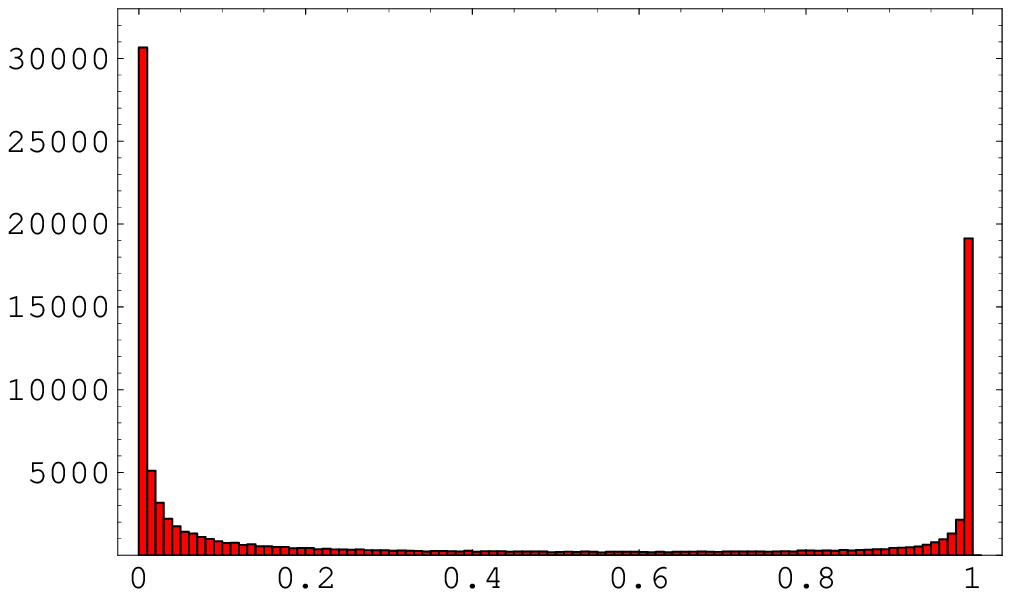} \\
$\theta_{12}/\pi$ & $\theta_{23}/\pi$ & $\sin\theta_{13}$ \\
\includegraphics[width=0.3\linewidth]{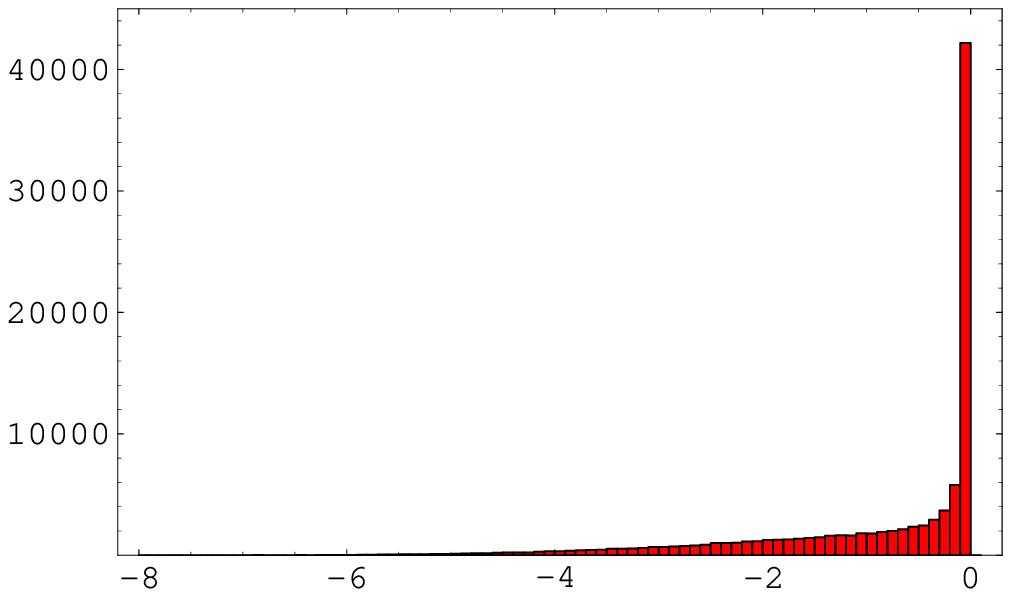} &
\includegraphics[width=0.3\linewidth]{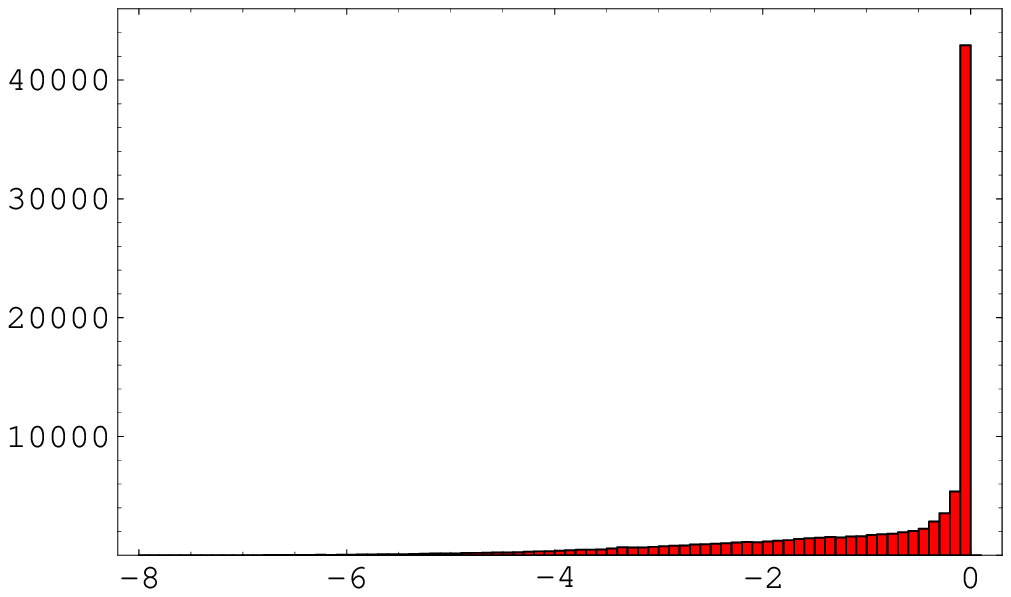} &
\includegraphics[width=0.3\linewidth]{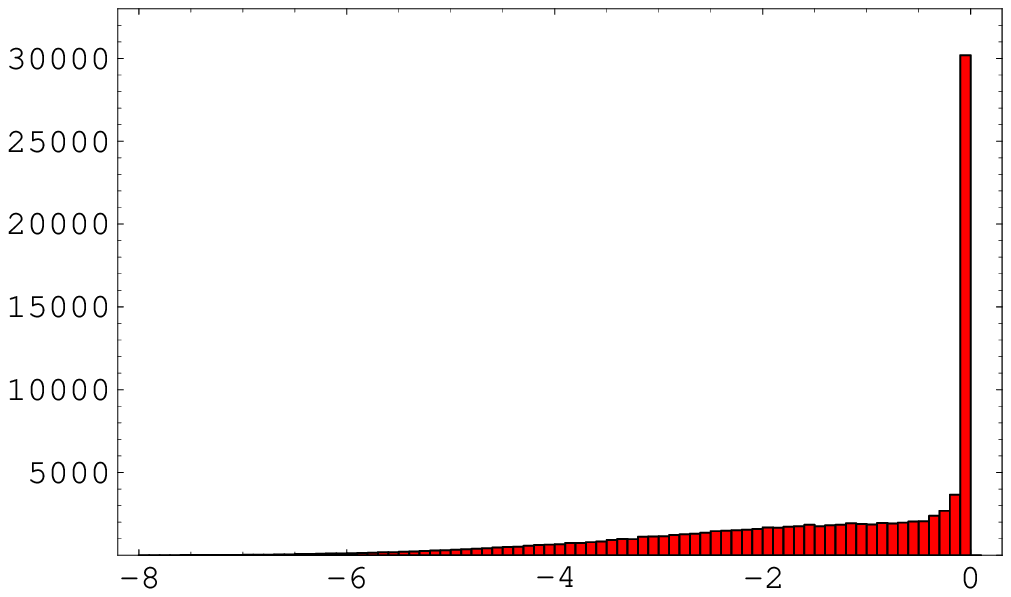} \\
$\log_{10}(2\theta_{12}/\pi)$ & $\log_{10}(2\theta_{23}/\pi)$ & 
$\log_{10}\sin\theta_{13}$ \\
\end{tabular}
\caption{\label{fig:Donoghue-mix} Distributions of the three mixing angles 
of the CKM matrix that results from the distribution 
(\ref{eq:nocorr-distr}).  The upper left panel is similar to Figure 11 
of~\cite{DDR}; however in~\cite{DDR} $\delta=1.16$ is used and the 
distribution is displayed only within the range 
$0\leq\theta_{ij}\leq \pi/4$.}
\end{center}
\end{figure}

Let us now study the mixing angles.  Figure~\ref{fig:Donoghue-mix} 
shows the distributions of mixing angles in the quark sector that result 
from a numerical simulation where each element of both the up-type 
and down-type Yukawa matrices is assumed to follow the distribution 
(\ref{eq:nocorr-distr}) independently.\footnote{Using  
(\ref{eq:distr-x3}--\ref{eq:distr-x1}) we find these 
distribution functions to be given by the approximate analytic form:  
\begin{eqnarray}
dP(t) \!&\simeq&\! \frac{3}{50}(10 - 15 t + 6 t^4 - t^9)
(5 - 8 t^3 + 3 t^8)\, dt\,, 
\qquad {\rm where~} t \equiv \frac{\ln\sin\theta_{12}}
{\ln (\lambda_{\rm min}/\lambda_{\rm max})}\,, \label{eq:dnht12} \\
dP(t) \!&\simeq&\! \frac{9}{32}(8 - 9 t + t^9)(1 - t^8)\, dt\,, 
\qquad\qquad\qquad\qquad\,\,\, {\rm where~} 
  t \equiv \frac{\ln\sin\theta_{23,13}}
  {\ln (\lambda_{\rm min}/\lambda_{\rm max})}\,. \label{eq:dnht23}
\end{eqnarray}
} 
The probability distribution functions of the mixing angles are shown 
against the axes of $d\theta_{12}$, $d\theta_{23}$ and 
$d(\sin \theta_{13})$ because the invariant measure\footnote{The 
importance of the invariant measure is emphasized in~\cite{HM}.}
of SO(3) mixing matrices is 
$d\theta_{12}\wedge d\theta_{23}\wedge d(\sin \theta_{13})$.  Since 
in this model the Yukawa couplings are all real- and positive-valued, 
the CKM quark mixing matrices are SO(3) matrices.

The prominent feature of these distributions is the twin peaks at 
$\theta_{ij}=0$ and $\theta_{ij}=\pi/2$ for all three mixing angles.  
This feature is straightforward to understand. Suppose that the 
randomly generated Yukawa matrices are of the form 
\begin{equation}
\lambda^u_{ij} \sim 
\left(\begin{array}{ccc}
* & {\lambda'}^{u}_2 & \bullet \\
* & \bullet & {\lambda'}^{u}_1 \\
{\lambda'}^{u}_3 & * & *
\end{array}\right), 
\qquad 
\lambda^d_{kj} \sim 
\left(\begin{array}{ccc}
\bullet & * & {\lambda'}^{d}_2 \\
* & {\lambda'}^{d}_3 & * \\
{\lambda'}^{d}_1 & * & \bullet 
\end{array}\right),
\label{eq:updown}
\end{equation}
where the $*$'s are assumed to be less than ${\lambda'}^{u}_3$ or 
${\lambda'}^{d}_3$, and the $\bullet$'s less than ${\lambda'}^{u}_2$ or 
${\lambda'}^{d}_2$. Ignoring the seesaw contributions to 
eigenvalues, i.e. when $(**/\lambda_3') \ll \lambda_2'$ and 
$(\bullet \bullet/\lambda_2') \ll \lambda_1'$, we find that 
the three left-handed quark doublets $q_j$ 
($j=1,2,3$) are approximately
\begin{equation}
 q_1 = (t_L, d_L), \qquad q_2 = (c_L, b_L), \qquad q_3 = (u_L,s_L), 
\label{eq:pairing}
\end{equation}
where $t_L$, $c_L$ and $u_L$ stand for left-handed components of the 
heaviest, middle and lightest mass eigenstates of up-type quarks.  The 
down-type mass eigenstates $b_L$, $s_L$ and $d_L$ are defined similarly.  
The CKM matrix for the up-type and down-type Yukawa matrices 
(\ref{eq:updown}) is roughly 
\begin{equation}
V_{CKM} \sim 
\left(\begin{array}{ccc}
& 1 & \\ & & 1 \\ 1 & & 
\end{array}\right), 
\label{eq:exCKM}
\end{equation}
corresponding to $\theta_{12} \sim \theta_{23} \sim \pi/2$ and 
$\theta_{13}\sim 0$. This explains one of the $2^3$ combinations of peaks 
in Figure~\ref{fig:Donoghue-mix}. When the ${\lambda'}^{u/d}_{3,2,1}$ are 
found in different entries of the $3\times3$ Yukawa matrices, other peak 
combinations are obtained.  Continuous distributions connecting 
$\theta_{ij}\sim 0$ to $\theta_{ij}\sim\pi/2$ originate from the seesaw 
contributions that we have ignored.  Thus the distributions become more 
and more localized around the peaks as 
$\lambda_{\rm max}/\lambda_{\rm min}$ is increased and seesaw 
contributions become less important.  Therefore the small mixing angles 
of the observed CKM matrix are not atypical of distributions with large 
values of $\lambda_{\rm max}/\lambda_{\rm min}$.

How then do we interpret the peaks at $\pi/2$?  The flavor structure
of the quark sector of the Standard Model is characterized by three 
general features.  On the one hand, the quark masses have a hierarchical 
structure, and this is successfully reproduced.  In addition, the 
$W$-boson current approximately connects three distinct pairs of 
quarks---we refer to this as ``pairing structure,'' and it is also 
found in this landscape.  Finally, the $W$-boson current connects pairs 
such that the lightest up-type quark is approximately paired with
the lightest down-type quark, the middle up-type quark is approximately 
paired with the middle down-type, etc.  We refer to this as the 
``generation structure'' of the Standard Model.  Mixing angles near 
$\pi/2$ maximally violate this generation structure; for example the set 
of angles $\theta_{12}\sim\theta_{23}\sim \pi/2$ and $\theta_{13}\sim 0$ 
of (\ref{eq:exCKM}) corresponds to the quark pairings in 
(\ref{eq:pairing}).  There are $3!$ combinations in forming three pairs,
only one of which has what we call the generation structure.\footnote{
When $\sin\theta_{13}\sim 1$, $\theta_{12}\sim\theta_{23}\sim 0$
and $\theta_{12}\sim\theta_{23}\sim\pi/2$ result in the same quark 
combinations. Likewise, $(\theta_{12},\theta_{23})\sim (0,\pi/2)$ and 
$(\theta_{12},\theta_{23})\sim (\pi/2,0)$ are the same.  Therefore there 
are only $3!$ physically different combinations, even though there are 
$2^3$ different ways to pick three peaks from 
Figure~\ref{fig:Donoghue-mix}.}

\subsection{Problems} 

Is the existence of mixing angle peaks about 
$\theta_{ij}=\pi/2$ really a problem?  The landscape that we have
discussed so far may reproduce the generation structure of mass 
eigenstates in the $W$-boson current (when $\theta_{12}\sim\theta_{23}
\sim\theta_{13}\sim0$), but more often it does not.  Although it might 
be argued that this is just a $1/3!$ coincidence problem, it is still 
difficult to accept that the generation structure of flavor is not 
revealing something important about the underlying theory.  It is also 
tempting to try read something deeper from the observed hierarchy between 
$V_{ub}\sim\theta_{13} \sim 4\times 10^{-3}$ and the Cabibbo angle 
$\theta_{12}\sim 0.2$.  In this landscape this hierarchy is just a random 
statistical fluctuation (c.f. Figures~\ref{fig:nocorr-egval-analytic} 
and~\ref{fig:Donoghue-mix}), and it does not appear this shortcoming can 
be overcome by a more ideal choice of $\lambda_{\rm max}/\lambda_{\rm min}$
(c.f. Figure~\ref{fig:Donoghue-egval}).

An even bigger problem is to understand how the probability distribution 
(\ref{eq:nocorr-distr}) arises or what is the correct distribution to 
replace it.  That is, although the phenomenology of the landscape that we 
have considered may be deemed acceptable, we do not have a solid 
theoretical ground upon which to base it.\footnote{Such an attempt is 
made in~\cite{DonoghueIIA}.  However, we consider that correlation among 
Yukawa couplings and the number of extra dimensions are crucial ingredients 
in understanding flavor physics, and these are missing 
in~\cite{DonoghueIIA}. The intersecting D6--D6 system mentioned 
in~\cite{DonoghueIIA} is dual to a $T^3$-fibered compactification of  
Heterotic string theory and is simulated by the $D=3$ Gaussian landscape 
models of this article. Although \cite{DonoghueIIA} guesses that the 
scale invariant (or nearly scale invariant power-law) distribution might 
arise from the intersecting D-brane systems, we conclude otherwise.  
Specifically, we find that the scale-invariant distribution 
(\ref{eq:nocorr-distr}) is derived from the $D=1$ Gaussian landscape
model, but not a $D=3$ model.  We also find that correlation among 
various elements of the Yukawa matrices is crucial to understanding 
flavor structure and the origin of generations.}  
It is also obscure how the phenomenology of a lepton sector with large 
mixing angles and that of quark sector with small mixing angles can be 
accommodated within a single theoretical framework.  In the remaining 
sections of this paper we analyze some landscape models that 
successfully reproduce the phenomenology of hierarchy, pairing, and 
generation structure, while making progress on each of the four problems 
described above.

\section{A Gaussian Landscape for Quarks in One Extra Dimension}
\label{sec:toy1}

The landscape discussed in section~\ref{sec:indep} assumes that all 18 
elements of $\lambda^u$ and $\lambda^d$ are scanned independently.  Yet 
without any correlation between these two Yukawa matrices, generation 
structure will never be obtained.  For example, in order to ensure that 
the heaviest up-type quark $t_L$ is contained in the same SU(2)$_L$ 
doublet as the heaviest down-type quark $b_L$, we require the following.  
When an ($i$, $j$) element of the up-type Yukawa coupling 
$\lambda_{ij}^u\, \bar{u}_i\, q_j\, h$ is large, at least one of the 
three down-type Yukawa couplings 
$\lambda^d_{kj}\, \bar{d}_k\, q_j\, h^*$ ($k=1,2,3$) involving the same 
quark doublet $q_j$ should be large.  A landscape of vacua must realize 
such a correlation between the up-type and down-type Yukawa matrices in 
order to explain the generation structure.

Perhaps one of the simplest ideas to introduce such a correlation is to 
introduce an extra dimension.  If a large Yukawa coupling of 
$\lambda^u_{ij}\, \bar{u}_i\, q_j\, h$ is due to a substantial overlap 
of the wavefunctions of $q_j$ and $h$, then the down-type Yukawa 
couplings involving the same $q_j$ tend to be larger because of the 
overlap of $q_j$ and $h$.  At the same time, for localized 
wavefunctions the overlap of some triplets of $\bar{u}$, $q$ and $h$ 
can be very small, and so there is hope to explain the hierarchically 
small Yukawa couplings necessary to account for light quarks.

In this section we present a simple toy landscape based on Gaussian
wavefunctions spanning a circular extra dimension. Although a single 
extra dimension is introduced for simplicity, this model captures the 
essence of what one expects more generally from such ``Gaussian 
landscapes'' based on multiple extra-dimensional field theories.  
Through numerical simulation and an approximate analytical analysis, we 
find that the hierarchy, pairing, and generation structures of quarks is 
obtained statistically in this landscape.  No flavor symmetry is needed.

\subsection{Emergence of Scale-Invariant Distributions}
\label{ssec:FN}

We introduce a single extra dimension with the simplest geometry: $S^1$.  
The wavefunctions for all of the quarks and the Higgs boson are assumed 
to be Gaussian with a common width $d$, and centered at arbitrary points 
on $S^1$:
\begin{equation}
\varphi(y ; y_0) \simeq \frac{1}{\pi^{1/4}\sqrt{M_5 d}}\, 
   e^{- \frac{(y - y_0)^2}{2 d^2 } }\,. 
\label{eq:Gaussian}
\end{equation}
Here $y$ is the coordinate of $S^1$ and $M_5$ is the cut-off scale of 
the effective field theory in $4+1$ dimensions.  This wavefunction is 
normalized so that 
\begin{equation}
 M_5 \int_{0}^{L} dy \, \varphi^2(y) = 1,
\label{eq:normalization}
\end{equation}
where $L$ is the circumference of $S^1$.  The wavefunction 
(\ref{eq:Gaussian}) should be made periodic on $S^1$, while maintaining 
the normalization in (\ref{eq:normalization}).  Yet as long as the width 
of the Gaussian profile $d$ is parametrically smaller than the 
circumference $L$, the wavefunction is almost Gaussian. One should 
examine whether Gaussian wavefunctions arise as solutions to equations of 
motion of field theories in extra dimensions, but we defer this 
theoretical study to section~\ref{sec:SYM}, and first study whether the 
assumption of Gaussian wavefunctions on extra dimensions leads to a 
successful explanation of the physics of quark and lepton masses and
mixing angles.

We calculate the up-type and down-type Yukawa matrices with the overlap 
integrals
\begin{eqnarray}
\lambda^u_{ij} & = & g M_5 \int_{S^1} dy\,\varphi^{\bar{u}}_i(y; a_i)\,
          \varphi^{q}_j(y; b_j)\,\varphi^h(y; y^h)\,,\nn\\
\lambda^d_{kj} & = & g M_5 \int_{S^1} dy\,\varphi^{\bar{d}}_k(y; c_k)\,
          \varphi^{q}_j(y; b_j)\,\varphi^h(y; y^h)\,,\label{eq:overlap}
\end{eqnarray}
where $g$ is an overall constant.\footnote{We will see in 
section~\ref{sec:SYM} that these interactions may originate from super 
Yang--Mills interactions on a higher dimensional spacetime, and then $g$ 
derives from the Yang--Mills coupling constant.  Despite this origin, 
$g$ can be different in different sectors in effective descriptions such 
as those using just one extra dimension. For simplicity we set the 
constant $g$ to be the same for both the up and down sectors.}
$\varphi^{q}_j(y)$, $\varphi^{\bar{u}}_i(y)$, $\varphi^{\bar{d}}_k(y)$ 
and $\varphi^h(y)$ are wavefunctions of left-handed quark doublets $q_j$ 
($j=1,2,3$), right-handed up-quarks $\bar{u}_i$ ($i=1,2,3$), right-handed 
down-quarks $\bar{d}_k$ ($k=1,2,3$) and of the Higgs boson, respectively, 
all of the Gaussian form (\ref{eq:Gaussian}).\footnote{In 
section~\ref{sec:SYM} we study higher dimensional supersymmetric field 
theories, where at short distances the up-type and down-type Yukawa 
couplings involve different Higgs doublets, $h_{1,2}$.  Our results are 
independent of whether or not supersymmetry survives to the weak scale.  
If a supersymmetry survives compactification, we assume that the $h_{1,2}$ 
zero modes have the same localization.  If no supersymmetry survives then 
we assume a single Higgs zero mode with a Gaussian profile.  There is a 
``$\tan\beta$'' factor between the up and down sectors that we have set to 
unity.  With weak scale supersymmetry this factor arises from the ratio of 
Higgs vacuum expectation values, while with high scale supersymmetry 
breaking it arises from the composition of the light Higgs boson.} 
The center coordinates of these wavefunctions are $b_j$, $a_i$, $c_k$ and 
$y^h$, respectively.  The matrices $\lambda^{u,d}$ are real, so that CP is 
conserved in this toy landscape. Complex Gaussian wavefunctions will be 
studied in section~\ref{sec:Lepton}, where the effects of phases on the 
distributions of quark masses and mixing angles is found to be small.  We 
assume that the center coordinates $b_j$, $a_i$, $c_k$ and $y^h$ are 
scanned freely and independently from one another on $S^1$.  Because of 
the translational symmetry of $S^1$, only the relative difference between 
these center coordinates affects observables. Thus the effective number 
of scanning parameters is $n_S=9$.  On the other hand, there are nine 
observables determined from the Yukawa matrices in the quark sector: six 
mass eigenvalues and three mixing angles. Thus the scanning parameters 
cover the space of observables and no precise prediction among the 
observables is available.  However, since this Gaussian landscape covers 
the space of observables, our vacuum is unlikely to be missed in this 
ensemble.

The other model parameters, namely the width $d$, circumference $L$, 
cut-off scale $M_5$, and coupling $g$, are treated as fixed.  This 
treatment is quite arbitrary;  among the myriad of other possibilities are 
to scan some or all of these parameters, to allow the up-sector and the 
down-sector to have different values of $g$, to choose different widths $d$ 
for different wavefunctions, etc.  An extreme version of the landscape 
would allow everything to scan, leaving no fixed parameters to be input by 
hand.  Our treatment---namely four fixed parameters---is equivalent to 
slicing a specific subset out of a possibly much larger landscape.  This
allows us to identify a phenomenologically successful subset of the 
landscape and at the same time more easily explore its properties.  It is 
a separate question whether this subset is typical of the full landscape, 
or if it is highly weighted because of cosmological evolution and/or 
environmental selection.  Indeed, it is not implausible that the 
distributions of these four fixed parameters could be sharply peaked in 
the full landscape, since some toy landscapes predict Gaussian 
distributions for some parameters and cosmological evolution can provide 
exponentially steep weight factors.  We consider that practical progress 
can be made by dividing the full problem into simpler parts.

Note that out of these four fixed parameters there are only two independent 
combinations that affect the Yukawa matrices. First, only the three 
dimensionless combinations, $g$, $M_5d$, and $M_5L$, are physical; the
value of $M_5$ simply sets the scale for dimensionful parameters.  Second,
even as we change the ``volume'' of the extra dimension $M_5L$, the 
Yukawa couplings remain the same if the width parameter $M_5d$ and the 
coupling $g$ are scaled so that the ratio $M_5d/M_5L=d/L$ and the 
effective coupling, 
\begin{equation}
g_{\rm eff}=\frac{g}{\sqrt{M_5L}}\,,
\end{equation} 
remain the same.  Therefore the Yukawa couplings are effectively controlled 
by only these two parameters, $d/L$ and $g_{\rm eff}$.  (Note however that 
the volume $M_5L$ does affect the low-energy value of Newton's constant.)

The Yukawa couplings, given by the overlap integrals (\ref{eq:overlap}), 
can be expressed more explicitly in terms of the underlying parameters in 
a restricted region of the parameter space.  Suppose that $d/L\ll 1$.  Then 
the compactness of $S^1$ is not important in the calculation of the Yukawa 
couplings, as long as the center coordinates of quarks, $a_i$ (or $c_k$) 
and $b_j$, are close to that of the Higgs boson $y^h$ (which, by 
translational invariance, we set as the origin of the coordinate $y$). 
For such a vacuum, the Yukawa couplings are given by 
\begin{equation}
\lambda^u_{ij} \simeq g_{\rm eff}
               \left(\frac{4}{9\pi}\frac{L^2}{d^2}\right)^{\frac{1}{4}}
               e^{-\frac{1}{3d^2}(a_i^2+b_j^2-a_i b_j)}\,, \qquad 
\lambda^d_{kj} \simeq g_{\rm eff}
               \left(\frac{4}{9\pi}\frac{L^2}{d^2}\right)^{\frac{1}{4}} 
               e^{-\frac{1}{3d^2}(c_k^2+b_j^2-c_k b_j)}\,.
\label{eq:FN-matrix}
\end{equation}
Let us compare this result to the form for the Yukawa
couplings that results from approximate Abelian Flavor Symmetries
(AFS).  In the most general AFS scheme there is a small symmetry
breaking factor associated with each quark field,
$\epsilon^{q,\bar{u},\bar{d}}$, which leads to Yukawa matrix elements
\begin{equation}
\lambda^u_{ij} = g_{ij} \, \epsilon_i^{\bar{u}}\epsilon_j^q \,,
 \qquad \lambda^d_{kj} = g'_{kj} \, \epsilon_k^{\bar{d}}\epsilon_j^q \,,
\label{eq:AFS}
\end{equation}
where the $g_{ij}$ and $g'_{kj}$ are all of order unity.  A mass hierarchy 
among the generations is realized by imposing 
$\epsilon_3\gg\epsilon_2\gg\epsilon_1$ in the left, right, or both sectors.  
Models with fewer parameters can be constructed and then the symmetry 
breaking parameters are not all independent; consider for example a single 
Abelian symmetry with a symmetry breaking parameter $\epsilon$ that 
appears in different entries with different powers due to generation 
dependent charges. Generation charges (0,2,3) then give 
$\epsilon_3:\epsilon_2:\epsilon_1=1:\epsilon^2:\epsilon^3$. No matter how 
the model is arranged, the mass hierarchy arises because the first 
generation feels much less flavor symmetry breaking than the third.  Note 
that AFS theories are very flexible---any hierarchical pattern of fermion 
masses can be described by an appropriate AFS.

The result (\ref{eq:FN-matrix}), which involves no flavor symmetry, has 
some similarities with the form (\ref{eq:AFS}).  First notice that $a_i$, 
$b_j$, and $c_k$ can be both positive and negative and therefore the 
factor of $a_i b_j$ (and $c_k b_j$) in (\ref{eq:FN-matrix}) is 
statistically neutral.  Anticipating this statistical averaging, we cast 
(\ref{eq:FN-matrix}) into the form of (\ref{eq:AFS}) with the 
identification 
\begin{equation}
\epsilon^{\bar{u}}_i = e^{-\frac{a_i^2}{3d^2}}\,, \qquad
\epsilon^q_j = e^{-\frac{b_i^2}{3d^2}}, \qquad 
\epsilon^{\bar{d}}_k = e^{-\frac{c_k^2}{3d^2}}\,.
\label{eq:AFS-supp}
\end{equation}
An important feature is that the AFS factor $\epsilon^q_j$
is shared by all elements of both the up-type and the down-type Yukawa 
couplings that involve the left-handed quark doublet $q_j$.  This 
introduces a correlation between the up-type and down-type Yukawa matrices.

We first study the probability distribution for a single entry in the 
Yukawa matrix, ignoring correlations with other entries. This allows us 
to determine the analogue of (\ref{eq:nocorr-distr}) for this Gaussian 
landscape.
\begin{figure}[t]
\begin{center}
\begin{tabular}{ccc}
\includegraphics[width=0.3\linewidth]{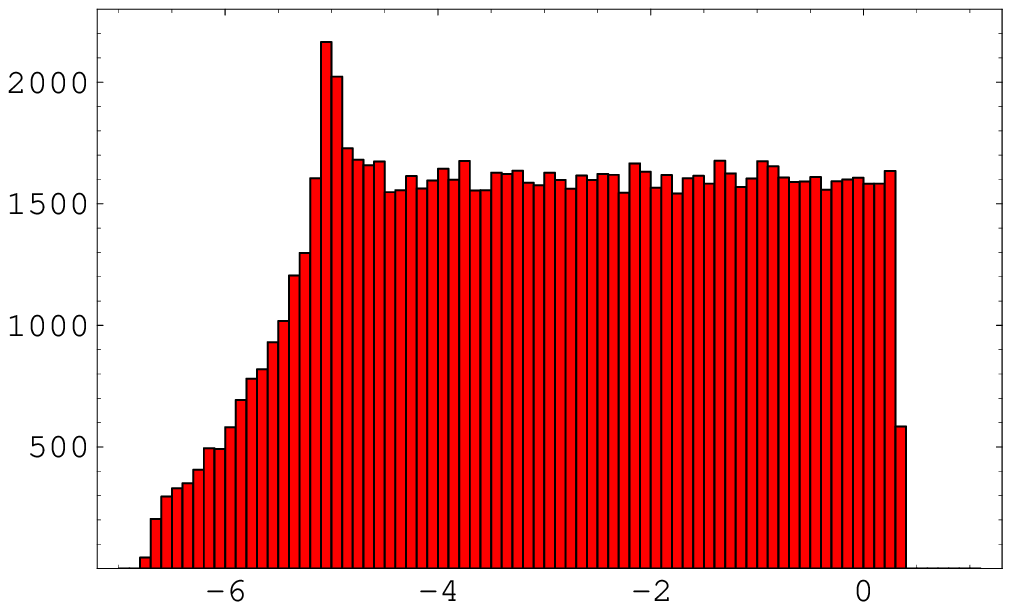} &
\includegraphics[width=0.3\linewidth]{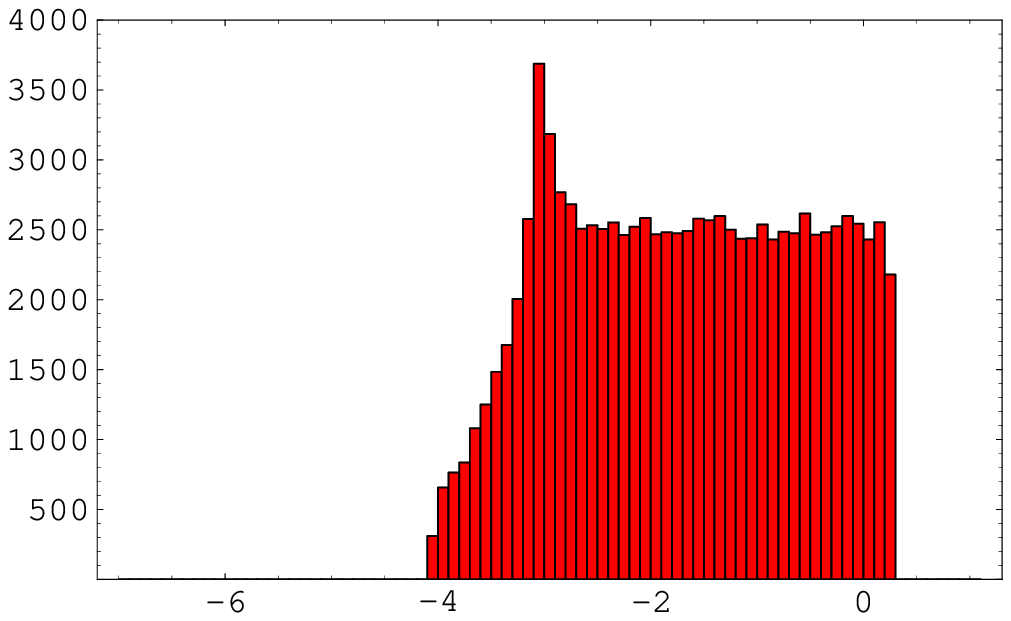} &
\includegraphics[width=0.3\linewidth]{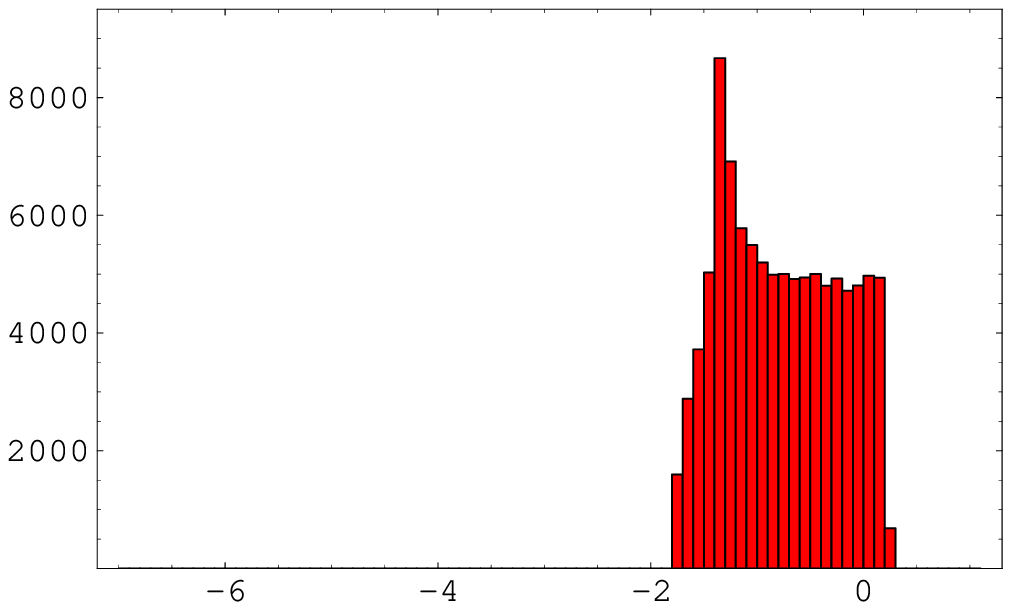} 
\end{tabular}
\caption{\label{fig:entrydistr} Distributions of $10^5$ Yukawa matrix 
elements, generated from the Gaussian landscape on $S^1$.  From left to 
right the panels correspond to $(d/L,g_{\rm eff})=(0.08,1)$, $(0.1,1)$, 
and $(0.14,1)$.}
\end{center}
\end{figure}
Here we do not have a distinction between the up-type and down-type 
Yukawa matrices because we assume that the center coordinates of both 
the $\bar{u}$ and $\bar{d}$ wavefunctions are distributed randomly over 
$S^1$.  The up-type Yukawa coupling in (\ref{eq:FN-matrix}) is a function 
on the two-dimensional parameter space 
$(a,b) \in [-L/2,L/2] \times [-L/2,L/2]$.  The probability that the 
Yukawa coupling is larger than some value $\lambda_0$ is proportional to 
the area of an ellipse, 
\begin{equation}
\frac{(a+b)^2}{4} + \frac{3(a-b)^2}{4} = 3 d^2 \ln 
      \left[ \frac{g_{\rm eff}}{\lambda_0}
      \left(\frac{4}{9\pi}\frac{L^2}{d^2}\right)^{1/4} \right].
\label{eq:ellipsearea}
\end{equation}
When the signs of $a$ and $b$ are opposite, the overlap of the three 
wavefunctions is small and the Yukawa coupling becomes small.  This is 
why the region of $\lambda>\lambda_0$ is short in the $(a-b)$ axis and 
long in the $(a+b)$ axis.  Comparing the area (\ref{eq:ellipsearea}) 
to the total area $L^2$, we see the probability that $\lambda>\lambda_0$ 
is given by 
\begin{equation}
P(\lambda > \lambda_0) \simeq 2\sqrt{3} \pi 
            \left(\frac{d}{L}\right)^2 
            \ln\left[ \frac{g_{\rm eff}}{\lambda_0}
            \left(\frac{4}{9\pi}\frac{L^2}{d^2}\right)^{1/4}\right] ,
\end{equation}
and hence the distribution is flat: 
\begin{equation}
\frac{dP(\lambda)}{d \ln \lambda} = 2\sqrt{3}\pi \left(\frac{d}{L}\right)^2 
 \simeq 11 \left(\frac{d}{L}\right)^2. \label{eq:flat}
\end{equation} 
The distribution may cease to be flat as $a$ or $b$ approaches $\pm L/2$, 
where the parameter space ends, because we ignored the periodic boundary 
condition in the calculation that led to (\ref{eq:FN-matrix}).  Setting
this point aside we see that the probability distribution of Yukawa 
couplings in this toy landscape is flat on the $\ln\lambda$ axis and has 
an approximate span of 
\begin{equation}
 \Delta \ln \lambda = 
 \ln \left(\frac{\lambda_{\rm max}}{\lambda_{\rm min}}\right) 
 \simeq \frac{1}{11} \left(\frac{L}{d}\right)^2, 
\label{eq:lgrange-lambda}
\end{equation}
arising from the inverse of the height of the distribution function 
(\ref{eq:flat}).  The overall hierarchy among Yukawa couplings 
$\Delta\ln\lambda=\ln(\lambda_{\rm max}/\lambda_{\rm min})$ is 
proportional to $(L/d)^2$; the narrower the wavefunctions become, the 
further the wavefunctions can be separated, and the smaller the Yukawa 
couplings can be.  As seen from (\ref{eq:FN-matrix}), the upper end of 
this scale-invariant distribution $\lambda_{\rm max}$ is roughly 
$g_{\rm eff}\sqrt{L/d}$.  Note that the scale-invariant distribution 
(\ref{eq:nocorr-distr}) was introduced almost by hand in~\cite{DDR} in 
order to account for the large hierarchy among Yukawa couplings.  It is 
interesting that this distribution is a natural prediction of our simple 
Gaussian landscape. 

We performed a numerical calculation to confirm the semi-analytical 
analysis above, taking account of the compactness of $S^1$ by making the 
wavefunction (\ref{eq:Gaussian}) periodic.  The center coordinates of 
the quark wavefunctions $a_i,b_j,c_k$ were generated randomly $10^5$ times, 
and the Yukawa coupling was calculated through (\ref{eq:overlap}). This 
process was repeated for three different sets of the $(d/L,g_{\rm eff})$ 
parameters: $(0.08,1)$, $(0.10,1)$ and $(0.14,1)$.  The resulting 
distributions, shown in Figure~\ref{fig:entrydistr}, are all roughly 
scale-invariant (flat on a logarithmic scale), with heights proportional
to $(d/L)^2$, just as we expected from the semi-analytical discussion.

\subsection{Quark-Sector Phenomenology of the Gaussian Landscape}
\label{ssec:phen}

\begin{figure}[t]
\begin{center}
\begin{tabular}{ccc}
\includegraphics[width=0.3\linewidth]{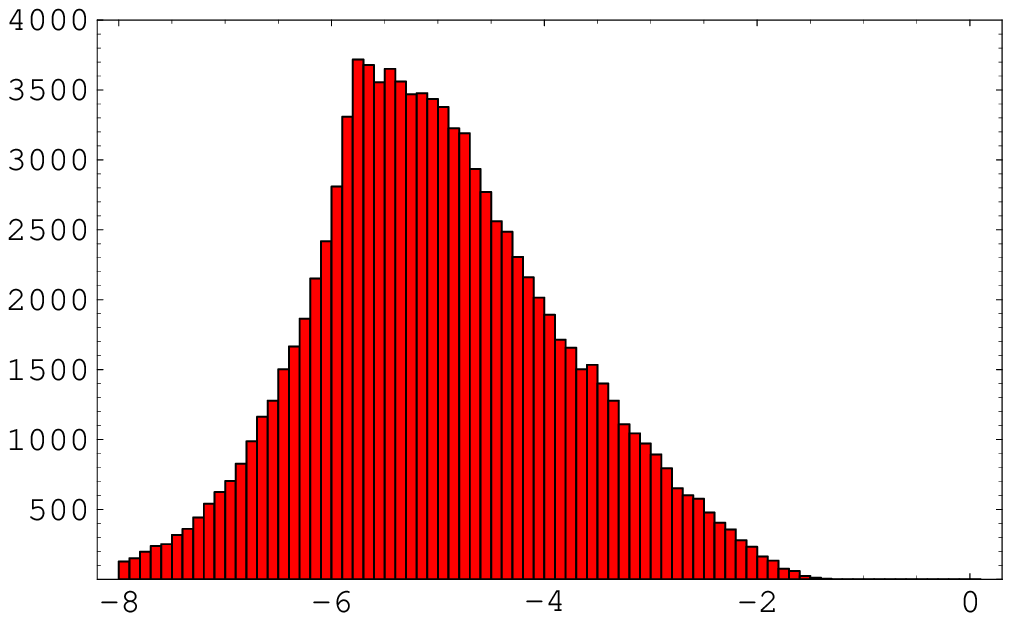} &
\includegraphics[width=0.3\linewidth]{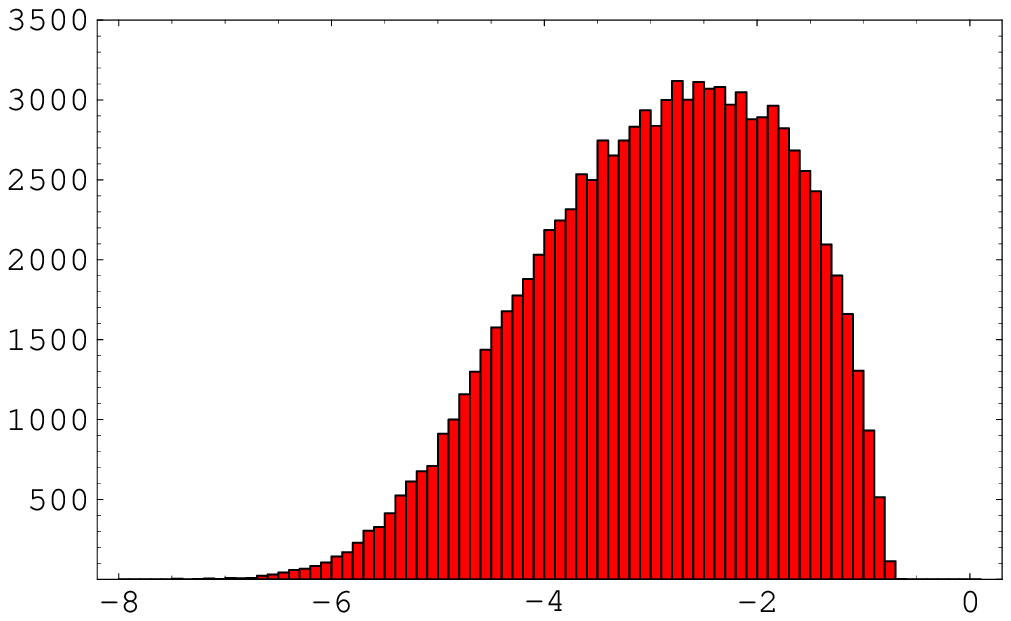} &
\includegraphics[width=0.3\linewidth]{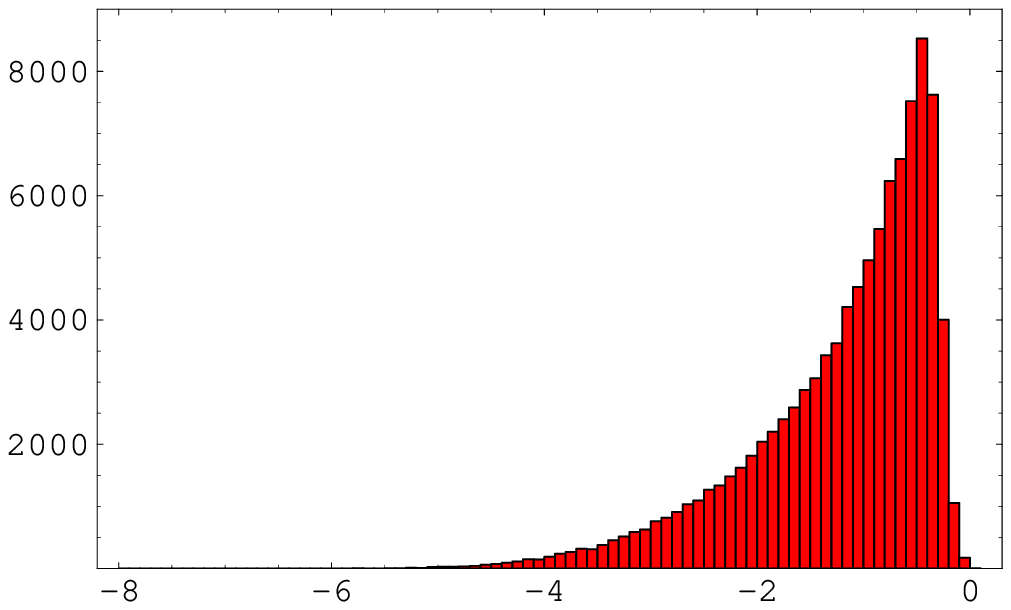} \\
$\log_{10}\lambda_{u,d}$ & $\log_{10}\lambda_{c,s}$ & 
$\log_{10}\lambda_{t,b}$ \\
\includegraphics[width=0.3\linewidth]{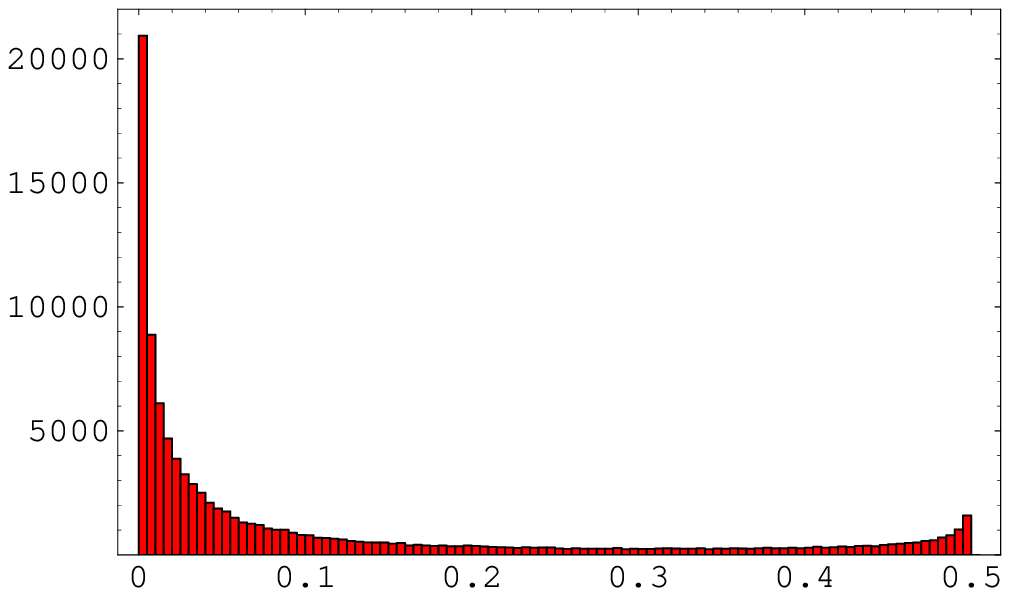} &
\includegraphics[width=0.3\linewidth]{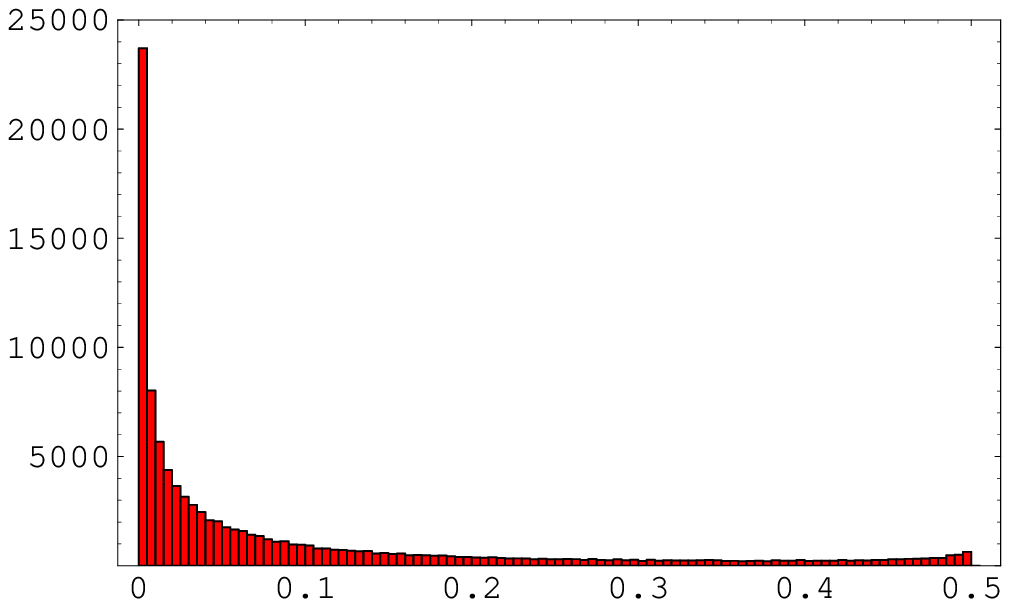} &
\includegraphics[width=0.3\linewidth]{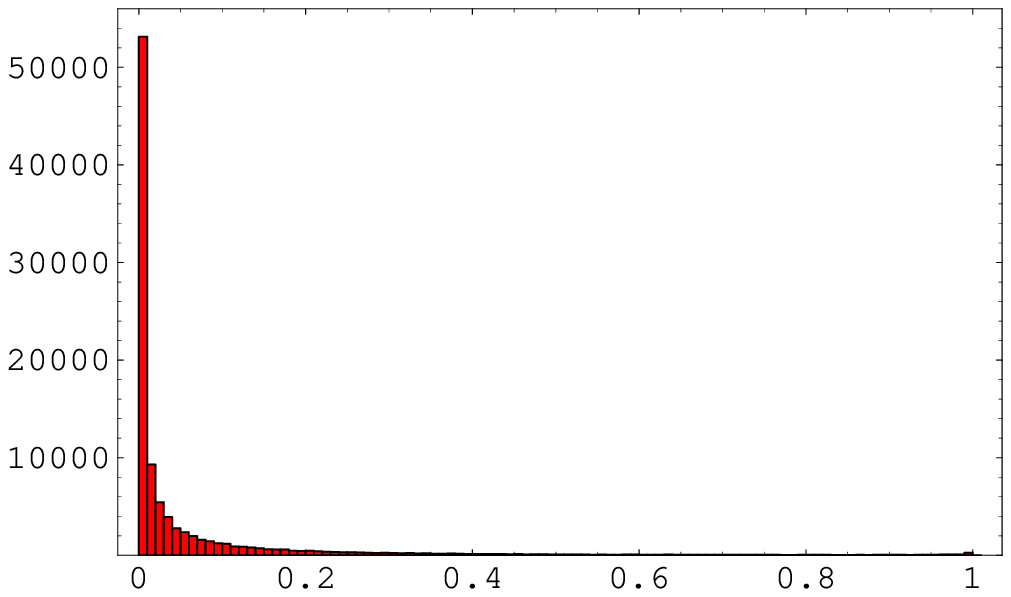} \\
$\theta_{12}/\pi$ & $\theta_{23}/\pi$ & $\sin\theta_{13}$ \\
\includegraphics[width=0.3\linewidth]{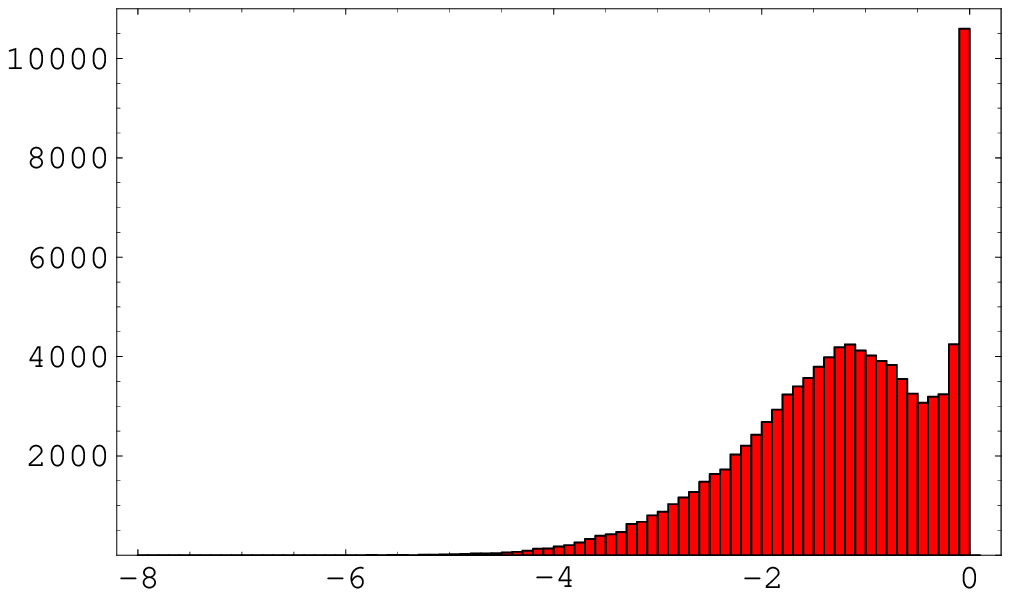} &
\includegraphics[width=0.3\linewidth]{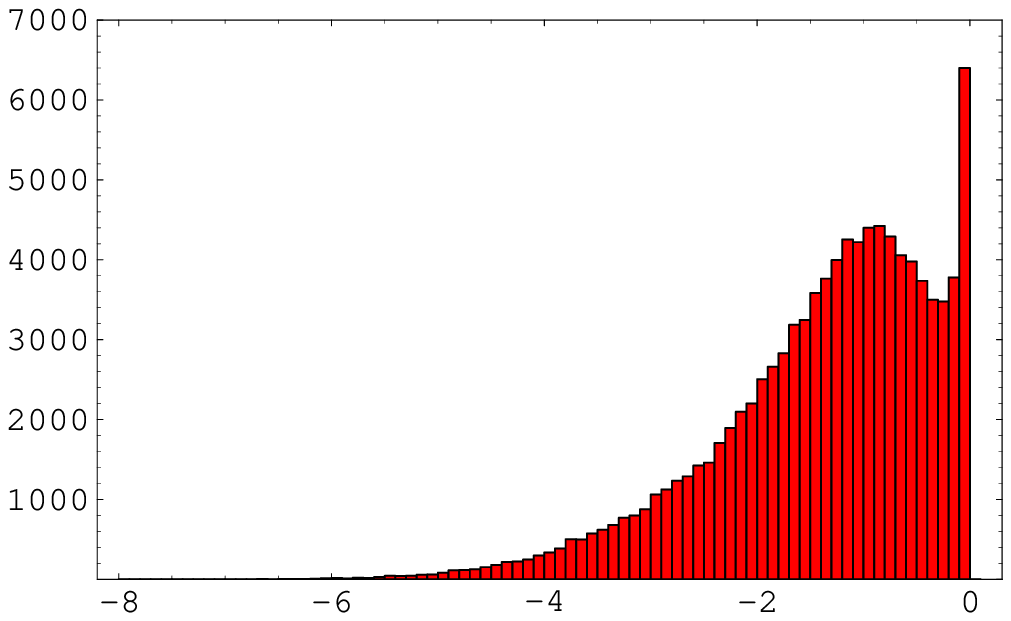} &
\includegraphics[width=0.3\linewidth]{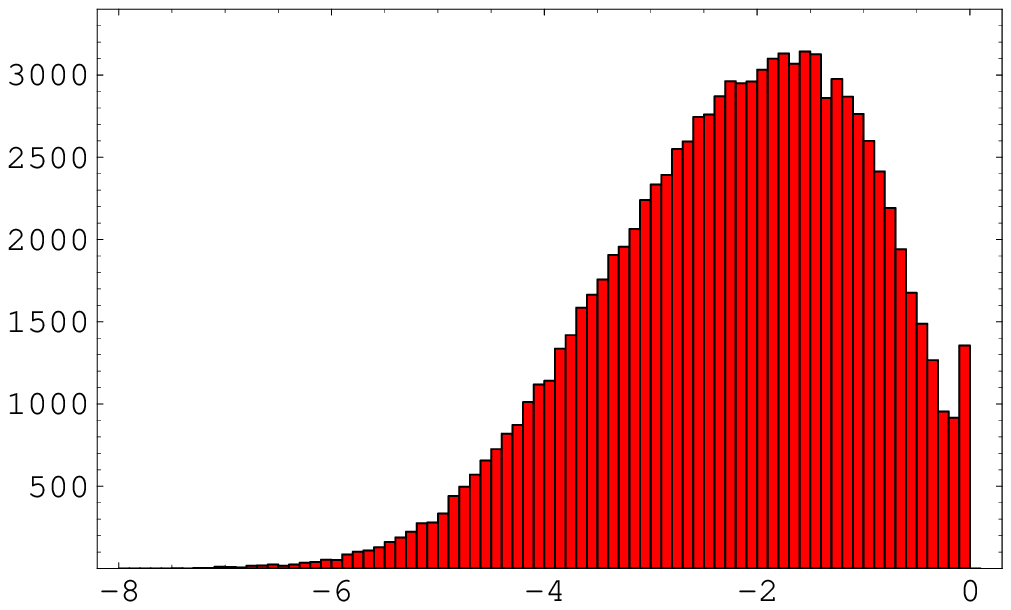} \\
$\log_{10}(2\theta_{12}/\pi)$ & $\log_{10}(2\theta_{23}/\pi)$ & 
$\log_{10}\sin\theta_{13}$ \\
\end{tabular}
\caption{\label{fig:D=1-nocut} Distributions of three quark Yukawa 
eigenvalues and mixing angles, based on a numerical simulation of the 
Gaussian landscape on $S^1$, using $(d/L,g_{\rm eff})=(0.08,0.2)$.}
\end{center}
\end{figure}

Let us now study the probability distributions of the mass eigenvalues 
and mixing angles.  Figure~\ref{fig:D=1-nocut} shows the result of a 
numerical simulation with $(d/L,g_{\rm eff})=(0.08, 0.2)$.
The distributions of Yukawa eigenvalues in Figure~\ref{fig:D=1-nocut}
are similar to those in Figure~\ref{fig:Donoghue-egval}, but with
narrower distributions of $\lambda_{u,d}$ and with the distributions of 
$\lambda_{c,s}$ shifted downward.  The prominent difference between the 
mixing-angle distributions in Figure~\ref{fig:D=1-nocut} and those in 
Figure~\ref{fig:Donoghue-mix} is the absence of the unwanted peaks at 
$\theta_{ij}\simeq \pi/2$.  Thus we find the generation structure of the 
quark sector follows from this Gaussian landscape; introducing 
correlations between the up-type and down-type Yukawa matrix elements 
works perfectly.  Moreover, in contrast to Figure~\ref{fig:Donoghue-mix} 
the distribution of $\theta_{13}$ in Figure~\ref{fig:D=1-nocut} has a 
clear peak at $\theta_{13} \ll {\cal O}(1)$ when displayed on a 
logarithmic scale.

The distributions of Yukawa eigenvalues and mixing angles in 
Figure~\ref{fig:D=1-nocut} can be understood analytically if we allow 
ourselves to make an approximation.
\begin{figure}[t]
\begin{center}
\begin{tabular}{ccc}
\includegraphics[width=0.3\linewidth]{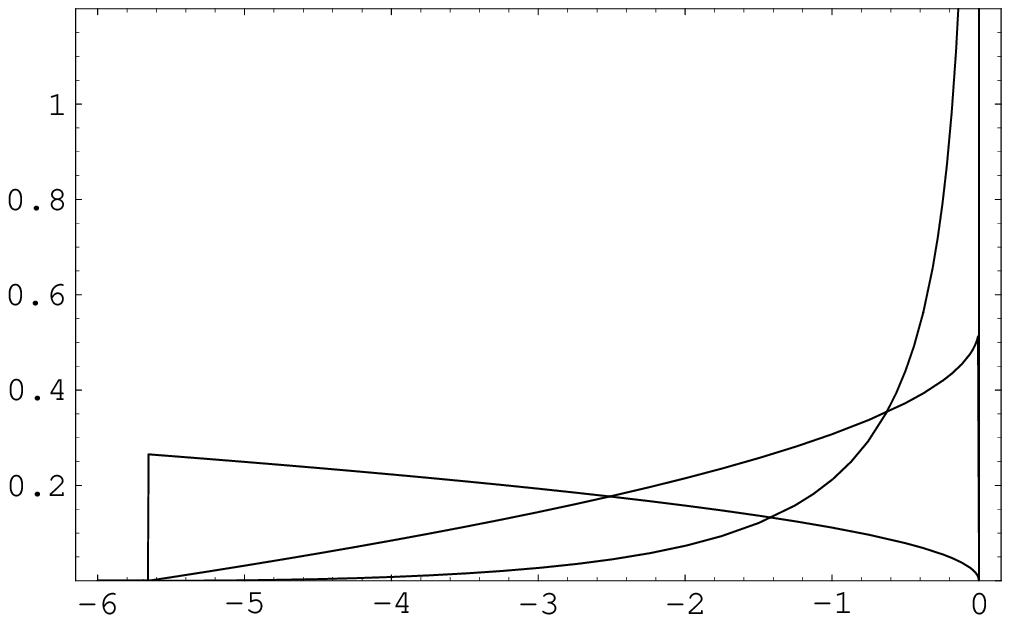} & 
\includegraphics[width=0.3\linewidth]{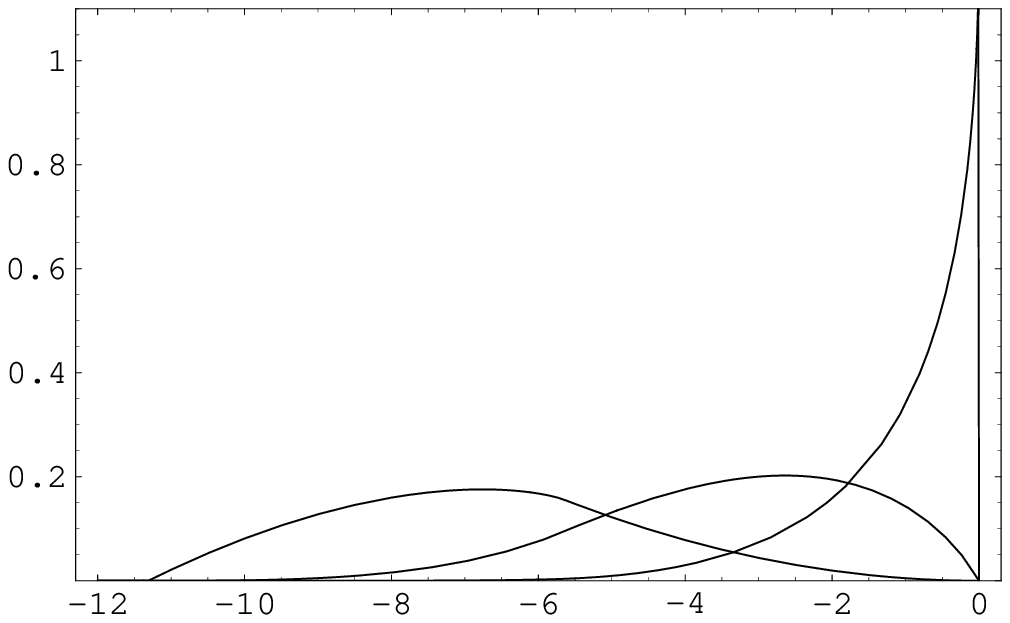} & 
\includegraphics[width=0.3\linewidth]{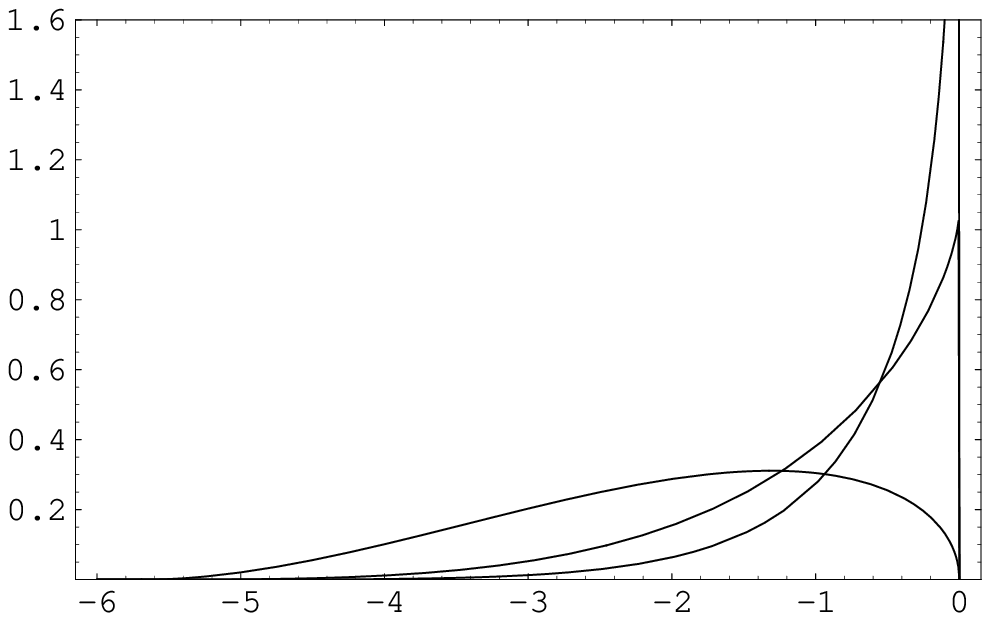} \\
$\log_{10}\epsilon_{1,2,3}$ & $\log_{10}\lambda_{1,2,3}$ &
$\log_{10}\sin\theta_{12,23,13}$ \\
\end{tabular}
\caption{\label{fig:fndistr} Distributions of the three AFS 
suppression factors, the three Yukawa eigenvalues, and the CKM
mixing angles; the latter calculated naively from (\ref{eq:mass-approx}) 
and (\ref{eq:Vus-approx}--\ref{eq:Vub-approx}) using the Gaussian 
landscape on $S^1$.  The CKM mixing angle $\theta_{23}$ is the one most 
sharply peaked at zero, while $\theta_{13}$ is most spread out.  These 
figures correspond to $d/L = 0.08$ and thus 
$\Delta\log_{10}\epsilon=5.65$.  Recall that these distribution functions 
are not reliable for small eigenvalues.}
\end{center}
\end{figure}
We have seen in section~\ref{ssec:FN} that both the up-type and 
down-type Yukawa matrices have an AFS structure.  Thus, we begin by 
determining the probability distribution of the AFS suppression factors 
$\epsilon^{q,\bar{u},\bar{d}}$ in (\ref{eq:AFS-supp}).  The value of 
$\epsilon^{q}$ ($\epsilon^{\bar{u},\bar{d}}$) is determined by the 
distance $|b|$ ($|a|$, $|c|$) of the left-handed (right-handed) quark 
wavefunction from the Higgs boson wavefunction.  Since the center 
coordinates are scanned randomly over the extra dimension $S^1$, the 
probability measure is 
\begin{equation}
 dP(b) = \frac{2}{L} d |b|\,, \qquad {\rm for~} 
0 \leq |b| \leq \frac{L}{2}.
\label{eq:FN-measure}
\end{equation}
The measure for the right-handed quarks is the same, and we only deal 
with the left-handed quarks hereafter.  Converting the variable from 
$|b|$ to $\ln \epsilon^q$ using (\ref{eq:AFS-supp}), we find 
\begin{equation}
dP(y) = \frac{dy}{2\sqrt{y}}\,,
\qquad {\rm for~} 0\leq y \leq 1\,, 
\label{eq:FN-distr}
\end{equation}
where $y\equiv\ln\epsilon^q/\Delta\ln\epsilon$ and we have defined 
$\Delta\ln\epsilon\equiv- (L/2d)^2/3=-(L/d)^2/12$.

The center coordinates of the three left-handed quark wavefunctions are 
chosen randomly, thus three AFS suppression factors follow 
(\ref{eq:FN-distr}) independently.  The smallest of these corresponds to 
the suppression factor $\epsilon_1$ for the lightest quark, while the 
middle factor $\epsilon_2$ and largest factor $\epsilon_3$ correspond to 
the middle and the heaviest quarks. The distribution of $\epsilon_{1,2,3}$ 
is given by 
\begin{equation}
dP(y_1,y_2,y_3) = \frac{3!}{2^3}
 \frac{ dy_1 \, dy_2 \, dy_3}{\sqrt{y_1 y_2 y_3}}\,
 \Theta(y_1 - y_2)\,\Theta(y_2 - y_3) \,, 
\qquad {\rm for~} 0\leq y_i\leq 1\,,
\label{eq:FN-combined}
\end{equation}
where $y_i\equiv\ln\epsilon_i/\Delta\ln\epsilon$.  Note that $y_1>y_2>y_3$.  
The distributions of the individual AFS suppression factors 
$\epsilon_{1,2,3}$ are obtained by integrating (\ref{eq:FN-combined}) with 
respect to the other two variables:
\begin{eqnarray}
dP(y_3) = \frac{3}{2}\frac{(1-\sqrt{y_3})^2}{\sqrt{y_3}}\, dy_3\,,\qquad 
dP(y_2) = 3\, (1-\sqrt{y_2})\, d y_2\,,\qquad 
dP(y_1) = \frac{3}{2}\sqrt{y_1} \, dy_1\,. 
\label{eq:FN}
\end{eqnarray}
These distribution functions are shown in Figure~\ref{fig:fndistr}.
Meanwhile, their averages are given by 
\begin{equation}
\frac{\vev{\ln\epsilon_3}}{\Delta\ln\epsilon} = \vev{y_3} = 0.1\,, \quad 
\frac{\vev{\ln\epsilon_2}}{\Delta\ln\epsilon} = \vev{y_2} = 0.3\,, \quad  
\frac{\vev{\ln\epsilon_1}}{\Delta\ln\epsilon} = \vev{y_1} = 0.6\,, \quad
\frac{\vev{\ln (\epsilon_1/\epsilon_2)}}
      {\vev{\ln (\epsilon_2/\epsilon_3)}} = 1.5\,.\,
\label{eq:FN-prediction}
\end{equation}
In a sense, this Gaussian landscape predicts the ratio of the AFS charges 
for the three generations: 6:3:1. However this ratio only describes the 
statistical ensemble, and the distribution functions (\ref{eq:FN}) 
contain more information.

The distributions of Yukawa eigenvalues follow from (\ref{eq:FN}) with 
the approximation 
\begin{equation}
\ln (\lambda_i/\lambda_{\rm max})\sim\ln\epsilon_i^q\epsilon_i^{\bar{q}}. 
\label{eq:mass-approx}
\end{equation}
Explicit expressions of the distribution functions derived in this 
way are found in the appendix, (\ref{eq:egval3-D1}--\ref{eq:egval1-D1}), 
and are plotted in Figure~\ref{fig:fndistr}.  From 
(\ref{eq:egval3-D1}--\ref{eq:egval1-D1}) we see that for small values of 
${z}_i\equiv\ln(\lambda_i/\lambda_{\rm max})/\Delta\ln\epsilon$ (these 
correspond to large eigenvalues), the distribution functions behave as 
$\neq 0$, $\propto {z}_2$, and $\propto {z}_1^2$, which is confirmed in 
the numerical simulation in Figure~\ref{fig:D=1-nocut}. Note that the 
distribution of $\lambda_{s,c}$ begins at around $10^{-1}$, and that of 
$\lambda_{u,d}$ at around $10^{-2}$, contrary to the approximate analytic 
distribution functions (\ref{eq:egval2-D1},\ref{eq:egval1-D1}), which 
begin at ${\cal O}(1)$.  This is probably due to the effects of 
diagonalizing the Yukawa matrices.  As is familiar in quantum mechanics, 
two degenerate eigenvalues split even in the presence of the slightest 
perturbation.  We later refer to this effect as the ``diagonalization 
effect.''

We should also note that the AFS approximation (\ref{eq:FN-matrix},
\ref{eq:AFS-supp}) breaks down for small values of Yukawa eigenvalues. 
The distributions based on the AFS approximation extend all the way down 
to $\ln(\lambda/\lambda_{\rm max})\sim 2\Delta\ln\epsilon$, while the 
numerical results cover a logarithmic range closer to 
(\ref{eq:lgrange-lambda}).  Since 
$\Delta\ln\epsilon\simeq-\Delta\ln\lambda$, the analytically derived 
range of $\ln\lambda$ covers twice the logarithmic scale that we expect.  
This discrepancy arises from the compactness of the extra dimension.  
That is, what really matters in the exponent of (\ref{eq:FN-matrix}) is  
\begin{equation}
{\rm min}\left[ (a+n L)^2 + (b+m L)^2 - (a+n L)(b+mL) \right]\,, 
\qquad {\rm for~} n\,, m \in\Z \,.
\label{FN-fullargument}
\end{equation}
As the center coordinates $a$ and $b$ approach $\pm L/2$, non-zero 
choices of $n$ and $m$ may become just as important as $n = m = 0$ in 
(\ref{eq:FN-matrix}).  Indeed, when $|a|\sim |b|\sim L/2$ integers $n$
and $m$ can be chosen so that the last term is negative.  When the 
compactness of the extra dimension is taken into account, the full 
expression (\ref{FN-fullargument}) cannot be larger than $(L/2)^2$.  This 
is why the distributions of Yukawa couplings and eigenvalues in 
Figure~\ref{fig:entrydistr} and \ref{fig:D=1-nocut} span over 
$\ln(\lambda/\lambda_{\rm max})\sim\Delta\ln\epsilon$ in the numerical 
results.  The AFS approximation breaks down because the coefficient 
$g_{ij}=e^{(a_i+nL)(b_j+mL)/3d^2}$ is not statistically neutral when 
$|a_i|$ and $|b_j|$ are around $L/2$.

Distribution functions of the mixing angles can also be obtained by 
pursuing the AFS approximation, along with the additional (crude) 
approximations: 
\begin{eqnarray}
 \ln V_{us} & \sim & \ln \left( {\rm max} \left\{ 
  (\epsilon^q_1/\epsilon^q_2)_{\rm u\mbox{-}sector},\, 
  (\epsilon^q_1/\epsilon^q_2)_{\rm d\mbox{-}sector} \right\} \right) 
     \label{eq:Vus-approx} \\
 \ln V_{cb} & \sim & \ln \left( {\rm max} \left\{ 
  (\epsilon^q_2/\epsilon^q_3)_{\rm u\mbox{-}sector},\, 
  (\epsilon^q_2/\epsilon^q_3)_{\rm d\mbox{-}sector} \right\} \right) 
    \label{eq:Vcb-approx} \\
 \ln V_{ub} & \sim & \ln \left( {\rm max} \left\{ 
  (\epsilon^q_1/\epsilon^q_3)_{\rm u\mbox{-}sector},\, 
  (\epsilon^q_1/\epsilon^q_3)_{\rm d\mbox{-}sector} \right\} \right).
    \label{eq:Vub-approx} 
\end{eqnarray}
Explicit expressions are found in (\ref{eq:th12-D1}--\ref{eq:th13-D1})
and are plotted in Figure~\ref{fig:fndistr}. These approximate analytic 
distribution functions capture qualitative features of the numerical 
results.  Note that the distribution function of $\sin\theta_{13}$ becomes 
zero at $\log_{10}\sin\theta_{13}\sim 0$ because $\sin\theta_{13}$ can be 
of order unity only when all three eigenvalues are almost degenerate in 
either the up- or down-sector, and the probability for this to occur is 
small.  The averages of the mixing angles in a logarithm scale are ordered  
\begin{equation}
 \vev{\theta_{13}} < \vev{\theta_{12}} < \vev{\theta_{23}} \,,
\end{equation}
both in the numerical simulation and in the analytic distributions; 
see (\ref{eq:tij}).  This is regarded as a consequence of the AFS charges 
in (\ref{eq:FN-prediction}); indeed in (\ref{eq:FN-prediction}) we have  
\begin{equation}
 \vev{\ln (\epsilon_1/\epsilon_3)} < \vev{\ln (\epsilon_1/\epsilon_2)} 
< \vev{\ln (\epsilon_2/\epsilon_3)} \,.
\end{equation}
Whether the assignments in (\ref{eq:FN-prediction}) are observationally 
acceptable or not is debatable, and we will return to this issue in 
section~\ref{sec:statistics}. We discuss in section~\ref{sec:Geometry} 
how the ``charge assignment'' changes when the geometry $S^1$ is replaced 
by other geometries.

To summarize, we see that the qualitative expectations from an AFS-type 
mass matrix hold true in this landscape.  In particular, the similarities 
between the landscape generated Yukawa couplings (\ref{eq:FN-matrix}) and 
those of (\ref{eq:AFS}) allow us to understand Gaussian landscapes, at 
least to some degree, using intuition based on models of AFS.  Of course, 
between these approaches the origin of small parameters is completely 
different: in the landscape they arise from small overlaps of 
wavefunctions well-separated in the extra dimension, whereas in AFS they 
arise from small symmetry breaking parameters.  Let us now emphasize this
distinction.

A crucial general feature of all AFS models is that the mass hierarchy 
between generations, $m_3 \gg m_2 \gg m_1$, arises because there is 
a hierarchy in the amount of symmetry breaking coupled to these 
generations.  This is true in the general Abelian case by the choice 
$\epsilon_3 \gg \epsilon_2 \gg \epsilon_1$.  In more restricted versions 
having $\epsilon_i \approx \epsilon^{Q_i}$, the hierarchy is imposed by a 
choice of charges $Q_1 > Q_2 > Q_3$.  If the flavor symmetry is non-Abelian, 
then there is a hierarchy of symmetry breaking, for example the rank may 
be broken from $i$ to $i-1$ with strength $\epsilon_i$. Thus AFS can in 
principle describe any flavor pattern, for example one heavy generation 
with Yukawas of order unity and two very light generations with Yukawas 
of order $10^{-10}$.  The situation with this Gaussian landscape is 
very different.  Each Yukawa coupling is a statistical quantity, with a 
probability distribution that is approximately scale invariant over a 
range determined by a single small parameter, $d/L$.  Relative to this 
range, the hierarchy of mass eigenvalues, including the typical 
inter-generational mass ratios and mixing angles, arises purely from 
statistics.  Unlike with the AFS parameters $\epsilon_i$, there is no 
sense in which the fundamental theory distinguishes among generations.  
Therefore unlike with AFS, this Gaussian landscape cannot accommodate 
one heavy generation and two very light generations of comparable mass.  
Within statistical uncertainties, the landscape determines the AFS charges.

\subsection{Environmental Selection Effects}
\label{ssec:environment}

It is a formidable task to understand all of the environmental effects 
that would be associated with a landscape scanning over the flavor 
parameters of the Standard Model.  Furthermore, without a specific theory 
for the landscape it is unclear whether certain qualitative features of 
the flavor sector arise due to environmental selection, due to systematic 
features of the landscape distributions, or due to accident.  Consider 
these examples.  In the Gaussian landscape of this section, 
$\lambda_1/\lambda_2$ tends to be smaller than $\lambda_2/\lambda_3$.  
Therefore if $d/L$ is chosen so as to explain the hierarchy 
$\lambda_2/\lambda_3$, then the relative lightness of the up and down 
quarks is explained.  The value of $d/L$ required to reproduce the quark 
masses we measure may be selected dynamically within the fundamental 
theory or it may be selected due to environmental effects associated with 
having light up and down quarks.  Likewise, the unexpected hierarchy 
$m_t/m_b$ may be due to the dynamical or accidental selection of 
different effective coupling constant $g_{\rm eff}$'s for the up and down 
sectors,\footnote{This hierarchy may also be due to weak-scale 
supersymmetry with a large $\tan\beta$; however this requires that the 
wavefunctions for the up-type and down-type Higgs are located at the same 
position in the extra dimension.} 
or environmental selection could favor a very large top mass as described 
in~\cite{FHW}.  

\begin{figure}[t]
\begin{center}
\begin{tabular}{ccc}
\includegraphics[width=0.3\linewidth]{Figures/D1L10d08/U1.eps} &
\includegraphics[width=0.3\linewidth]{Figures/D1L10d08/U2.eps} &
\includegraphics[width=0.3\linewidth]{Figures/D1L10d08/U3.eps} \\
$\log_{10}\lambda_{u,d}$ & $\log_{10}\lambda_{c,s}$ & 
$\log_{10}\lambda_{t,b}$ \\
\includegraphics[width=0.3\linewidth]{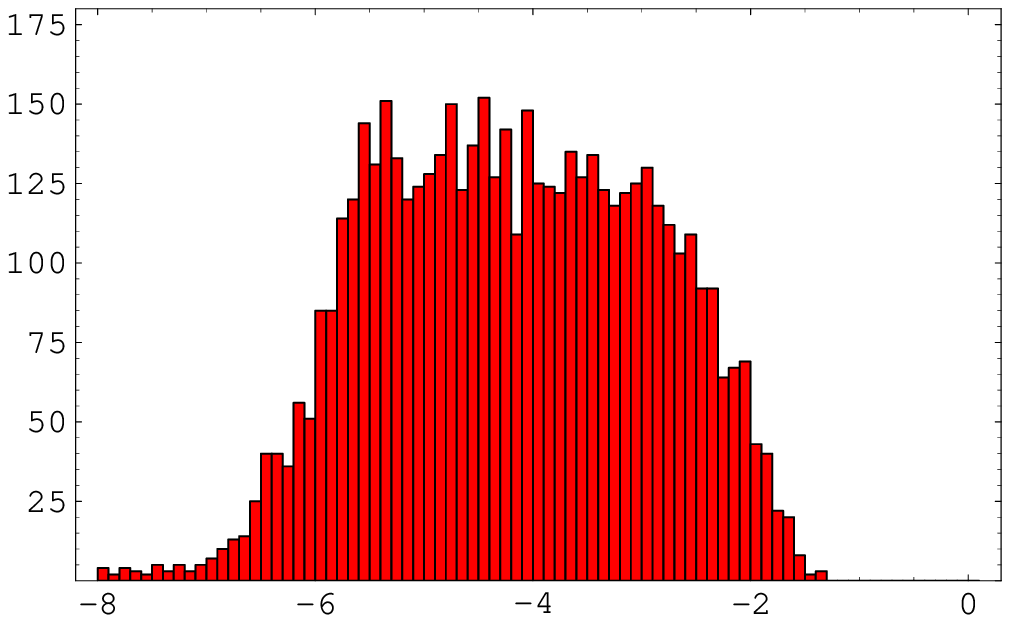} &
\includegraphics[width=0.3\linewidth]{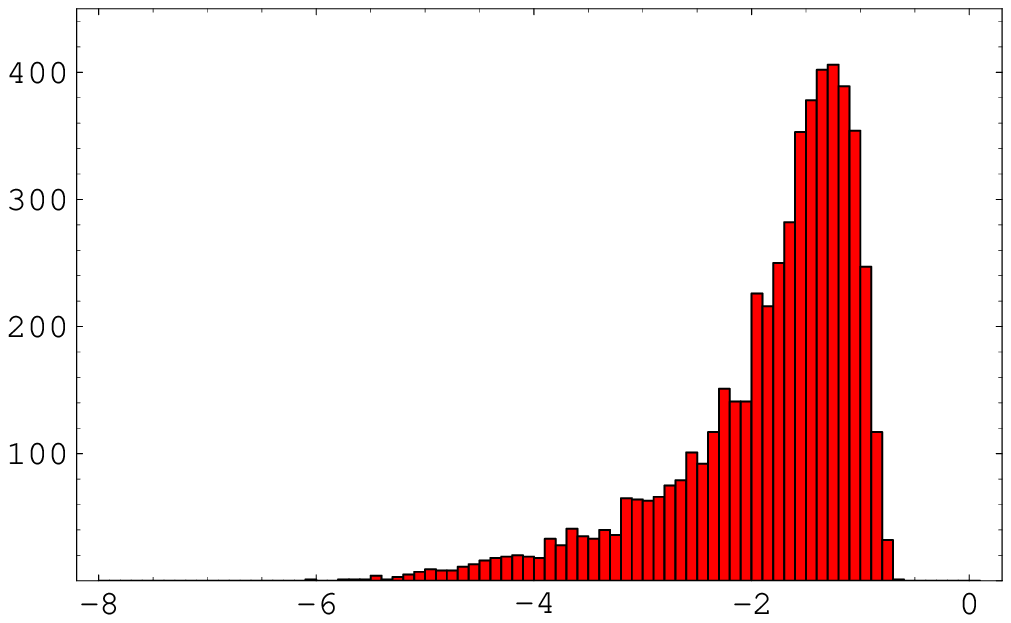} &
\includegraphics[width=0.3\linewidth]{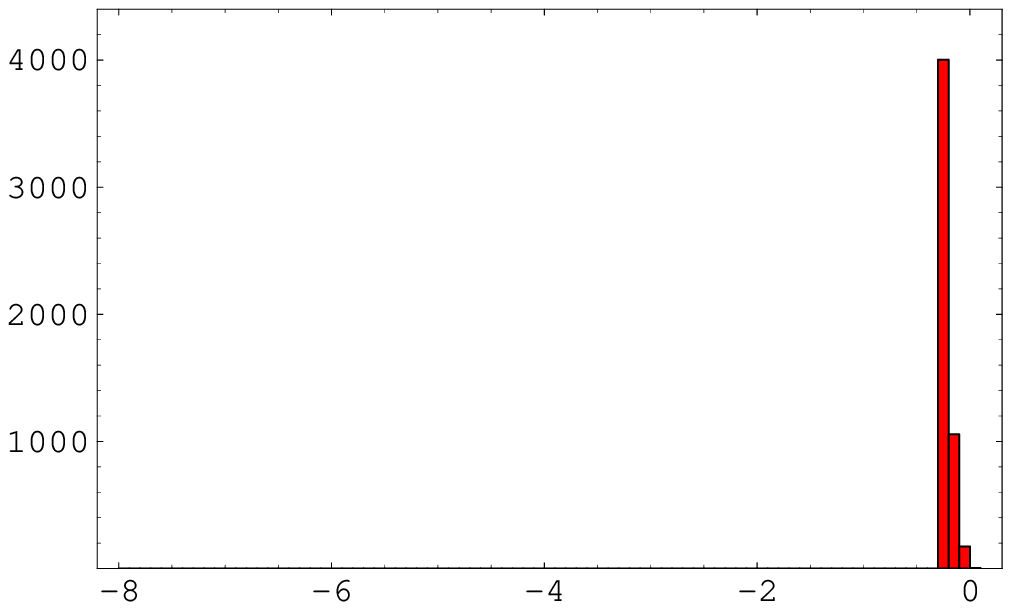} \\
$\log_{10}\lambda_u$ & $\log_{10}\lambda_c$ & $\log_{10}\lambda_t$ \\
\includegraphics[width=0.3\linewidth]{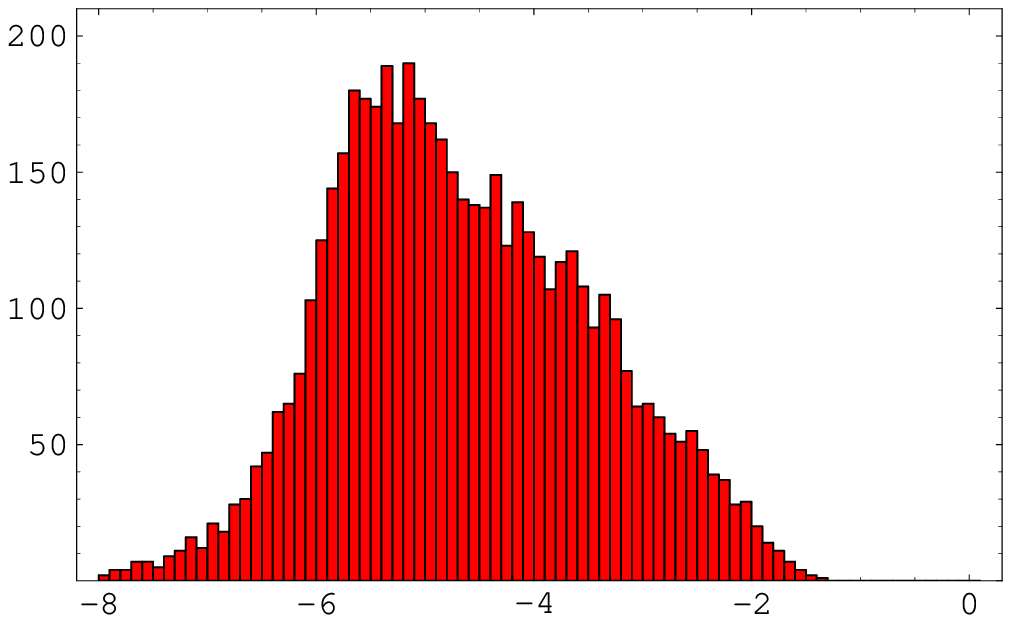} &
\includegraphics[width=0.3\linewidth]{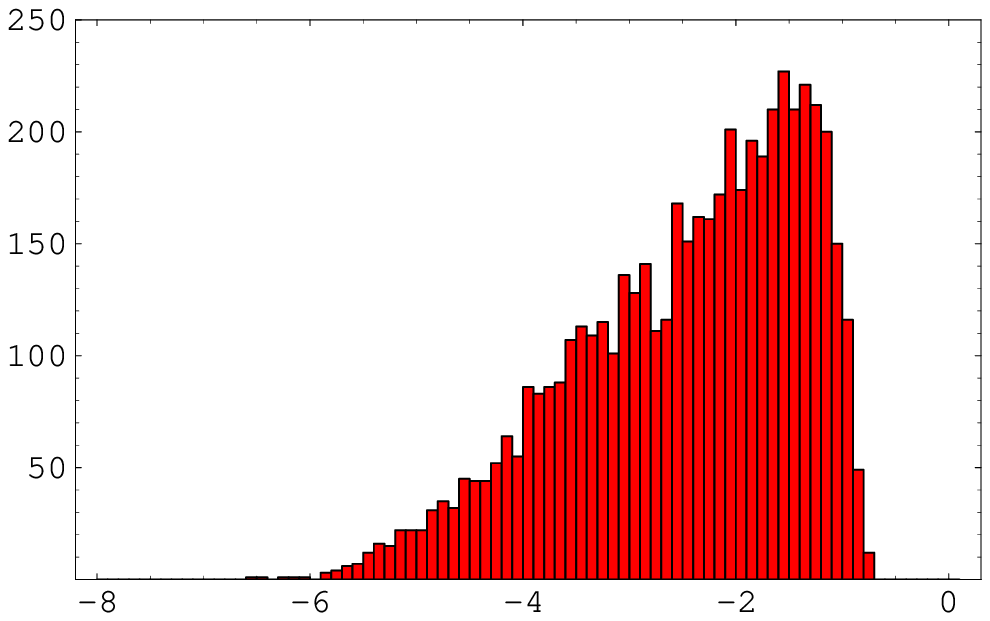} &
\includegraphics[width=0.3\linewidth]{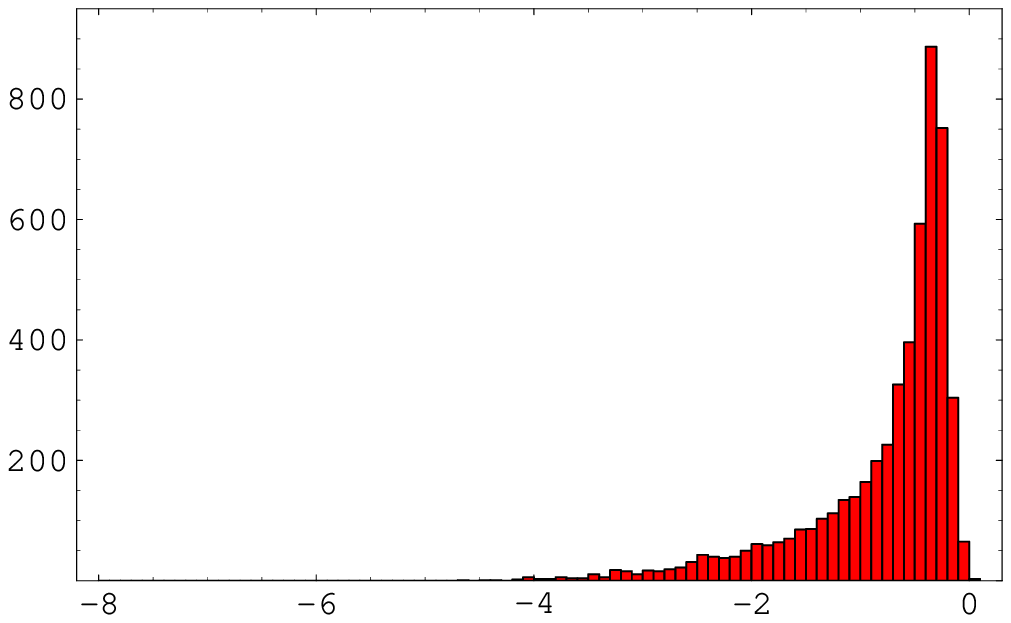} \\
$\log_{10}\lambda_d$ & $\log_{10}\lambda_s$ & $\log_{10}\lambda_b$
\end{tabular}
\caption{\label{fig:q-spec-tcut} Distributions of quark Yukawa 
eigenvalues based on the Gaussian landscape on $S^1$.  The first row 
shows the distributions from Figure~\ref{fig:D=1-nocut} before the 
$t$-cut is imposed.  The last two rows show the distributions of the 
roughly 5\% of Yukawa matrices that pass the $t$-cut.}
\end{center}
\end{figure}

We now describe qualitative effects associated with one possibility for
environmental selection, which is the selection of a large top mass to 
ensure the stability of our present Higgs phase~\cite{FHW}.  Consider a 
simple cut on the top Yukawa:
\begin{eqnarray}
\log_{10}\lambda_t \geq -\,0.3\,. \label{eq:cut-t}
\end{eqnarray}
We emphasize that we impose this $t$-cut to study qualitative effects; 
it is not intended to be precise. We first study the distributions of 
the quark Yukawa eigenvalues, which follow the distributions shown in 
Figure~\ref{fig:q-spec-tcut}. 
The three distributions in the first row correspond to Yukawa eigenvalues
before the environmental cut is imposed.  Since our landscape has not 
introduced any difference between the up- and down-type sectors, the three 
distributions are the same in both sectors.  Imposing the $t$-cut 
($\ref{eq:cut-t}$) reduces the sample to 5\% of its original size.  After 
the $t$-cut is imposed, the eigenvalue distributions are modified into 
those displayed in the second and third rows of Figure~\ref{fig:q-spec-tcut}.  
A notable effect of the $t$-cut is that the distributions of the other 
Yukawa eigenvalues are dragged upward.  This effect is more evident in the 
up-sector than in the down-sector, improving the fit of  
$\lambda_b/\lambda_t$, $\lambda_c/\lambda_b$, and $\lambda_s/\lambda_c$.

One might consider that this improvement is not enough. Certainly the
$\lambda_b/\lambda_t$ prediction is improved, but we find that the Standard 
Model value only moves from the 3rd to about the 6th percentile.
Furthermore, the typical value of $\lambda_b$ is large even without the 
$t$-cut, and the distribution of $\lambda_b$ is shifted upward after this 
cut is imposed.  There are a number of ways to modify the Gaussian 
landscape of this section to alleviate this problem.  Some of these have 
already been mentioned; others are presented in section~\ref{sec:Geometry}, 
where we study how the geometry of the extra dimensions affects the 
distributions of observables in Gaussian landscapes.

\begin{figure}[!t]
\begin{center}
\begin{tabular}{ccc}
\includegraphics[width=0.3\linewidth]{Figures/D1L10d08/T12log.eps} &
\includegraphics[width=0.3\linewidth]{Figures/D1L10d08/T23log.eps} &
\includegraphics[width=0.3\linewidth]{Figures/D1L10d08/S13log.eps} \\
$\log_{10}(2\theta_{12}/\pi)$ & $\log_{10}(2\theta_{23}/\pi)$ & 
$\log_{10}\sin\theta_{13}$ \\
\includegraphics[width=0.3\linewidth]{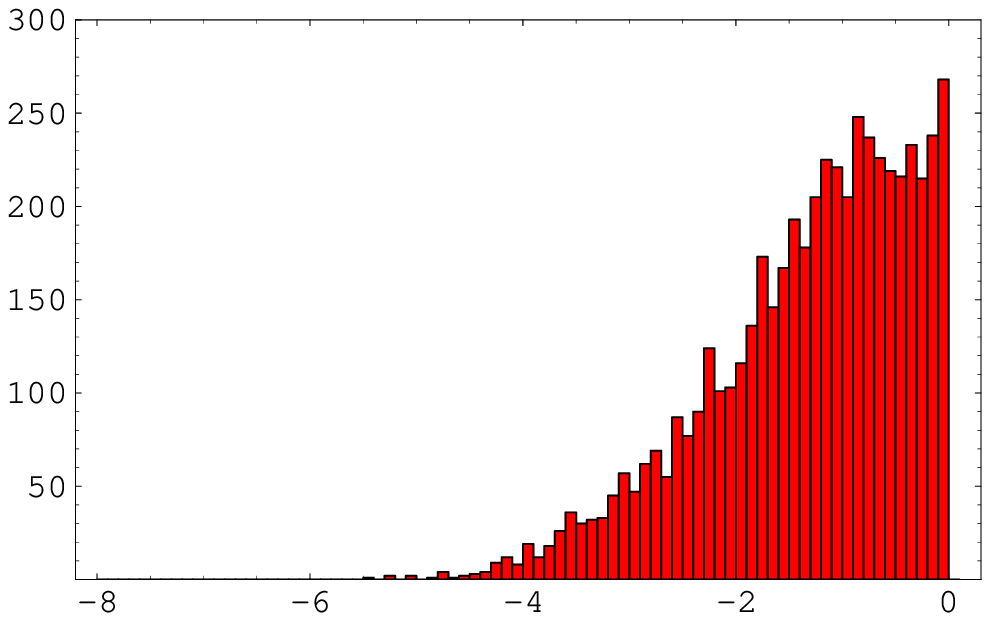} &
\includegraphics[width=0.3\linewidth]{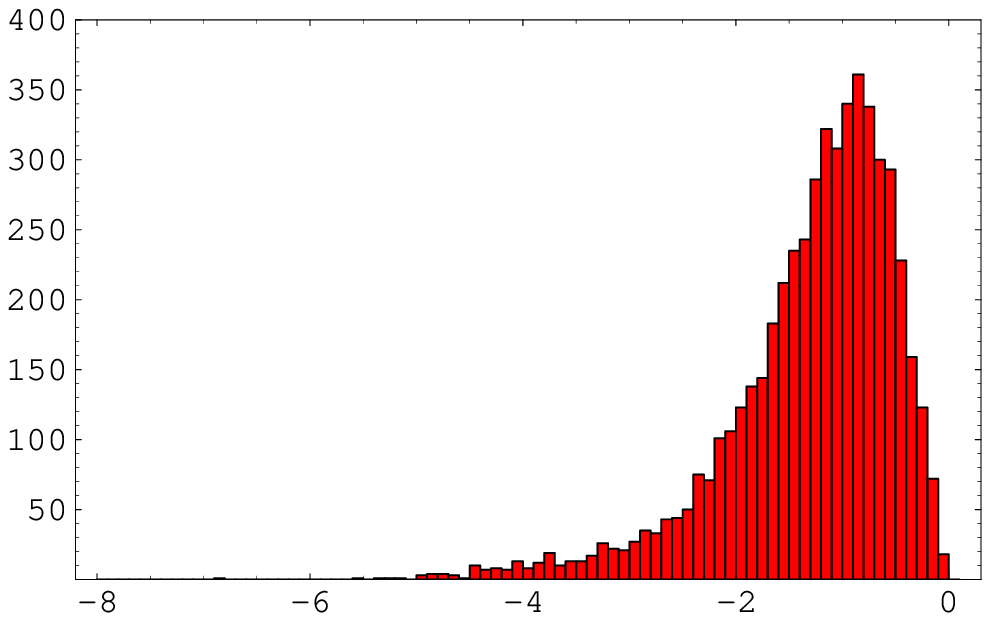} &
\includegraphics[width=0.3\linewidth]{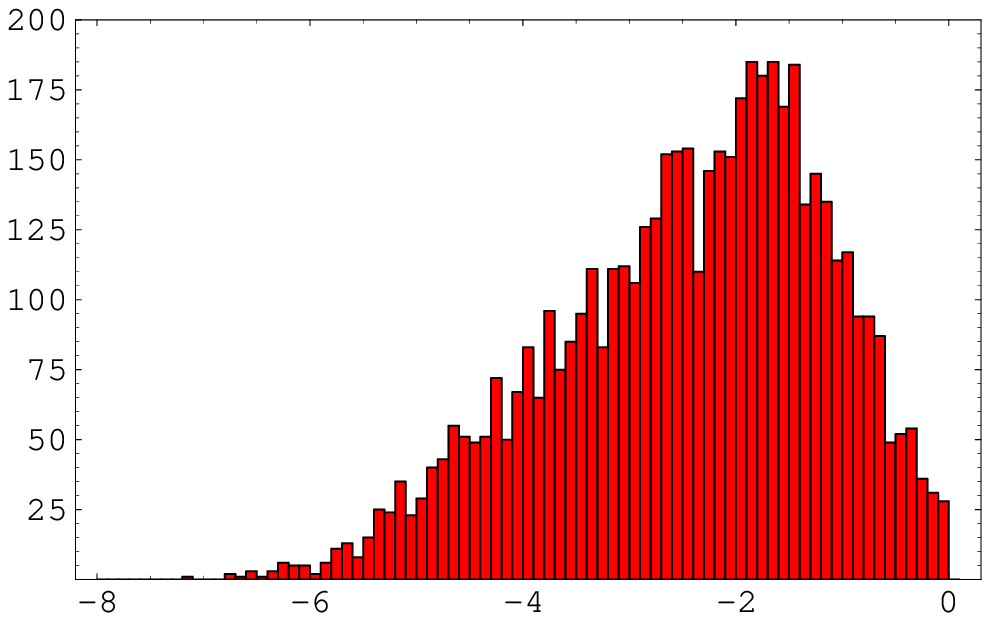} \\
$\log_{10}(2\theta_{12}/\pi)$ & $\log_{10}(2\theta_{23}/\pi)$ & 
$\log_{10}\sin\theta_{13}$ \\
\end{tabular}
\caption{\label{fig:q-mix-tcut} Distributions of CKM mixing angles, based
on the Gaussian landscape on $S^1$.  The first row shows the distributions 
before the $t$-cut is imposed, while the second row shows the distributions 
corresponding to the roughly 5\% of Yukawa matrices that pass the $t$-cut.}
\end{center}
\end{figure}

The effects of the $t$-cut are also seen in the distributions of the 
mixing angles; see Figure~\ref{fig:q-mix-tcut}.
The probability that $\theta_{23}\sim {\cal O}(1)$ is reduced 
significantly, which is understandable because $\lambda_t$ (and
hence $\lambda_b$) tends to be very close to ${\cal O}(1)$ in the sample 
passing the $t$-cut, while $\lambda_c$ and $\lambda_s$ are rarely larger 
than $10^{-1}$.  The $\theta_{23}$ distribution is reduced at small values 
as well, probably because the distributions of $\lambda_c$ and $\lambda_s$ 
are reduced at small values. Thus environmental selection on some flavor 
parameters modifies the distributions of other flavor parameters through a 
complicated chain of correlations.

\subsection{Summary}

We introduce a simple Gaussian landscape, based on a more microscopic 
description of the origin of flavor involving a single extra dimension, 
which predicts probability distributions for the CP-conserving quark sector 
flavor parameters.  Using only two free parameters, $g_{\rm eff}$ and $d/L$, 
this restrictive theory provides as good a fit to the flavor parameters as 
conventional theories of flavor based on flavor symmetries.  This is 
especially so when the Gaussian landscape is augmented by environmental 
selection in favor of large top Yukawa coupling.  This Gaussian landscape 
is characterized by the homogeneous scanning of the center coordinates of 
Gaussian wavefunctions in the extra dimension, which results in homogeneous 
scanning of Yukawa matrix elements on a logarithmic scale. The three 
eigenvalues of each Yukawa matrix then tend to be well separated from one 
another on the logarithmic scale. Thus, the hierarchical structure of quark 
mass eigenvalues is obtained.  The wavefunctions of the three quark doublets 
$q_j$ and the Higgs boson $h$ are assumed to be quite localized in the extra 
dimension, and the overlap of these wavefunctions introduces correlation 
between the up-type and down-type Yukawa matrices.  This is the essence 
behind (the microscopic explanation) for the generation structure of the 
quark sector, that is the phenomenon that the heaviest (middle and lightest) 
up-type quark is coupled to the heaviest (middle and lightest) down-type 
quark in the $W$-boson current.

\section{Goodness-of-fit Tests of Landscape Models}
\label{sec:statistics}

Before proceeding to generalize the Gaussian landscape, we here take an 
aside to discuss some possibilities for testing theories that assume a 
huge landscape of vacua.  A landscape theory that contains the observed
vacuum is not directly falsifiable in the convention sense, since we can
only compare one value of any ``variable'' (the value in our universe) to 
its predicted distribution.  On the other hand, with a statistical 
theoretical prediction we can at least calculate the probability to have 
measured a more atypical value for a landscape variable.  If this 
probability is small, the landscape can be excluded at a 
certain confidence level.  Indeed, the more numerous are the predictions 
of a landscape theory, the greater the possibility to exclude the theory.  
Consider for example quantum theory, where statistical theoretical 
predictions are well-tested using scattering experiments that count the 
event rates in a large number of bins/modes.  The parameterization of the 
Standard Model is dominated by the 22 flavor 
parameters, 20 of which have either been measured or may be measured in 
the near future.  Therefore it seems that the flavor structure of the 
Standard Model provides the best realm in which to test landscape theories.

In this section we study two statistical tests, the chi-square statistic 
and the p-value statistic, to illustrate how Gaussian landscapes may be
excluded by experiment.  As this section is intended to be only 
illustrative, we consider only the quark sector as described by the 
Gaussian landscape on $S^1$ (which in this section is referred to as the 
$S^1$ model) and we do not consider any weight factors that may arise due 
to cosmological evolution or environmental selection effects.  The 
existence of correlations among the distributions of flavor parameters 
in Gaussian landscapes allows for a rich statistical analysis, and the 
discussion below only scratches the surface.

\subsection{The Chi-Square Statistic}

Our first example is to use the so-called chi-square statistic as a 
measure of goodness-of-fit between the observed flavor parameters and 
their hypothetical landscape distributions.  The CP-conserving flavor 
parameters in the quark sector consist of the nine variables:
\begin{equation}
\{ X_I \} 
\equiv\log_{10}\{ \lambda_u,\, \lambda_c,\, \lambda_t,\, 
\lambda_d,\, \lambda_s,\, \lambda_b,\, \theta_{12},\, 
\theta_{23},\, \sin\theta_{13}\} \,, \quad {\rm for~} I=1,\cdots, 9 \,,
\label{qfex}
\end{equation}
with the logarithm acting on every element of the list. The CKM phase 
is not included in this list because the $S^1$ model we consider here 
is CP-conserving.  The chi-square statistic is
\begin{equation}
\chi^2=\sum_{I,\,J} \big( \hat{X}_I-\vev{X_I} \big)
\left(V^{-1}\right)_{IJ}
\big( \hat{X}_J-\vev{X_J} \big)\,,
\label{chisquare}
\end{equation}
where hats denote the measured value of a parameter.  The values listed 
in Table~\ref{table1} are used for $\{\hat{X}_I\}$.  Brackets denote the 
landscape average, and the co-variance matrix $V$ is given by
\begin{equation}
V_{IJ}={\rm Cov}(X_I,\,X_J)=\vev{X_I X_J}-\vev{X_I}\vev{X_J} \,.
\end{equation}
The chi-square statistic is invariant under any linear transformation 
among the variables $\{X_I\}$.
 
The correlation among the parameters in (\ref{qfex}) cannot be ignored. 
Specifically, for the three mass eigenvalues in the up sector, the 
correlation matrix, $\rho_{IJ}\equiv V_{IJ}/\sqrt{V_{II}V_{JJ}}$, is 
given by
\begin{equation}
 \rho_{IJ} = 
\left( \begin{array}{ccc}
         1. & 0.57 & 0.39 \\
	 0.57 & 1. & 0.58 \\
	 0.39 & 0.58 & 1. 
	\end{array}\right) ,\quad {\rm for~} I,J=1,2,3\,. 
\end{equation}
Note that the off-diagonal terms are not negligible compared to the
diagonal terms. The positive correlations are as expected from the combined 
probability distribution (\ref{eq:FN-combined}) of the AFS suppression 
factors $\log_{10} \epsilon^{q,\bar{u}}_i$.  Meanwhile, the correlation 
between the up and down sector mass eigenvalues, 
$\{ \log_{10}\lambda_u, \log_{10}\lambda_c, \log_{10}\lambda_t \}
\times \{\log_{10}\lambda_d, \log_{10}\lambda_s, \log_{10}\lambda_b\}$, 
is given by  
\begin{equation}
 \rho_{IJ} = \left(\begin{array}{ccc}
   0.32 & 0.27 & 0.20 \\ 0.27 & 0.39 &  0.29 \\ 0.20 & 0.29 & 0.43  
 \end{array}\right) ,\quad
{\rm for~} I=1,2,3\,\, {\rm and}\,\, J=4,5,6\,.
\end{equation}
This is also sizable and positive, confirming that the mass eigenvalues 
in the down sector are dragged upward under the $t$-cut.  The mixing 
angles are also correlated with mass hierarchy; the correlation matrix 
between 
$\log_{10} \{ \lambda_u/\lambda_c, \lambda_c/\lambda_t,
\lambda_u/\lambda_t \}$ and 
$\log_{10} \{\theta_{12}, \theta_{23}, \sin \theta_{13}\}$ is given by
\begin{equation}
\rho_{IJ} = \left( \begin{array}{ccc}
   0.34 & -0.14 & 0.28 \\ -0.21 & 0.42 & 0.14 \\ 0.16 & 0.21 & 0.38
      \end{array}\right), 
\end{equation}
where now $I=X_1-X_2,\,X_2-X_3,\,X_1-X_3$ and $J=7,8,9$.  We see that 
larger mass hierarchies are correlated with smaller mixing angles.

The covariance matrix $V_{IJ}$ determines the principal axes 
$Y_i=c_{iJ}X_J$, with the covariance matrix $V_{ij}$ being diagonal when 
the basis $\{X_I\}$ is changed to $\{Y_i\}$.  Since $V_{ij}$ is diagonal 
the observables $\{Y_i\}$ are independent at least up to second order.  
If the distributions of the $Y_i$'s were {\em completely} independent, 
and Gaussian, then the quantity $\chi^2$ would follow the chi-square 
distribution for random measurements of $\{Y_i=c_{iJ}X_J \}$.  This 
distribution has probability density
\begin{equation}
\frac{dP}{d\chi^2}= 2^{N/2}\Gamma(N/2)\,\chi^{N-2}e^{-\chi^2/2} \,,
\end{equation} 
where $N$ is the number of independent Gaussian random variables entering 
(\ref{chisquare}).  This distribution has a mean of $N$ and a standard 
deviation of $\sqrt{2N}$.  Thus the prediction 
$\chi^2\approx N\big(1\pm\sqrt{2/N}\big)$ is very sharp when a large 
number of observables are predicted by the landscape.  Note that too large 
a value of $\chi^2$ indicates that the observed flavor parameters are 
atypical of what is expected from the landscape distributions, while too 
small a value of $\chi^2$ indicates that the correct underlying theory 
has less randomness than is exhibited by the landscape in question.
Therefore, even though we can measure only one value for each observable, 
and hence measure only one value of $\chi^2$, the chi-square 
statistic can still be a powerful tool in testing landscape theories.    

Our $S^1$ model predicts distributions for nine flavor parameters in the 
quark sector, but it contains $n_F=2$ free parameters ($g_{\rm eff}$ and 
$d/L$) that can be fixed by hand and tuned\footnote{In fact the free 
parameters $g_{\rm eff}$ and $d/L$ have not been fully exploited, since 
we have not performed a maximum likelihood analysis to determine the 
values that minimize $\chi^2$ in (\ref{chisquare}).  However, the 
parameter values we use have been chosen to qualitatively fit our 
expectations for the measured values of $\{X_I\}$.} 
to fit two out of nine of the $\hat{Y}_i$'s, rendering two terms in 
$\sum_i (\hat{Y}_i-\vev{Y_i})^2V^{-1}_{ii}$ to vanish.  Thus we have 
$N = 9 - n_F = 7$ independent degrees of freedom.  Therefore if the 
distributions describing the $Y_i$ were Gaussian, we would expect 
$\chi^2\approx 7\pm 4$.  In fact the $S^1$ model does not predict 
Gaussian distributions for any of the flavor parameters. 
\begin{figure}[t]
 \begin{center}
\begin{tabular}{cc}
\includegraphics[width=0.3\linewidth]{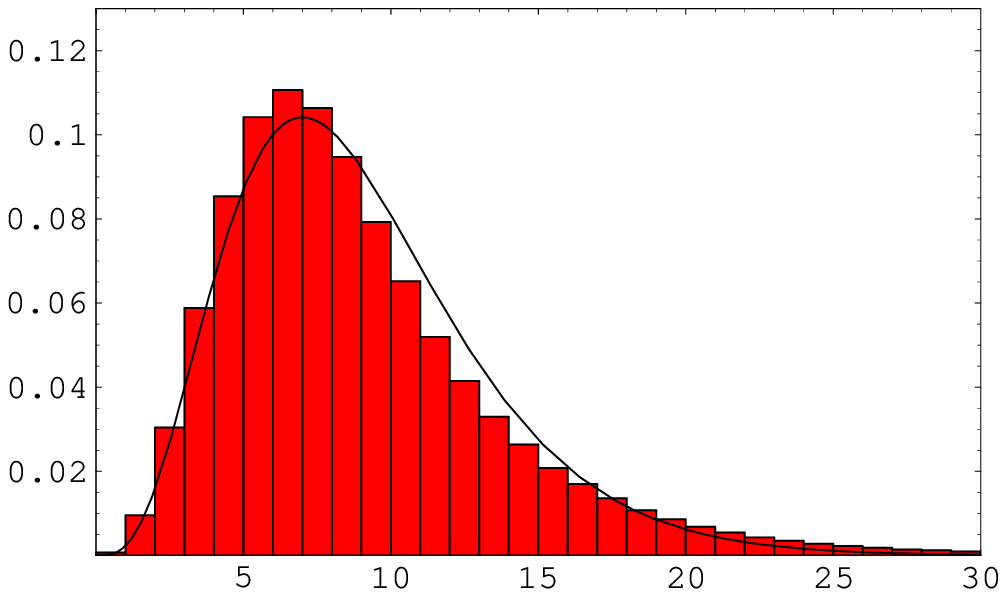} &
\includegraphics[width=0.3\linewidth]{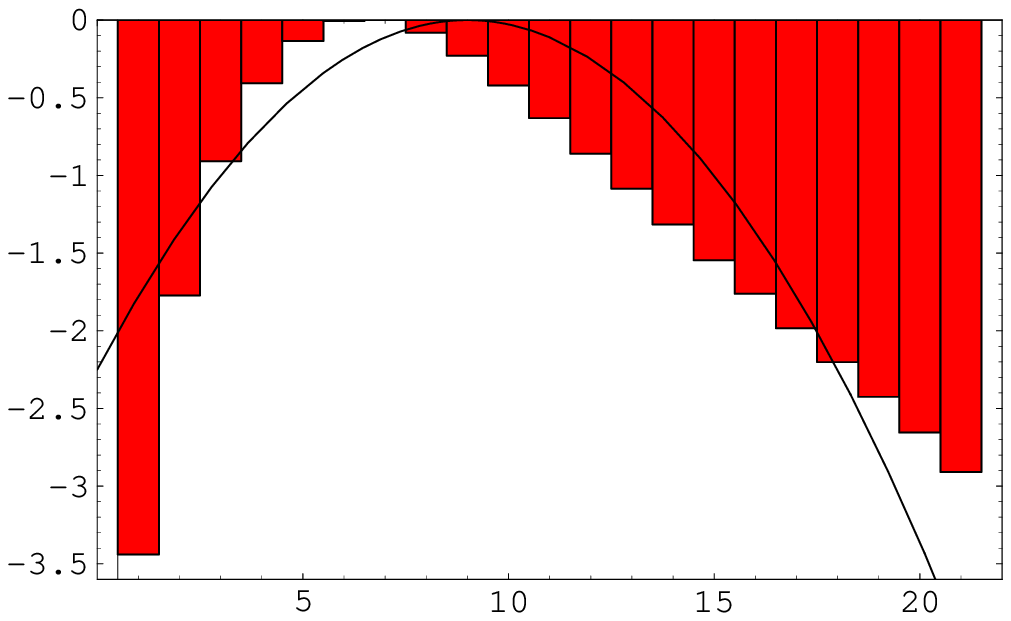} \\
$dP/d\chi^2$ & $\ln (dP/d\chi^2)$ 
\end{tabular}
\caption{\label{fig:chisquare} The left panel displays the probability 
distribution of $\chi^2$ obtained numerically (histogram) using data 
generated by the $S^1$ model, along with the actual chi-square distribution 
for $N=9$.  The right panel displays the logarithm of the numerical 
distribution of $\chi^2$ (histogram) and a normal distribution with a mean 
$N=9$ and standard deviation $\sqrt{2N}=\sqrt{18}$.}
\end{center}
\end{figure}
However, Figure~\ref{fig:chisquare} shows that the distribution of $\chi^2$ 
(\ref{chisquare}) using data generated by the $S^1$ model looks quite 
similar to the corresponding chi-square distribution, which has $N=9$.  
Furthermore, the central limit theorem guarantees that as $N$ is increased 
the distribution of $\chi^2$ approaches a normal distribution with a mean 
$N$ and a variance that grows in proportion to $N$.  $N$ can be as large 
as 20 for the observable flavor parameters of the Standard Model, assuming 
a landscape with at least 20 scanning parameters.

We calculate the chi-square statistic for the nine CP-conserving 
quark-sector flavor parameters using the $S^1$ model to calculate 
$\vev{X_I}$ and $V_{IJ}$, and using Table~\ref{table1} for $\hat{X}_I$.  
This gives $\chi^2 \approx 6$, which agrees very well with the prediction 
$\chi^2 \approx 7 \pm 4$.  That the obtained value is neither too large 
nor too small indicates that the observed flavor parameters are typical 
and that they exhibit the expected level of statistical fluctuation in 
this landscape.  The same analysis can be carried out for the $S^1$ model 
after a hypothetical environmental weight factor favoring a large top 
mass has been imposed.  Using only $S^1$ model data that pass the $t$-cut 
(\ref{eq:cut-t}), we find $\chi^2 \approx 8$, which also agrees well with 
the prediction.  Imposing the $t$-cut increases the value of $\chi^2$, 
but one should keep in mind that the parameters $g_{\rm eff}$ and $d/L$ 
were chosen to fit the observed flavor parameters before imposing the 
$t$-cut, without respect for the ensemble that passes the $t$-cut.

\subsection{The P-Value Statistic}

The chi-square statistic has a disadvantage as well.  Although the central 
limit theorem guarantees that for large $N$ that the distribution of  
$\chi^2$ approaches a normal distribution, regardless of most of the 
details of the actual distributions of the $\{ X_I\}$, $N$ is not more 
than 20 for the application to flavor parameters.  Moreover, some of 
distributions of the flavor observables have sharp cut-offs, such that the 
probability to obtain a large value of $(\hat{X}_I-\vev{X_I})^2$ may be 
much smaller than what is expected from a normal distribution with a mean 
$\vev{X_I}$ and a variance $V_{II}$.  The chi-square statistic does not 
completely account for such a situation.

On the other hand, the p-value statistic is capable of handling 
non-Gaussian distribution functions.  The p-value of a single flavor 
parameter $X_I$ is simply the fraction of the distribution that is more 
atypical than the measured value $\hat{X}_I$.  Specifically, we may 
consider the p-value to be the fraction of the distribution that is 
farther from the median $\overline{X}_I$ than is $\hat{X}_I$.  If we 
approximate the probability density as continuous with profile $f(X_I)$, 
then the p-value $p_I$ is
\begin{equation}
p_I(\hat{X}_I) = \int_{-\infty}^{\hat{X}_I} 2f(X_I)\,dX_I \quad {\rm for~}
\hat{X}_I < \overline{X}_I \,,\quad
p_I(\hat{X}_I) = \int_{\hat{X}_I}^\infty 2f(X_I)\,dX_I \quad {\rm for~}
\hat{X}_I > \overline{X}_I \,.\,\,
\label{p1}
\end{equation}
Note that like $\chi^2$, $p_I$ is a random variable for every independent
measurement $\hat{X}_I$.  However, whereas $\chi^2$ is distributed according 
to the chi-square distribution, $p_I$ is distributed uniformly between zero 
and one.  The p-value is also the probability that the Kolmogorov--Smirnov 
statistic (the value of the $D$ function) of one sampling of the variable 
$X_I$ could have been larger than what is calculated from $\hat{X}_I$. 
Hence a measurement giving p-value $p_I$ excludes the corresponding theory 
at a confidence level of max$\{p_I,\,1-p_I\}\times 100$\%.\footnote{P-values
very near zero indicate that the measured values $\{\hat{X}_I\}$ are 
collectively atypical of their predicted distributions; p-values very 
near unity indicate that the measured values do not exhibit the randomness
expected from the predicted distributions.}

Reference~\cite{dGM} used the p-value statistic to test the hypothesis 
of Neutrino Anarchy. There, the probability distribution of the lepton 
sector flavor parameters is factorized into those of mass eigenvalues, 
of each individual mixing angle, and of the CP phase~\cite{dGM}. Then 
the p-value statistic can be applied separately to each of the three mixing 
angles.  In the case where the statistical variables $X_I$ are independent 
of each other, $\hat{k}\equiv \prod_I p_I(\hat{X}_I)$ represents the 
fraction of the distribution of $\{X_I\}$ for which each element $X_I$ is 
more atypical than the measured value $\hat{X}_I$.  The probability 
that $k\equiv \prod_I p_I(X_I)$ could have been smaller than 
$\hat{k}$ is given by (e.g.~\cite{dGM}) 
\begin{equation}
p=\int^{k\leq\hat{k}}\prod_I d p_I
= \hat{k}\sum_{a=0}^{N-1}\frac{1}{a!}\left( -\ln\hat{k}\right)^a\, .
\label{pvalue}
\end{equation}
Here $a$ is simply a summation index used to simplify the last expression, 
and $N$ is the number of degrees of freedom in $\{X_I\}$.  For a random
measurement of a set of independent variables $\{X_I\}$, the p-value 
calculated through (\ref{pvalue}) is distributed uniformly from zero to
one.  Thus a random measurement giving p-value $p$ can exclude a landscape
theory with a confidence level max$\{p,\,1-p\}\times 100$\%.  If we treat
the nine CP-conserving flavor observables in the quark sector as if they
were independent, then the data $\hat{X}_I$ and the hypothetical 
distributions $dP(X_I)/dX_I$ of the $S^1$ model can be used to calculate
the p-value through (\ref{pvalue}).  Here, $N=7$ is used in (\ref{pvalue})
because the values of $g_{\rm eff}$ and $d/L$ can be tuned to bring two
of the $X_I$ to be the same as their medians.  We find $p\simeq0.66$, 
suggesting a good fit with observation.   

In fact, however, the statistical variables $\{X_I\}$ are not independent. 
To the best of our knowledge, there is no standard definition of the 
p-value statistic for correlated multi-variable distributions.  One 
possibility to account for correlation is to calculate $\hat{k}$ by 
actually counting the fraction of randomly generated parameter sets 
$\{X_I\}$ that have each $X_I$ more atypical than that of the measured 
set $\{\hat{X}_I\}$.  This is as opposed to using 
$\hat{k}=\prod_I p_I(\hat{X}_I)$, which gives this fraction in the 
absence of correlations.  In a sample of $5\times 10^6$ data sets 
generated using the numerical simulation of the $S^1$ model, we find 52 
such sets, giving $\hat{k}=2^9\times\left(\frac{52}{5\times10^6}\right)$ 
and hence $p\simeq 0.73\pm0.03$.  The uncertainty is obtained by assuming 
the variance in counting $N_{\rm atyp}$ atypical data points is of the 
order of $N_{\rm atyp}$.  Yet another attempt at a ``p-value like'' 
measure of the goodness-of-fit may be to judge atypicality by using in 
(\ref{p1}) the distributions of the principal axes $Y_i$, their measured 
values $\hat{Y}_i$, and their medians 
$\overline{Y}_i=c_{iJ}\overline{X}_J$---as opposed to using $X_I$, 
$\hat{X}_I$, and $\overline{X}_I$---since the statistical variables $Y_i$ 
do not have correlations, at least up to second order.  The fraction of 
$\{ Y_i\}$ more atypical than the measured set $\{\hat{Y}_i\}$ can be 
calculated using $\hat{k}=\prod_i p_i(\hat{Y}_i)$; using (\ref{pvalue})
we find this gives $p\simeq 0.42$.  If there were truly no correlations
among the $Y_i$ this would give the same result as calculating $\hat{k}$
by actually counting the fraction of randomly generated parameter sets 
$\{ Y_i \}$ that have each $Y_i$ more atypical than that of the measured 
set $\{\hat{Y}_i\}$.  However, in our ensemble of $5 \times 10^6$ data 
sets we find just one for which each element is more atypical than its 
corresponding measured value, giving $p\simeq 0.19^{+0.07}_{-0.19}$.

\section{Geometry Dependence}
\label{sec:Geometry}

In section~\ref{sec:toy1} we introduced and analyzed a Gaussian landscape 
based on a single extra dimension, and found that it could provide the
hierarchy, pairing, and generation structures of the observed quark 
sector.  However, the number and geometry of extra dimensions need not 
correspond to the $S^1$ Gaussian landscape.  Therefore, we initiate a 
study into how the compactification geometry of extra dimensions affects 
the probability distributions of observables.  We find that the 
qualitative results of section~\ref{sec:toy1} can be achieved by Gaussian 
landscapes in other geometries of extra dimensions and that the 
distribution functions of flavor observables are largely insensitive to 
the details of these geometries.  However, the behavior of the distribution 
functions of mass eigenvalues at large values is affected by the number of 
extra dimensions.  In section~\ref{sec:SYM}, we argue that the geometries 
used in Gaussian landscapes are related to base manifolds of torus-fibered 
geometries used in the compactification of string theory.

\subsection{Gaussian Landscapes on $T^2$ and $S^2$}
\label{ssec:toyT2}

To investigate the robustness of the Gaussian landscape on $S^1$, we
study other geometries.  We first look at the simplest extension to more 
than one dimension, $T^2 = S^1 \times S^1$, and focus on a ``square torus'' 
where the two periods of the torus are both $L$, and the directions of 
the two $S^1$ are orthogonal. We assume that each of the quarks and the 
Higgs have rotation-symmetric Gaussian wavefunctions of the form
\begin{equation}
\varphi(\vec{y};\vec{y}_0) \propto e^{-\frac{|\vec{y}-\vec{y}_0|^2}{2d^2}},
\end{equation}
where the center coordinates $\vec{y}_0$ of each particle are randomly 
scanned over the internal space $T^2$.  The up-type and down-type Yukawa 
matrices are calculated by the overlap integration (\ref{eq:overlap}), 
which is naturally generalized to integration on $T^2$.  We defer to 
section~\ref{sec:SYM} a discussion of the extent to which these 
assumptions result from a dynamical field theory on extra dimensional 
spacetime, and for the moment focus on the phenomenology of this landscape.

\begin{figure}[t]
\begin{center}
\begin{tabular}{ccc}
\includegraphics[width=0.3\linewidth]{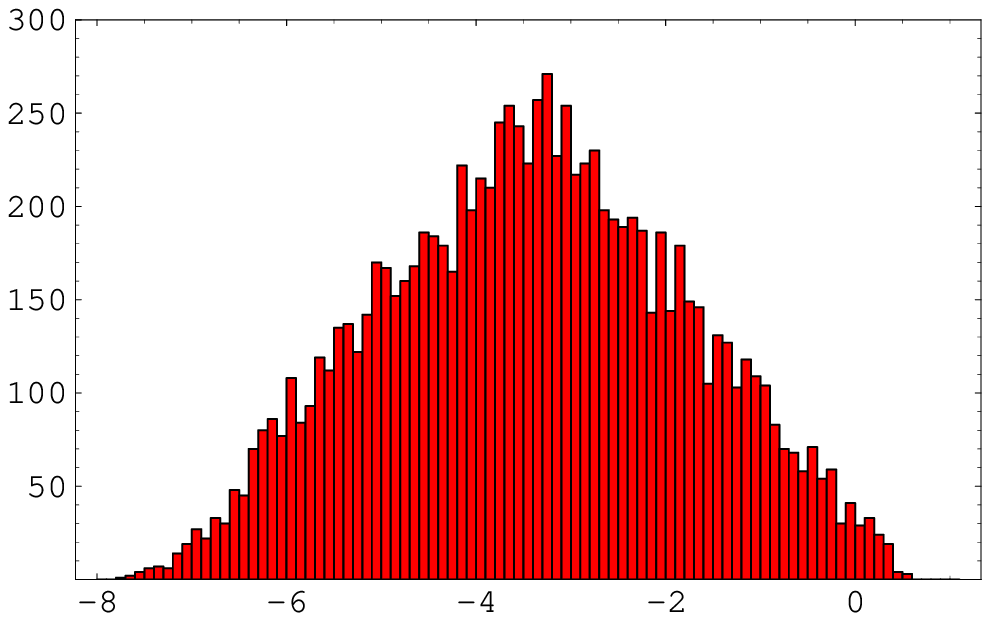} &
\includegraphics[width=0.3\linewidth]{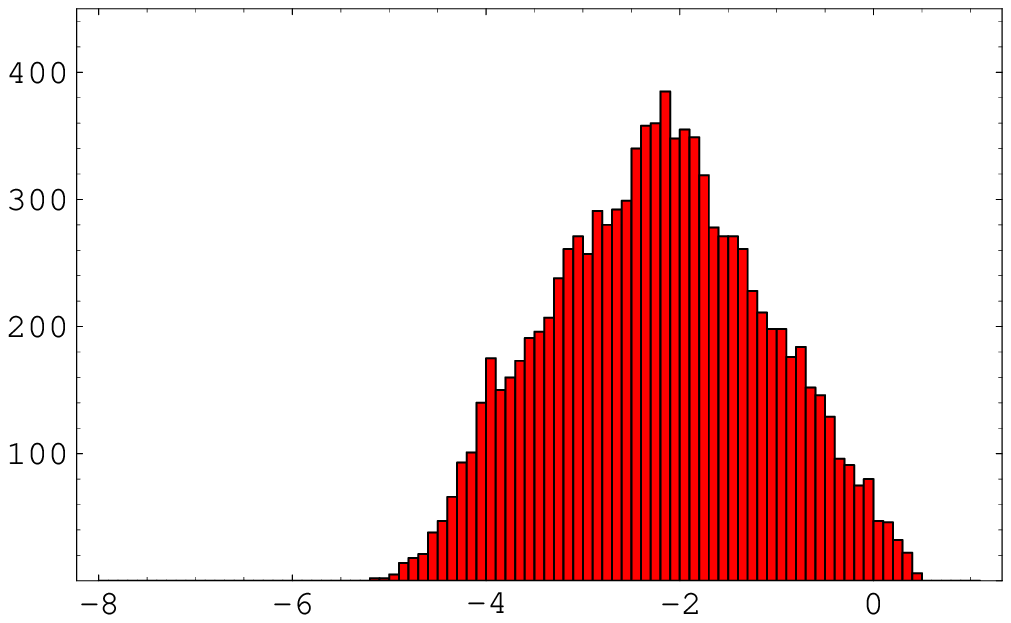} &
\includegraphics[width=0.3\linewidth]{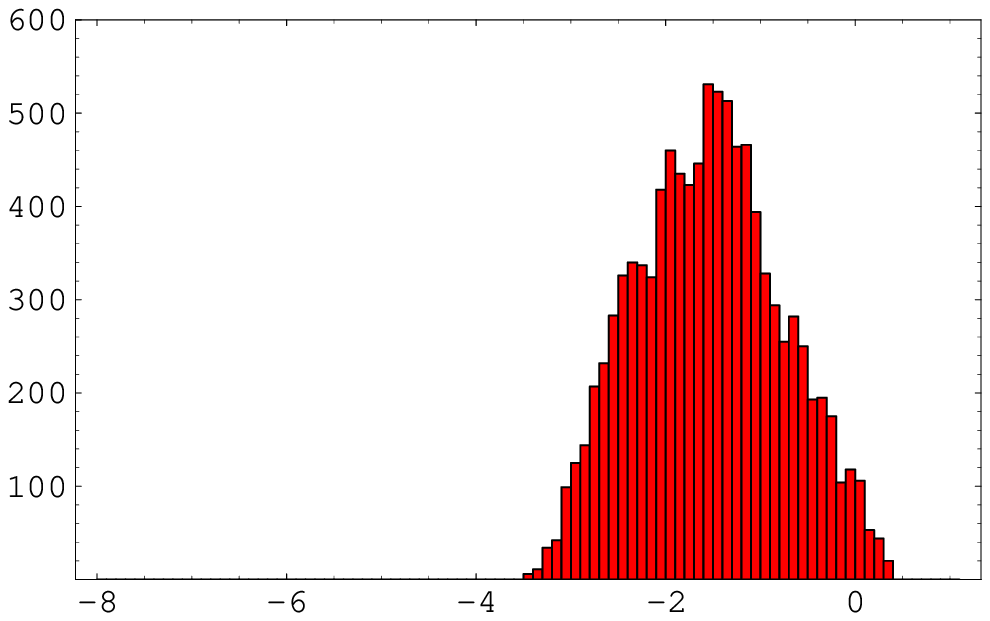}
\end{tabular}
\caption{\label{fig:entry-distr-T2} Distributions of $10^4$ Yukawa matrix 
elements, generated from the Gaussian landscape on $T^2$.  From left to 
right the panels correspond to $(d/L,g_{\rm eff})=(0.1,1)$, $(0.12,1)$ and 
$(0.14,1)$.}
\end{center}
\end{figure}

\begin{figure}[t]
\begin{center}
\begin{tabular}{ccc}
\includegraphics[width=0.3\linewidth]{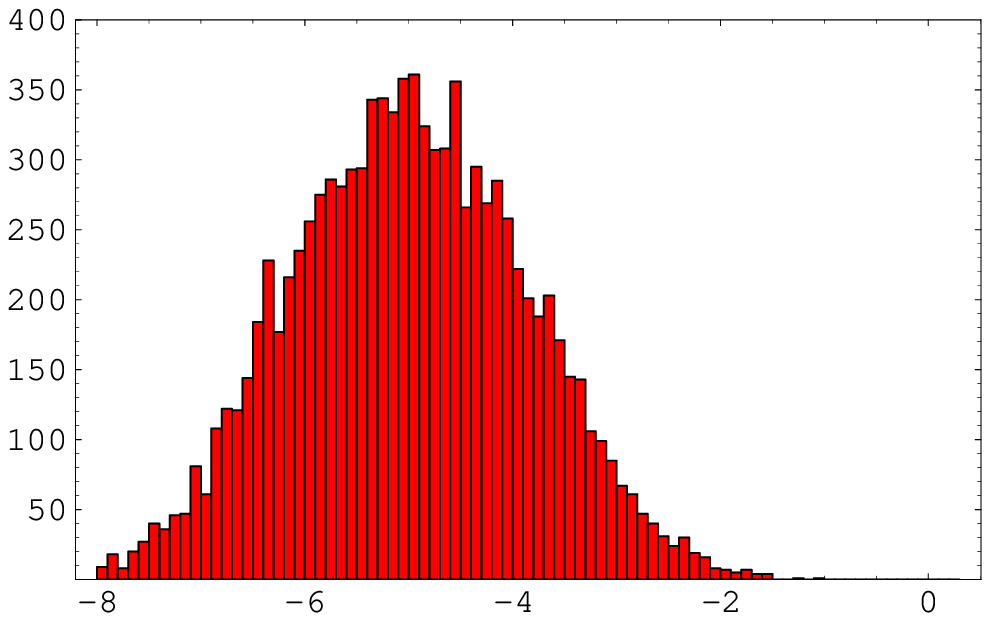} &
\includegraphics[width=0.3\linewidth]{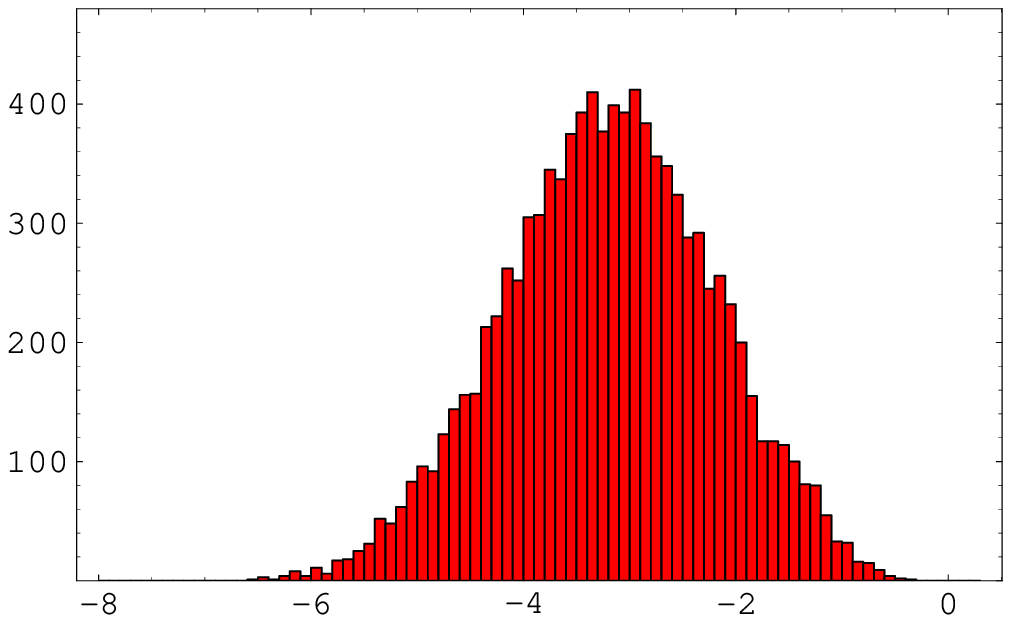} &
\includegraphics[width=0.3\linewidth]{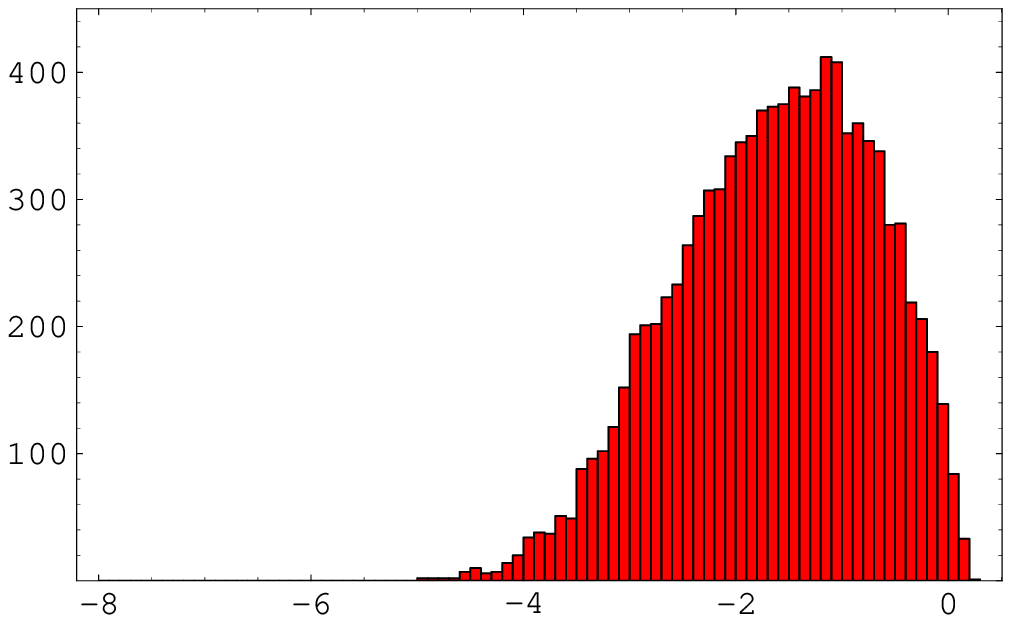} \\
$\log_{10}\lambda_{u,d}$ & $\log_{10}\lambda_{c,s}$ & 
$\log_{10}\lambda_{t,b}$ \\
\includegraphics[width=0.3\linewidth]{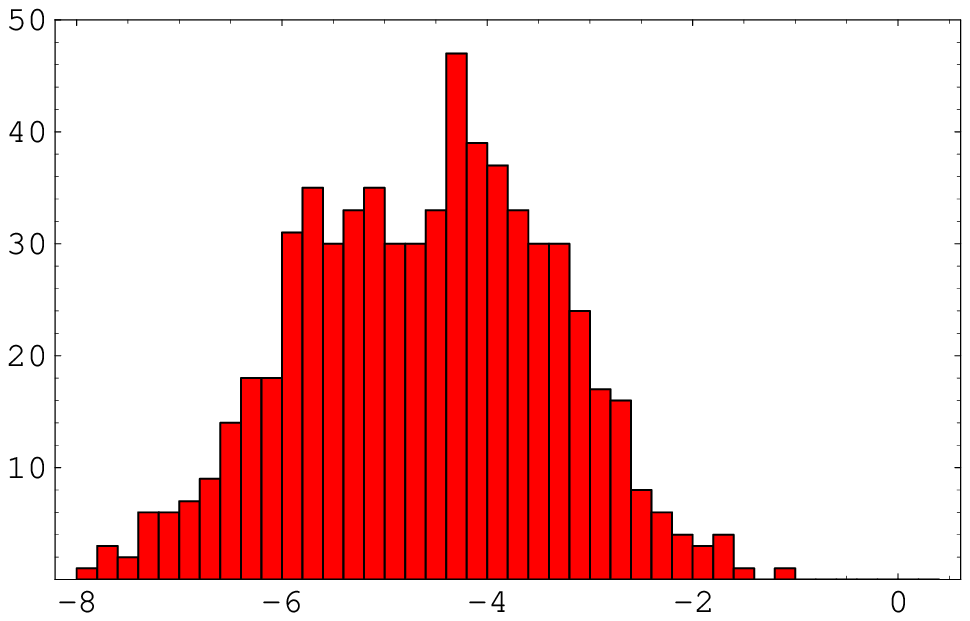} &
\includegraphics[width=0.3\linewidth]{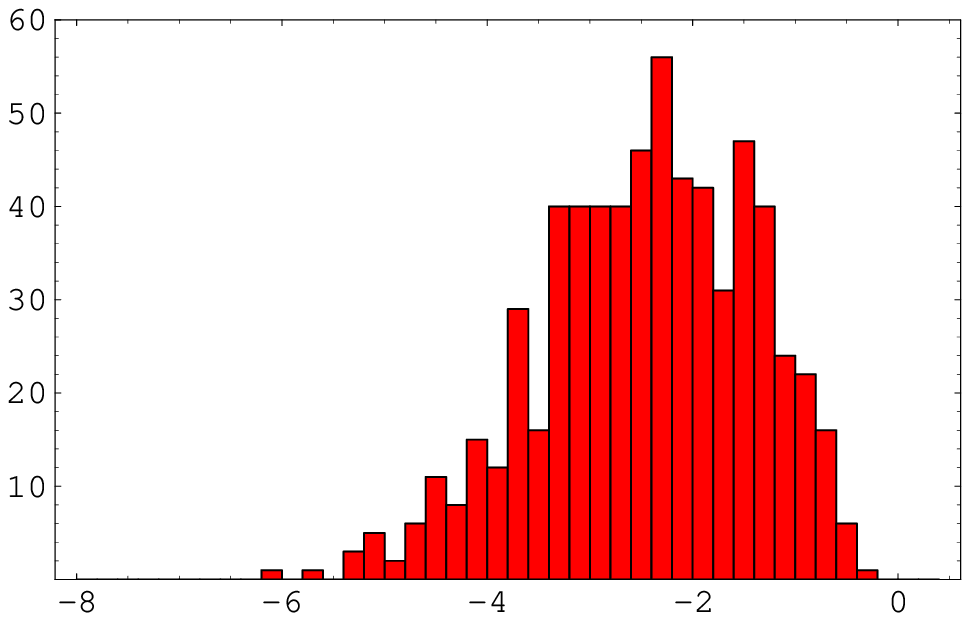} &
\includegraphics[width=0.3\linewidth]{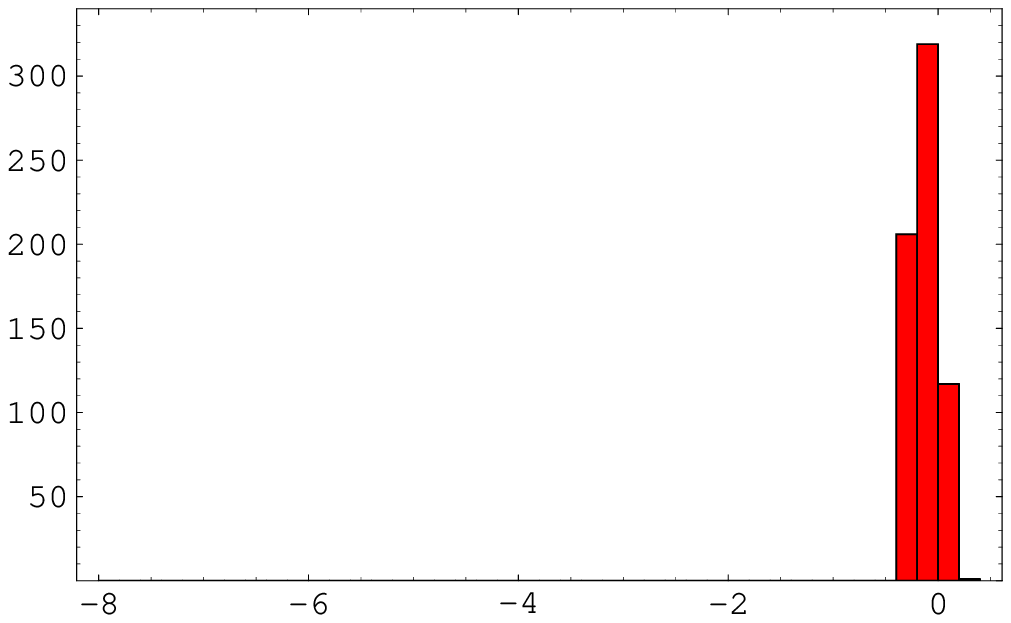} \\
$ \log_{10}\lambda_u$ & $\log_{10}\lambda_c$ & $\log_{10}\lambda_t$ \\
\includegraphics[width=0.3\linewidth]{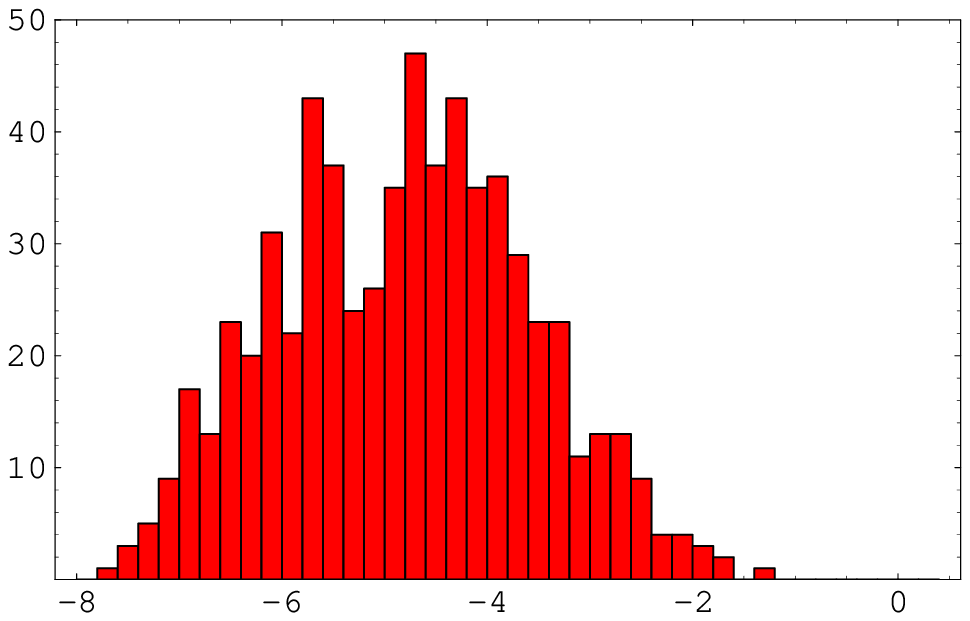} &
\includegraphics[width=0.3\linewidth]{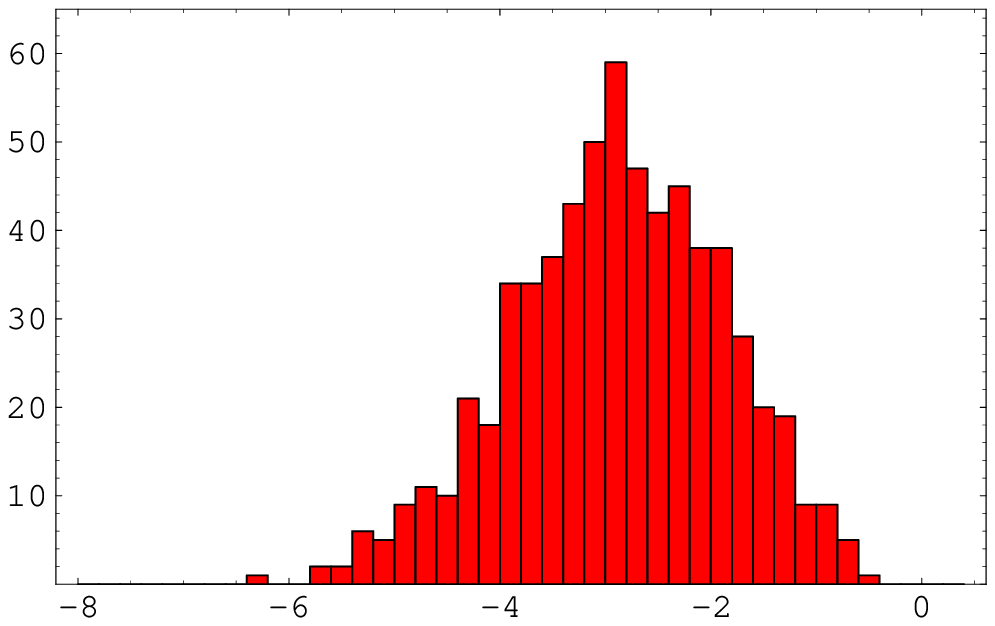} &
\includegraphics[width=0.3\linewidth]{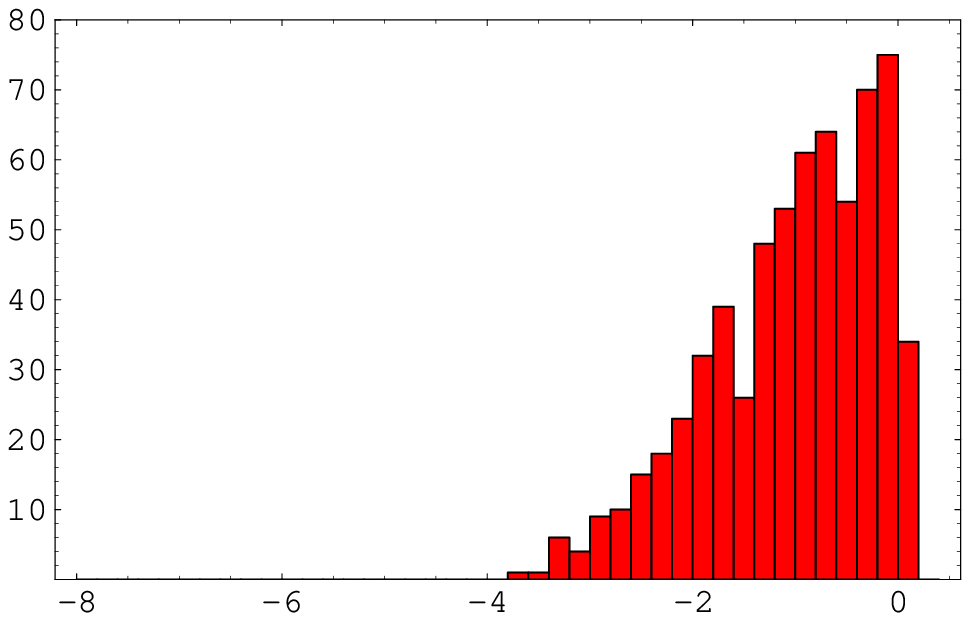} \\
$ \log_{10}\lambda_d$ & $\log_{10}\lambda_s$ & $\log_{10}\lambda_b$
\end{tabular}
\caption{\label{fig:T2YukawaEgval} Distributions of Yukawa eigenvalues 
in the Gaussian landscape on $T^2$, based on a numerical simulation 
with $(d/L,g_{\rm eff})=(0.1,0.4)$.  The first row shows the distribution 
of the three eigenvalues of the up (and down) sector.  The second and third
rows display the eigenvalues of the roughly 6\% of matrices that survive 
the $t$-cut of section~\ref{ssec:environment}.}
\end{center}
\end{figure}

\begin{figure}[t]
\begin{center}
\begin{tabular}{ccc}
\includegraphics[width=0.3\linewidth]{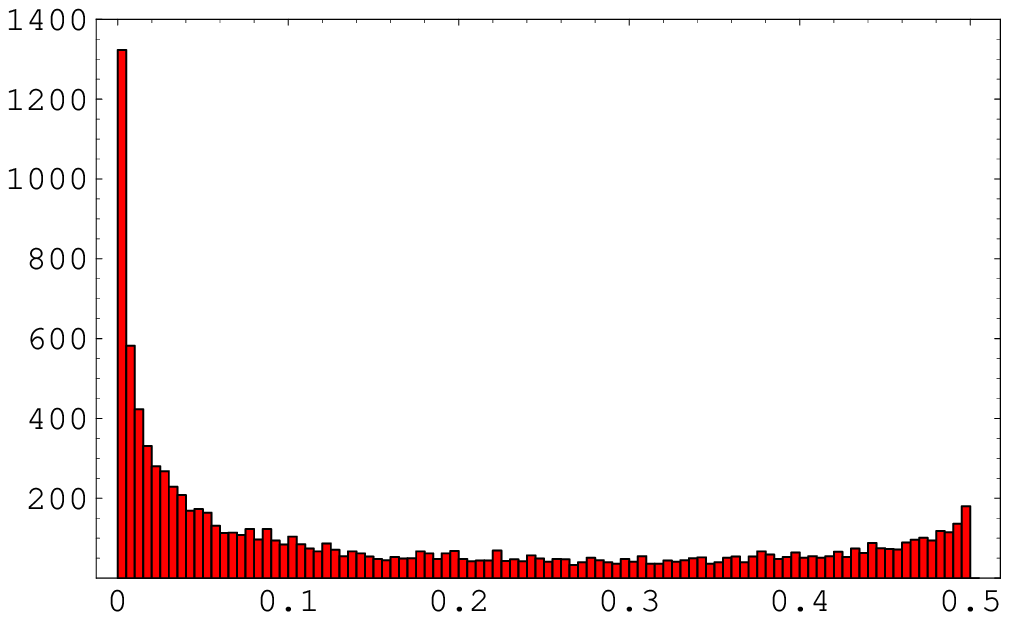} &
\includegraphics[width=0.3\linewidth]{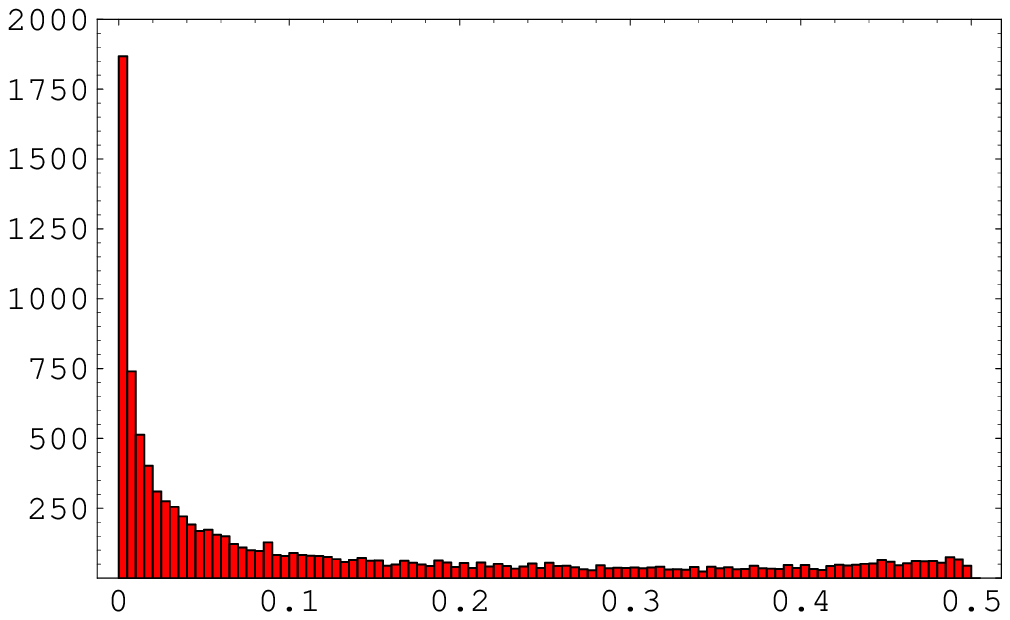} &
\includegraphics[width=0.3\linewidth]{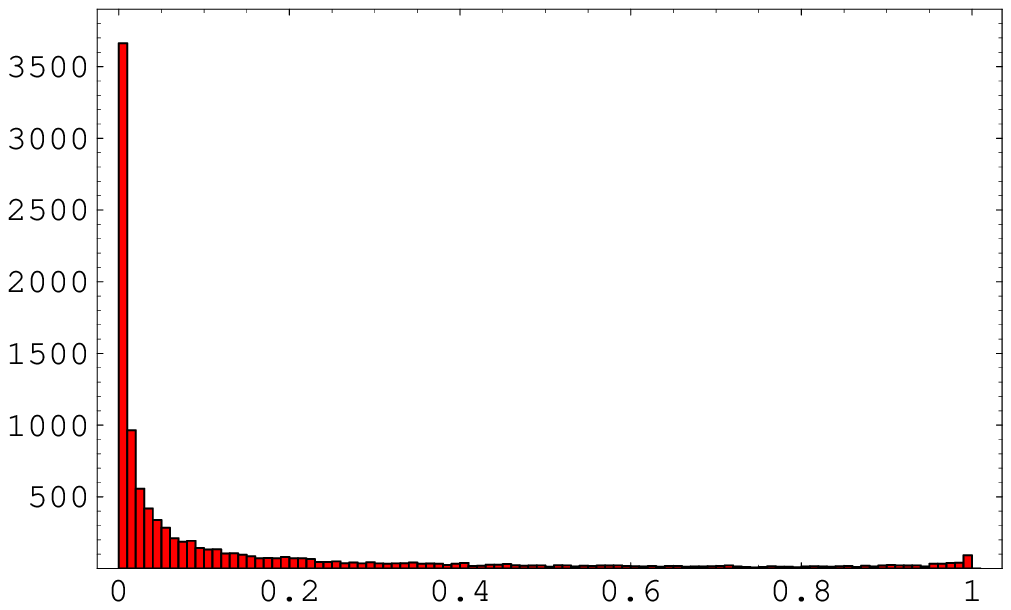} \\
$\theta_{12}/\pi$ & $\theta_{23}/\pi$ & $\sin\theta_{13}$ \\
\includegraphics[width=0.3\linewidth]{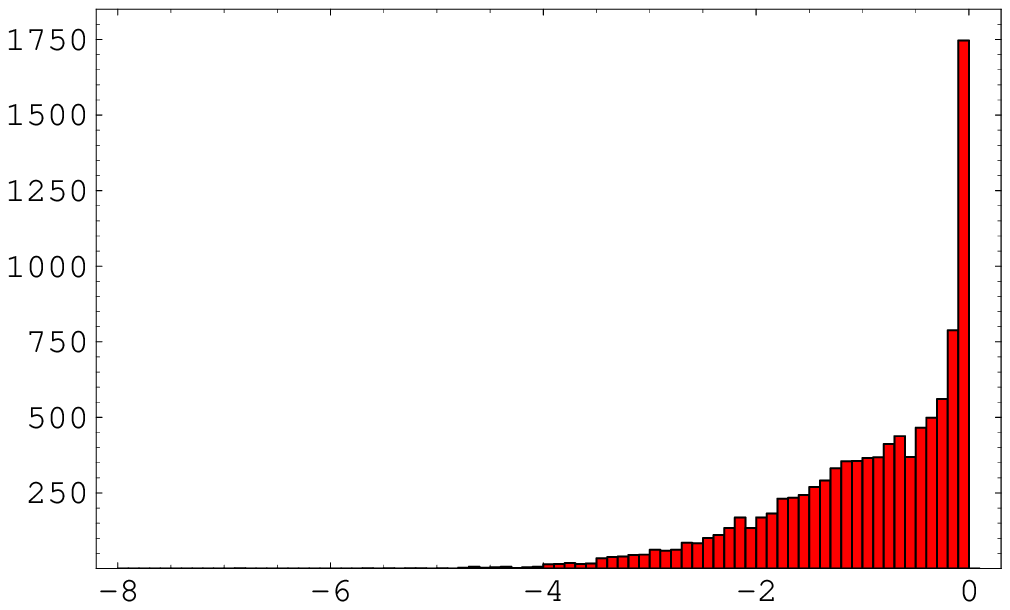} &
\includegraphics[width=0.3\linewidth]{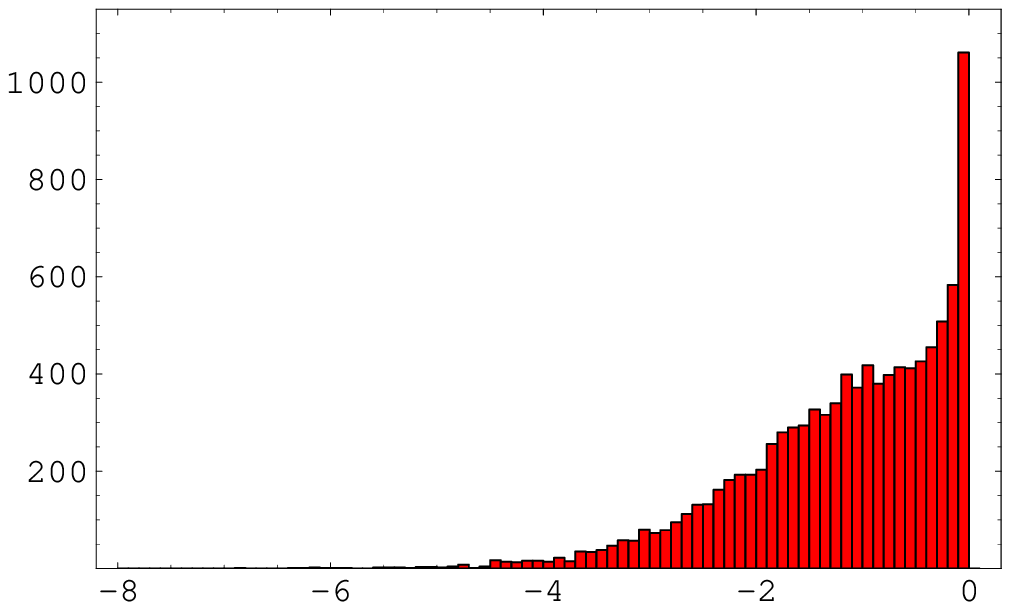} &
\includegraphics[width=0.3\linewidth]{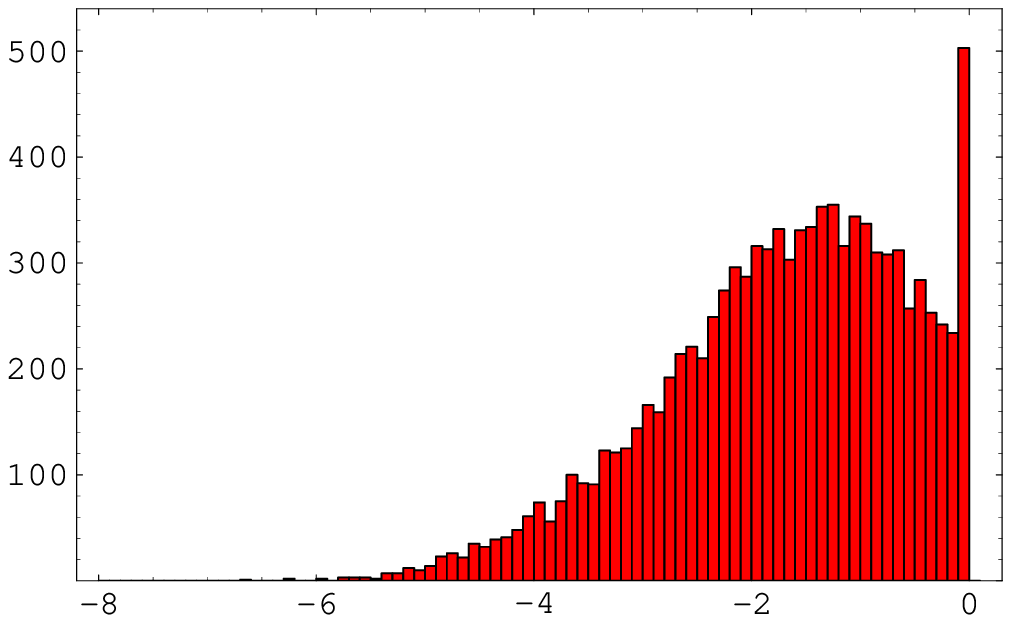} \\
$\log_{10}(2\theta_{12}/\pi)$ & $\log_{10}(2\theta_{23}/\pi)$ & 
$\log_{10}\sin\theta_{13}$ \\
\includegraphics[width=0.3\linewidth]{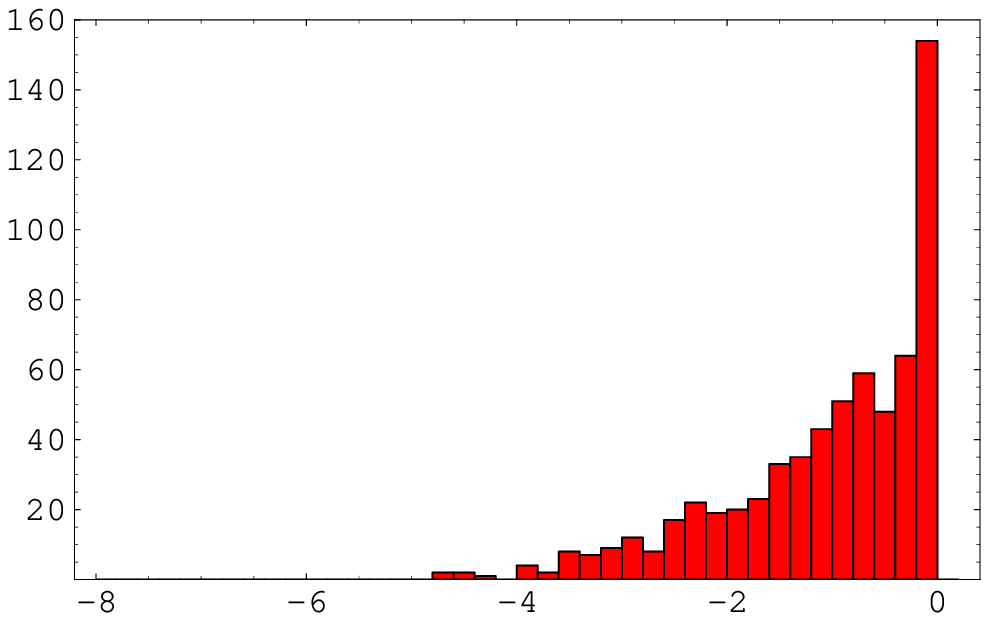} &
\includegraphics[width=0.3\linewidth]{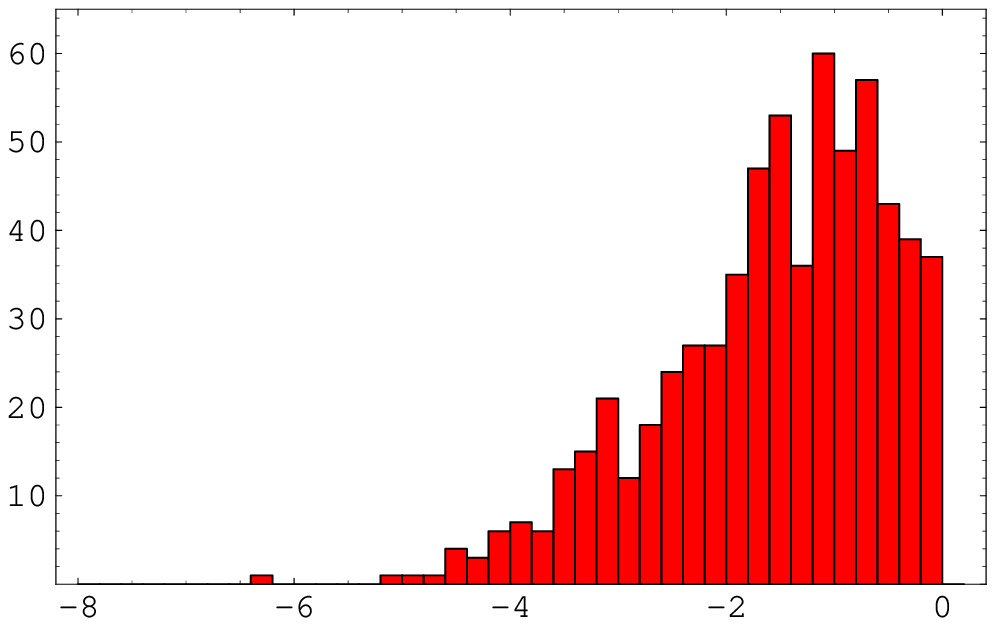} &
\includegraphics[width=0.3\linewidth]{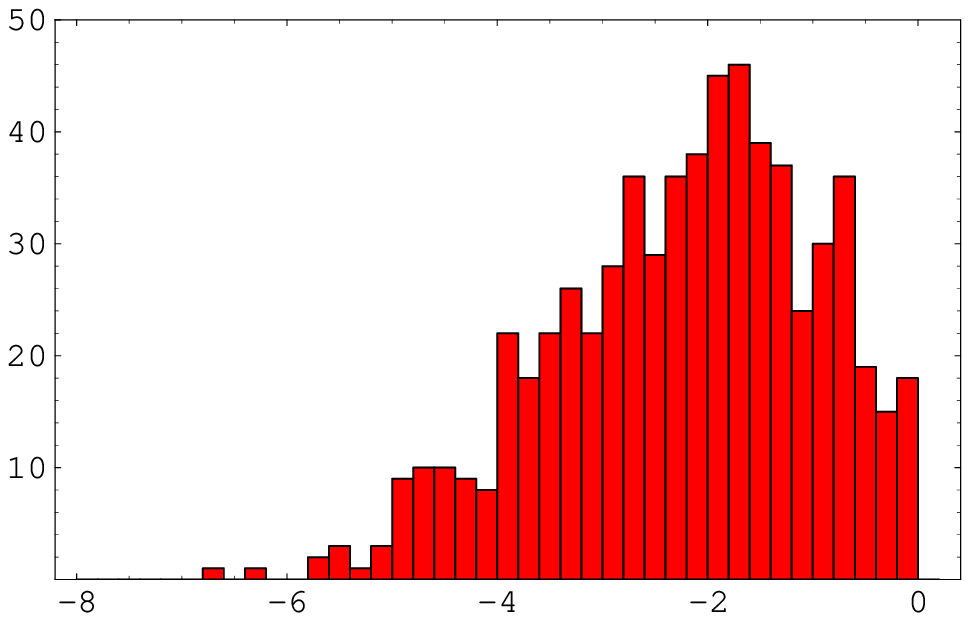} \\
$\log_{10}(2\theta_{12}/\pi)$ & $\log_{10}(2\theta_{23}/\pi)$ & 
$\log_{10}\sin\theta_{13}$ \\
\end{tabular}
\caption{\label{fig:T2mixangle} Distributions of CKM mixing angles  
in the Gaussian landscape on $T^2$, based on a numerical simulation 
with $(d/L,g_{\rm eff})=(0.1,0.4)$.  The bottom row displays the mixing 
angles of the roughly 6\% of matrices that survive the $t$-cut of section 
\ref{ssec:environment}.}
\end{center}
\end{figure}

Figure~\ref{fig:entry-distr-T2} shows the distribution of Yukawa matrix
elements for a numerical simulation of this model. 
As in the $S^1$ model, the only parameters relevant to these distributions 
are $g_{\rm eff}=g/(M_6L)$ and $d/L$, where now $M_6$ is the cut-off scale
of the effective theory in 5+1 dimensions.  Furthermore, a larger hierarchy 
is generated when the wavefunctions are more localized, i.e. when $d/L$ is 
smaller. The key difference from $S^1$ is that the distribution of Yukawa 
matrix elements is no longer scale invariant.  Instead, on $T^2$ the 
probability density for the largest and the smallest matrix elements is 
depleted.  With regard to the Yukawa eigenvalues and mixing angles, we 
find that a hierarchical pattern of Yukawa eigenvalues is generated 
(Figure~\ref{fig:T2YukawaEgval}) 
and that the distributions of mixing angles are peaked at $\theta_{ij}=0$ 
but not at $\theta_{ij} = \pi/2$ (Figure~\ref{fig:T2mixangle}).  Thus the 
flavor structure of the quark sector follows from the Gaussian landscape 
on $T^2$.
Despite the apparent difference between the distribution of the Yukawa 
matrix elements of the two landscapes (comparing Figure~\ref{fig:entrydistr} 
and Figure~\ref{fig:entry-distr-T2}), we see that the distribution of 
masses and mixing angles are roughly the same when we compare 
Figures~\ref{fig:T2YukawaEgval} and~\ref{fig:T2mixangle} to 
Figures~\ref{fig:q-spec-tcut} and~\ref{fig:q-mix-tcut}.  In particular, 
these distributions all come with a width of about an order of magnitude, 
and the differences between the distributions from the two toy landscapes 
is not, statistically, very significant compared to this width.  This 
demonstrates that the flavor structure in the quark sector is a robust 
feature of Gaussian landscapes.

The biggest difference between the distribution of observables in 
the two toy landscapes is a more left/right symmetric probability 
distribution for $\log_{10}\lambda_{b,t}$ and $\log_{10}\lambda_{s,c}$ on
$T^2$.  This is a consequence of the difference in the distribution of 
individual Yukawa matrix elements, Figure~\ref{fig:entry-distr-T2} vs. 
Figure~\ref{fig:entrydistr}. The depleted probability density of the 
largest Yukawa matrix elements in the Gaussian landscape on $T^2$ results 
in reduced probability for largest values of $\lambda_{b,t}$ and 
$\lambda_{s,c}$. The distribution of $\lambda_{b,t}$ becomes much
broader and less peaked at the largest possible value, 
$\lambda_{\rm max}$.  This allows the bottom Yukawa coupling to become as 
small as its measured value rather easily, while still accommodating the 
measured value of the top Yukawa coupling. 

The effects of a possible environmental selection for a large top 
Yukawa coupling are studied in Figure~\ref{fig:T2YukawaEgval} and in
Figure~\ref{fig:T2mixangle}, using the cut condition (\ref{eq:cut-t}) 
as a crude approximation to the environmental selection effect.
The distributions of the mixing angles $\theta_{23}$ and $\theta_{13}$ 
are shifted toward smaller angles, just as on $S^1$.  Since the cut 
condition is in favor of a larger hierarchy between the lighter quarks 
and the heaviest quarks, smaller $\theta_{23}$ and $\theta_{13}$ are 
natural consequences.  The inequality $\vev{\lambda_c}>\vev{\lambda_s}$ 
also follows from the cut, just like in section~\ref{sec:toy1}.  The most 
important difference between the two landscapes may be in the distribution 
of the bottom Yukawa coupling after the cut is imposed.  It was rare that 
$\lambda_b$ be less than $10^{-2}$ in the lower-right distribution of 
Figure~\ref{fig:q-spec-tcut}, but a significant fraction is below 
$10^{-2}$ in Figure~\ref{fig:T2YukawaEgval}. Therefore the observed 
hierarchy $\lambda_t/\lambda_b$ may be understood within the context of a 
Gaussian landscape with an appropriately chosen geometry, especially when 
there is environmental selection for a large top Yukawa coupling.

For comparison, we also perform a numerical study of a Gaussian landscape 
defined on $S^2$.  As before, the quarks and Higgs are represented by 
localized wavefunctions with width $d$.  Specifically, on $S^2$ a 
wavefunction centered at $\theta=0$ is given by\footnote{Although this 
wavefunction is not smooth at $\theta=\pi$, this is not of present concern.  
That is, the purpose of this numerical simulation is not to determine the 
distribution precisely but to study its qualitative aspects.}
\begin{equation}
\varphi(\theta,\phi) \propto e^{-\frac{\theta^2}{2(d/R)^2}} \,.
\end{equation}
Of course, to generate an ensemble of Yukawa matrices the central
coordinates of each wavefunction are scanned independently and uniformly 
over the geometry $S^2$.  Note that like the previously described 
landscapes, the Gaussian landscape on $S^2$ is characterized by two
free parameters, $d/R$ and $g_{\rm eff}=g/(\sqrt{\pi}M_6R)$.
  
\begin{figure}[t]
\begin{center}
\begin{tabular}{ccc}
\includegraphics[width=0.3\linewidth]{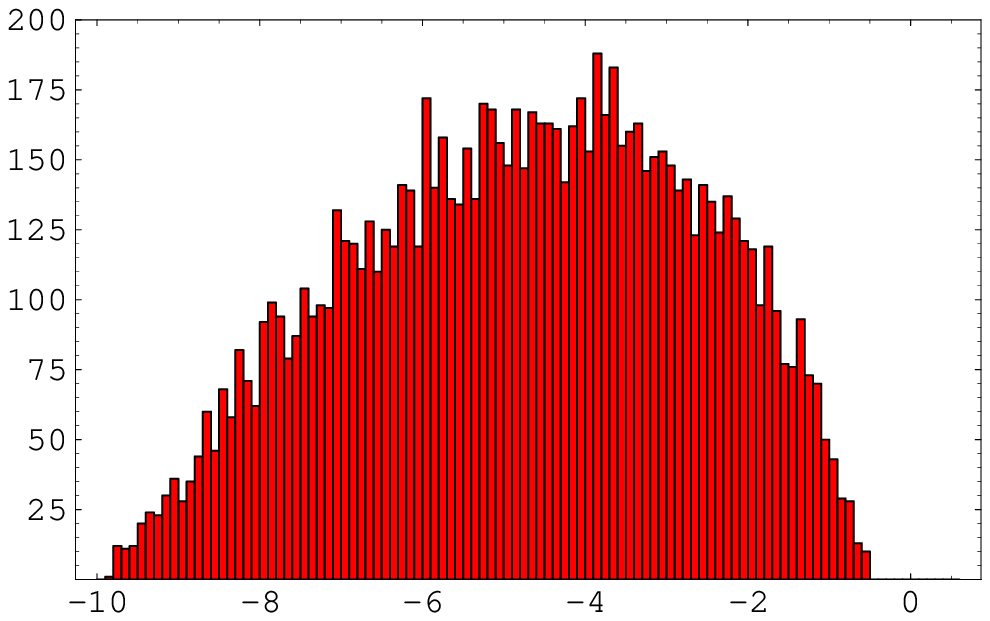} &
\includegraphics[width=0.3\linewidth]{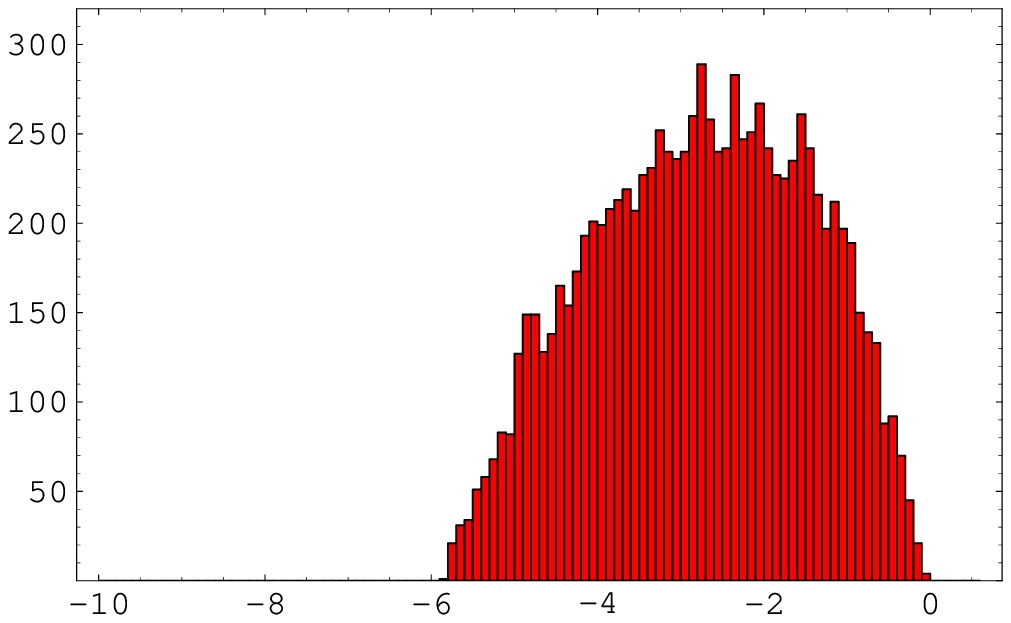} &
\includegraphics[width=0.3\linewidth]{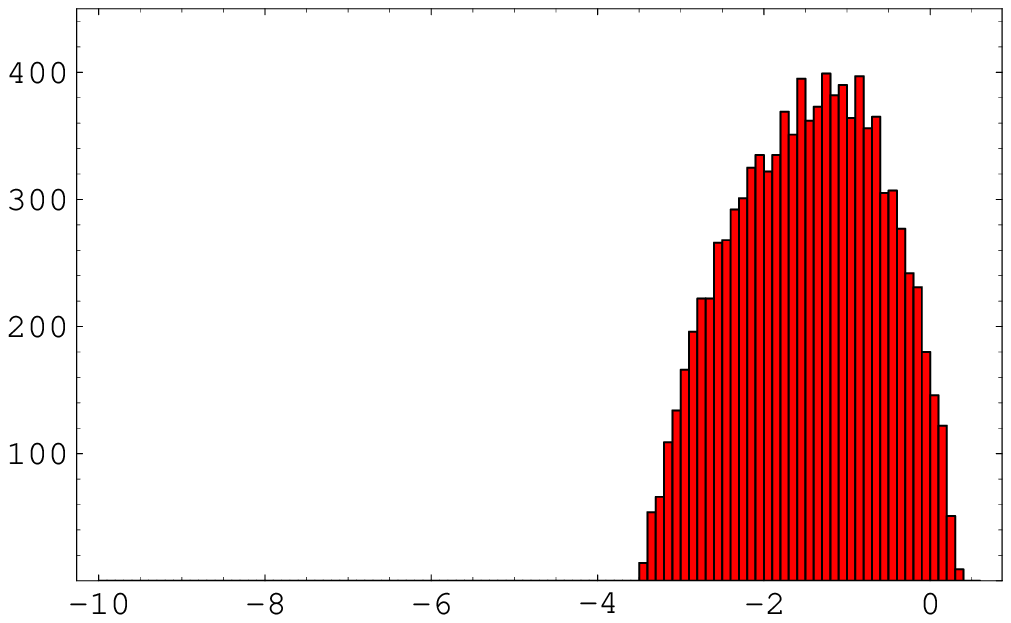} \\
\end{tabular}
\caption{\label{fig:entry-distr-S2} Distributions of $10^4$ Yukawa matrix 
elements, generated from the Gaussian landscape on $S^2$.  From left to 
right the panels correspond to $(d/R,g_{\rm eff})=(0.4,1)$, $(0.5,1)$ and 
$(0.6,1)$.}
\end{center}
\end{figure}

\begin{figure}[t]
\begin{center}
\begin{tabular}{ccc}
\includegraphics[width=0.3\linewidth]{Figures/D2L10d10/U1.eps} &
\includegraphics[width=0.3\linewidth]{Figures/D2L10d10/U2.eps} &
\includegraphics[width=0.3\linewidth]{Figures/D2L10d10/U3.eps} \\
$\log_{10}\lambda_{u,d}$ & $\log_{10}\lambda_{c,s}$ 
& $\log_{10}\lambda_{t,b}$ \\
\includegraphics[width=0.3\linewidth]{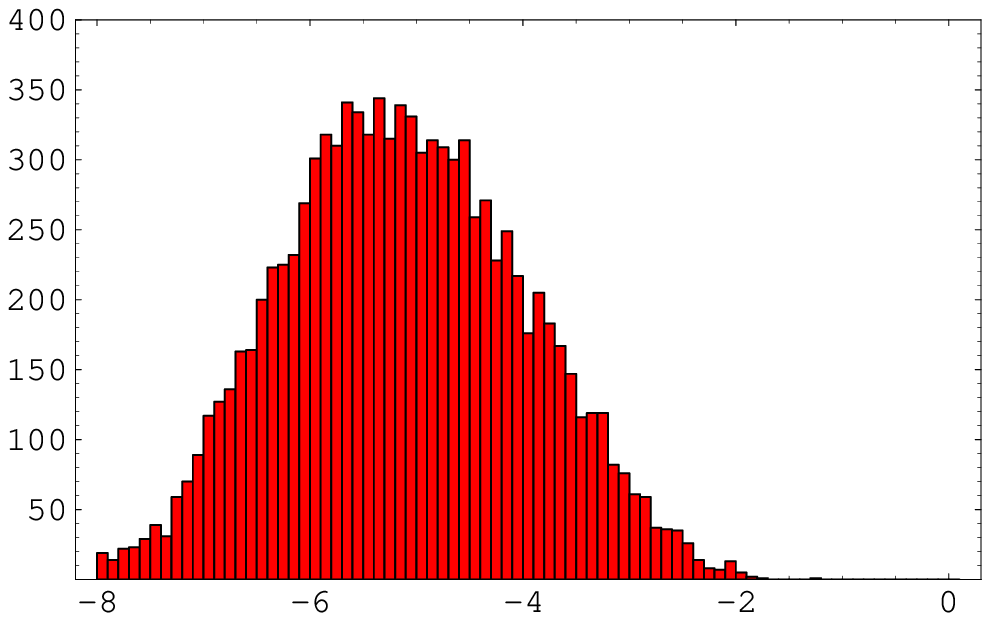} &
\includegraphics[width=0.3\linewidth]{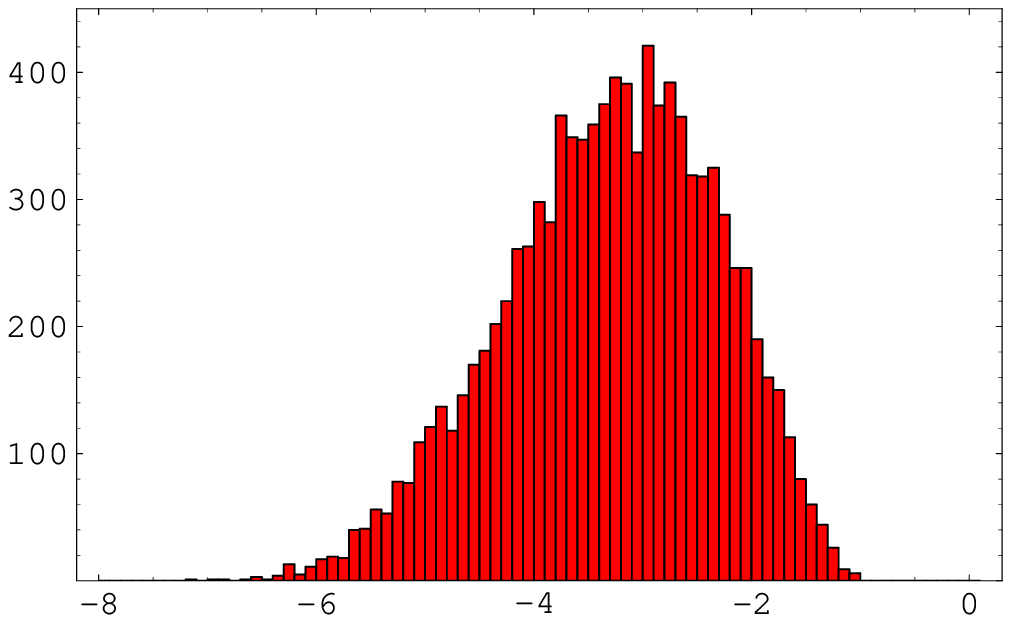} &
\includegraphics[width=0.3\linewidth]{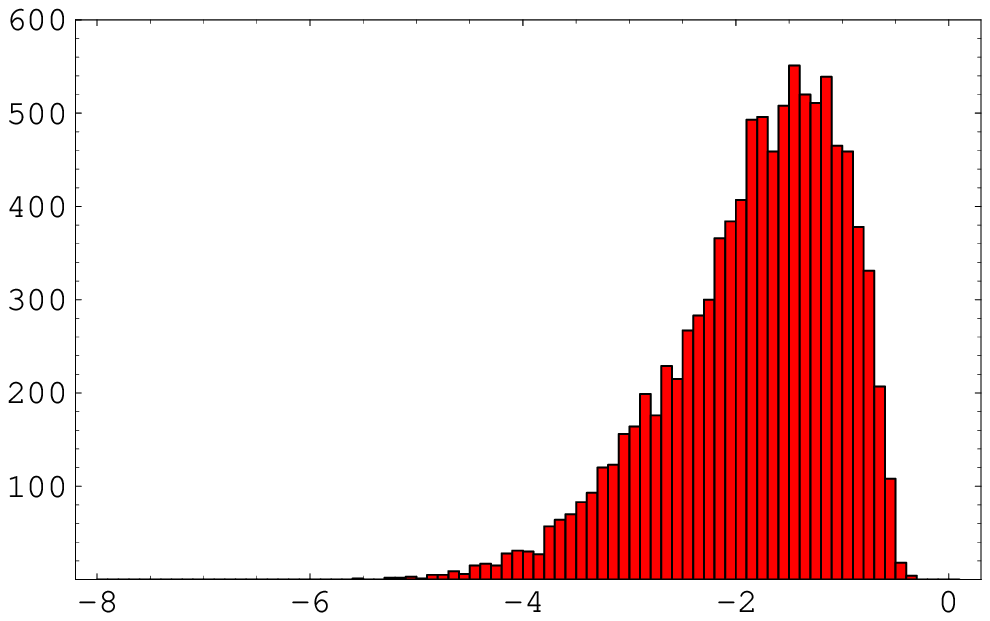} \\
$\log_{10}\lambda_{u,d}$ & $\log_{10}\lambda_{c,s}$ 
& $\log_{10}\lambda_{t,b}$ \\
\end{tabular}
\caption{\label{fig:compareT2S2} Comparison between the distributions 
of Yukawa eigenvalues from the Gaussian landscape on $T^2$ with 
$(d/L,g_{\rm eff})=(0.1,0.4)$ (top row) and the Gaussian landscape on 
$S^2$ with $(d/R,g_{\rm eff})=(0.45,0.4)$ (bottom row).}
\end{center}
\end{figure}

Figure~\ref{fig:entry-distr-S2} shows the distribution of Yukawa matrix 
elements for this model.
Note that the overall shapes of the distributions are remarkably similar 
to those in Figure~\ref{fig:entry-distr-T2}, which correspond to $T^2$.  
Figure~\ref{fig:compareT2S2} displays the distributions of the three 
mass eigenvalues for the Gaussian landscapes on $S^2$ and $T^2$.
These distributions are quite similar between $S^2$ and $T^2$.  We 
find that the other phenomenological aspects of the Gaussian landscape,
discussed above with respect to the geometry $T^2$, hold true on $S^2$ 
as well.

\subsection{Dependence on Dimensionality and Geometry Independence}
\label{ssec:Ndim}

Having seen that the hierarchical mass eigenvalues and the generation 
structure of the quark sector are robust predictions of Gaussian 
landscapes, the next subject of interest is to understand to what 
extent the details of distribution functions depend on the internal 
geometry.  Already we have seen slight differences between the Gaussian
landscape on $S^1$ and the landscapes defined on $T^2$ and $S^2$, yet
remarkable similarity between the distributions coming from $T^2$ and 
$S^2$.  We would like to understand where these differences and 
similarities come from, without having to run simulations on all possible 
choices of internal geometry.  To do this, we recall the analysis of 
sections~\ref{ssec:FN} and~\ref{ssec:phen}, where the distributions of
flavor observables on $S^1$ were understood {\it analytically} within the 
context of the AFS approximation.  As we show in this section, this 
analysis can be generalized to Gaussian landscapes on any geometry of 
extra dimensions.\footnote{There is also a practical motivation for this 
approach.  The numerical integration time involved with performing overlap 
integrals on extra dimensions grows very large as the number of extra 
dimensions is increased; meanwhile generating ensembles of flavor 
observables involves performing large numbers of these integrals.}

Let us consider a $D$-dimensional internal space with local coordinates 
$\vec{y}$. Then in the limit where we can ignore the finite size 
of the internal space the Yukawa coupling matrix is given by 
\begin{equation}
\lambda^{u}_{ij} \propto e^{-\frac{1}{3d}
(|\vec{a}_i|^2 + |\vec{b}_j|^2 - \vec{a}_i\cdot \vec{b}_j)},
\label{eq:approx-overlap-mD}
\end{equation}
which replaces $\lambda^u_{ij}$ in (\ref{eq:FN-matrix}) for the Gaussian 
landscape on $S^1$.  Here $\vec{a}_i$ and $\vec{b}_j$ are the central 
coordinates of the quark wavefunctions, relative that of the Higgs.  The 
Yukawa matrix again has the AFS form (\ref{eq:AFS}), with suppression 
factors  
\begin{equation}
\epsilon^q_j = e^{- \frac{|\vec{b}_j|^2}{3d^2}}\,, \qquad
\epsilon^{\bar{u}}_i = e^{- \frac{|\vec{a}_i|^2}{3d^2}}\,,
\label{eq:AFS-supp-mD}
\end{equation}
replacing (\ref{eq:AFS-supp}). In the extra factor 
$g_{ij} = e^{\vec{a}_i\cdot\vec{b}_j/3d^2}$ (which is not necessarily
of order unity) the vectors $\vec{a}_i$ and $\vec{b}_j$ are sometimes 
unaligned, sometimes parallel, and sometimes anti-parallel.  This 
generates a random coefficient to each Yukawa coupling that is 
statistically neutral in the AFS approximation.  We note again that this 
analysis is valid only when the compactness of the internal space is 
unimportant and the local geometry can be approximated as a flat 
$D$-dimensional space.  This is equivalent to focusing on only the 
largest Yukawa matrix elements.

Ignoring the statistically neutral factor 
$e^{\vec{a}_i\cdot\vec{b}_j/3d^2}$, the Yukawa matrix elements are roughly  
\begin{equation}
\lambda \sim e^{- \frac{|\vec{\bf r}|^2}{3d^2}}\,, \qquad 
-\ln\lambda = \frac{|\vec{\bf r}|^2}{3d^2}\,, 
\end{equation}
with $\vec{\bf r} = (\vec{a},\vec{b})$ scanning a $2D$-dimensional space. 
The natural probability measure is 
\begin{equation}
dP \propto \frac{d^{2D}\vec{\bf r}}{L^{2D}} 
 \propto \frac{|\vec{\bf r}|^{2D-2}\,d |\vec{\bf r}|^2}{L^{2D}}
 \propto \left(\frac{d}{L}\right)^{2D}
    (-\ln\lambda)^{D-1}d|\ln \lambda| \, ,
\label{eq:Yukawa-distr-Ddep}
\end{equation}
where $L$ is the typical size of the extra dimensions.  Indeed, the 
distribution of the Gaussian landscape on $S^1$ ($D-1=0$) is flat, c.f. 
Figure~\ref{fig:entrydistr}, and those of the Gaussian landscapes on $T^2$
and $S^2$ ($D-1=1$) are linear in $|\ln \lambda|$ at their upper 
end.\footnote{The distributions on $S^2$ are not as precisely 
linear for large matrix elements as those on $T^2$. This can be
understood in terms of geometry:  $S^2$ has positive curvature 
while $T^2$ is flat.}  
These results are independent of the value of $d/L$.  The logarithmic 
range of the distribution of Yukawa couplings scales as $(L/d)^{2}$.

\begin{figure}[t!]
\begin{center}
\begin{tabular}{cccc}
$S^1$ & $T^2$ & $S^2$ & $S^3$ \\
\includegraphics[width=0.23\linewidth]{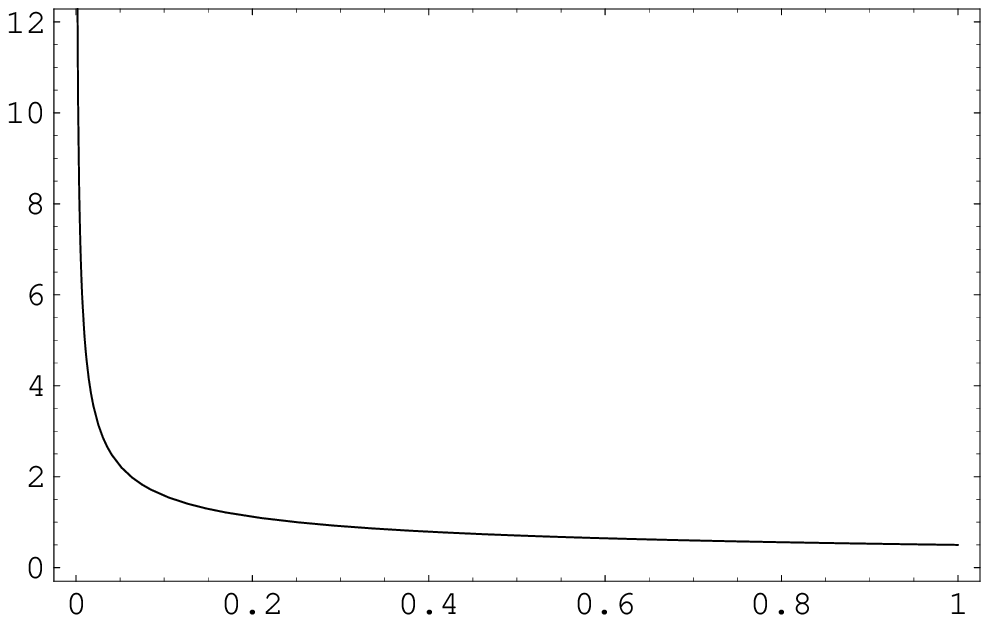} &
\includegraphics[width=0.23\linewidth]{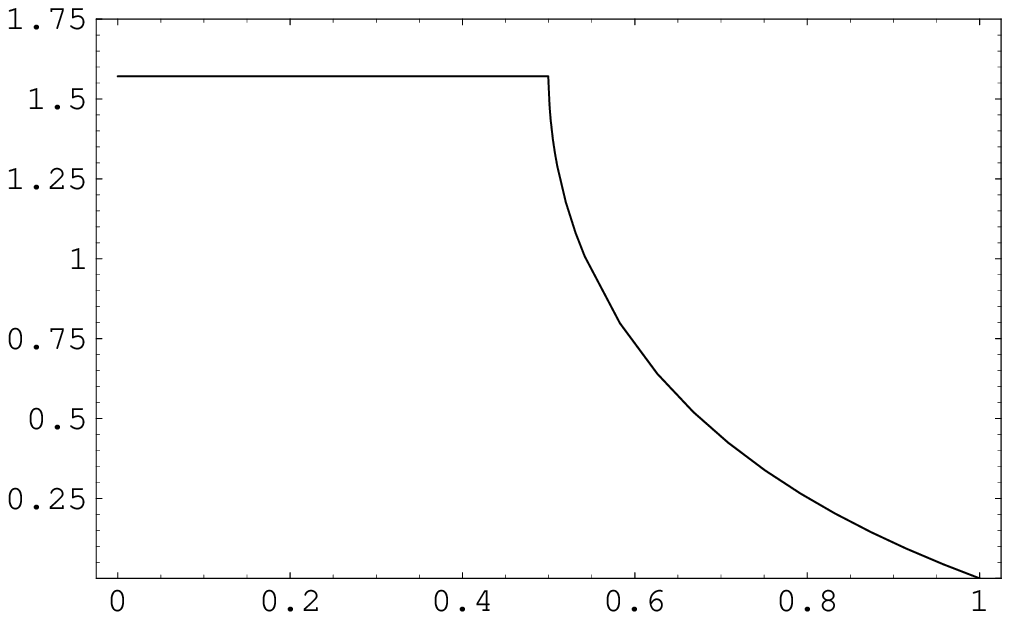} &
\includegraphics[width=0.23\linewidth]{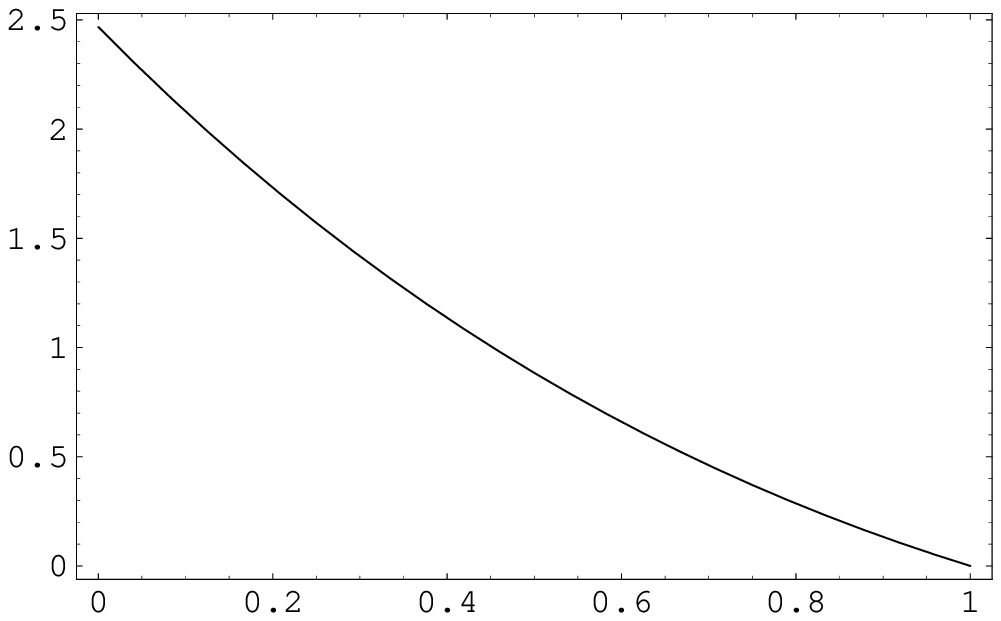} &
\includegraphics[width=0.23\linewidth]{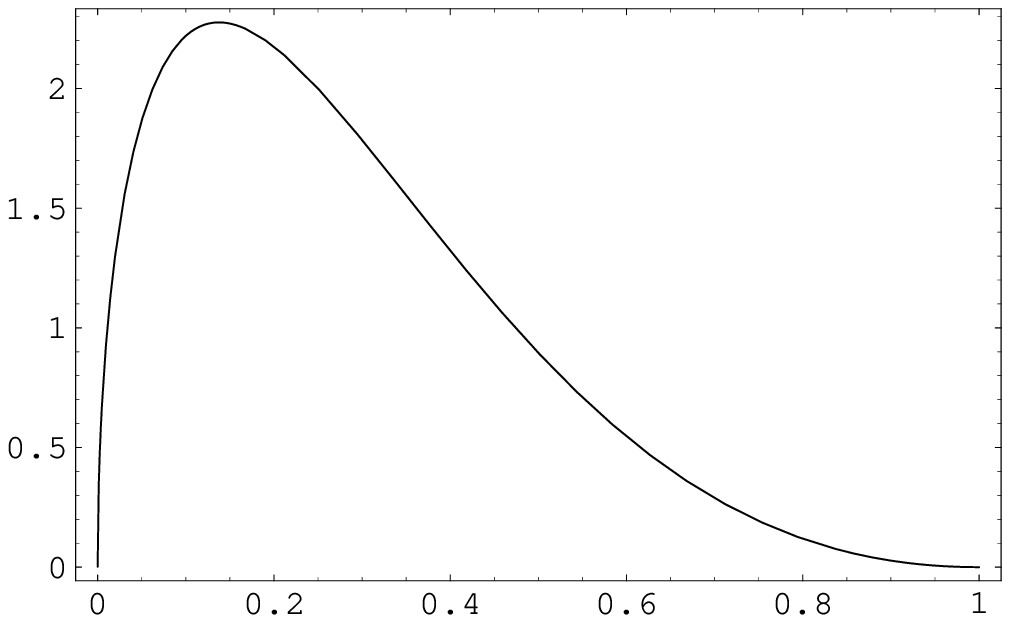} \\
$f_{S^1}(y)$ & $f_{T^2}(y)$ & $f_{S^2}(y)$ & $f_{S^3}(y)$ \\
\includegraphics[width=0.23\linewidth]{Figures/FN/fnS1.eps} &
\includegraphics[width=0.23\linewidth]{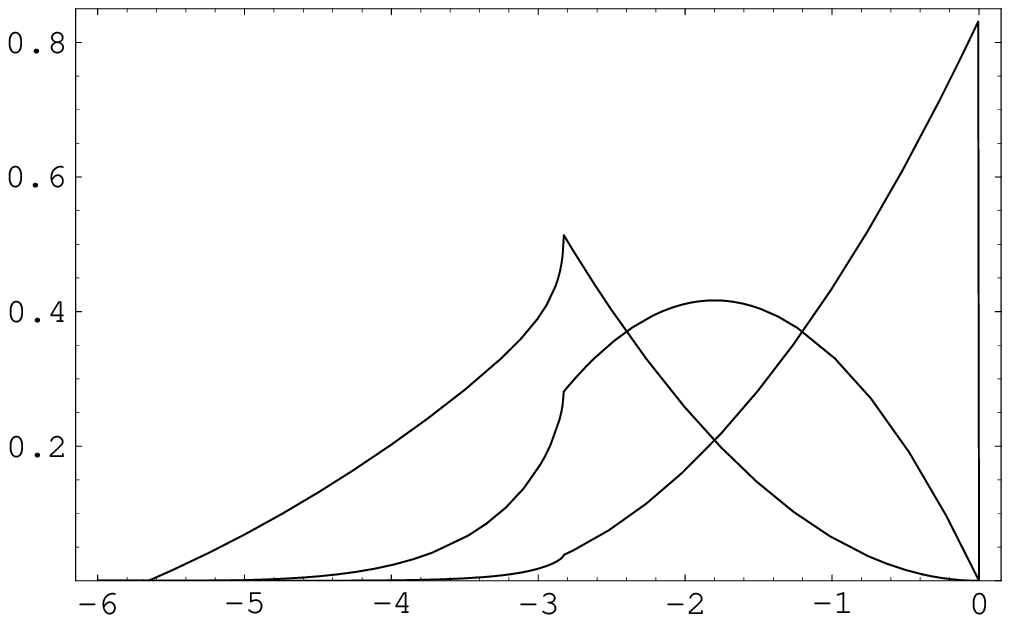} &
\includegraphics[width=0.23\linewidth]{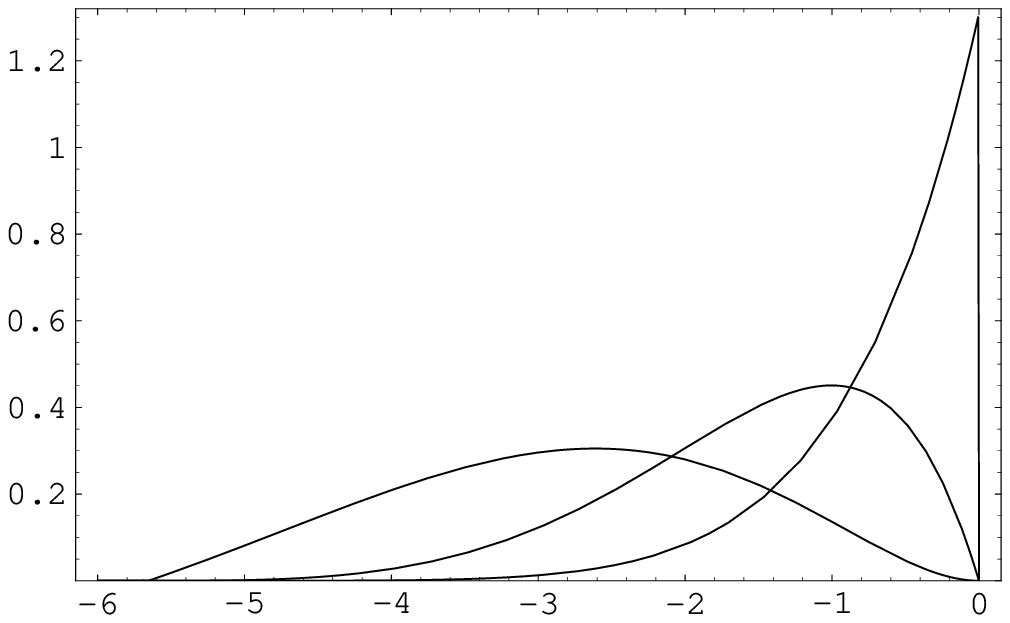} &
\includegraphics[width=0.23\linewidth]{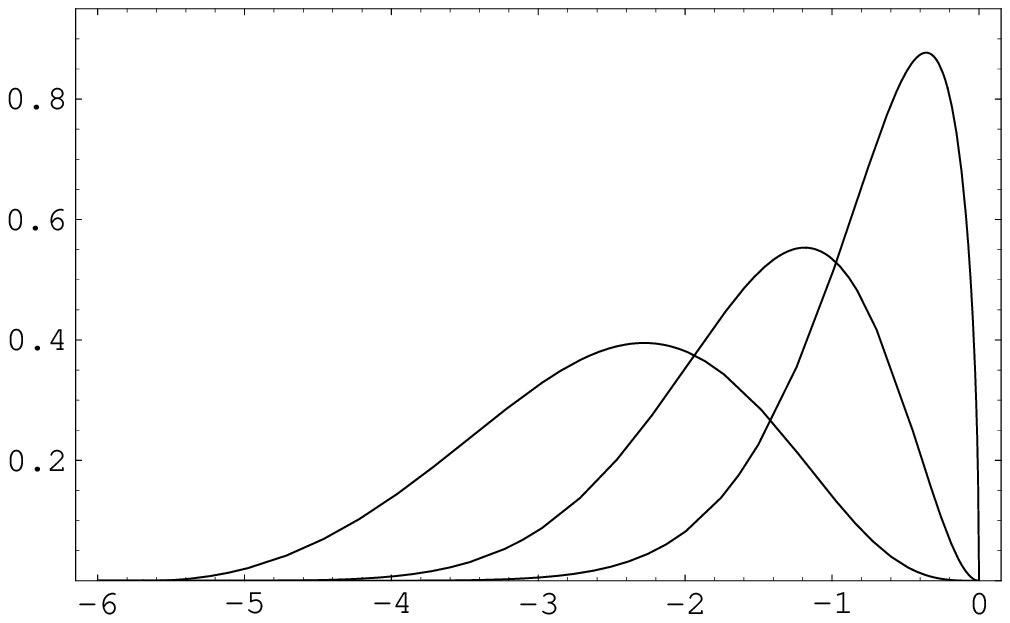} \\
$\log_{10} \epsilon_{1,2,3}$ & $\log_{10} \epsilon_{1,2,3}$ & 
$\log_{10} \epsilon_{1,2,3}$ & $\log_{10} \epsilon_{1,2,3}$ \\
\includegraphics[width=0.23\linewidth]{Figures/FN/massS1.eps} &
\includegraphics[width=0.23\linewidth]{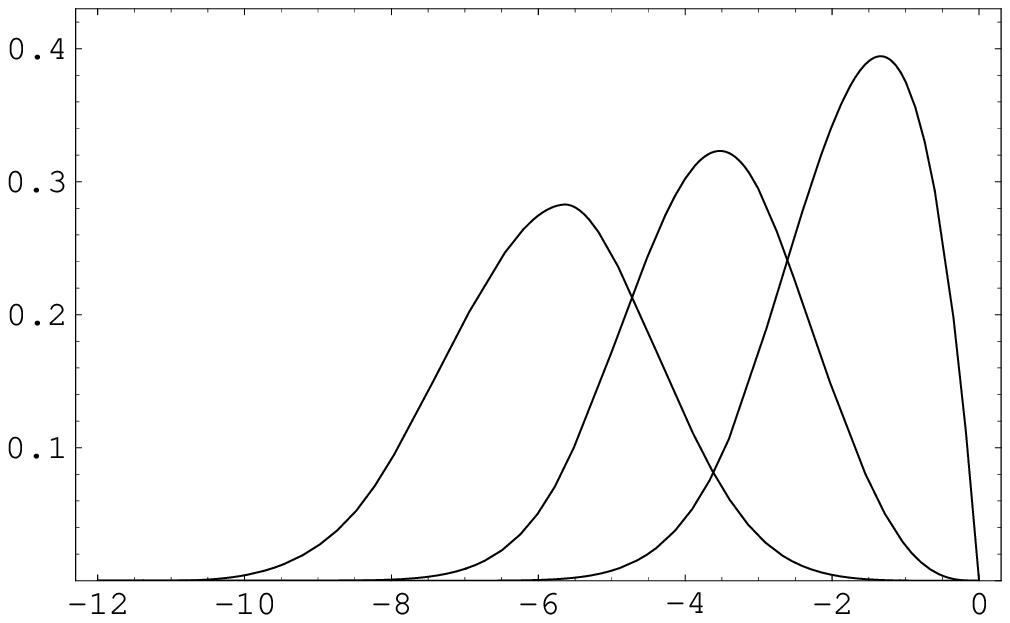} &
\includegraphics[width=0.23\linewidth]{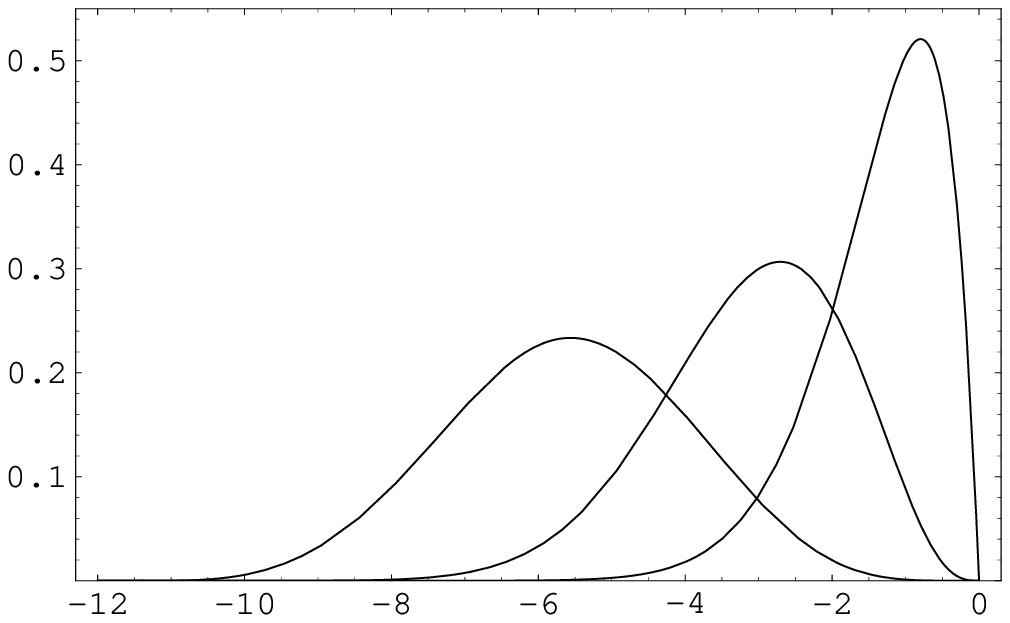} &
\includegraphics[width=0.23\linewidth]{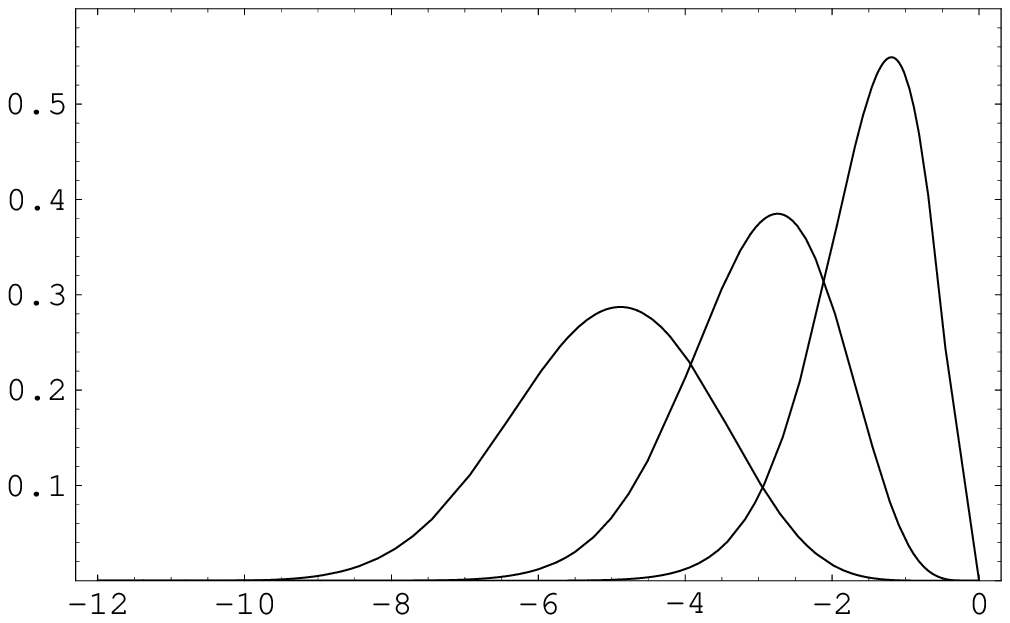} \\
$\log_{10} \lambda_{1,2,3}$ & $\log_{10} \lambda_{1,2,3}$ & 
$\log_{10} \lambda_{1,2,3}$ & $\log_{10} \lambda_{1,2,3}$ \\
\includegraphics[width=0.23\linewidth]{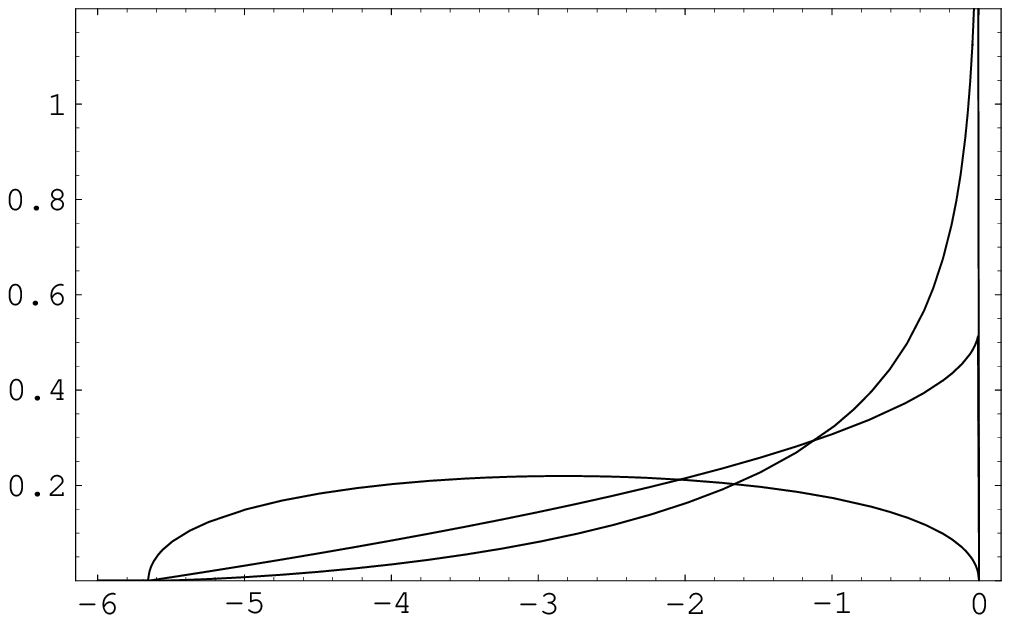} &
\includegraphics[width=0.23\linewidth]{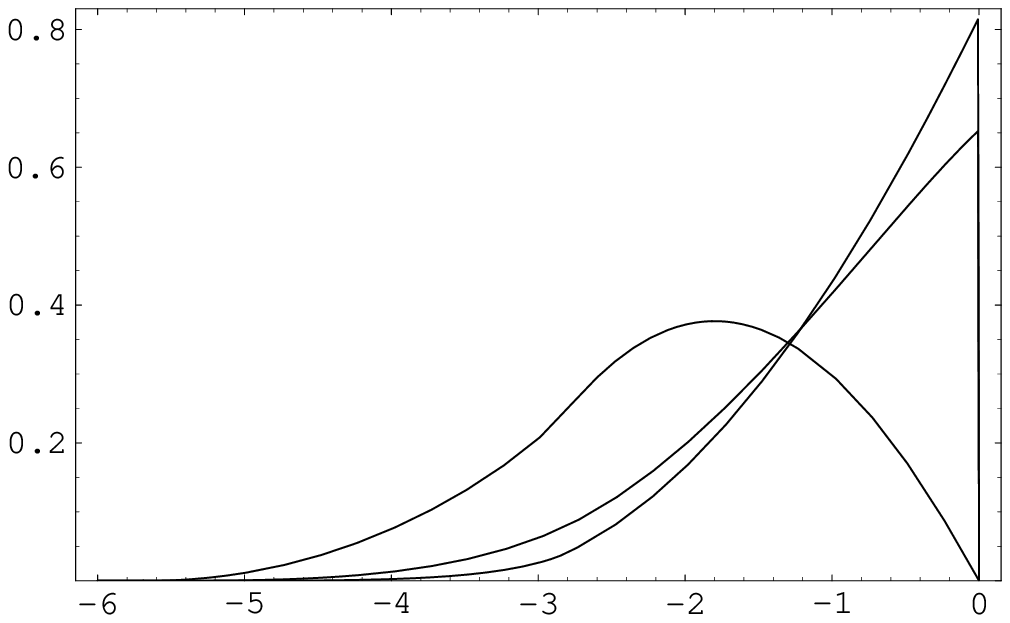} &
\includegraphics[width=0.23\linewidth]{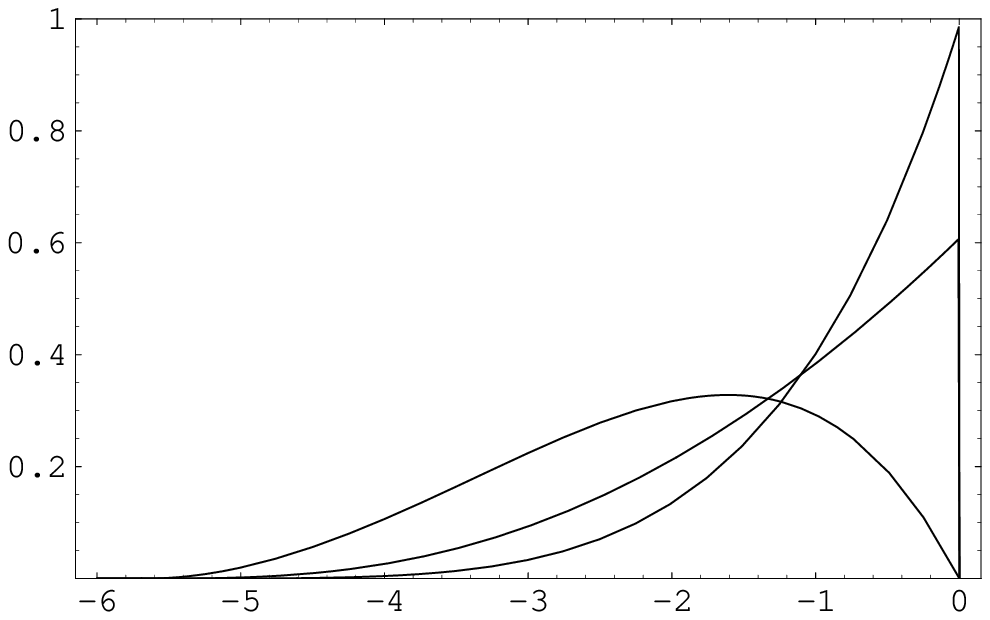} &
\includegraphics[width=0.23\linewidth]{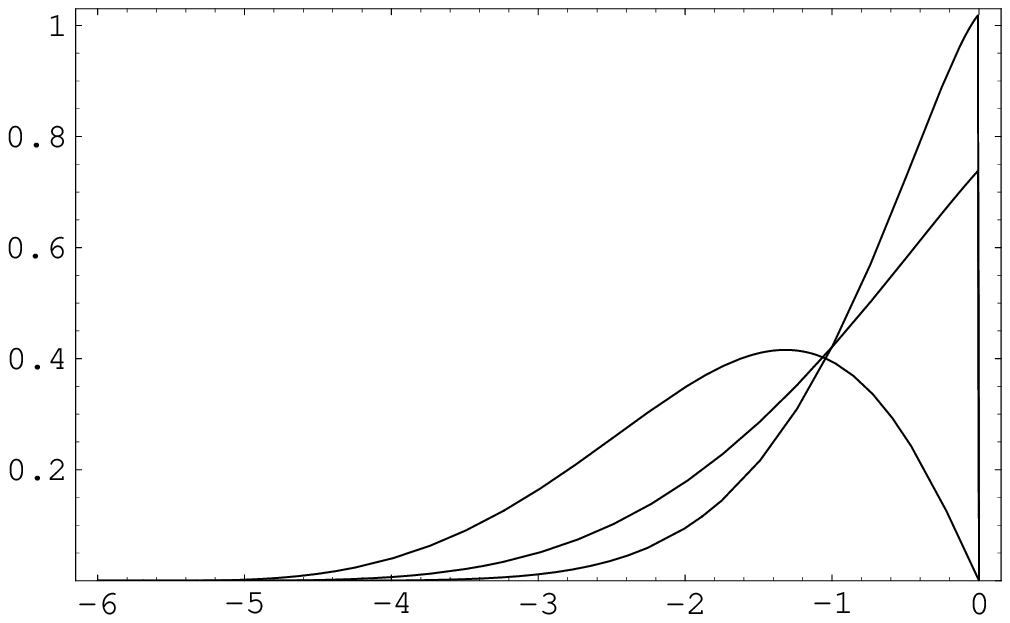} \\
$\log_{10} (\epsilon_i/\epsilon_j)$ &
$\log_{10} (\epsilon_i/\epsilon_j)$ &
$\log_{10} (\epsilon_i/\epsilon_j)$ &
$\log_{10} (\epsilon_i/\epsilon_j)$ \\ 
\includegraphics[width=0.23\linewidth]{Figures/FN/mixS1.eps} &
\includegraphics[width=0.23\linewidth]{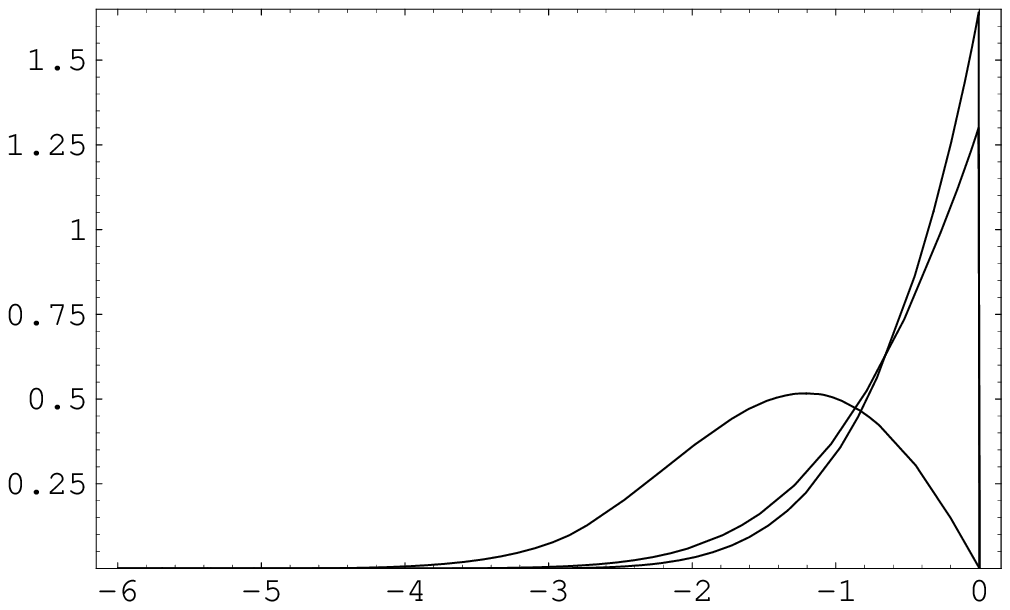} &
\includegraphics[width=0.23\linewidth]{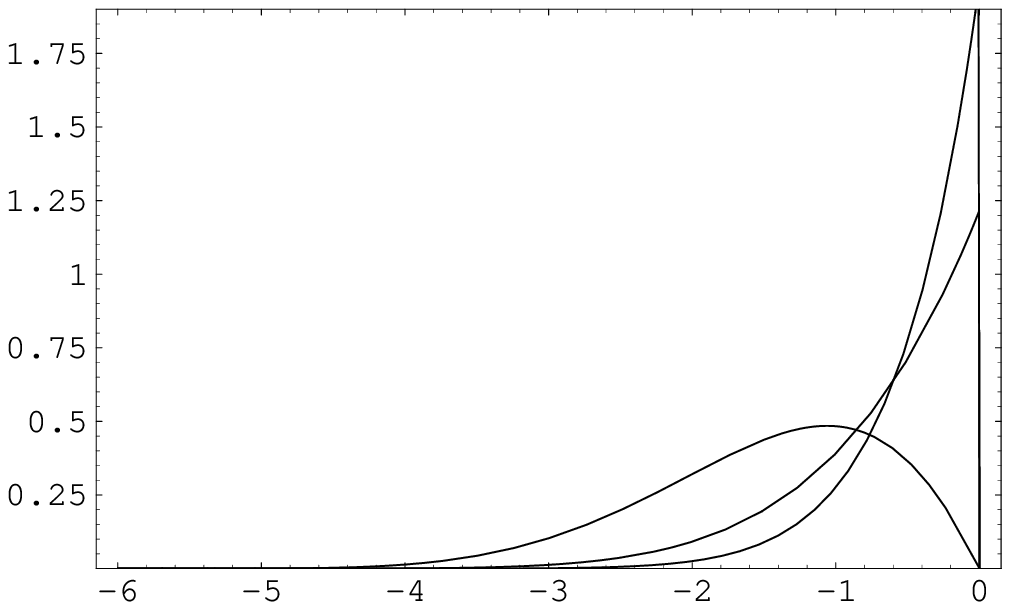} &
\includegraphics[width=0.23\linewidth]{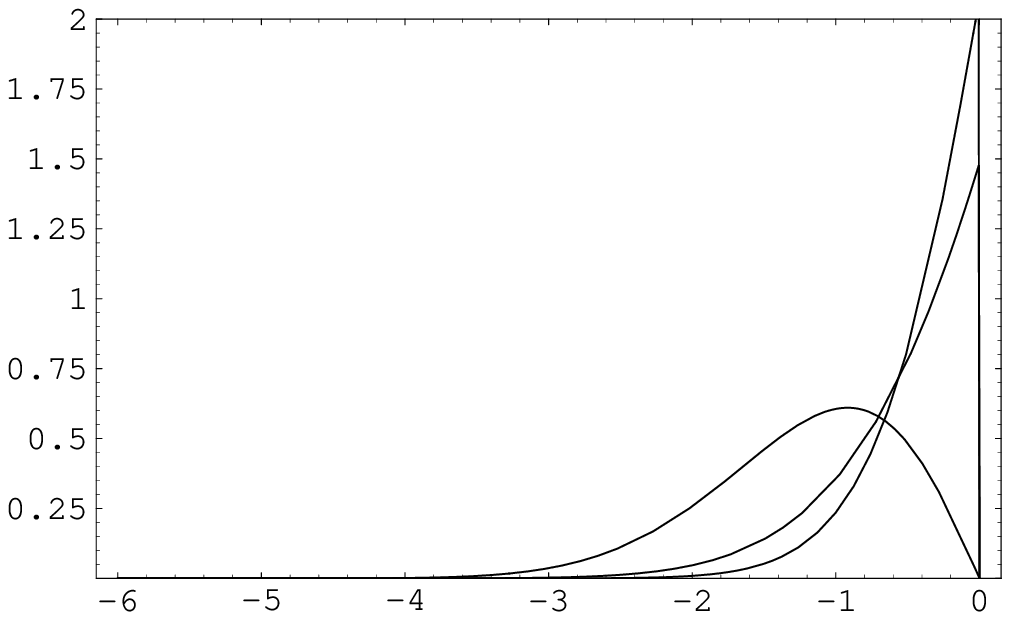} \\
$\log_{10} \sin \theta_{12,23,13}$ & 
$\log_{10} \sin \theta_{12,23,13}$ &
$\log_{10} \sin \theta_{12,23,13}$ &
$\log_{10} \sin \theta_{12,23,13}$ 
\end{tabular}
\caption{\label{fig:FN-D23} The volume distribution functions 
$f(y)$, AFS suppression factors $\log_{10}\epsilon_{1,2,3}$, the 
Yukawa eigenvalues $\log_{10}\lambda_{1,2,3}$, the diagonalization 
angles $\log_{10}(\epsilon_i/\epsilon_j)$ ($i<j$), and the CKM mixing 
angles $\log_{10}\sin\theta_{ij}$ for $S^1,\,T^2,\,S^2$ and $S^3$.
In all cases the broadest mixing angle distribution corresponds to 
$\theta_{13}$, while the distribution most sharply peaked at zero 
corresponds to $\theta_{23}$.  These use approximations 
(\ref{eq:mass-approx}) and (\ref{eq:Vus-approx}--\ref{eq:Vub-approx}), 
which are not reliable for small Yukawa eigenvalues.  For clear 
comparison the variables $y_i$, $z_i$, and $t_{ij}$ have been 
converted to observables using a common logarithmic scale,
$\Delta\log_{10}\epsilon = -5.66$, that corresponds to the scale of
the Gaussian landscape on $S^1$ with $d/L = 0.08$.}
\end{center}
\end{figure}

The distribution of the AFS suppression factors, $\epsilon^q_j$, can be 
obtained as in section~\ref{sec:toy1}.  In more than one dimension the 
measure (\ref{eq:FN-measure}) is generalized to 
\begin{equation}
dP(b)\sim\frac{d^D b}{L^D} \sim L^{2-D}\frac{d V(b)}{db^2}
\frac{d b^2}{L^2}\,, 
\label{eq:FN-measure-gen}
\end{equation}
where $b\equiv|\vec{b}|$ and $V(b)$ is the volume enclosed within a 
distance $b$ from a given point.  Using (\ref{eq:AFS-supp-mD}), it can 
be converted into a distribution of the AFS suppression factors,
\begin{equation}
dP(y) = f(y)\, dy\,, \qquad
f(y) = \frac{b_{\rm max}^2}{V_{\rm tot}}
       \frac{d V(b)}{d b^2}\bigg|_{b=b_{\rm max}\sqrt{y}} \,.
\label{eq:defoff}
\end{equation}
Here, $\ln \epsilon$ is normalized by 
$\Delta \ln \epsilon=- \frac{1}{3} (b_{\rm max}/d)^2$, so that 
$y\equiv\ln\epsilon/\Delta\ln\epsilon=(b/b_{\rm max})^2$ 
ranges from zero to one.  For example, $b_{\rm max}=L/2$ for $S^1$, 
$b_{\rm max}=L/\sqrt{2}$ for $T^2$ and $b_{\rm max}=\pi R$ for $S^2$.
The ratio $(b_{\rm max}/d)^2$ determines the overall logarithmic range 
of hierarchy, and the volume distribution function $f(y)$ controls the 
shape of the distributions within $y \in \left[ 0, 1 \right]$.  It is 
straightforward to find $f(y)$ for a given geometry of extra dimensions;  
for example, for $S^1$, $S^2$ and $S^3$,  
\begin{equation}
f_{S^1}(y) = \frac{1}{2\sqrt{y}}\,, \qquad
f_{S^2}(y) = \frac{\pi}{4}\frac{\sin (\pi \sqrt{y})}{\sqrt{y}}\,,\qquad 
f_{S^3}(y) = \frac{\sin^2 (\pi \sqrt{y})}{\sqrt{y}}\, ,
\label{eq:fS123}
\end{equation}
and for the ``square torus'' $T^2$,
\begin{equation}
f_{T^2}(y) = \frac{\pi}{2}-2\,{\rm arcsin} \left(
\sqrt{1-\frac{1}{2y}}\right)\Theta(y-1/2) \,.
\end{equation}
These volume distribution functions are displayed in the first row of 
Figure~\ref{fig:FN-D23}.

The distribution of the smallest, middle, or largest AFS suppression 
factor is obtained by integrating the other two variables out of the 
probability distribution 
\begin{equation}
 dP(y_1, y_2,y_3) = 3! f(y_1) f(y_2) f(y_3) 
 \Theta(y_1-y_2) \Theta(y_2-y_3)\, dy_1 dy_2 dy_3 \,,
\label{eq:distr-eps-D}
\end{equation}
where $y_i\equiv\ln\epsilon_i/\Delta\ln\epsilon$ and we remind that in 
this notation $y_1>y_2>y_3$.  The AFS suppression factors $y_i$ for 
$S^{1,2,3}$ are displayed in the second row of Figure~\ref{fig:FN-D23}.  
The mass eigenvalues $\lambda^{u(d)}_i$ are approximated by 
$\epsilon^{q}_i \epsilon^{\bar{u}(\bar{d})}_i$, and hence the distribution 
of $z_i\equiv\ln(\lambda^{u(d)}_i/\lambda_{\rm max})/\Delta\ln\epsilon$ is 
given by a convolution of the distribution function of 
$y_i=\ln \epsilon_i/\Delta\ln\epsilon$ (third row of 
Figure~\ref{fig:FN-D23}).  The diagonalization matrices for the up and 
down sectors are determined by three diagonalization angles, which are 
approximately equal to $\epsilon_i/\epsilon_j$, $i<j$. Thus, the 
distribution of 
$\Delta_{ij}\equiv\ln(\epsilon_i/\epsilon_j)/\Delta\ln\epsilon=y_i-y_j$ is 
obtained by integrating the variables $y_{1,2,3}$ out of 
(\ref{eq:distr-eps-D}) while keeping the distance $\Delta_{ij}=y_i-y_j$ 
(fourth row of Figure~\ref{fig:FN-D23}).  Finally, the CKM mixing angles 
are obtained from the diagonalization angles by approximating the mixing 
angles to be larger one of $\Delta_{ij}$ in the up sector and the down 
sector.  These are displayed in the fifth row of Figure~\ref{fig:FN-D23}.
Note that the analytical results for the mass eigenvalue distributions 
(the third row of Figure~\ref{fig:FN-D23}) capture the qualitative 
features of the numerical results for the Gaussian landscapes on $S^1$, 
$T^2$ and $S^2$ very well.

It is now possible to understand why the distributions of flavor 
observables are quite similar between the Gaussian landscapes on $T^2$ 
and $S^2$.  All of these distributions are calculated using only the 
volume distribution function $f(y)$.  Although $f(y)$ is quite different
for these two geometries, as is seen in the first row of 
Figure~\ref{fig:FN-D23}, a number of consecutive integrations is required 
to obtain distribution functions for observable flavor parameters.  For 
any AFS suppression factor $y_i\propto\ln\epsilon_i$, the volume 
distribution function $f(y_i)$ in (\ref{eq:distr-eps-D}) is not integrated, 
and thus singularities in the original $f(y_i)$ remain in the distribution 
of $\log_{10}\epsilon_i$.  Nevertheless, the difference in $f(y)$ between
$T^2$ and $S^2$ is already less evident in the distributions of 
$\log_{10}\epsilon_i$, (second row of Figure~\ref{fig:FN-D23}), because the 
integration of two variables out of (\ref{eq:distr-eps-D}) takes a certain 
average of $f(y)$.  Meanwhile, to obtain the distribution functions of 
the mass eigenvalues, diagonalization angles, and the CKM mixing angles, 
each $f(y_i)$ is integrated at least once, and the geometry-dependent 
information contained in $f(y)$ is further smeared out.  This is why the 
distributions of the Yukawa eigenvalues are quite similar between $T^2$ 
and $S^2$ in Figures~\ref{fig:compareT2S2} and~\ref{fig:FN-D23}.
Since the distributions of CKM mixing angles involve a greater number of
integrations, these distributions are even less sensitive to the 
underlying geometry.

The analytical discussion so far explains why the distributions of flavor 
observables are similar for the Gaussian landscapes on $T^2$ and $S^2$, 
but it remains unseen why the distributions of mass eigenvalues of the 
$S^1$ Gaussian landscape are somewhat different from those of $T^2$ and 
$S^2$. Therefore, we now focus on how the number of dimensions in the 
internal geometry affects the distributions of flavor observables.  As 
we have already seen in the discussion surrounding 
(\ref{eq:Yukawa-distr-Ddep}), the number of extra dimensions directly 
affects the shape of the volume distribution function.  Specifically, 
we find $f(y) \propto y^{D/2-1}$ for $y$ greater than but near zero.  
Thus for very small $y_i$, 
\begin{equation}
dP(y_1)\propto y_1^{3D/2-1}dy_1\,, \qquad 
dP(y_2)\propto y_2^{D-1}dy_2\,, \qquad 
dP(y_3)\propto y_3^{D/2-1}dy_3\,. 
\label{eq:Ddep-AFSfactors}
\end{equation}
Distribution functions of the masses and mixing angles can be obtained 
by using the approximations (\ref{eq:mass-approx}) and 
(\ref{eq:Vus-approx}--\ref{eq:Vub-approx}), analogous to the analysis 
earlier in this section.  In the limit of large eigenvalues and large 
mixing angles these distribution functions behave as
as
\begin{equation}
\begin{array}{l l l}
\displaystyle\frac{dP(z_1)}{dz_1}\propto z_1^{3D-1}\,, 
\phantom{\Bigg(\Bigg)} &
\displaystyle\frac{dP(z_2)}{dz_2}\propto z_2^{2D-1}\,, 
\phantom{\Bigg(\Bigg)} &
\displaystyle\frac{dP(z_3)}{dz_3}\propto z_3^{D-1}\,, \\
\displaystyle\frac{dP(t_{12})}{dt_{12}}\neq 0\,, 
\phantom{\Bigg(\Bigg)} & 
\displaystyle\frac{dP(t_{23})}{dt_{23}}\neq 0\,, 
\phantom{\Bigg(\Bigg)} &   
\displaystyle\frac{dP(t_{13})}{dt_{13}}\propto t_{13}\,,
\end{array}
\label{eq:Ddep-observables}
\end{equation}
where again $z_i\equiv\ln(\lambda_i/\lambda_{\rm max})/\Delta\ln\epsilon$ 
and $t_{ij}\equiv\ln \sin\theta_{ij} /\Delta\ln\epsilon$.  Note that the
behavior of the distribution functions of the mixing angles does not depend 
on the number of dimensions.  The $D$-dependence above is qualitatively 
confirmed in Figure~\ref{fig:FN-D23}.  Because of the slow rising of the 
$z_3=\ln(\lambda_{t,b}/\lambda_{\rm max})/\Delta\ln\epsilon$ distribution 
function in higher dimensions, the weight of the distribution of $z_3$ is 
shifted toward larger values, allowing for smaller values of $\lambda_b$.

As is evident from Figure~\ref{fig:FN-D23}, however, the approximation 
$f(y)\propto y^{D/2-1}$ is valid for only a very narrow range of $y$ near 
zero. Using the explicit form of $f(y)$ instead, we can extract another 
systematic $D$-dependence.  The statistical average of 
$y_i=\ln\epsilon_i/\Delta\ln\epsilon$ plays the role of an AFS charge of 
the $i$th generation, and $y_i-y_j$ determines the hierarchy between the 
$i$th and $j$th generations.  For sequences of geometries with different
numbers of extra dimensions---$S^1$, $S^2$, and $S^3$, and $T^1 = S^1$, 
$T^2$, and $T^3=S^1\times S^1\times S^1$---an explicit calculation reveals 
\begin{eqnarray}
 \frac{\vev{y_1 - y_2}}{\vev{y_2 - y_3}} & = &
   1.50 \; (S^1), \qquad 1.44 \; (S^2),\qquad 1.38 \; (S^3)\,, \\
 \frac{\vev{y_1 - y_2}}{\vev{y_2 - y_3}} & = &
   1.50 \; (T^1), \qquad 1.20 \; (T^2), \qquad 1.14 \; (T^3)\,. 
\end{eqnarray}
This ratio clearly decreases in Gaussian landscapes with a greater
number of extra dimensions, meaning that the middle eigenvalue becomes 
(statistically) closer to the smallest eigenvalue on a logarithmic scale. 
This also means that in in the AFS approximation the diagonalization angle 
(and hence the mixing angle) between the first and second generations 
becomes more significant compared with that between the second and third 
generations.  This tendency is confirmed in Figure~\ref{fig:FN-D23}, 
where the $\theta_{12}$ distribution becomes closer to the $\theta_{23}$ 
distribution for larger $D$.  Given that we measure 
$\lambda_d/\lambda_s>\lambda_s/\lambda_b$ and $\theta_{12}>\theta_{23}$, 
this might be regarded as an indication that $D > 1$.  Note however that 
the compactness of extra dimensions affects the distributions of the 
smallest eigenvalues, and therefore the distributions of observables based 
on the AFS approximation cannot reliably be used to infer the number of 
extra dimensions.

Let us now summarize the conclusions of this section.  The basic features 
of flavor in the Standard Model follow from Gaussian landscapes on any 
geometry of extra dimensions.  Specifically, the distribution of Yukawa 
matrix elements is generally broad and thus there is always hierarchy 
among Yukawa mass eigenvalues.  Furthermore, the up-type and down-type
Yukawa matrices are correlated due to the common overlaps of 
left-handed quark doublets and the Higgs, so that generation structure 
is always realized.  Even in the details of the distribution functions of 
flavor observables, there is not much geometry dependence.  We understand
this in terms of the AFS approximation, where all of the geometry 
dependence is encoded in the volume distribution function $f(y)$.  This
function is integrated and convoluted many times to obtain the 
distributions of observables, thus smearing out the original geometry
dependence.  This means that we cannot learn very much about the 
geometry of extra dimensions from the observed masses and mixing angles.  
On the other hand, it appears that we can understand the qualitative 
pattern of masses and mixing angles without knowing much about the details 
of the underlying geometry.  The number of extra dimensions, however, 
leaves its footprint on some distribution functions near $y\sim 0$, 
because the boundary at $y=0$ remains a boundary in some convolutions.  We 
also find that the mass eigenvalues of the second generation become closer 
(statistically) to those of the first generation as we increase the number
of extra dimensions in Gaussian landscapes.

For various geometries, the volume distribution functions $f(y)$ are 
usually moderately varying functions of $y$ over the range 
$y\in\left[ 0, 1\right]$.  Since the distribution functions of the 
AFS suppression factors $y_i\propto\ln\epsilon_i$, the mass eigenvalues 
$z_i\propto\ln\lambda_i$, and the mixing angles 
$t_{ij}\propto\ln\sin\theta_{ij}$ are all derived from $f(y)$, these 
distribution functions are moderately varying functions of the 
logarithmic variables $y_i$, $z_i$, and $t_{ij}$.  The widths of 
these statistical distributions are quite broad; to achieve otherwise
would require an exponentially steep $f(y)$, which does not happen in 
Gaussian landscapes.  Therefore, the Gaussian 
landscapes in this article are different from those in~\cite{A-HDK}, 
where all of the dimensionless coupling constants of the Standard Model 
have narrow-width Gaussian distributions.  This feature of our toy 
landscapes is traced back to our assumption that the localized 
wavefunctions become exponentially small as one moves away from the 
centers of localization.

\section{Including the Lepton Sector} 
\label{sec:Lepton}

Yukawa couplings generated from simple Gaussian landscapes are in good 
qualitative agreement with the mass hierarchies and small mixing angles 
in the quark sector.  On the other hand, the lepton sector is 
characterized by large mixing angles and very small neutrino masses.  We 
now turn our attention to how these qualitatively distinct features might 
be explained within a single landscape model.  With the challenges of 
describing the lepton sector comes the opportunity to predict the 
probability distributions of three yet-to-be measured observables in the 
lepton sector:  the mixing angle $\theta_{13}$, the CP phase in neutrino 
oscillations and the mass parameter $m_{\beta\beta}$ of neutrinoless 
double beta decay.

In order to accommodate large mixing angles in the lepton sector, in 
section~\ref{ssec:broad-5bar} the Gaussian landscape is extended to 
include different widths for wavefunctions of particles in different 
representations.  We will see in section~\ref{ssec:mixcp} that the mixing 
angles can be large indeed, and we also learn that complex (CP violating) 
phases play a crucial role in determining the distribution of mixing 
angles (and vice versa in section~\ref{ssec:prediction}).  Before 
studying mixing angles we study the charged lepton mass spectrum in 
section~\ref{ssec:chargedleptons}.  Small neutrino masses are assumed 
to be generated via the seesaw mechanism, and in section~\ref{ssec:seesaw}
we describe how to generate the statistics of the right-handed neutrino 
Majorana mass terms in Gaussian landscapes.  Finally, in 
section~\ref{ssec:prediction} we impose some cuts on the statistics 
generated by a particular version of the Gaussian landscape.  This is 
done to obtain a general feeling for how the probability distributions of 
yet-to-be measured observables can be affected when distributions 
conditional on measurements already performed in this universe are 
considered.  

Note that our goal in this section is to find a single theoretical 
framework that can describe the various flavor structures that are 
observed; i.e.\! we aim to identify what subset of landscapes share key 
qualitative features with the observed flavor structure.  Ultimately, 
the success of the landscape picture will depend on both the existence 
of phenomenologically viable subsets to the landscape and that these 
subsets are not too atypical of what is expected from the full landscape, 
after cosmological and environmental selection effects are accounted.

\subsection{Landscapes with Delocalized $\bar{\bf 5} = (\bar{d},l)$}
\label{ssec:broad-5bar}

In the traditional AFS approach, the large mixing angles of neutrino 
oscillation and a mild charged lepton mass hierarchy result if the three 
lepton doublets are not strongly distinguished by the flavor symmetry.   
For example, an AFS may be broken by a single parameter, with the three 
$q\subset {\bf 10}=(q, \bar{u},\bar{e})$'s strongly distinguished by the 
symmetry charges while the three $l\subset\bar{\bf 5}=(\bar{d}, l)$'s are 
not.  Since Gaussian landscapes can mimic an AFS structure in the Yukawa 
couplings, the idea of different strengths of flavor symmetry breaking 
for $\bar{\bf 5}$ and ${\bf 10}$ can be translated into the framework of 
Gaussian landscapes. 

As is seen in sections~\ref{sec:toy1} and~\ref{sec:Geometry}, the ratio 
of effective AFS charges for the three fermions in a given representation 
is determined (statistically) by the Gaussian landscape; we have no 
freedom to choose these by hand.  On the other hand, the overall 
hierarchy depends on the parameter $\Delta\ln\epsilon\propto-(L/d)^2$.  
So far, for simplicity the parameter $d$ has been chosen to be the same 
for all of the Standard Model wavefunctions;  however this parameter 
can be different for fields in different representations.  If the 
${\bf 10}$ and $\bar{\bf 5}$ fields have different width parameters 
$d_{\bf 10}$ and $d_{\bar{\bf 5}}$, then the AFS suppression factors 
associated with these fields are also different.  In such a landscape the 
single parameter $d$ is replaced by three: $d_H$, $d_{\bf 10}$ and 
$d_{\bar{\bf 5}}$.

If the wavefunctions of the fermions in the $\bar{\bf 5}$ representation 
are not particularly localized, the overlaps between the wavefunctions of 
the $\bar{\bf 5}$ fields and the Higgs boson do not vary hierarchically 
with the peak locations of the $\bar{\bf 5}$ Gaussian wavefunctions.  
Therefore the Yukawa couplings associated with the three $\bar{\bf 5}$ 
fields are not hierarchically separated when $d_{\bar{\bf 5}}$ is not much 
less than $L$.  This is along the line of the idea in~\cite{HMR}.
On the other hand, such a choice of $d_{\bar{\bf 5}}$ maintains the 
hierarchical structure of the masses and mixing angles in the quark sector, 
when $d_{\bf 10}/L$ and $d_H/L$ are chosen the same as before.  The main 
exception is that the hierarchy of the down-type quark masses becomes a 
little smaller, as we see below, which is actually in good agreement with 
observation.

\subsection{Charged Lepton Mass Spectrum}
\label{ssec:chargedleptons}

The masses of the charged leptons derive from the Yukawa interaction
\begin{equation}
 {\cal L} = \lambda^e_{ai} \bar{e}_a\,l_i\, h^* \,,
\end{equation}
where $\lambda^e_{ai}$ is generated in analogy to (\ref{eq:overlap}), 
except with a large width $d_{\bar{\bf 5}}$ to the wavefunction of $l_i$. 
In the limit $d_{\bar{\bf 5}}\gg d_{\bf 10},d_H$ but still 
$d_{\bar{\bf 5}}\approx L$, 
$\lambda^e_{ai}$ obtains an AFS structure 
\begin{equation}
 \lambda^e_{ai} \sim \varphi^l_{i} 
   \left(y = \frac{d_H^2}{d_{\bf 10}^2 + d_H^2}
    y^{\bar{e}}_a; 
y^l_i \right) \; 
   e^{- \frac{(y^{\bar{e}}_a)^2}{2(d_{\bf 10}^2+d_H^2)}}\,,
\label{eq:lopsided}
\end{equation}
where $y^{\bar{e}}_a$ is the center coordinate of the $\bar{e}_a$ 
wavefunction relative that of the Higgs, and $\varphi^l_{i}(y; y^l_i)$ is 
the broad-width wavefunction of $l_i$, centered at $y^l_i$.   Scanning 
$y^l_i$ and $y^{\bar{e}}_a$, the first factor becomes a random 
coefficient of order unity for all elements of the $3\times 3$ Yukawa 
matrix, while the second factor determines the flavor suppression. As the 
three eigenvalues are roughly the same as the three flavor suppression 
factors, the distributions of the three eigenvalues, $\lambda^e_{1,2,3}$, 
should be roughly the same as those of the three AFS suppression factors.

\begin{figure}[t]
\begin{center}
\begin{tabular}{ccc}
\includegraphics[width=0.3\linewidth]{Figures/D1L10d08/U1.eps} &
\includegraphics[width=0.3\linewidth]{Figures/D1L10d08/U2.eps} &
\includegraphics[width=0.3\linewidth]{Figures/D1L10d08/U3.eps} \\
$\log_{10}\lambda_u$ & $\log_{10}\lambda_c$ & $\log_{10}\lambda_t$ \\
\includegraphics[width=.3\linewidth]{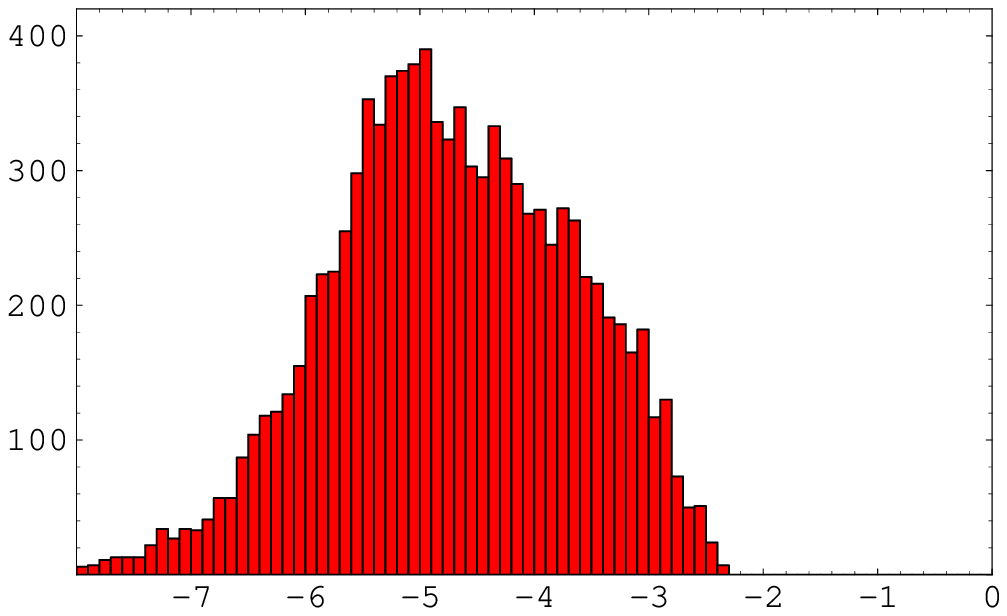} &
\includegraphics[width=.3\linewidth]{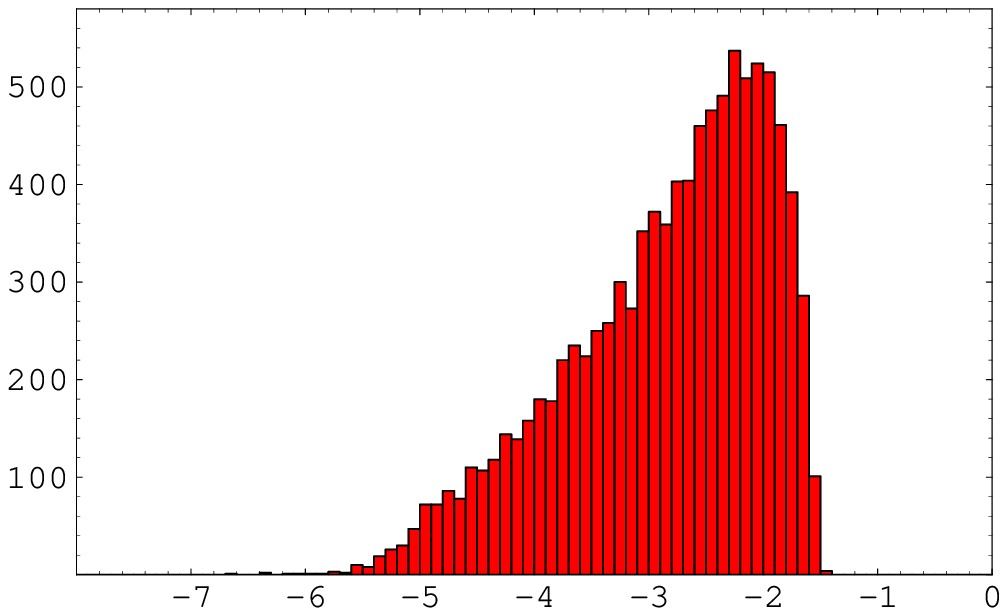} &
\includegraphics[width=.3\linewidth]{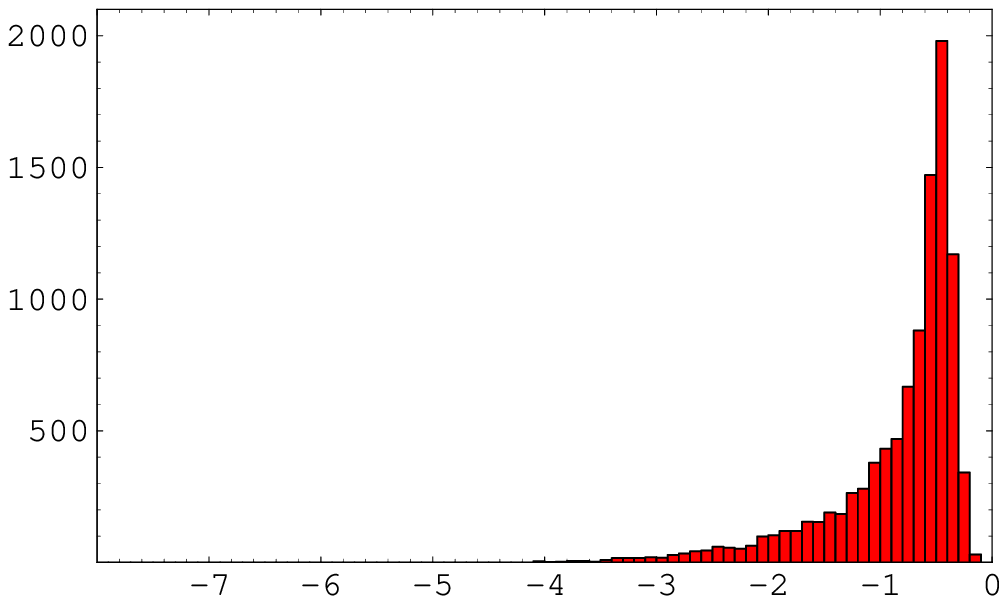} \\
$\log_{10}\lambda_{d,e}$ & $\log_{10}\lambda_{s,\mu}$ & 
$\log_{10}\lambda_{b,\tau}$  
\end{tabular}
\caption{\label{fig:5vs10} Distributions for Yukawa eigenvalues from
a Gaussian landscape on $S^1$:  $g_{\rm eff}=0.2$ for all overlap integrals 
and the Gaussian widths are $d/L = 0.08$ for all fields except for 
$\bar{d}$ and $l$, which have $d_{\bar{\bf 5}}/L = 0.3$.}
\end{center}
\end{figure}

Therefore, in the Gaussian landscape on $S^1$ the shape of the 
distribution functions of the charged lepton Yukawa eigenvalues should be 
like those in the left panel of Figure~\ref{fig:fndistr}.  We will 
discuss the overall range of the hierarchy shortly.  If we use a common 
width parameter for ${\bf 10}=(q,\bar{u},\bar{e})$ and a different common 
width for $\bar{\bf 5}=(\bar{d}, l)$,\footnote{It is an interesting 
theoretical question whether higher dimensional field theories can give 
rise to independent scanning of the centers of each of the 
$q,\bar{u},\bar{d},l,\bar{e}$ wavefunctions while preserving the 
SU(5)$_{\rm GUT}$ symmetric widths.} 
then the distributions of eigenvalues for the down quark sector should be 
the same as those of the charged lepton sector.  On the other hand, the 
up sector Yukawa couplings involve two fields in the ${\bf 10}$ 
representation, so the distribution of eigenvalues should be closer to 
those in the central panel of Figure~\ref{fig:fndistr} (see 
section~\ref{sec:toy1} regarding the limitations of applying the AFS 
approximation to small eigenvalues).  The results of a numerical 
simulation, displayed in Figure~\ref{fig:5vs10}, confirm that the AFS 
approximation captures the width-parameter dependence of the distribution 
functions. 
Although we do not present numerical results for Gaussian landscapes
other than on $S^1$, the approximate analytic distribution functions
for the AFS suppression factors on $T^2$, $S^2$, and $S^3$ are presented 
in Figure~\ref{fig:FN-D23}.  In principle one can generalize the analytic 
discussion to many geometries.    

In the limit of broad $\bar{\bf 5}$ wavefunctions, 
$d_{\bar{\bf 5}}\gg d_{\bf 10},d_{H}$ (but $d_{\bar{\bf 5}}\approx L$ so 
that values of $\varphi^l_{i}$ remain random), the logarithmic range 
covered by the charged lepton Yukawa eigenvalues is 
\begin{equation}
\Delta\ln\epsilon = -\frac{L^2}{8(d_{\bf 10}^2 + d_H^2)}\,.
\end{equation}
For simplicity we choose $d_{\bf 10}$ and $d_H$ to be the same, giving
$\Delta\ln\epsilon = -(L/d)^2/16$. This is $3/4$ of the result 
$\Delta\ln\epsilon = -(L/d)^2/12$ when all of the three widths are the 
same.  In the numerical results of Figure~\ref{fig:5vs10}, the overall 
hierarchy in the up sector is slightly larger than that of the 
down/charged-lepton sectors, consistent with these analytic expectations.

In Gaussian landscapes with $d_{\bar{\bf 5}}\gg d_{{\bf 10}}, d_{H}$, 
$\lambda_{b,\tau}$ tend to be predicted larger than $\lambda_t$.  This 
is of course a problem if there is no factor such as $\tan\beta$ in the
two-Higgs doublet model.  This problem can be solved in two different 
ways. One is that there may be environmental selection in favor of a 
large top Yukawa coupling, which makes the observed top Yukawa coupling at
the far upper end of its prior distribution, while the bottom and tau 
Yukawa couplings are not.  This scenario does not work very well in the 
Gaussian landscape on $S^1$, but it may work in Gaussian landscapes in 
higher dimensions, since the distributions of the largest eigenvalues 
receive a high-end tail in $D>1$ extra dimensions (see 
section~\ref{sec:Geometry}).  The other solution is that the coefficient 
$g_{\rm eff}$ of overlap integration is smaller for the 
down-type/charged-lepton sectors than that for the up-type sector.  
Since the overall hierarchy for the up-type sector is larger than in 
the down and charged lepton sectors, and because the distributions for
the smallest eigenvalues are very broad, this can be done while still 
maintaining the order of magnitude agreement between $\lambda_{d,e}$ 
and $\lambda_u$.

It is interesting to note a difference between Gaussian landscapes with 
large $d_{\bar{\bf 5}}$ and the AFS models that accomplish neutrino 
anarchy with very weak AFS for the lepton doublets.  In these AFS models, 
the logarithmic range of the hierarchy in the down-type/charged-lepton 
sectors is half of that of the up-type sector.  On the other hand, this 
ratio is typically about $3/4$ in the Gaussian landscape on $S^1$ 
(when $d_{\bf 10}=d_H$).  Of course the precise number $3/4$ may be 
different for different numbers of dimensions, geometry, and choice of 
$d_{\bf 10}/d_H$.  The difference between AFS and Gaussian landscapes 
stems from the fact that in the latter the AFS suppression factors 
associated with ${\bf 10}$ are different for the up-type and 
down-type/charged-lepton Yukawa matrices,
\begin{equation}
\epsilon^{q, \bar{u}({\rm u\mbox{-}sector})} \sim 
e^{- \frac{d_{\bf 10}^2 + d_H^2}{d_{\bf 10}^2 + 2 d_H^2}
        \frac{y_a^2}{2 d_{\bf 10}^2  } }, \qquad 
\epsilon^{q,\bar{e}({\rm d/e\mbox{-}sector})} \sim 
          e^{- \frac{y_a^2}{2(d_{\bf 10}^2+d_H^2)}}\,.
\end{equation}
In other words, the flavor suppression factors are not determined by 
symmetry charges intrinsically assigned to fields in the {\bf 10} 
representation, but depend on the nature of the other fields.  In our 
universe, the down-type/charged-lepton sector hierarchy is not as small 
as half of that of the up-sector, and Gaussian landscapes have enough 
flexibility to accommodate this situation.

\subsection{Neutrino Mass Hierarchy}
\label{ssec:seesaw}

Very small neutrino masses are obtained via the seesaw mechanism.  
Specifically, we assume left-handed neutrino masses derive from the 
effective dimension-five operators, 
\begin{equation}
 {\cal L}_{\rm eff} = \frac{C_{ij}}{M}\,l_i\,l_j\,h\,h\,,
\end{equation}
which are generated after integrating out heavy right-handed neutrinos, 
with interactions  
\begin{eqnarray}
{\cal L} =
M c_{\alpha\beta}\,\overline{\nu}_{R\alpha}\,\overline{\nu}_{R\beta}\, 
+\lambda^{\nu}_{\alpha i}\,\overline{\nu}_{R\alpha}\,l_i\,h\,;
\qquad 
C_{ij} = \left( \lambda^{\nu\, T} \, c^{-1} \, \lambda^\nu \right)_{ij} \,.
\label{eq:seesaw}
\end{eqnarray}
In a Gaussian landscape the neutrino Yukawa couplings 
$\lambda_{\alpha i}^\nu$ are generated by scanning the center coordinates 
of Gaussian wavefunctions, analogous to the generation of 
$\lambda_{ij}^{u,d}$.  Yet the low energy observables in the neutrino 
sector also depend on $Mc_{\alpha\beta}$, so we need to introduce some 
additional assumptions regarding how the Majorana mass terms of the 
right-handed neutrinos are generated.  The traditional approach has been 
to assume an AFS pattern in the mass matrix $Mc_{\alpha\beta}$.  However 
our Gaussian landscapes are not based on the breaking of any flavor 
symmetry, but on the overlap integration of localized wavefunctions.  
Therefore we incorporate this principle into the generation of 
right-handed Majorana mass terms. 

Right-handed neutrinos are singlets under the Standard Model gauge group, 
but they are charged in more unified gauge groups.\footnote{The term 
``unified gauge group'' is here used in a loose sense.  For instance, 
we do not assume that the Standard Model gauge group is unified as an 
effective field theory in 3+1 dimensions.}  
For example, unified gauge groups such as $\SU(4)_C \times \SU(2)_L 
\times \SU(2)_R$, $\SO(10)$, and $E_6\to E_8$ contain a $B-L$ symmetry, 
and the right-handed neutrinos carry non-vanishing charges under 
this symmetry.  In order for charged particles to have Majorana mass 
terms, some scalar fields on 3+1 dimensions have to be inserted to form 
gauge-invariant operators, and the expectation values of these scalar 
fields convert the operators into Majorana mass terms:
\begin{equation}
{\cal L} = \vev{\phi_{SB}}\,\overline{\nu}_R\,\overline{\nu}_R\,, 
\qquad {\rm ~or~} \qquad 
{\cal L} =
\vev{\phi'_{SB}\phi'_{SB}}\,\overline{\nu}_R\,\overline{\nu}_R\,. 
\label{eq:MRmass}
\end{equation}
The scalar fields $\phi_{SB}$ or $\phi'_{SB}$ are singlets of the 
Standard Model gauge groups, but carry charges under whatever symmetry 
the right-handed neutrinos are charged. In other words, they 
are moduli fields describing the symmetry breaking of the more unified 
symmetry.

Since the Majorana mass terms (\ref{eq:MRmass}) involve vacuum expectation 
values of scalar fields, the coefficients of the mass terms will involve 
overlap integrations:
\begin{equation}
Mc_{\alpha\beta}\propto\int dy\,\varphi^{SB}\,  
\varphi^{\overline{\nu}_{R}}_\alpha \varphi^{\overline{\nu}_{R}}_\beta\,,
\label{eq:overlap-NR}
\end{equation}
where $\varphi^{\overline{\nu}_{R}}_\alpha$ is the zero-mode wavefunction 
of the $\alpha$ copy of the right-handed neutrino and $\varphi^{SB}$ 
represents all of the other effects of localization in the extra
dimension(s), including the wavefunctions of the symmetry-breaking moduli
fields $\phi_{SB}$ and/or $\phi'_{SB}$.  These fields have nothing to 
do with the Higgs boson, and hence there is no reason to believe that 
$\varphi^{SB}$ is localized at the same place in the extra dimensions as 
the Higgs boson wavefunction.  Thus while the AFS structure of all of the 
Yukawa couplings $\lambda^{u,d,e,\nu}$ is due to an overlap between fermion 
wavefunctions and that of the Higgs, any flavor structure in the Majorana 
mass term follows from an overlap involving $\varphi^{SB}$.  In particular, 
unless there is a strong correlation between the wavefunctions of 
$\varphi^{SB}$ and the Higgs, the flavor structure of the mass matrix 
$Mc_{\alpha\beta}$ is statistically independent of that of the Yukawa 
matrices.

We study the consequences of this new flavor structure in 
sections~\ref{sssec:narrow} and~\ref{sssec:broad}.  However, before we 
begin some comments are in order.  Our first remark concerns the overall 
mass scale $M$ of the right-handed neutrinos.  The overlap integration 
in (\ref{eq:overlap-NR}) sets the flavor structure of $Mc_{\alpha\beta}$, 
but does not say anything about the overall scale of the symmetry 
breaking.  Thus an extra assumption has to be introduced in order to set 
the distribution of the overall scale $M$.  Even if we know this 
distribution from theoretical considerations, it may be modulated by 
environmental selection related to leptogenesis.  Therefore we set aside 
this issue and instead focus only on the hierarchy among neutrino masses 
and mixing angles in the lepton sector, which can be determined 
independent of the overall scale of the neutrino masses.  This is 
equivalent to studying a fixed $M$ cross section of the full landscape.  
Our second comment concerns the form of the wavefunctions of the 
right-handed neutrinos and the symmetry breaking source $\varphi^{SB}$.  
To date we have very little knowledge about these fields; yet in a 
Gaussian landscape their most crucial aspect will be whether in the extra 
dimensions their wavefunctions are localized or not.  Therefore we set 
aside the theoretical origin of these fields and simply represent them 
with Gaussian wavefunctions (\ref{eq:Gaussian}), while considering the 
width parameters $d_N$ and $d_{SB}$ as unknown.  By choosing $d_N$ and 
$d_{SB}$ to be large or small, we can simulate various possibilities.  
For example, Majorana mass terms generated by world-sheet instantons 
wrapped on topological cycles may be mimicked by a (possibly 
multi-centered) Gaussian wavefunction with a small $d_{SB}$.

It turns out that the mass matrices of the neutrino sector are quite 
different depending on whether the right-handed neutrinos have 
localized wavefunctions or not.  Therefore sections~\ref{sssec:narrow} 
and \ref{sssec:broad} are separately devoted to these two possibilities.
In either case, to generate a statistical ensemble of 
$\lambda^\nu_{\alpha i}$ and $c_{\alpha\beta}$ the center coordinates 
of wavefunctions are scanned randomly and independently for 
$\varphi^{SB}$, the right-handed neutrinos, lepton doublets, and the 
Higgs.

\subsubsection{Narrow Right-Handed Neutrino Wavefunctions}
\label{sssec:narrow}

If right-handed neutrinos have narrow wavefunctions, then the neutrino 
Yukawa matrix has a structure similar to that of charged leptons.  That 
is, we have 
\begin{equation}
\lambda^\nu_{\alpha i} \sim \varphi^l_{i} 
\left( y = \frac{d_H^2}{d_N^2 + d_H^2} y^{\overline{\nu}_R}_\alpha ; 
y^l_i \right)
e^{-\frac{(y^{\overline{\nu}_R}_\alpha)^2}{2(d_N^2 + d_H^2)}}
\equiv g_{\alpha i}\,\epsilon^{D}_\alpha\,. 
\label{eq:nuYukawa-str}
\end{equation}
This is (\ref{eq:lopsided}) with $d_{\bf 10}$ replaced by $d_N$; and again 
the $g_{\alpha i}$ are effectively random coefficients of order unity.  
The range of the hierarchy is 
$\Delta\ln\epsilon^D = -L^2/[8(d_N^2 + d_H^2)]$.

\begin{figure}[t]
\begin{center}
\begin{tabular}{ccc}
\includegraphics[width=0.3\linewidth]{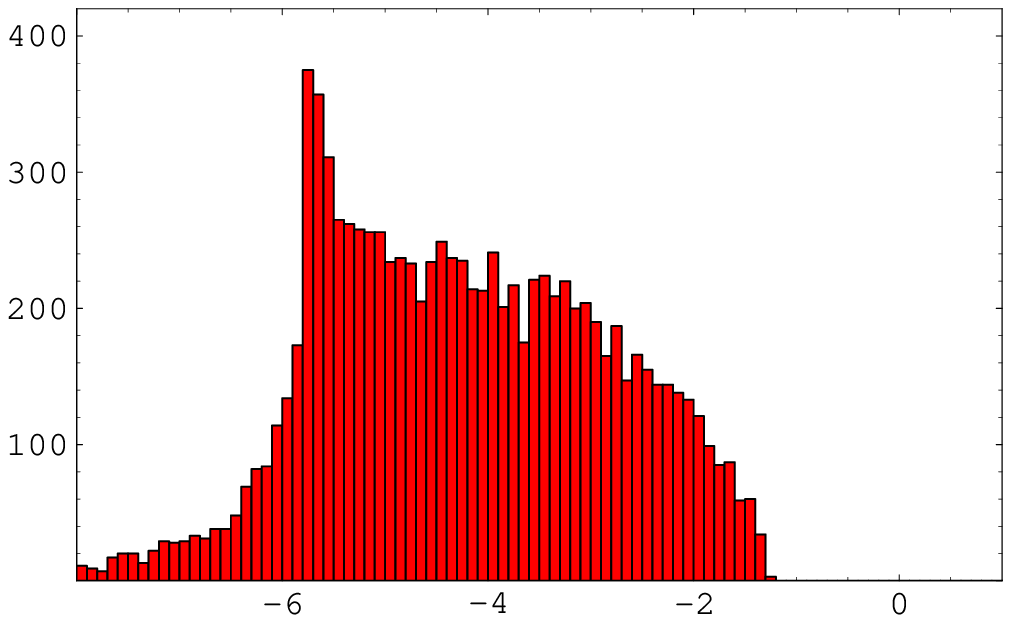} &
\includegraphics[width=0.3\linewidth]{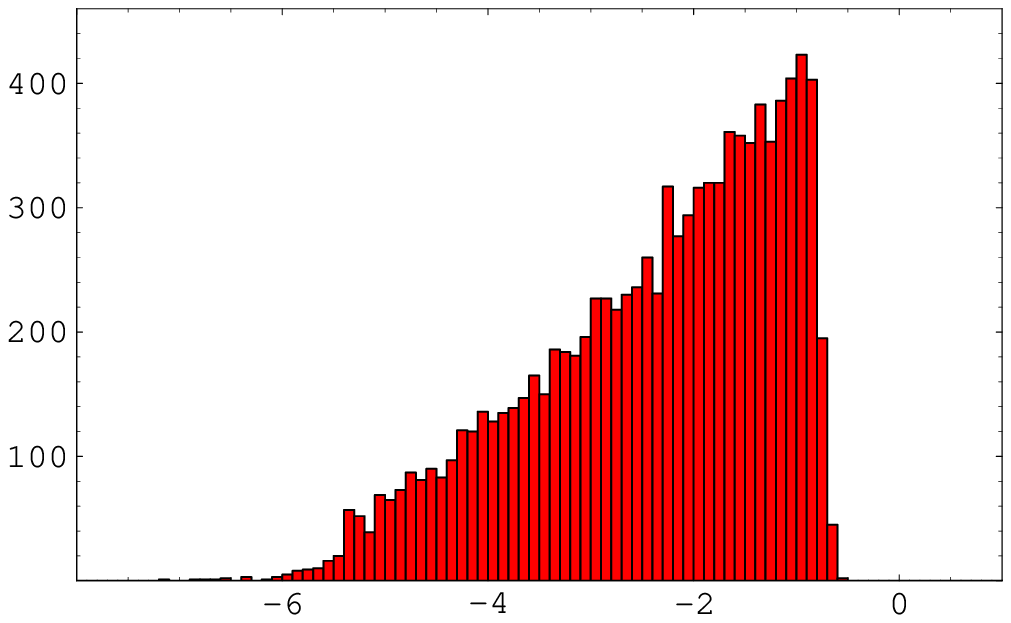} &
\includegraphics[width=0.3\linewidth]{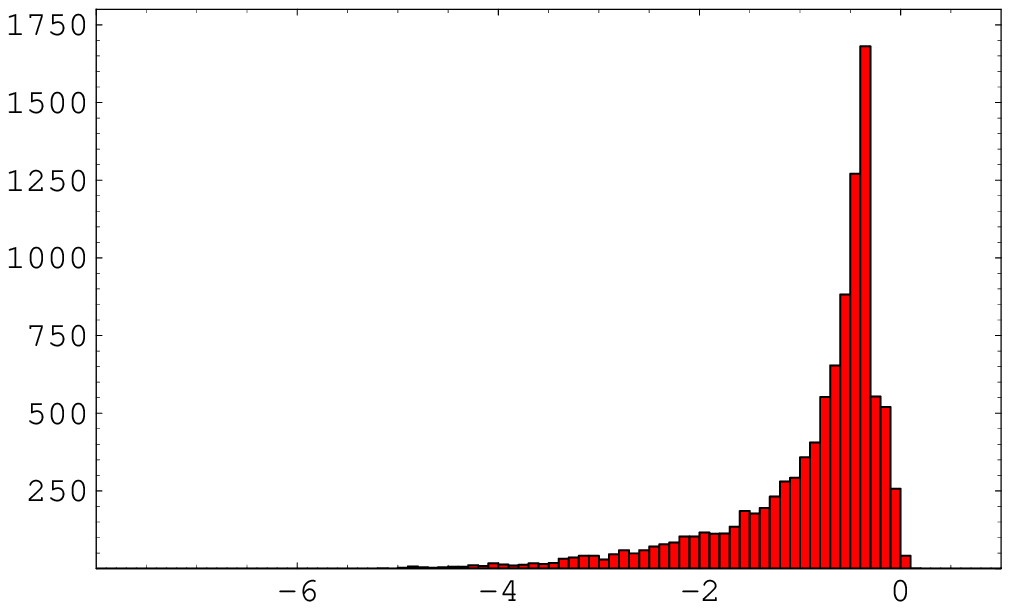} \\
$\log_{10}(M_1/M)$ & $\log_{10}(M_2/M)$ & $\log_{10}(M_3/M)$ \\
\includegraphics[width=0.3\linewidth]{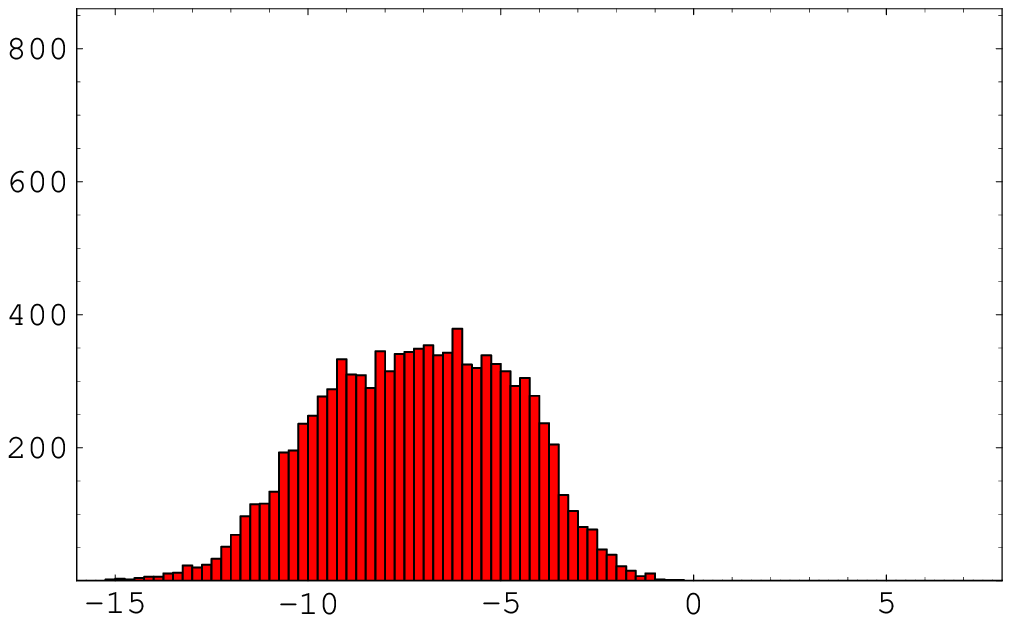} &
\includegraphics[width=0.3\linewidth]{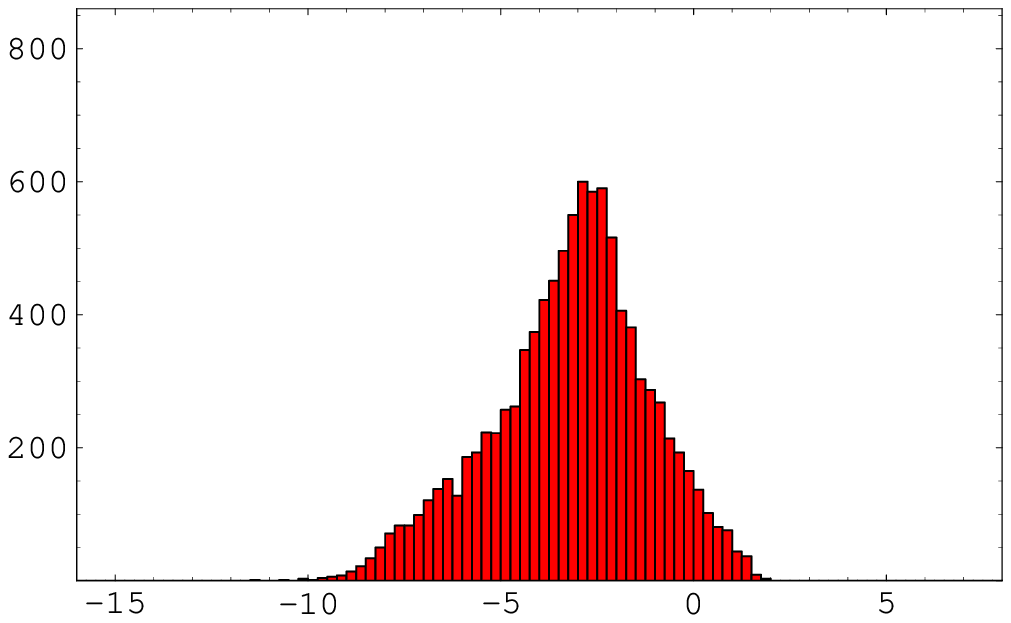} &
\includegraphics[width=0.3\linewidth]{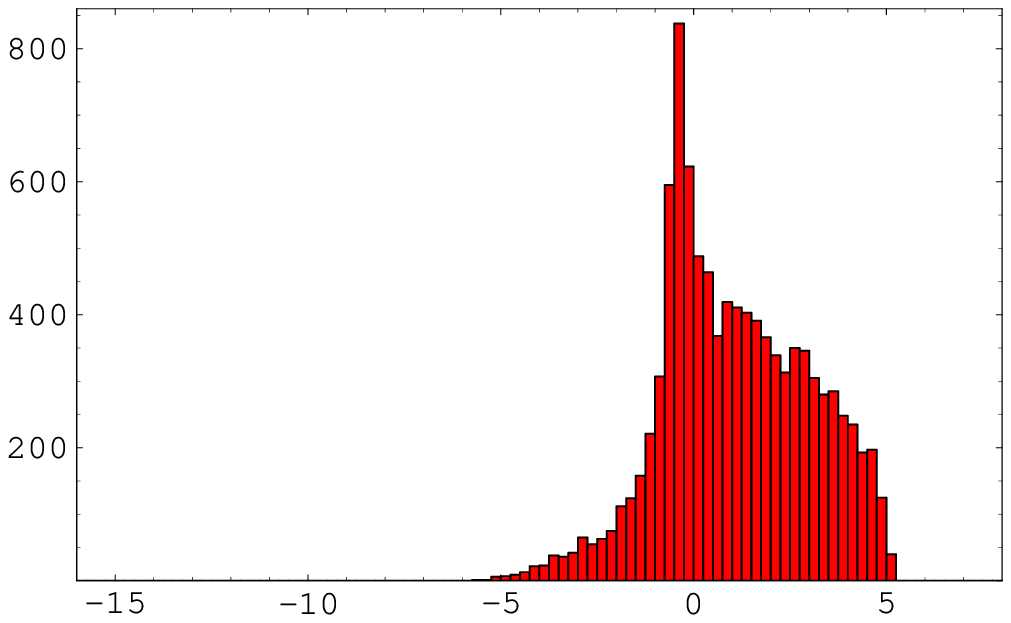} \\
$\log_{10}(m_1/m_\nu)$ & $\log_{10}(m_2/m_\nu)$ & $\log_{10}(m_3/m_\nu)$ 
\end{tabular}
\caption{\label{fig:narrow-narrow} Distributions of the eigenvalues of 
the right-handed neutrino mass matrix and the left-handed seesaw Majorana 
mass matrix, normalized by $M$ and $m_\nu\equiv\vev{h}^2/M$, respectively. 
The Gaussian landscape on $S^1$ is used for this simulation, with 
$d/L=0.08$ for the wavefunctions of $\overline{\nu}_{R\alpha}$, 
$\varphi^{SB}$, and the Higgs while $d/L=0.3$ for the lepton doublets.}
\end{center}
\end{figure}

Meanwhile, the Majorana mass matrix $c_{\alpha\beta}$ of right-handed 
neutrinos is almost diagonal, because the off-diagonal entries are 
suppressed due to the (statistically) small overlap of wavefunctions 
between different right-handed neutrinos.  If the symmetry-breaking source 
$\varphi^{SB}$ is also very localized, then the Majorana mass matrix 
has the approximate structure 
\begin{equation}
 c_{\alpha\beta} \sim \delta_{\alpha\beta}\, 
  e^{- \frac{(y^{\overline{\nu}_R}_\alpha - y^{SB})^2}{2 d_{SB}^2 + d_N^2}} 
  \equiv \delta_{\alpha\beta}\,\epsilon_\alpha^{M}\,, 
\label{eq:RHNmassFN}
\end{equation}
where $d_{SB}$ is the width of Gaussian wavefunction of $\varphi^{SB}$ and
$y^{SB}$ is the center coordinate of this wavefunction.  The positions 
$y^{\overline{\nu}_R}_\alpha$ and $y^{SB}$ are scanned randomly and the 
largest, middle and smallest $\epsilon_\alpha$ become the three 
eigenvalues of the right-handed neutrino mass matrix.  In particular, 
their distribution functions should be like those of the AFS suppression 
factors $\epsilon_{1,2,3}$ in Figure~\ref{fig:fndistr} or~\ref{fig:FN-D23}.  
For Gaussian landscapes on $S^1$ we have 
$\Delta\ln\epsilon^M = -L^2/[4(2d_{SB}^2+ d_N^2)]$.  Results of a 
numerical simulation on the $S^1$ landscape are shown in the first row of 
Figure~\ref{fig:narrow-narrow}, confirming the theoretical expectations 
so far. 
If the symmetry-breaking source $\varphi^{SB}$ does not have a localized 
wavefunction, then we can take $d_{SB}$ to be very large.  Then  
$\Delta\ln\epsilon^M\approx 0$, meaning that there is not much hierarchy 
among the mass eigenvalues of right-handed neutrinos. 

\begin{figure}[t]
\begin{center}
\begin{tabular}{ccc}
\includegraphics[width=0.3\linewidth]{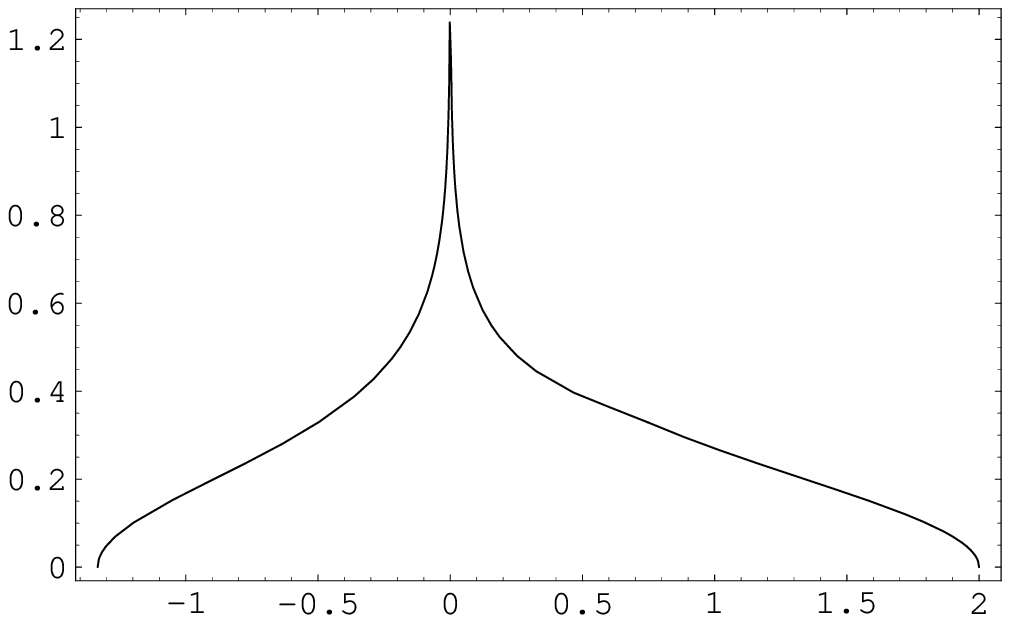} &
\includegraphics[width=0.3\linewidth]{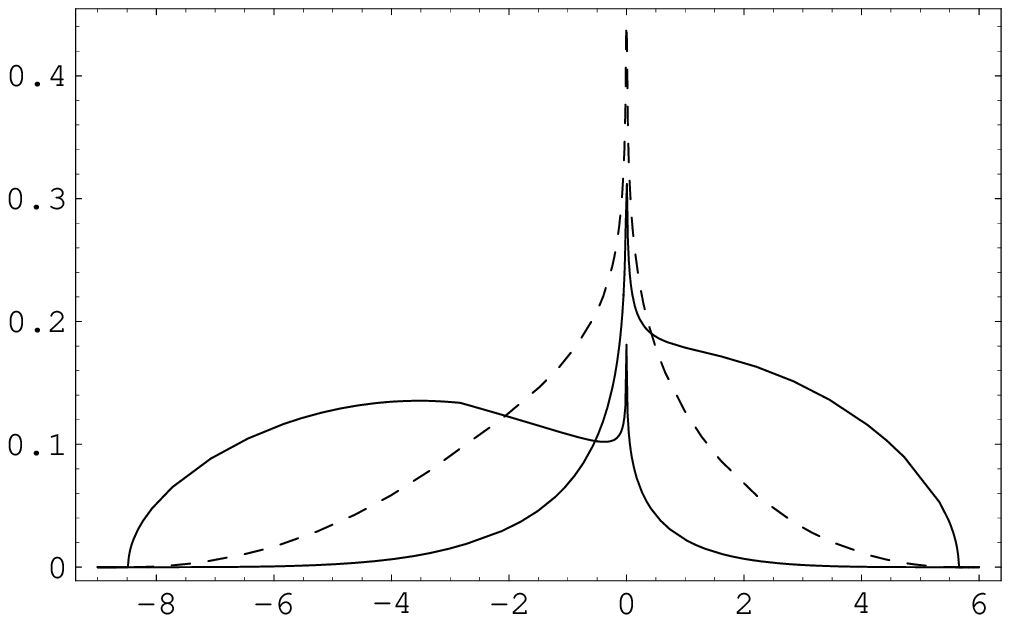} &
\includegraphics[width=0.3\linewidth]{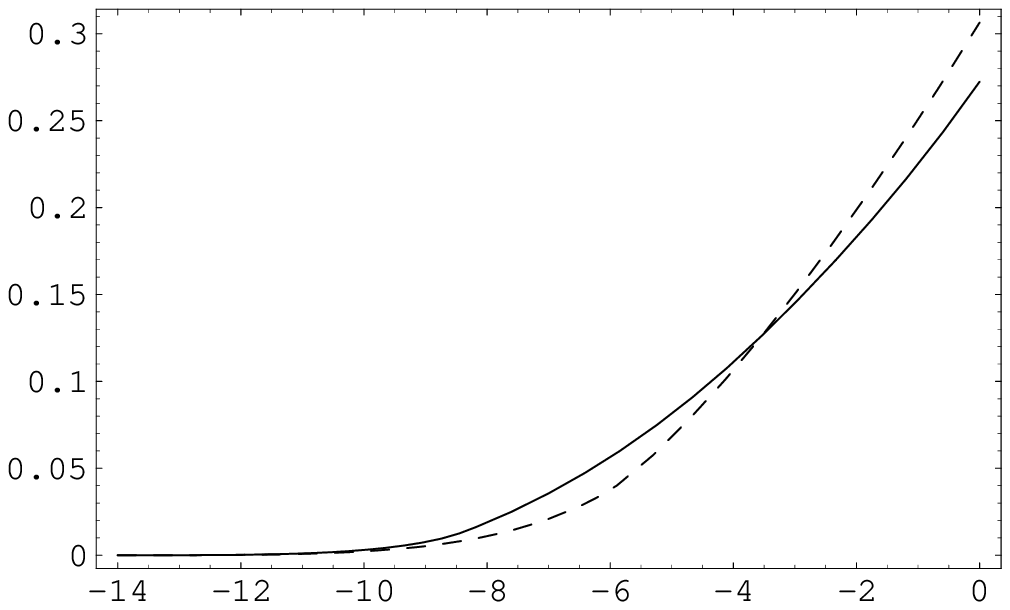} \\
$f^{S^1}_{\rm ss}(z^{\rm ss})$ &
$\log_{10}(m_{1,2,3}/m_\nu)$ &
$\log_{10}(m_1/m_2)$, $\log_{10}(m_2/m_3)$ \\
\includegraphics[width=0.3\linewidth]{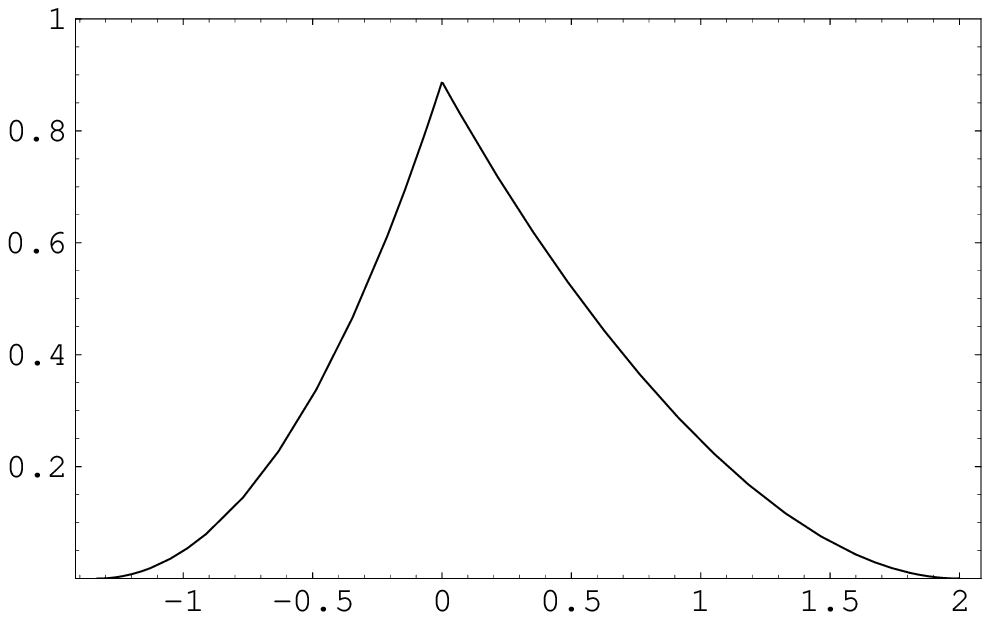} &
\includegraphics[width=0.3\linewidth]{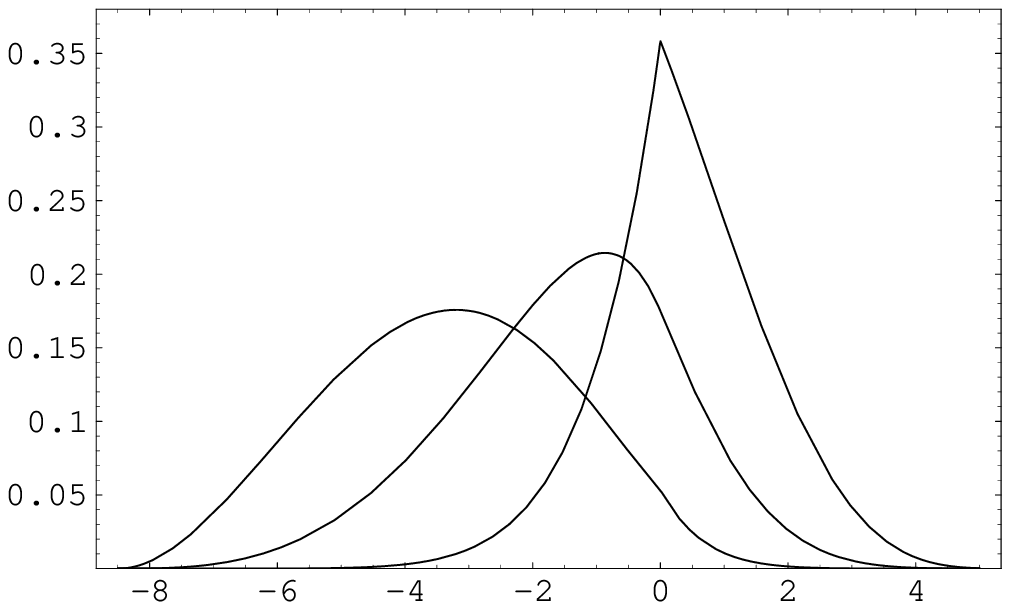} &
\includegraphics[width=0.3\linewidth]{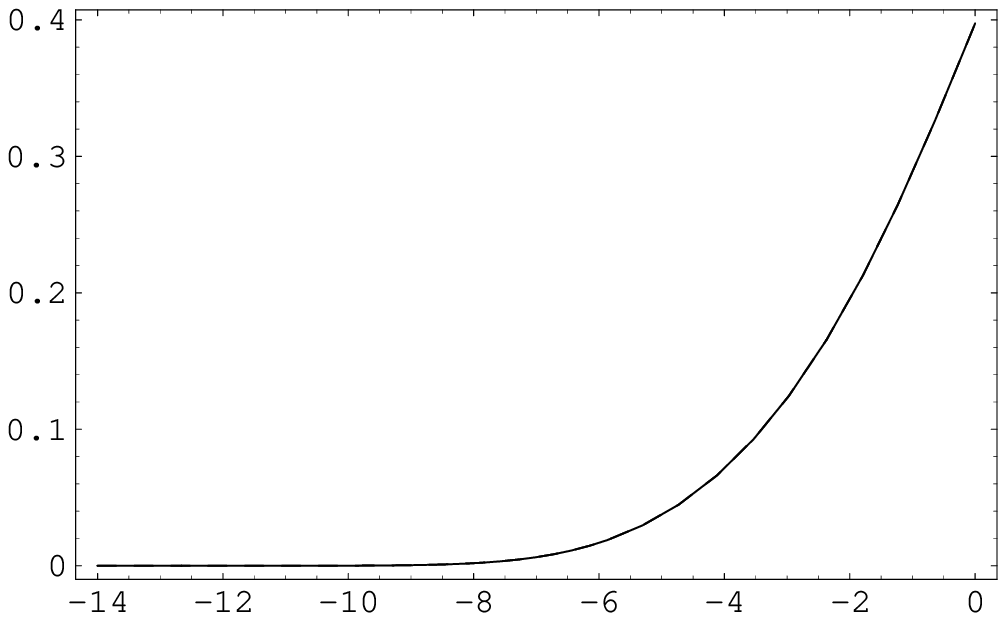} \\
$f^{S^2}_{\rm ss}(z^{\rm ss})$ &
$\log_{10}(m_{1,2,3}/m_\nu)$ &
$\log_{10}(m_1/m_2)$, $\log_{10}(m_2/m_3)$
\end{tabular}
\caption{\label{fig:neutrinoegval-analytic} Distributions of density 
functions $f_{\rm ss}(z^{\rm ss})$ (left column), low-energy left 
handed neutrino eigenvalues (middle column), and the mass ratios 
$(m_1/m_2)$ and $(m_2/m_3)$ (right column).  The top row is for $S^1$
while the bottom row is for $S^2$ (to reduce computation time we have
approximated $f_{S^2}$ with a polynomial that is always within a part in
a thousand of the corresponding expression in (\ref{eq:fS123})).  In the 
top row the distributions for the middle eigenvalue $m_2/m_\nu$ and the 
ratio $m_2/m_3$ are displayed with a dashed line.  The two mass ratios 
are the same on $S^2$.  For these curves $d_H/L=d_N/L=d_{SB}/L=0.08$.}
\end{center}
\end{figure}

From (\ref{eq:seesaw}), (\ref{eq:nuYukawa-str}), and (\ref{eq:RHNmassFN}) 
the mass matrix of low-energy neutrinos is approximately  
\begin{equation}
C_{ij}\sim\sum_\alpha g_{\alpha i} g_{\alpha j} 
\frac{(\epsilon^D_\alpha)^2}{\epsilon^M_\alpha} \,. 
\label{eq:CijAFS}
\end{equation}
Since we introduce no correlation between the Higgs and $\varphi^{SB}$, 
we do not expect net statistical cancellations in the ratio 
$(\epsilon^D_\alpha)^2/\epsilon_\alpha^M$.  This contrasts sharply with 
flavor theories that assume an AFS structure for the Majorana mass matrix 
of right-handed neutrinos.  Let the distributions of $\ln \epsilon^D$ and 
$\ln \epsilon^M$ be denoted by
\begin{equation}
dP(y^D) = f_D(y^D)\,dy^D\,, \qquad 
dP(y^M) = f_M(y^M)\,dy^M\,, \
\end{equation}
where we introduce the new variables 
$y^{D,M}\equiv\ln\epsilon^{D,M}/\Delta\ln\epsilon^D$ ($y^D$ is normalized 
to run from zero to one; note that the variable $y^M$ does not necessarily
run from zero to one).   Because of the form of $\epsilon^D$ and 
$\epsilon^M$ in (\ref{eq:nuYukawa-str}) and (\ref{eq:RHNmassFN}), $f_D(y)$ 
and $f_M(y)$ are given by the volume distribution function $f(y)$ of 
(\ref{eq:defoff}), after re-normalizing the distribution function and 
rescaling the argument, if necessary.  To approximate the distributions of 
eigenvalues of $\ln C_{ij}$, we first introduce the variable 
$z^{\rm ss}\equiv\ln\left[(\epsilon^D)^2/\epsilon^M\right]
/\Delta\ln\epsilon^D$.  The distribution of $z_{\rm ss}$ follows from 
appropriately convoluting the distributions of $\ln\epsilon^D$ and 
$\ln\epsilon^M$, 
\begin{equation}
\frac{dP(z^{\rm ss})}{dz^{\rm ss}}\equiv f_{\rm ss}(z^{\rm ss}) = 
\int\frac{1}{2}\, f_D\left(z/2\right) f_M(z-z^{\rm ss})\, dz\,. 
\label{eq:ss-DM}
\end{equation}
Note that the variable $z^{\rm ss}$ runs from 
$-2(d_H^2+d_N^2)/(2d_{SB}^2+d_N^2)$ to 2 on $S^1$.  The distribution 
function $f_{\rm ss}$ can be calculated explicitly once the underlying
volume distribution function $f(y)$ is known; the results for the Gaussian 
landscapes on $S^1$ and $S^2$ are shown in 
Figure~\ref{fig:neutrinoegval-analytic}.   
Using the volume distribution function $f_{\rm ss}(z_{\rm ss})$, we 
obtain the distributions of the three eigenvalues of $C_{ij}$ by 
integrating the other two variables out of the combined 
distribution\footnote{The combined probability distribution of neutrino 
mass eigenvalues is discussed in~\cite{DDR} for the case with $f_D$ and 
$f_M$ proportional to a $\delta$-function.  Reference~\cite{HM} deals 
with the neutrino anarchy scenario, where the diagonalization effects on 
the eigenvalue distribution are also taken into account.} 
\begin{equation}
 dP(z^{\rm ss}_3,z^{\rm ss}_2,z^{\rm ss}_1) = 3! 
f_{\rm ss}(z^{\rm ss}_3) f_{\rm ss}(z^{\rm ss}_2) f_{\rm ss}(z^{\rm ss}_1)
\,\Theta(z^{\rm ss}_1-z^{\rm ss}_2)\,\Theta(z^{\rm ss}_2-z^{\rm ss}_3) 
\,dz^{\rm ss}_3 dz^{\rm ss}_2 dz^{\rm ss}_1\,,
\end{equation}
where $z^{\rm ss}_{1,2,3} = - 8[(d_H^2+d_N^2)/L^2]\ln (m_{1,2,3}/m_\nu)$,
with $m_\nu\equiv\vev{h}^2/M$.  Distribution functions of other 
observables such as a ratio of low-energy neutrino masses are also 
calculated from this combined probability distribution. Some examples 
are displayed in Figure~\ref{fig:neutrinoegval-analytic}.  Note that the 
low-energy neutrino masses tend to have a very large hierarchy, because 
the distribution function $f_{\rm ss}$ covers the enormous logarithmic 
range 
\begin{equation}
\Delta \ln m_\nu = 2 \Delta \ln \epsilon^D + \Delta \ln \epsilon^M 
= - \frac{1}{4} \left(\frac{L^2}{d_N^2 + d_H^2} +
    \frac{L^2}{2d_{SB}^2 + d_N^2} \right).
\label{eq:nu-hierarchy}
\end{equation} 
Specifically, the hierarchies of neutrino Yukawa matrix and the Majorana 
mass matrix add rather than cancel, c.f.~\cite{HMW, HM}.  Even when 
$d_{SB}$ is very large, low-energy neutrino masses have a hierarchy twice 
as large as that of eigenvalues of the neutrino Yukawa matrix. 

The low-energy neutrino eigenvalues of Figure~\ref{fig:narrow-narrow} 
show a remarkable similarity\footnote{The peak of the distribution of 
$\log_{10} m_2/m_\nu$ in Figure~\ref{fig:narrow-narrow} is near $-2$, as 
opposed to the expected location $0$, shown in 
Figure~\ref{fig:neutrinoegval-analytic}.  The peak of $m_2/m_\nu$ being 
lower than that of $m_3/m_\nu$ is not surprising. We have seen similar 
phenomena in Figures~\ref{fig:D=1-nocut}, \ref{fig:T2YukawaEgval},
\ref{fig:compareT2S2}, and~\ref{fig:5vs10}.  As mentioned earlier, these 
are likely due to the diagonalization effect.} 
in shape with the corresponding distributions in 
Figure~\ref{fig:neutrinoegval-analytic}, confirming that the theoretical 
derivation of distribution functions is fairly reliable.  Although 
results of a numerical simulation on the Gaussian landscape on $T^2$ are 
not presented here, we have confirmed that the distributions of 
$\ln(m_i/m_\nu)$ are similar to corresponding distributions in 
Figure~\ref{fig:neutrinoegval-analytic}. This success arises because the 
overlap integration for $\lambda^e_{a i}$, $\lambda^\nu_{\alpha i}$ and 
$c_{\alpha\beta}$ involves only two very localized wavefunctions.  
Although we have seen around (\ref{FN-fullargument}) that the analytic 
distribution functions based on the volume distribution function have a 
systematic error due to the compactness of the extra dimension(s), the 
error occurs only for overlap integrations involving {\em three} 
localized Gaussian wavefunctions.  Therefore these analytic results can 
be used to compute distribution functions of flavor observables in the 
lepton sector in other geometries for which numerical simulation is 
costly. 

The underlying geometry is reflected in the volume distribution functions 
$f_D$ and $f_M$. However, these are integrated once to obtain 
$f_{\rm ss}$, which smears the effects of geometry. The original $f$'s 
are integrated at least once in the distribution functions of 
$\ln (m_i/m_\nu)$, like those of the up-sector eigenvalues. Moreover, the 
distribution functions of observables such as $\ln(m_1/m_2)$ and 
$\ln(m_2/m_3)$ involve two integrations over the $f$'s, such that very 
little geometry dependence is left in the distribution of the 
observable.\footnote{The cuspy peaks in the distributions of $f_{\rm ss}$ 
and $\ln (m_i/m_\nu)$ on $S^1$ are only logarithmic; the $y^{-1/2}$ 
singularity of the original $f_{S^1}$ is integrated at least once in 
each of these.  We chose a fine binning in Figure~\ref{fig:narrow-narrow} 
in order to accentuate the logarithmic singularity and emphasize the
agreement between the simulation and analytic derivation. When a coarser
binning is chosen, the logarithmic singularities are smeared out. Thus, 
there is not much practical difference in the distributions of 
$\ln (m_i/m_\nu)$'s whether $S^1$ or $T^2$ or any other geometry is used.}

\subsubsection{Broad Right-Hand Neutrino Wavefunctions}
\label{sssec:broad}

\begin{figure}[t]
\begin{center}
\begin{tabular}{ccc}
\includegraphics[width=0.3\linewidth]{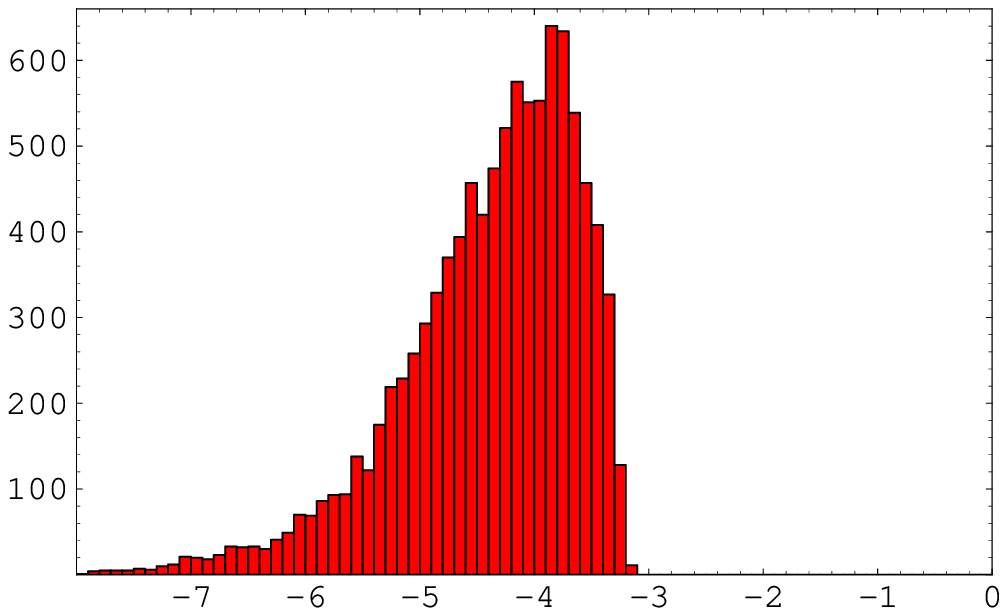} &
\includegraphics[width=0.3\linewidth]{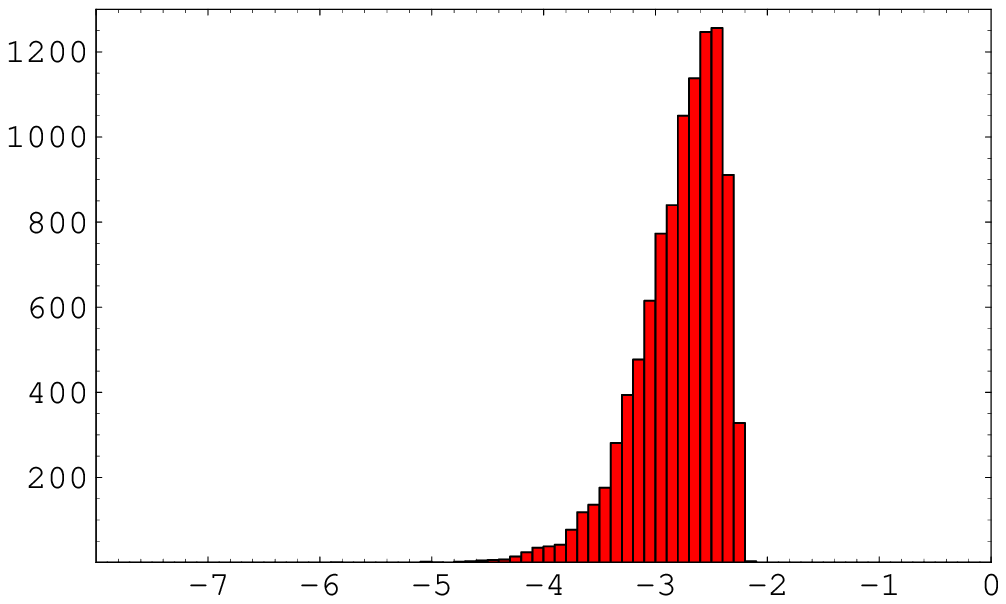} &
\includegraphics[width=0.3\linewidth]{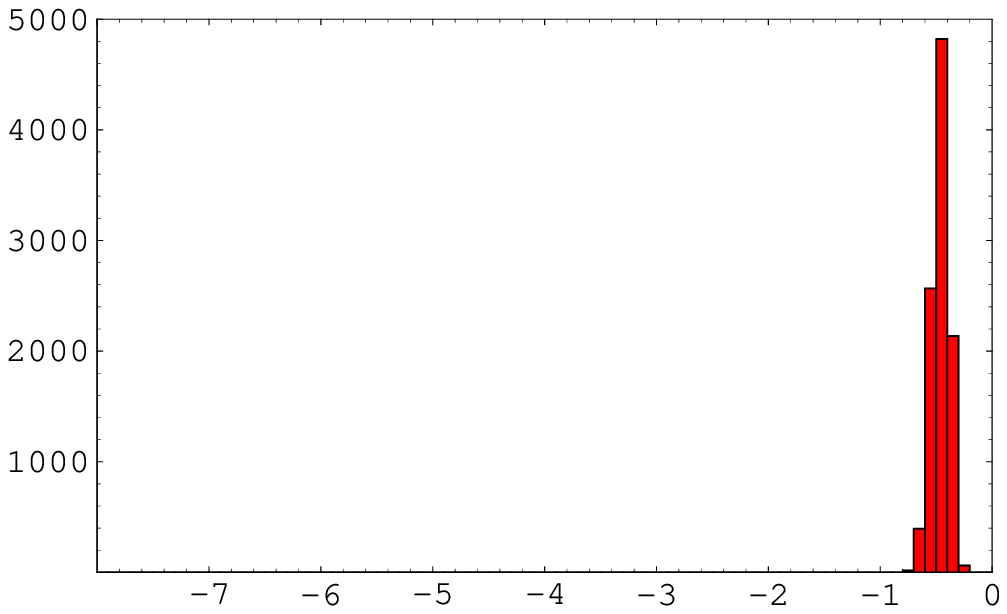} \\
$\log_{10}\lambda^\nu_1$ & $\log_{10}\lambda^\nu_2$ &
$\log_{10}\lambda^\nu_3$ \\
\includegraphics[width=0.3\linewidth]{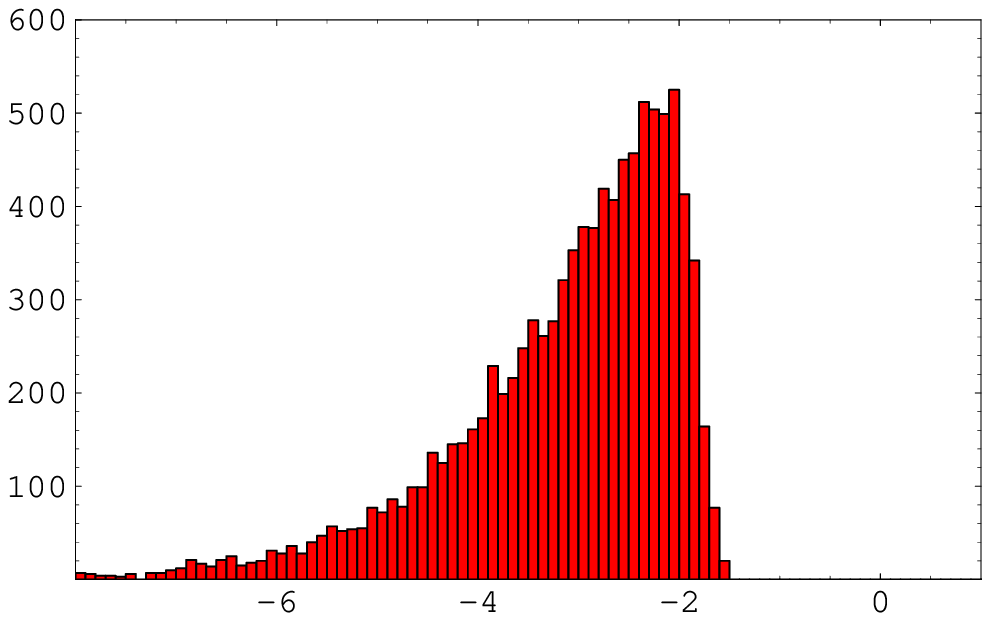} &
\includegraphics[width=0.3\linewidth]{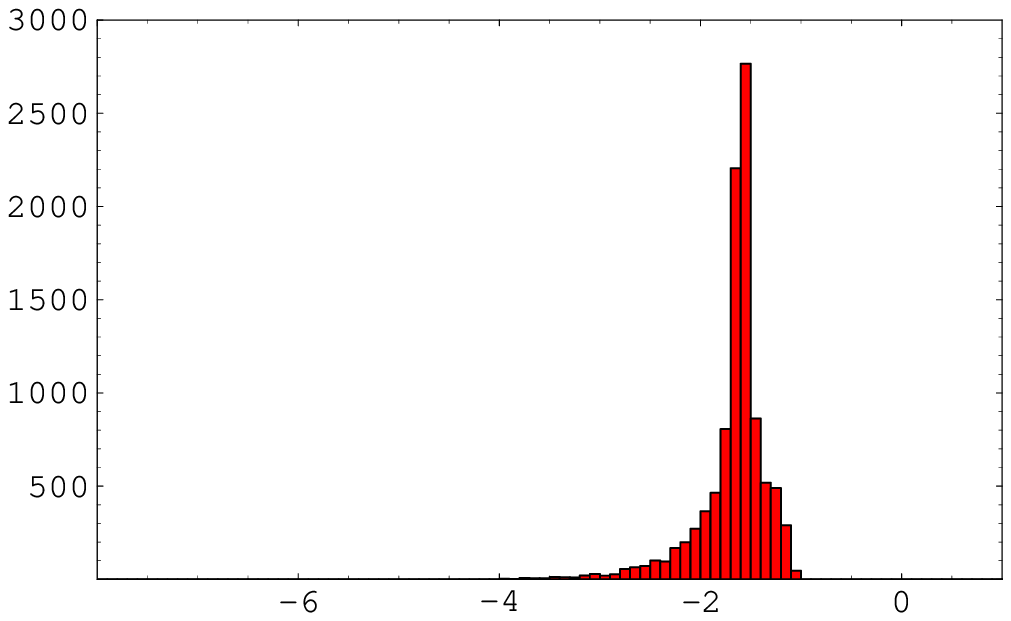} &
\includegraphics[width=0.3\linewidth]{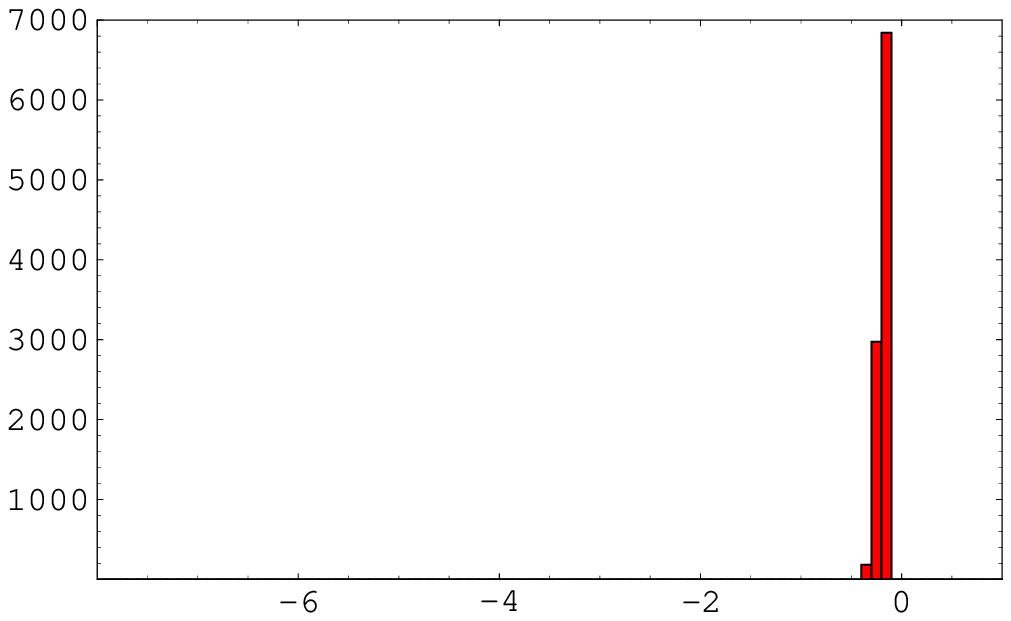} \\
$\log_{10}(M_1/M)$ & $\log_{10}(M_2/M)$ & $\log_{10}(M_3/M)$ \\
\includegraphics[width=0.3\linewidth]{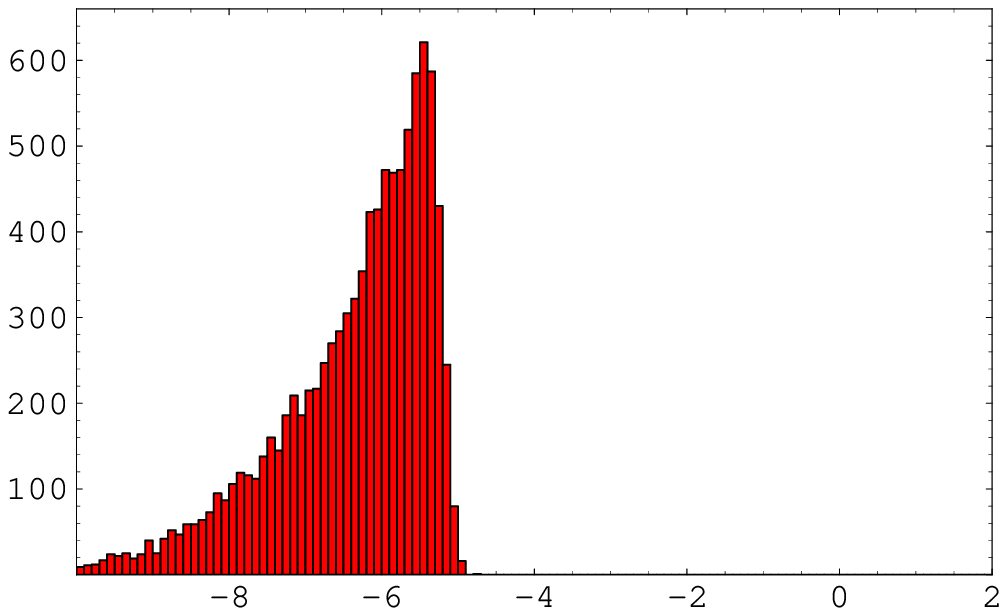} &
\includegraphics[width=0.3\linewidth]{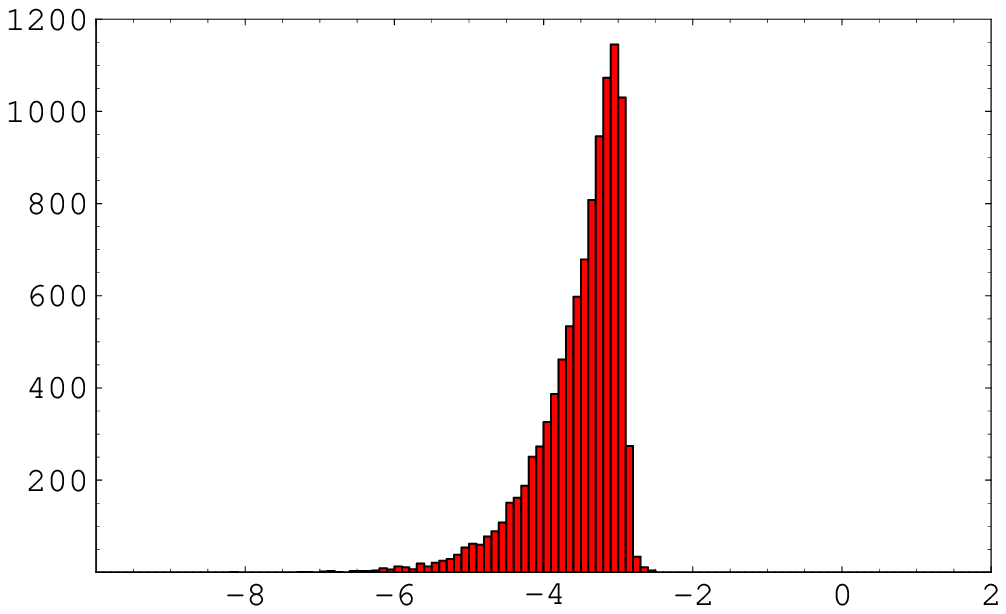} &
\includegraphics[width=0.3\linewidth]{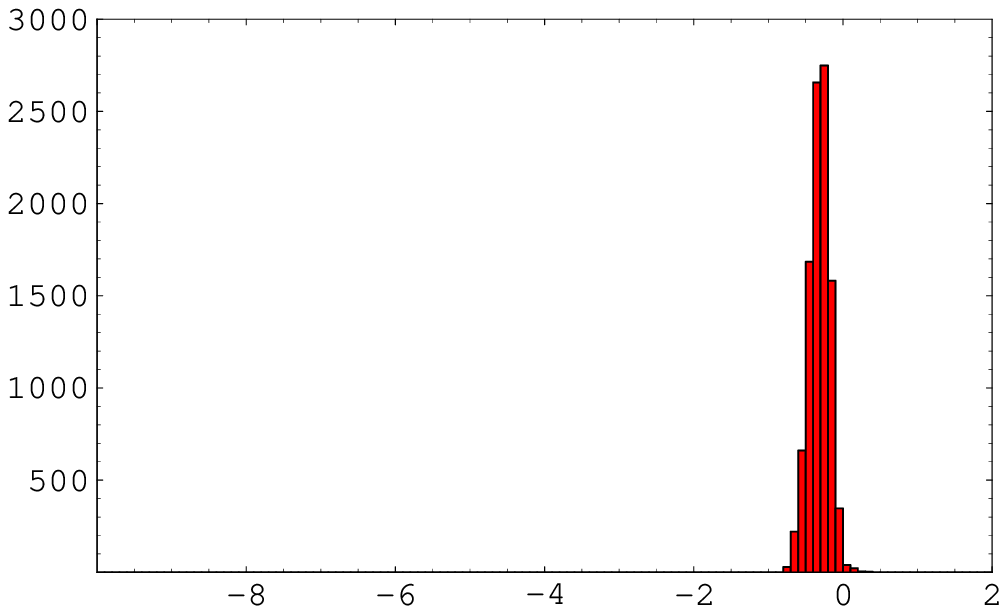} \\
$\log_{10} (m_1/m_\nu)$ & $\log_{10} (m_2/m_\nu)$ & $\log_{10} (m_3/m_\nu)$
\end{tabular}
\caption{\label{fig:broad-broad} Distributions of the eigenvalues of the
left-handed neutrino Yukawa matrix, the right-handed Majorana mass matrix,
and the low-energy left-handed Majorana mass matrix.  The numerical 
simulation uses a Gaussian landscape on $S^1$, with $d_H/L=0.08$ and 
$d_{SB}/L=d_N/L=d_{\bar{\bf 5}}/L=0.3$.}
\end{center}
\end{figure}

If the zero-mode wavefunctions of right-handed neutrinos are not
localized, i.e. the width parameter $d_N$ is not much less than $L$, 
then in the neutrino Yukawa coupling only the Higgs has a localized 
wavefunction.  Thus the overlap integral is evaluated around the peak of 
the Higgs wavefunction. Expanding the other wavefunctions,
\begin{eqnarray}
\varphi^l_{j}(y) & = & \varphi^l_j(y^h) + {\varphi^l_j}' (y^h)(y-y^h)
  + \frac{1}{2} {\varphi^l_j}'' (y^h) (y-y^h)^2\,, \\
\varphi^{\overline{\nu}_{R}}_\alpha (y) & = &
  \varphi^{\overline{\nu}_R}_\alpha (y^h) +
  {\varphi^{\overline{\nu}_R}_\alpha}' (y^h)(y-y^h)
  + \frac{1}{2} {\varphi^{\overline{\nu}_R}_\alpha}'' (y^h) (y-y^h)^2\,,
\end{eqnarray}
we see that the neutrino Yukawa matrix is of the form
\begin{equation}
\lambda^\nu_{i\alpha}\propto\varphi^l_{i}\varphi^{\overline{\nu}_R}_\alpha
+ \frac{1}{2}\left[ {\varphi^l_i}'' \varphi^{\overline{\nu}_R}_\alpha
+ \varphi^l_i {\varphi^{\overline{\nu}_R}_\alpha}'' 
+ 2{\varphi^l_i}' {\varphi^{\overline{\nu}_R }_\alpha}' \right] d_H^2
+ \ldots \,,
\end{equation}
where all of the wavefunctions are evaluated at $y^h$. Note that the first
term is a rank one matrix while the additional terms are suppressed by
$(d_H/{\rm min}\{d_N,d_{\bar{\bf 5}}\})^2$.  Thus we obtain a neutrino
Yukawa matrix that has the form of the ``democratic'' mass matrix ansatz, 
i.e. it is rank one at leading order~\cite{democratic}.  Deviations from 
absolute democracy (i.e. from being rank one) result from the 
wavefunctions for $l_i$ and $\overline{\nu}_{R\alpha}$ not being 
absolutely flat.  The first row of Figure~\ref{fig:broad-broad} shows the 
distributions of the three eigenvalues of the neutrino Yukawa matrix.  
Note that $\varphi$, $\varphi'$ and $\varphi''$ have upper limits and 
there are no effects enhancing the middle eigenvalue; hence the sharp 
cut-off of the distribution of $\ln \lambda^\nu_2$.
An interesting conclusion is that the neutrino Yukawa matrix has a 
certain amount of hierarchy, no matter how large or small the width 
parameter $d_N$ of the right-handed neutrinos is.  Small Yukawa couplings  
result from small overlap integrals for localized right-handed neutrinos, 
or from rank reduction for non-localized right-handed neutrinos. 

The Majorana mass matrix of right-handed neutrinos is not hierarchical 
when the source of symmetry breaking $\varphi^{SB}$ is not localized.
Thus in the second row of Figure~\ref{fig:broad-broad} the distribution
of $M_2/M$ is roughly order unity, and the hierarchy $M_2/M_3$ is not more 
than what is expected from the diagonalization effect.\footnote{We do not 
have a clear explanation why the distribution of $M_1/M$ has a long tail 
toward smaller values.} 
On the other hand, the Majorana masses of right-handed neutrinos become 
hierarchical when $\varphi^{SB}$ is localized. Then $c_{\alpha\beta}$ is 
rank one at leading order, and deviation from being rank one should come 
at the order of $(d_{SB}/d_N)^2$.

The third row of Figure~\ref{fig:broad-broad} shows the
distribution of mass eigenvalues of the low-energy seesaw neutrinos for 
$d_N/L = d_{SB}/L = 0.3$. The distribution of $\ln (m_2/m_\nu)$ is sharply 
cut off at its largest value; this cut-off is presumably traced back to 
the similarly sharp cut-off of the distribution of the second largest 
eigenvalue of the neutrino Yukawa matrix $\lambda^\nu_2$.  As opposed to 
when the right-handed neutrinos have localized wavefunctions, it is 
not as straightforward to develop an analytical understanding of the 
distribution of seesaw neutrino mass eigenvalues when the right-handed  
neutrinos have non-localized wavefunctions.  Instead, we run numerical
simulations with the Gaussian landscape on $S^1$, using a range of 
$d_N/L \in \left[ 0.2, 0.5 \right]$ and 
$d_{SB}/L \in \left[ 0.08, 0.5\right]$.  We find that the distributions 
of $\ln (m_i/m_\nu)$ remain qualitatively the same as in the last row 
of Figure~\ref{fig:broad-broad}. It is only when $d_{SB}$ is much
smaller than $d_N$ that there is a qualitative difference; then the 
Majorana mass matrix of right-handed neutrinos has hierarchical 
eigenvalues, and lighter right-handed neutrinos give rise to enhanced 
values of $m_2/m_\nu$, erasing the sharp cut-off in the distribution of 
$\ln (m_2/m_\nu)$.

\subsubsection{Neutrino Mass Hierarchy in Gaussian Landscapes}

It is a robust consequence of Gaussian landscapes on any geometry that 
the seesaw masses are hierarchical.  If the right-handed neutrinos have 
localized wavefunctions, then the neutrino Yukawa matrix obtains 
statistically hierarchical mass eigenvalues just like in the analysis of 
sections~\ref{sec:indep} and~\ref{sec:toy1}.  On the other hand, if the 
wavefunctions are not particularly localized over the extra dimension(s), 
then the Yukawa matrix always has small eigenvalues due to rank
reduction.  Because of the absence of correlation between the Majorana 
mass terms of right-handed neutrinos and the neutrino Yukawa couplings, 
the hierarchical structure of the neutrino Yukawa matrix remains in the 
seesaw mass eigenvalues.  As a consequence, an inverted hierarchy is 
unlikely.  In addition, it is likely that $m_1 \ll m_2$ and 
\begin{equation}
 \frac{m_2}{m_3} \simeq
 \sqrt{\frac{\Delta m^2_{\odot}}{\Delta m^2_{\rm atm}}} \,. 
\label{eq:ignorem1}
\end{equation}

\begin{figure}[t]
\begin{center}
\begin{tabular}{cccc}
\includegraphics[width=0.23\linewidth]{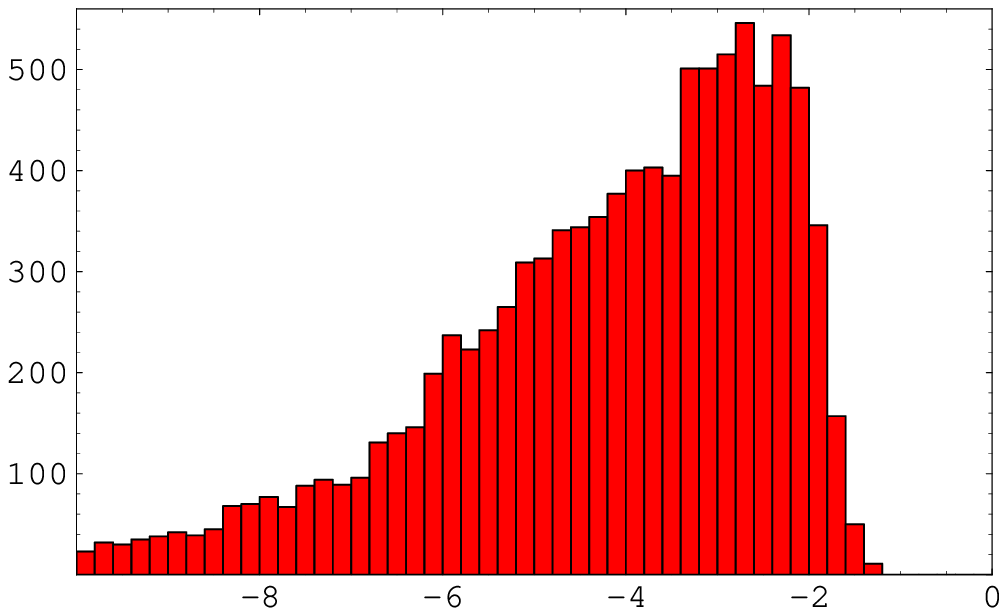} &
\includegraphics[width=0.23\linewidth]{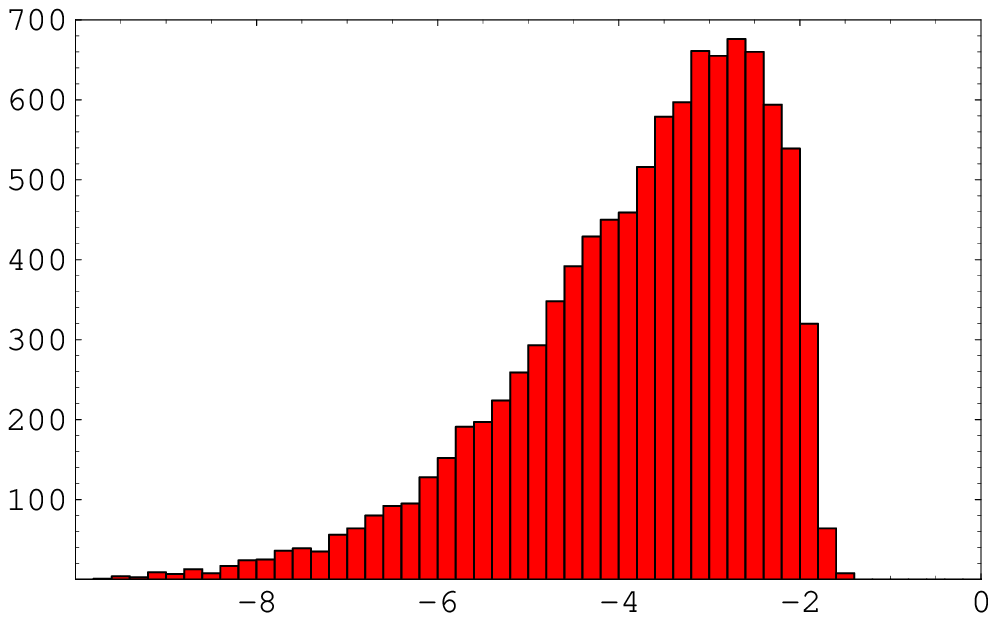} &
\includegraphics[width=0.23\linewidth]{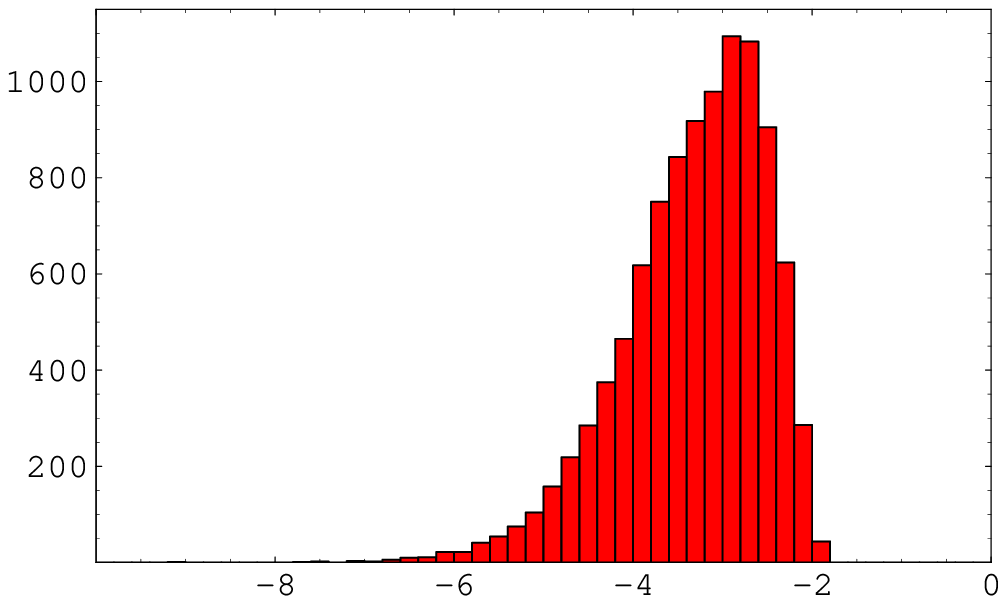} &
\includegraphics[width=0.23\linewidth]{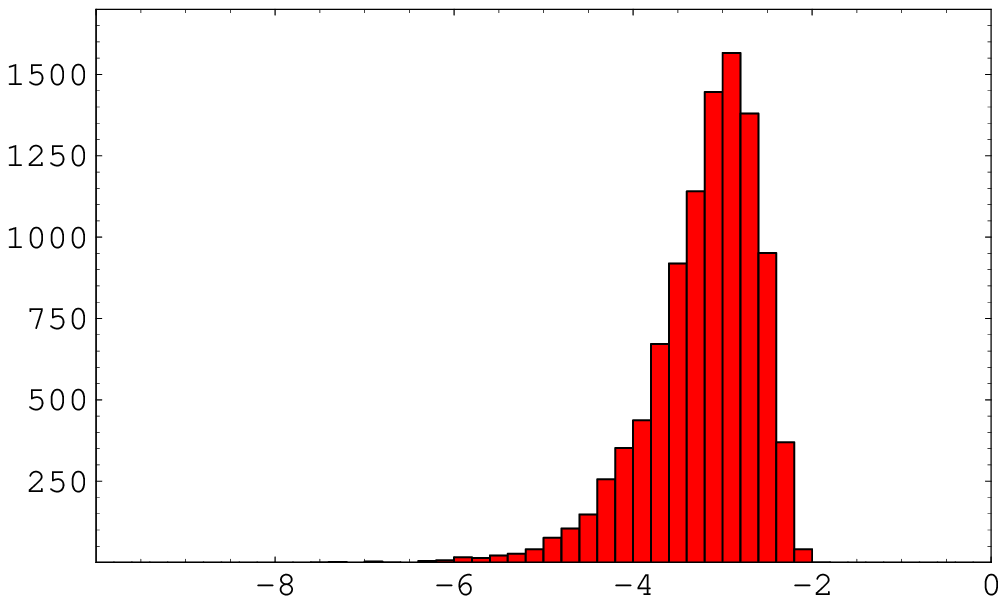} \\
(a) & (b) & (c) & (d) \\
\includegraphics[width=0.23\linewidth]{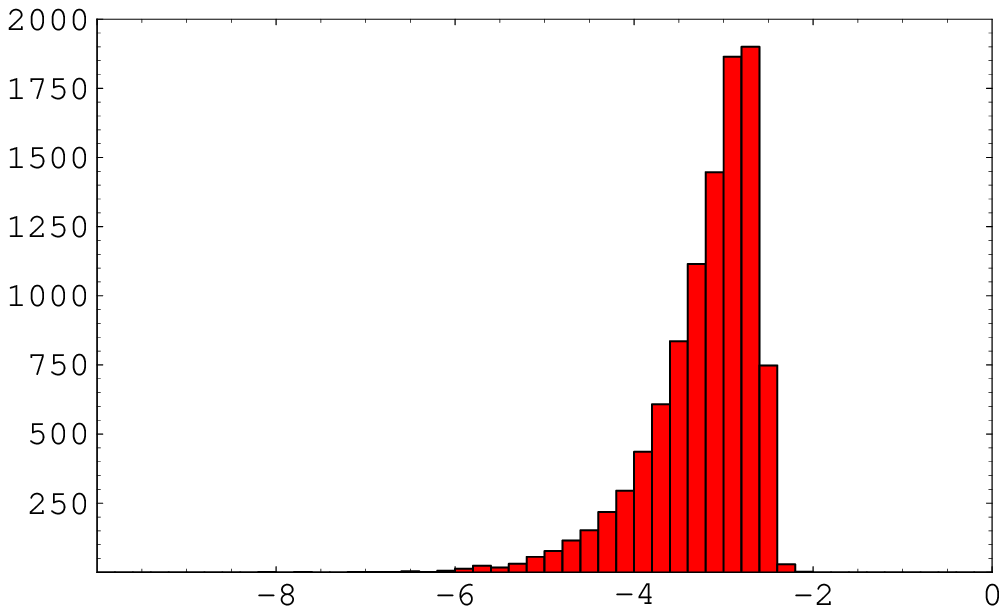} &
\includegraphics[width=0.23\linewidth]{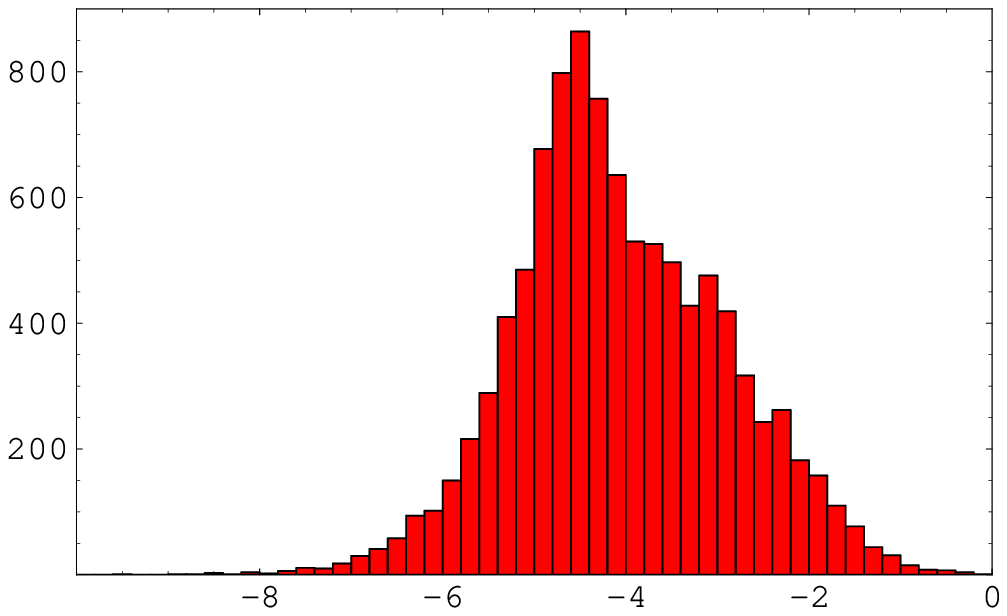} &
\includegraphics[width=0.23\linewidth]{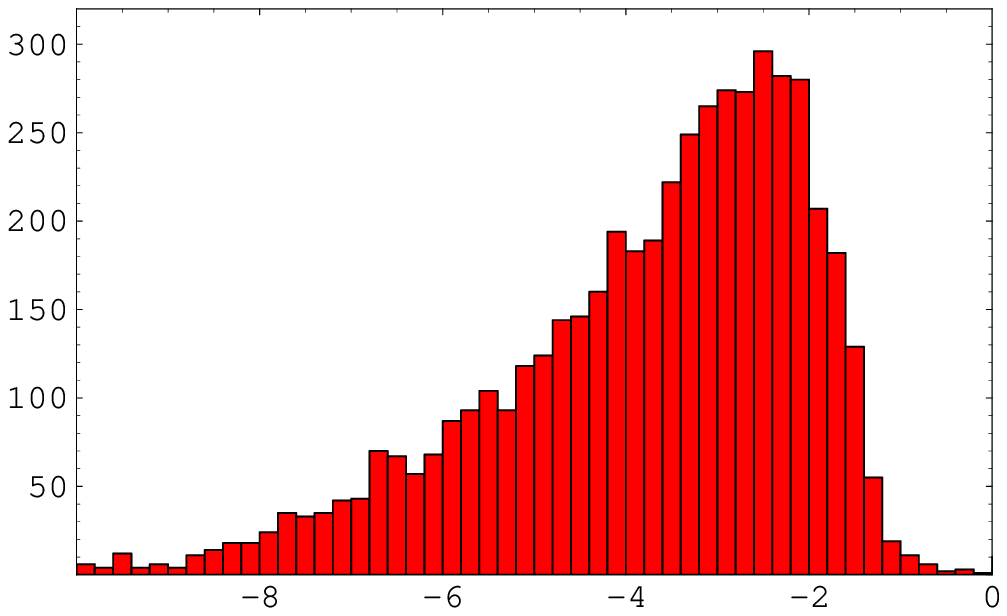} & 
\includegraphics[width=0.23\linewidth]{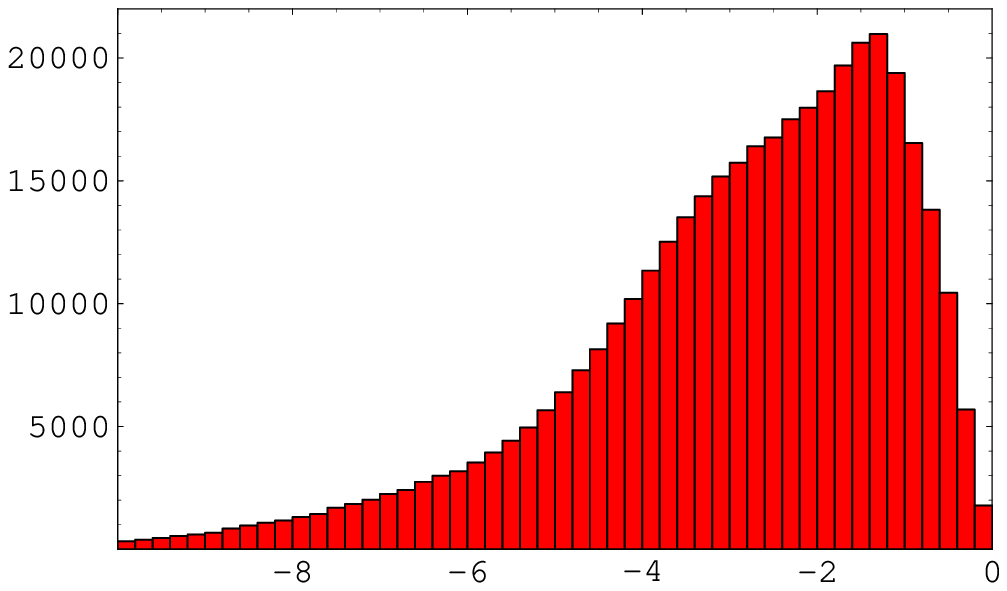} \\
(e) & (f) & (g) & (h) 
\end{tabular}
\begin{tabular}{ c l c l} 
(a): & $d_N/L = d_{SB}/L =0.08$ \phantom{space} & 
(e): & $d_N/L = d_{SB}/L = 0.3$ \\
(b): & $d_N/L = d_{SB}/L = 0.10$ & 
(f): & $d_N/L = 0.3$, $d_{SB}/L = 0.02$, $d_H/L = 0.1$ \\
(c): & $d_N/L = d_{SB}/L = 0.15$ & 
(g): & $d_N/L = d_{SB}/L = d_H/L=d_{\bf 10}/L= 0.1$, on $T^2$ \\ 
(d): & $d_N/L = d_{SB}/L = 0.2$ & 
(h): & $d_N/L = d_{SB}/L =0.08$, $r=3.0$ \\ 
\end{tabular}
\caption{\label{fig:compare23} Distributions of $\log_{10}(m_2/m_3)$ for
various choices of parameters.  The Gaussian landscape on $S^1$ was used 
for all simulations except (g), for which $T^2$ was used.
In (f), $d_{SB}$ is chosen to be very small so that $d_H^4/d_{SB}^2$ is 
large, while (h) involves complex Gaussian profiles (see 
section~\ref{ssec:mixcp}).  Unless otherwise specified, 
$d_{\bf 10}/L = d_H/L = 0.08$ and $d_{\bar{\bf 5}}/L = 0.3$.}
\end{center}
\end{figure}

Although Gaussian landscapes predict that the seesaw masses are 
hierarchical {\it on average}, the {\it distribution} of mass ratios 
from Gaussian landscapes are qualitatively different, depending on the 
choice of the width parameters $d_N$ and $d_{SB}$. Consider the 
distribution of $m_2/m_3$ for various choices of these width parameters,
displayed in Figure~\ref{fig:compare23}.
When neither the right-handed neutrino nor the symmetry-breaking 
wavefunction(s) are localized, the distribution of $m_2/m_3$ is sharply 
cut off at around $10^{-2}$, seen in Figure~\ref{fig:compare23} (d) 
and (e).  In this case there is no chance to accommodate the observed 
value, 
\begin{equation}
 \sqrt{\frac{\Delta m^2_{\odot}}{\Delta m^2_{\rm atm}}} 
    \simeq (1.6\mbox{--}2.0) \times 10^{-1} \,.
\label{eq:s2a}
\end{equation}

If $d_N$ is large and the wavefunction $\varphi^{SB}$ is strongly localized 
in the extra dimension(s), then the distribution of $m_2/m_3$ has a tail 
extending upward as in Figure~\ref{fig:compare23} (f).  This is because of 
the previously mentioned enhancement in the seesaw mass matrix---due to 
small eigenvalues in the Majorana mass terms of the right-handed 
neutrinos---which erases the sharp upper bound on $\lambda^\nu_2$. 
The distribution in Figure~\ref{fig:compare23} (f) barely covers the 
observed value (\ref{eq:s2a}), but the width parameters may be chosen 
a little differently to provide better coverage in the tail of the 
distribution.  Yet even in this scenario, the distribution of $m_2/m_3$ 
is peaked at a value much smaller than the observed value (\ref{eq:s2a}).
If environmental selection factors (such as those associated with 
leptogenesis) and weight factors associated with cosmological evolution 
are in favor of such an outlier, then this scenario may be acceptable. 
Otherwise, one should conclude that this scenario is not a good 
approximation to the actual landscape governing the flavor structure of
the multiverse.

It is left to consider localized wavefunctions for the right-handed 
neutrinos. The approximate distribution of $m_2/m_3$ is derived 
analytically in this scenario, and plotted in the right column of 
Figure~\ref{fig:neutrinoegval-analytic}.  Regardless of the geometry 
this distribution covers a broad logarithmic range, with the greatest 
weight toward the largest value, $m_2/m_3 \sim 1$.  These analytical 
distributions capture the qualitative features of 
Figure~\ref{fig:compare23} (a), (b), and (g) very well, with one important
difference:  the distributions in the numerical results fall off to almost 
zero for $m_2/m_3 \simgt 10^{-2}\mbox{--}10^{-1}$.  This behavior of the 
numerical results is presumably due to the diagonalization effect, a 
phenomenon first mentioned in section~\ref{ssec:phen} and subsequently 
referred to with regard to Figures~\ref{fig:D=1-nocut}, 
\ref{fig:T2YukawaEgval}, \ref{fig:compareT2S2}, \ref{fig:5vs10}, 
and~\ref{fig:narrow-narrow}.  This reduced weight in the probability 
distribution functions at $m_2/m_3 \simgt 10^{-1}$ means that this 
scenario is also in conflict with the observation (\ref{eq:s2a}). 

There is an important caveat, however, which is that the diagonalization
effect changes as the real-valued matrices that we have dealt with so far
are generalized to complex-valued matrices.  See the appendix of~\cite{HM}, 
for example.  Figure~\ref{fig:compare23} (h) is based on an extended 
version of the Gaussian landscape on $S^1$ (presented in 
section~\ref{ssec:mixcp}) that includes complex phases. The diagonalization 
effect clearly has much less impact on the distribution function of 
$m_2/m_3$, and this distribution extends almost all the way up to 
$m_2/m_3 \simeq 1$.  Thus we find the scenario with localized wavefunctions
for the right-handed neutrinos is compatible with the large value observed
for $\Delta m^2_{\odot}/\Delta m^2_{\rm atm}$, assuming the Gaussian 
landscape contains complex phases.

\subsection{Mixing Angles and CP Phases}
\label{ssec:mixcp}

\begin{figure}[t]
\begin{center}
\begin{tabular}{ccc}
\includegraphics[width=0.3\linewidth]{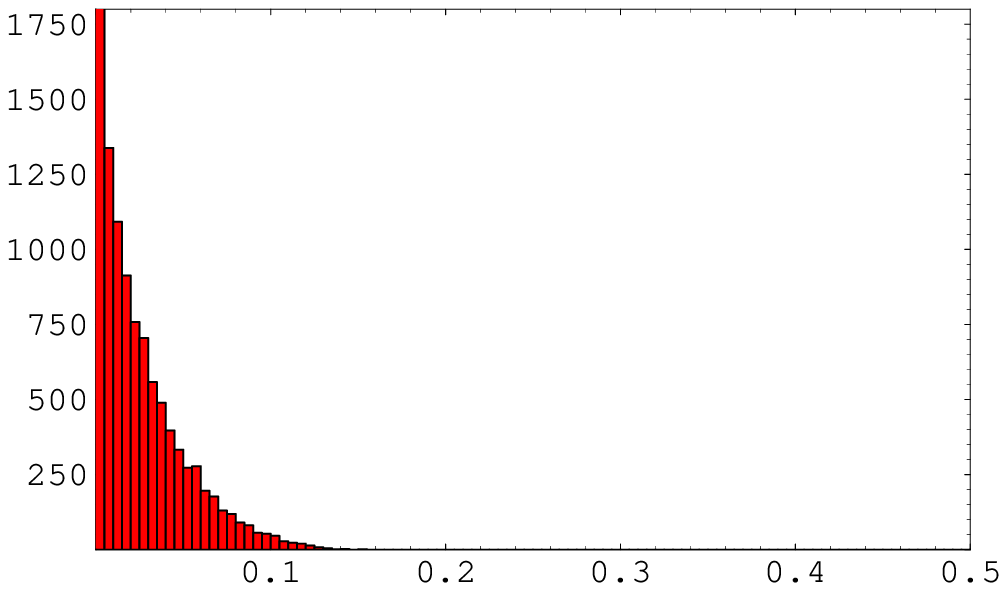} &
\includegraphics[width=0.3\linewidth]{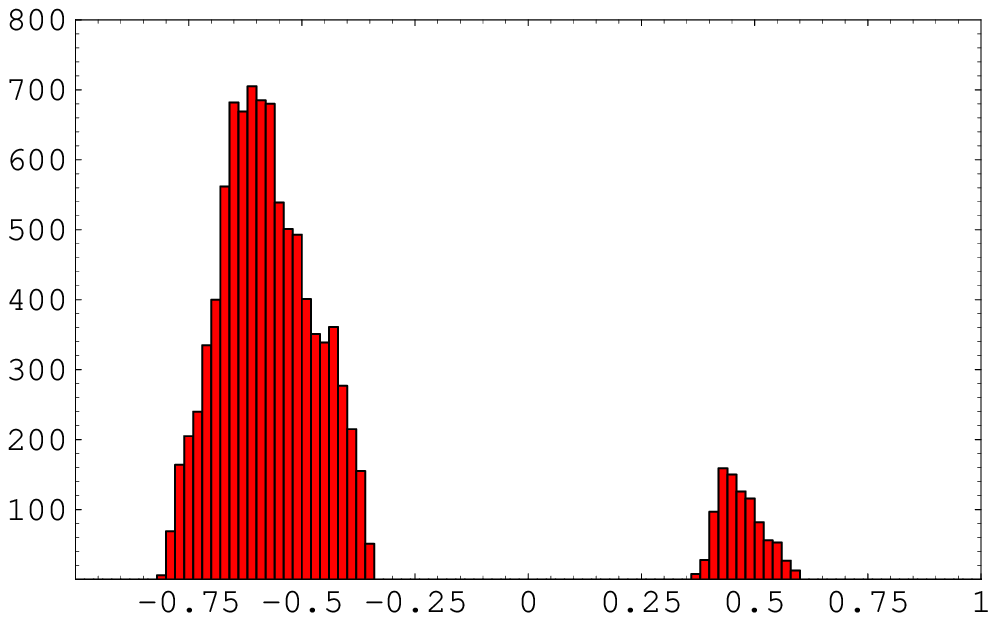} &
\includegraphics[width=0.3\linewidth]{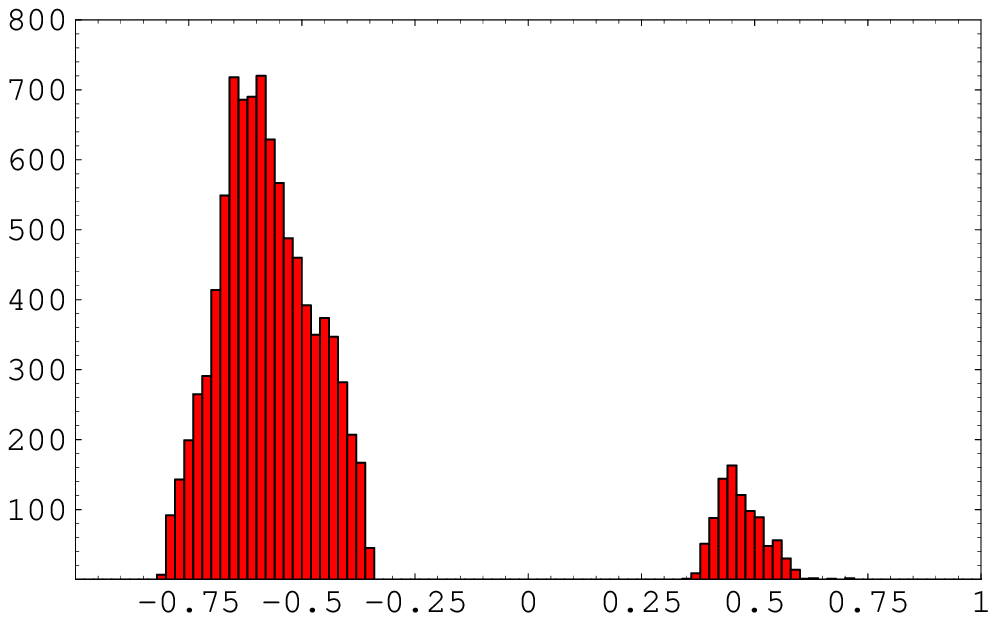} \\
$\theta_{23}/\pi$ & $V^e_{23}$ & $V^\nu_{23}$
\end{tabular}
\caption{\label{fig:fail2get23} The distributions of the PMNS mixing angle
$\theta_{23}$, the $(2,3)$ element of the diagonalization matrix $V^e$, 
and that element of $V^\nu$.  The distributions result
from a numerical simulation of the Gaussian landscape on $S^1$ with 
$d/L=0.08$ for all fields except $d_{\bar{\bf 5}}/L = 0.3$.}
\end{center}
\end{figure}

We introduced non-localized wavefunctions for the fields in 
$\bar{\bf 5}$ because we expected this to result in large leptonic 
mixing angles.  However numerical simulation reveals that the mixing 
angles, particularly $\theta_{23}$, are still very small, as can be 
seen in the left panel of Figure~\ref{fig:fail2get23}. 
\begin{figure}[t]
\begin{center}
\begin{tabular}{cc}
\includegraphics[width=0.25\linewidth]{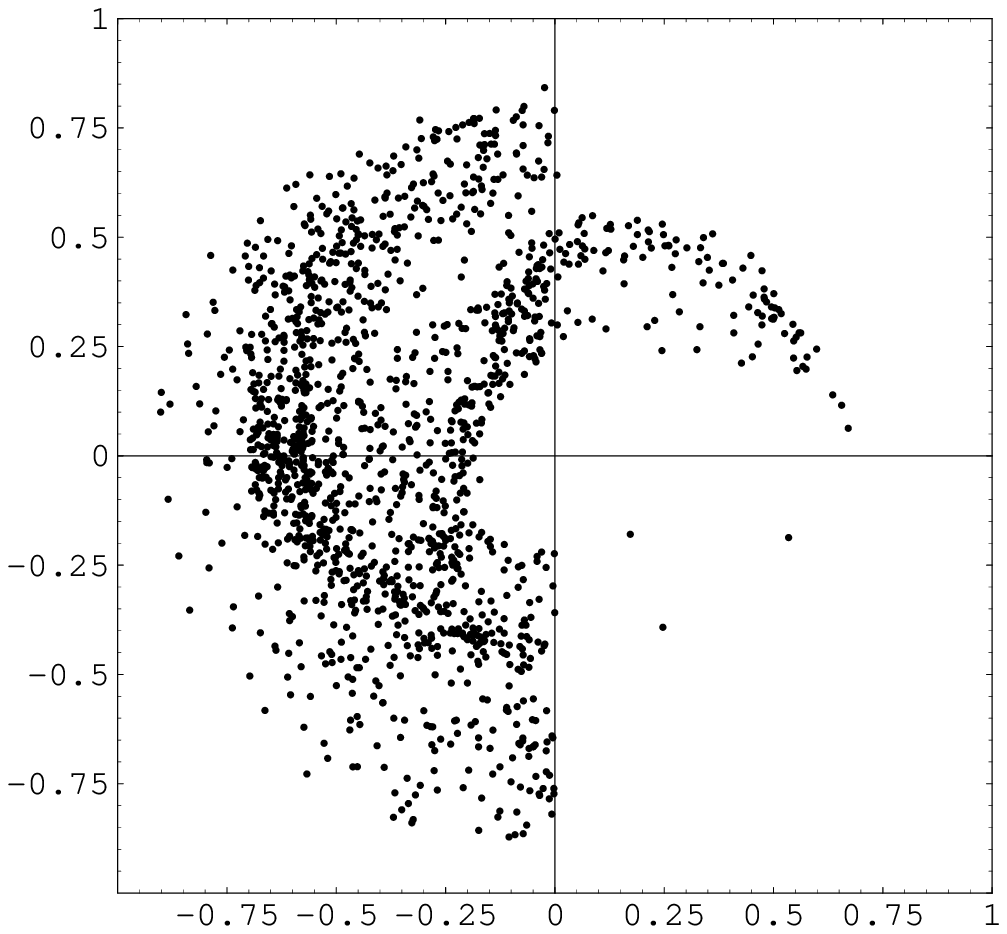} &
\includegraphics[width=0.25\linewidth]{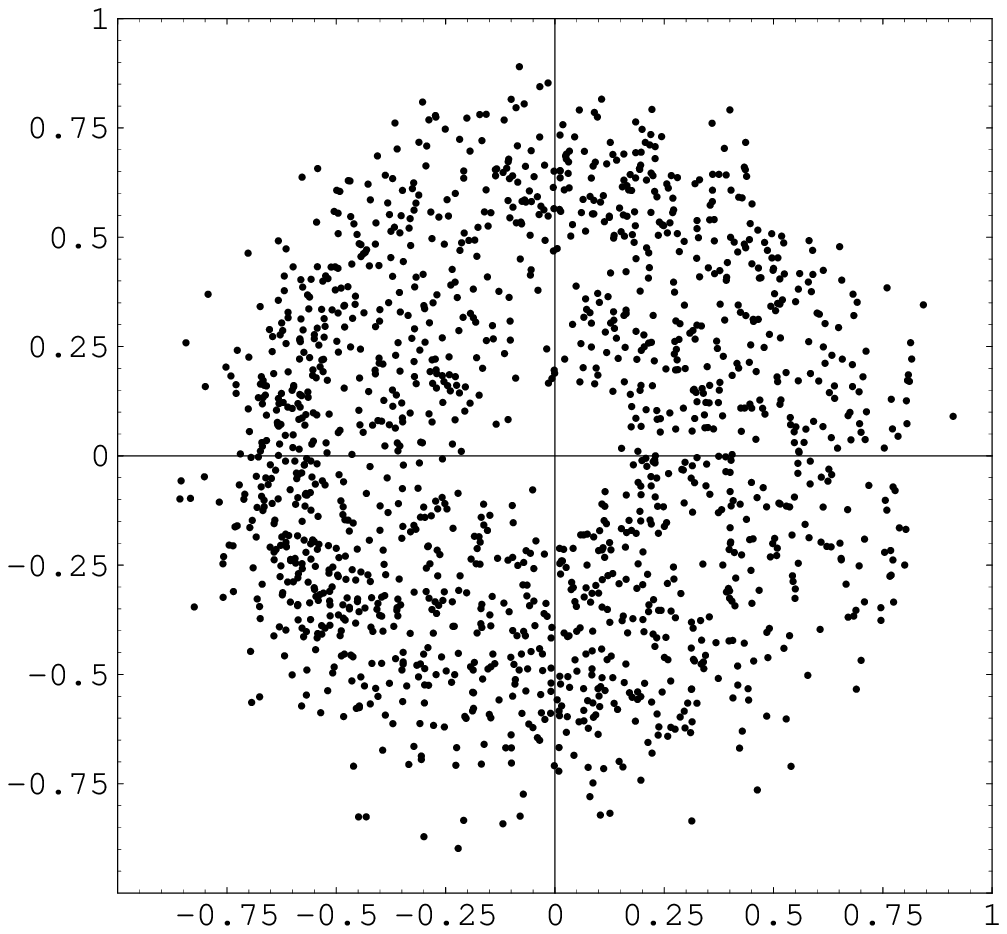} \\ 
$V^e_{23}$ & $V^\nu_{23}$ \\
\end{tabular}
\caption{\label{fig:rotation-angle} Distribution of the $(2,3)$ 
component of the diagonalization matrices $V^e$ and $V^\nu$ in the 
complex plane.  The simulation uses the Gaussian landscape on $S^1$, 
with $d/L=0.08$ for all fields except $d_{\bar{\bf 5}}/L = 0.3$; 
while $r=3$.  The distributions of $|V^e_{23}|$ and $|V^\nu_{23}|$ remain 
much the same as in Figure~\ref{fig:fail2get23}, but the phase angle 
prevents precise cancellations in $\theta_{23}$.}
\end{center}
\end{figure}
The figure presented here is a simulation using small $d_N/L$ and 
$d_{SB}/L$; but the $\theta_{23}$ distribution is confined near 
$\theta_{23}\sim 0$ no matter how small or large these parameters are 
taken to be.

The distributions of $V^e_{23}$ and $V^\nu_{23}$ in
Figure~\ref{fig:fail2get23} explain why, contrary to our expectation, 
we failed to get large $\theta_{23}$. An SO(3) matrix $V^e$ acts on 
the left-handed charged leptons $e_{Li} \subset l_i$ when diagonalizing 
the charged lepton Yukawa matrix $\lambda^e_{ai}$, and $V^\nu$ acts on 
the left-handed neutrinos $\nu_{Li} \subset l_i$ when diagonalizing the 
seesaw mass matrix $C_{ij}/M$.  Their (2,3) components are large, just 
as we expected. Yet the mixing angle of atmospheric neutrino oscillation, 
the (2,3) component of $U =(V^e)^{-1} V^\nu$, is small due to a 
precise cancellation between the large $V^e_{23}$ and $V^\nu_{23}$.
This cancellation is not a new phenomenon. The democratic mass matrix 
ansatz, originally applied to the quark sector, yields small mixing 
angles through this precise cancellation. 

To obtain large mixing angles, some modification is needed to prevent 
this cancellation.  We replace the real-positive valued Gaussian 
wavefunctions in (\ref{eq:Gaussian}) by Gaussian wavefunctions with a 
winding complex phase:\footnote{This form for the wavefunction is not 
without motivation, c.f. the discussion surrounding (\ref{eq:complexGs7}).  
However, we use this wavefunction only as one of the simplest means to 
introduce complex phases into Gaussian landscapes.  There are more 
complicated, more sophisticated, and possibly better justified ways to 
introduce complex phases into Gaussian landscapes, or landscapes based on 
extra dimensions in general.  It would be interesting to explore such 
models that include complex phases to study the correlation between the 
induced CP violation and the distribution of other observables.  However 
this subject is beyond the scope of this paper.  We restrict attention to 
(\ref{eq:cpxGaussian}) and find that even this simple model is sufficient 
to obtain large leptonic mixing angles, when $d_{\bar{\bf 5}}$ is large.}
\begin{equation}
\varphi(y ; y_0) \propto e^{-(1+r i) \frac{(y-y_0)^2}{2d^2}}\,, 
\label{eq:cpxGaussian}
\end{equation}
introducing a new parameter $r$.  For simplicity we use the same value of
$r$ for all of the wavefunctions in the overlap integrations,
$\varphi^{q,\bar{u},\bar{d},l,\bar{e},\overline{\nu},SB,h,h^*}$.  The
wavefunction of $h^*$ does not have to be the complex conjugate of that of
$h$, since for example in the framework of section~\ref{sec:SYM} these 
wavefunctions are not just scalars but have more complicated internal 
structure.\footnote{We have checked that one can obtain results very similar 
to those presented in this section, but taking $\varphi^h=(\varphi^{h^*})^*$, 
if one also allows for non-universal values of $r$.} 

As seen in Figure~\ref{fig:rotation-angle}, using this form of wavefunction
makes $V^e_{23}$ and $V^\nu_{23}$ complex, and their phases decrease the 
likelihood of a precise cancellation in $\theta_{23}$.  Furthermore, these 
complex Gaussian wavefunctions introduce CP violation into the flavor 
physics.  Meanwhile, this introduction of complex phases has little effect
on the distributions of charged fermion mass eigenvalues. This can 
be understood analytically by tracing the discussion in the preceding 
sections with the wavefunction (\ref{eq:cpxGaussian}).  We also confirmed 
this result by running numerical simulations.  The diagonalization effect 
is not taken into account in the theoretical arguments, but in practice 
this matters only for the $m_2/m_3$ neutrino mass distribution, and we 
have seen that this improves the agreement with observation. 

\begin{figure}[!t]
\begin{center}
\begin{tabular}{ccc}
\includegraphics[width=0.3\linewidth]{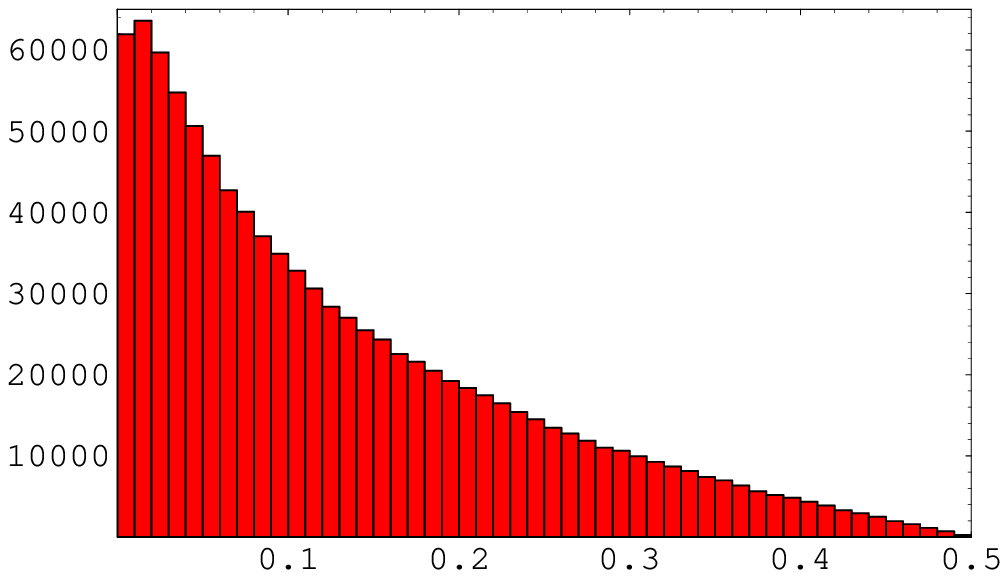} &
\includegraphics[width=0.3\linewidth]{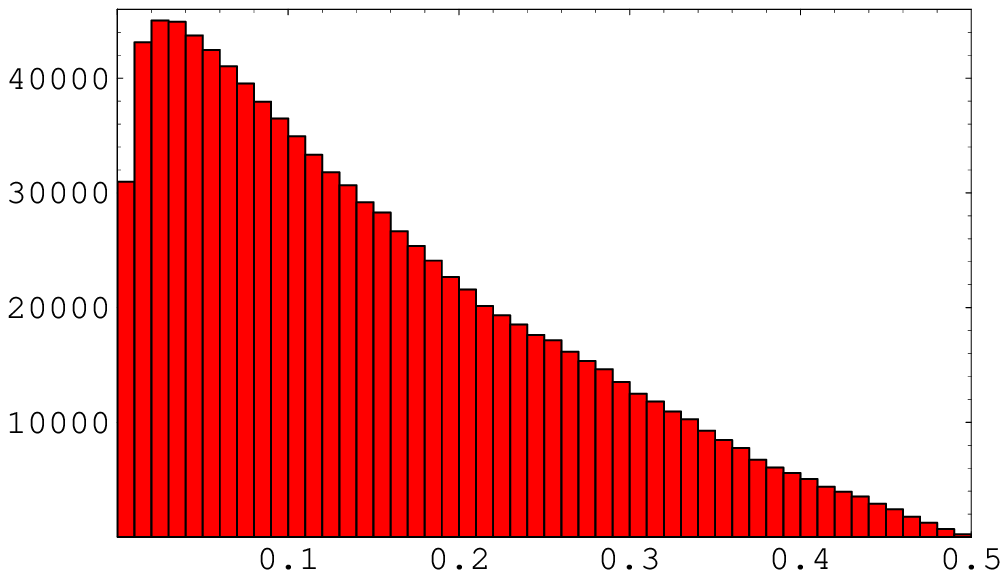} &
\includegraphics[width=0.3\linewidth]{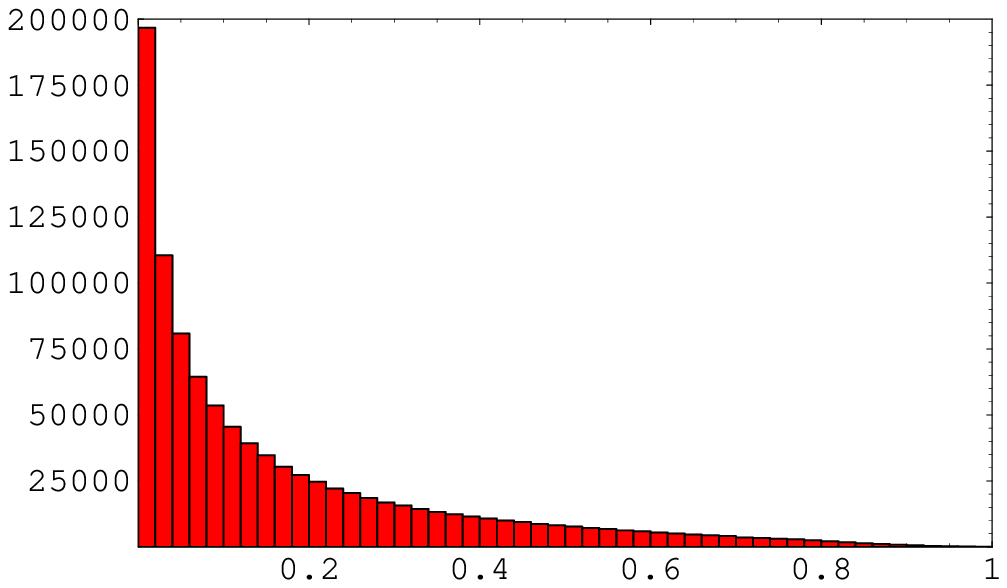} \\
$\theta_{12}/\pi$ & $\theta_{23}/\pi$ & $\sin \theta_{13}$ \\
\includegraphics[width=0.3\linewidth]{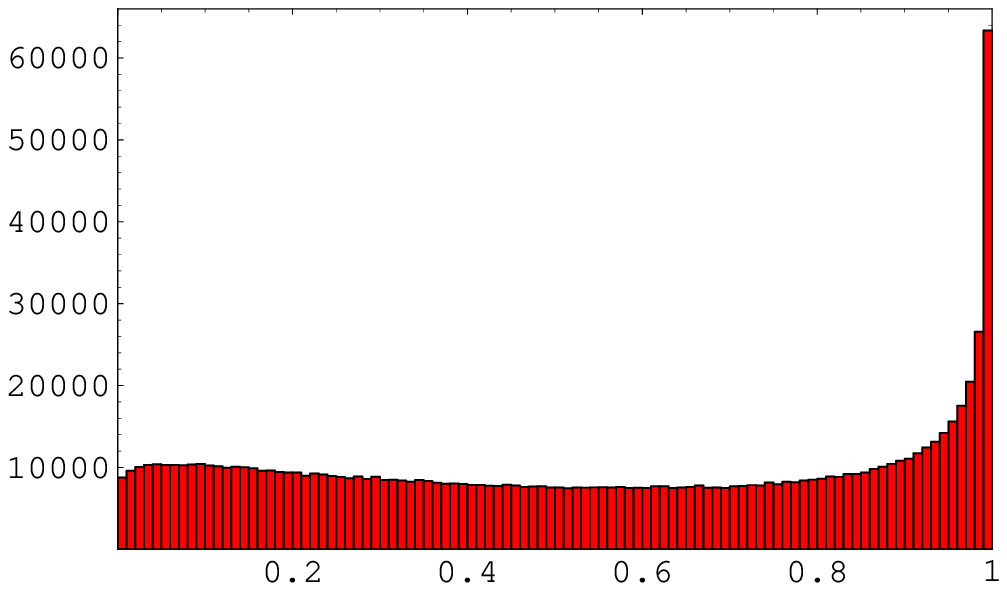} &
\includegraphics[width=0.3\linewidth]{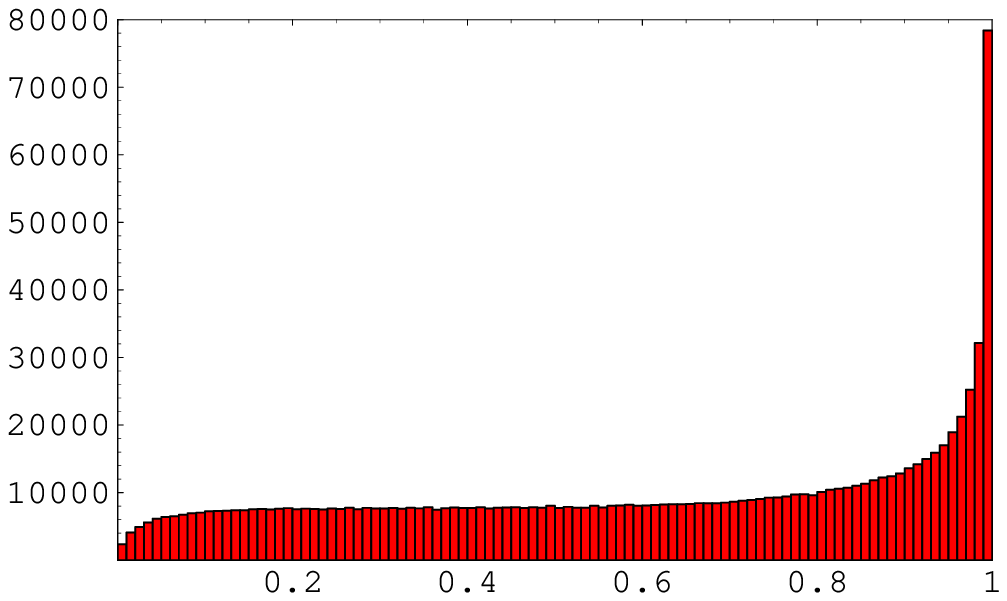} & 
\includegraphics[width=0.3\linewidth]{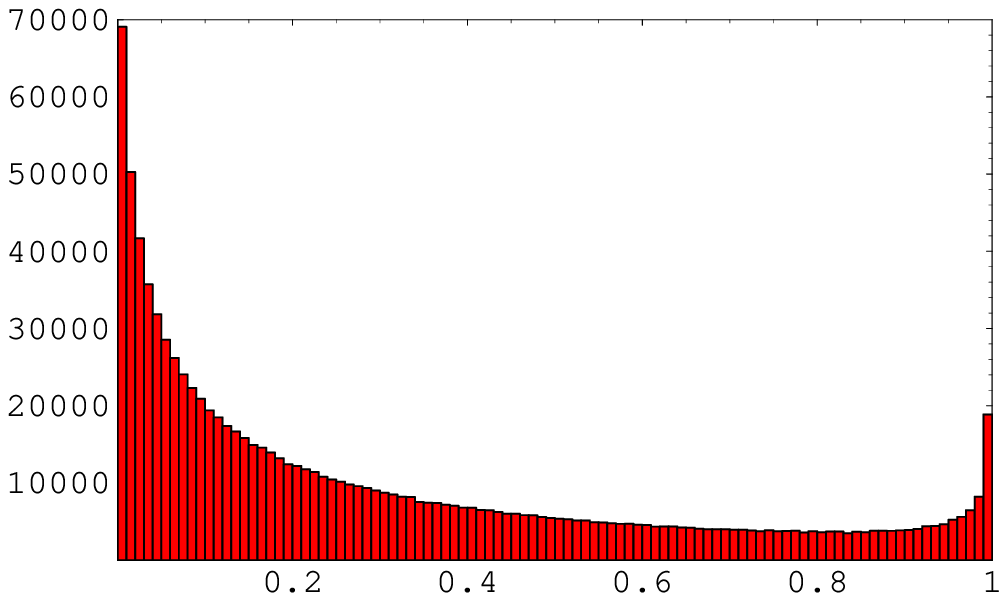} \\
$\sin (2\theta_{12})$ & $\sin (2\theta_{23})$ & $\sin (2\theta_{13})$\\
\includegraphics[width=0.3\linewidth]{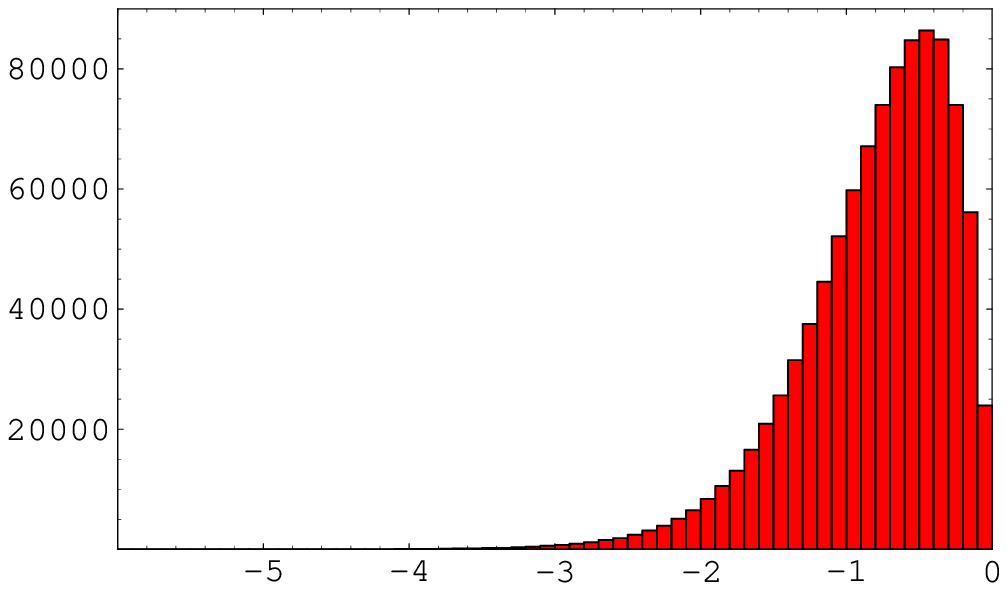} &
\includegraphics[width=0.3\linewidth]{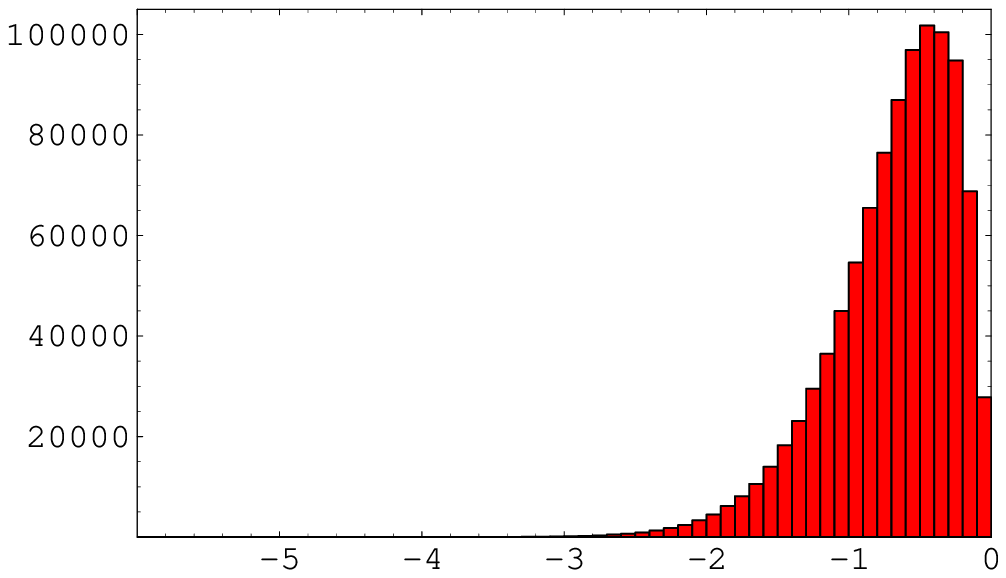} &
\includegraphics[width=0.3\linewidth]{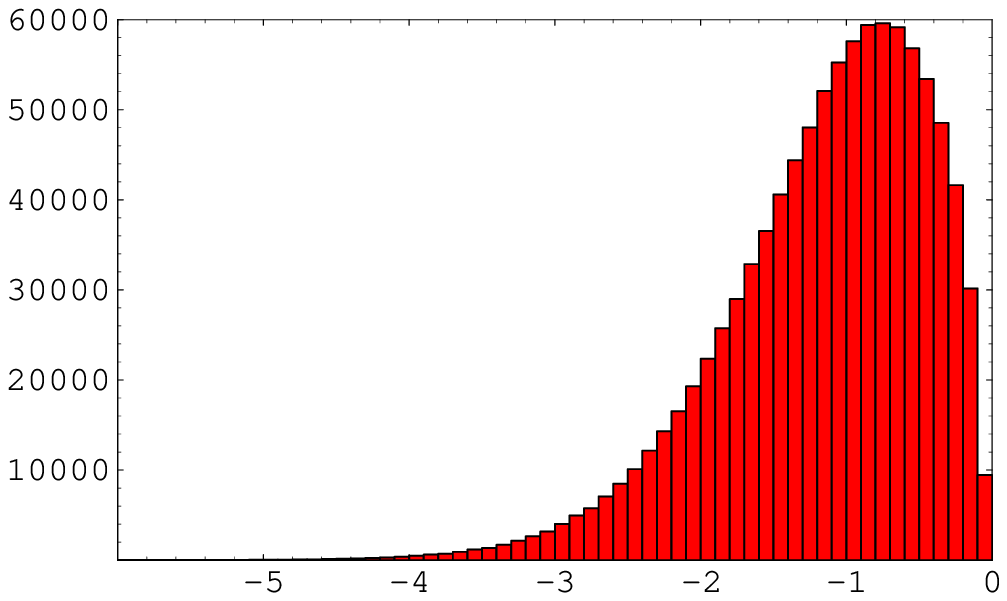} \\
$\log_{10} (2\theta_{12}/\pi)$ & $\log_{10} (2\theta_{23}\pi)$ & 
$\log_{10} \sin \theta_{13}$ 
\end{tabular}
\caption{\label{fig:cpx-lepton-mix} Distribution of the leptonic mixing 
angles in the Gaussian landscape on $S^1$, with complex-valued 
wavefunctions with $r=3$ and $d/L = 0.08$ for all fields except  
$d_{\bar{\bf 5}}/L = 0.3$.}
\end{center}
\end{figure}

On the other hand, introducing complex phases dramatically changes the 
distributions of mixing angles.  Figure~\ref{fig:cpx-lepton-mix} shows 
the distributions of the three leptonic mixing angles resulting from a 
numerical simulation.  
\begin{figure}[t]
\begin{center}
\begin{tabular}{ccc}
\includegraphics[width=0.3\linewidth]{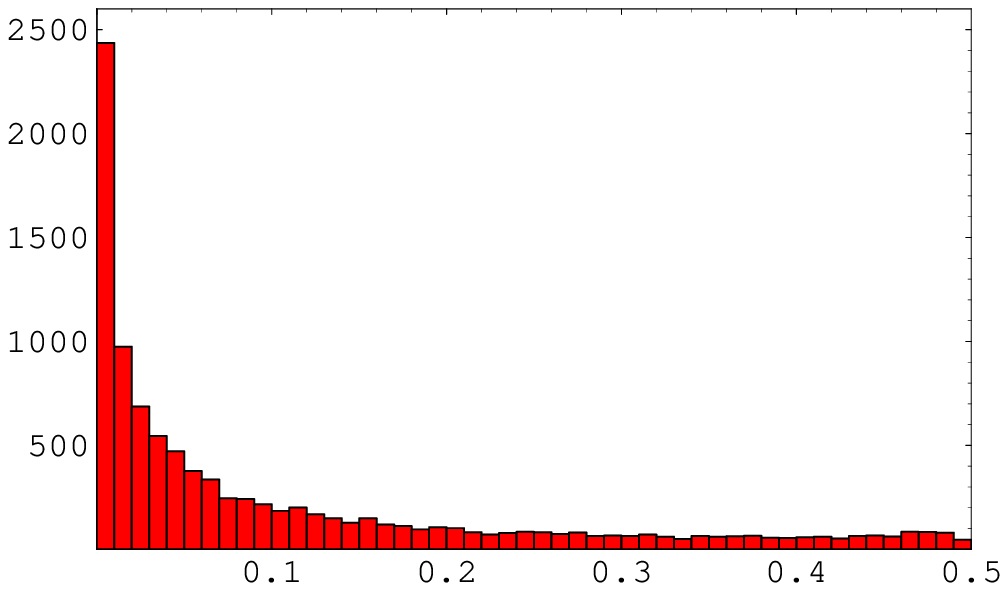} &
\includegraphics[width=0.3\linewidth]{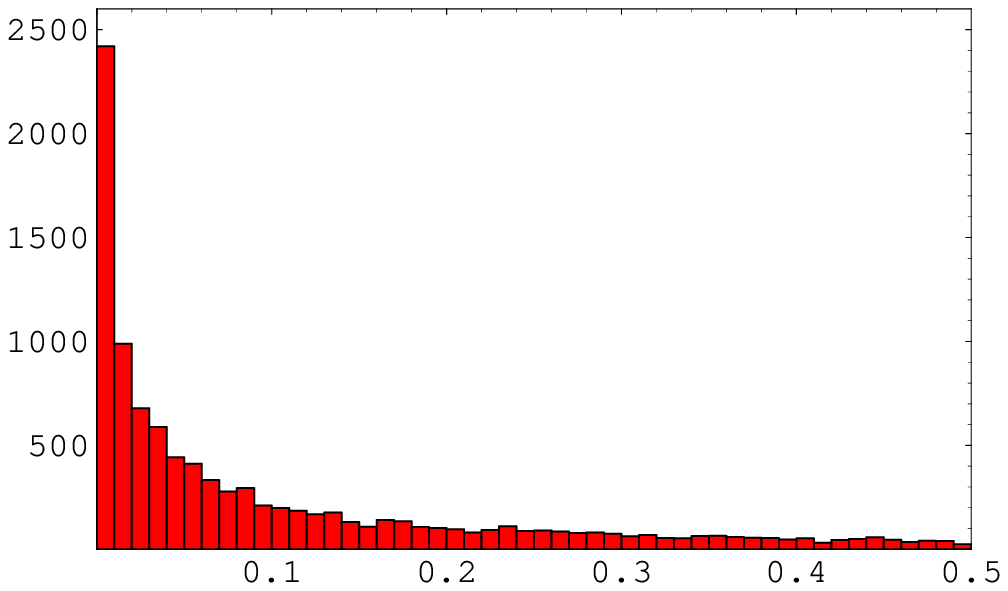} &
\includegraphics[width=0.3\linewidth]{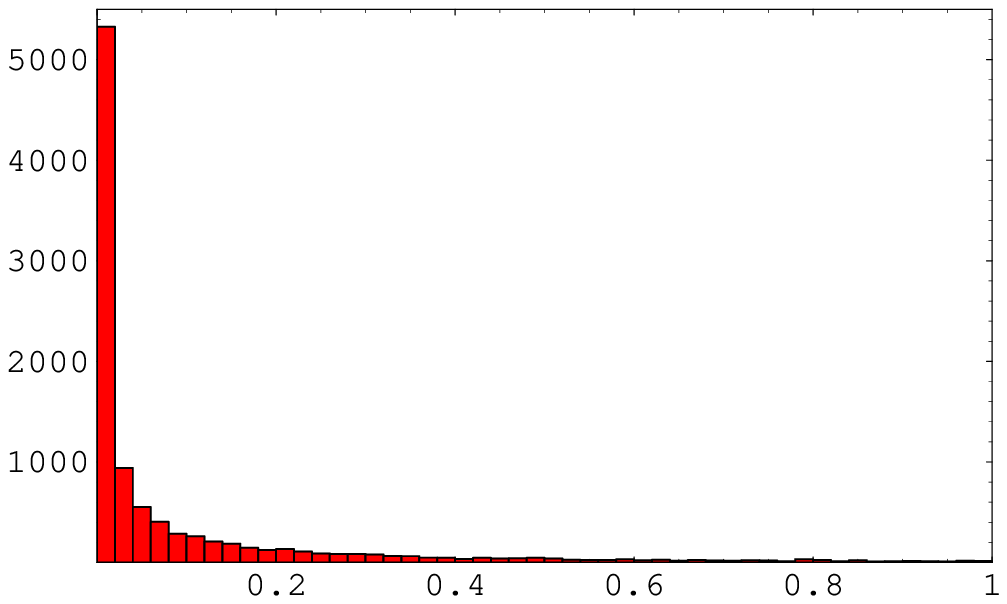} \\
$\theta_{12}/\pi$ & $\theta_{23}/\pi$ & $\sin \theta_{13}$ \\
\includegraphics[width=0.3\linewidth]{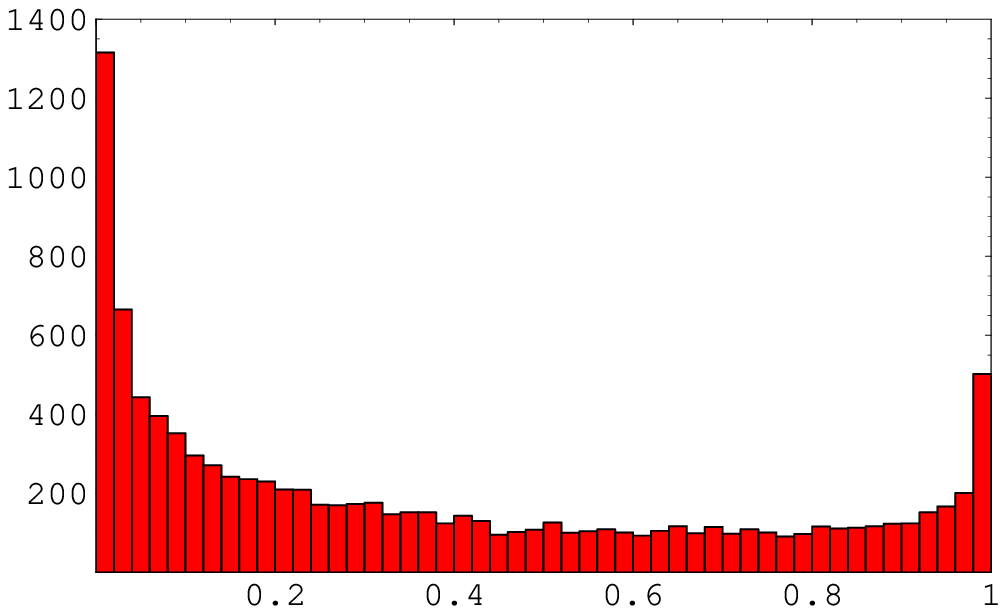} &
\includegraphics[width=0.3\linewidth]{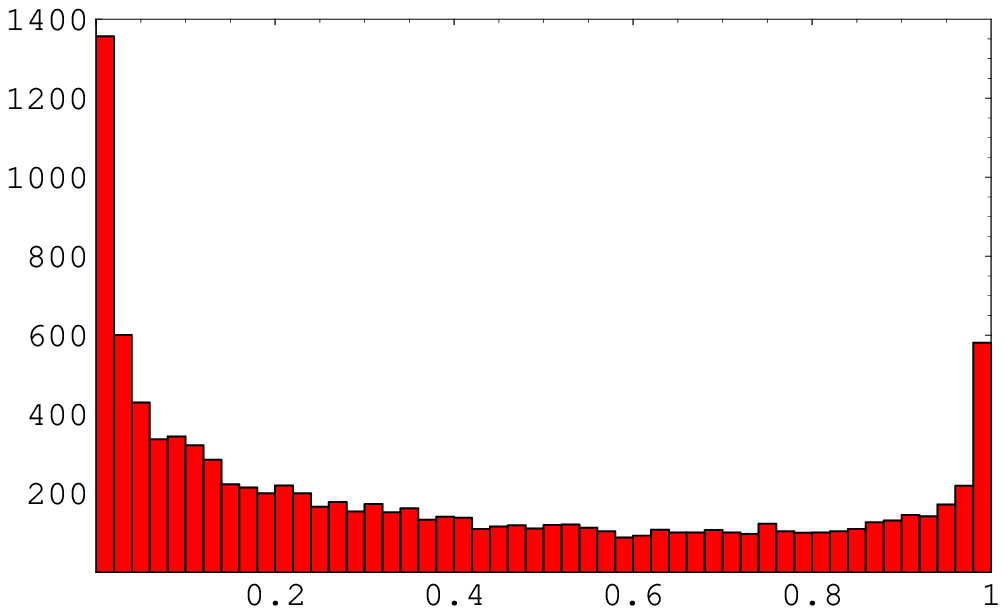} &
\includegraphics[width=0.3\linewidth]{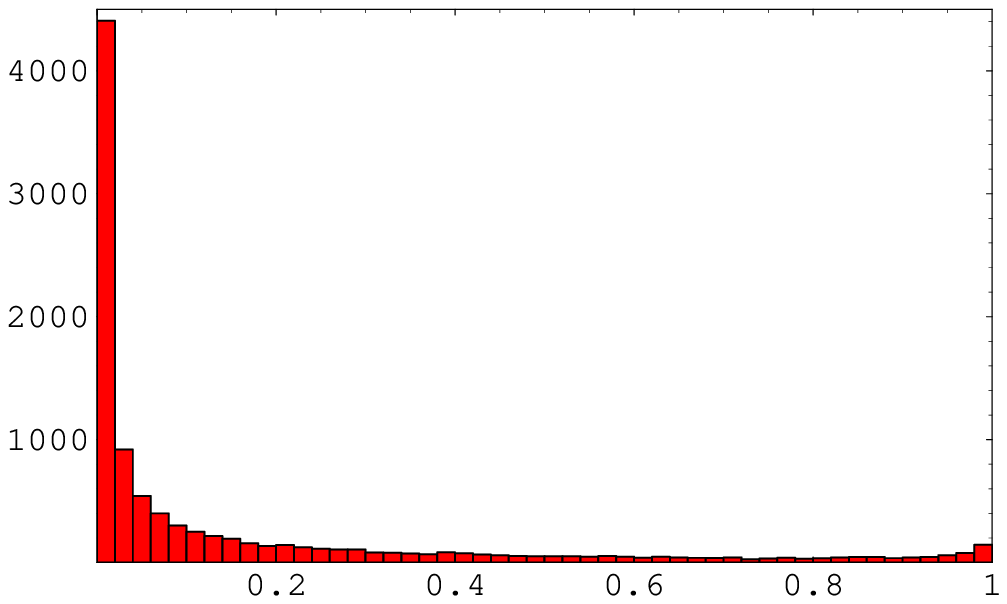} \\
 $\sin (2\theta_{12})$ & $\sin (2\theta_{23})$ & $\sin (2\theta_{13})$ \\
\includegraphics[width=0.3\linewidth]{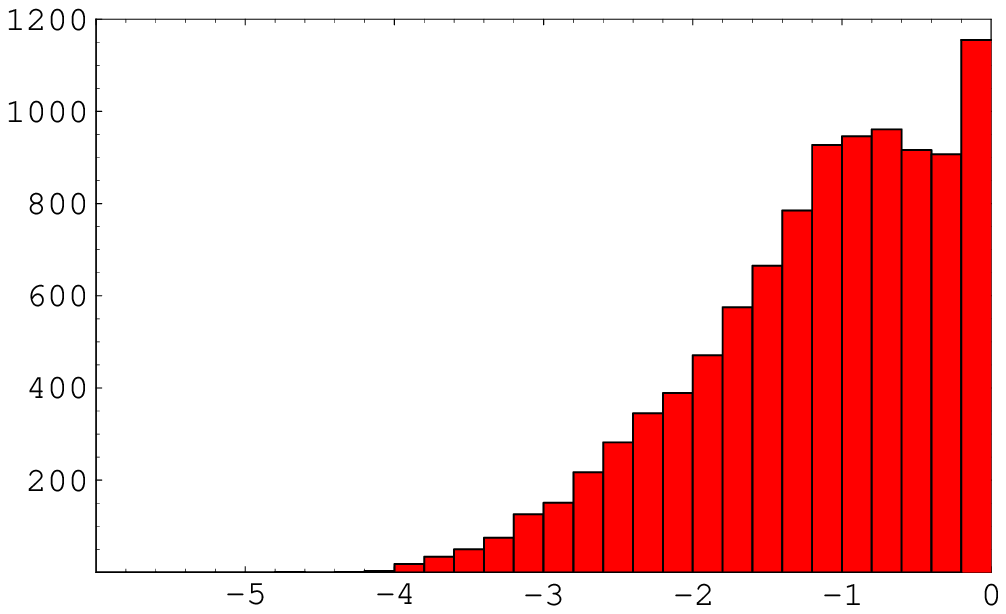} &
\includegraphics[width=0.3\linewidth]{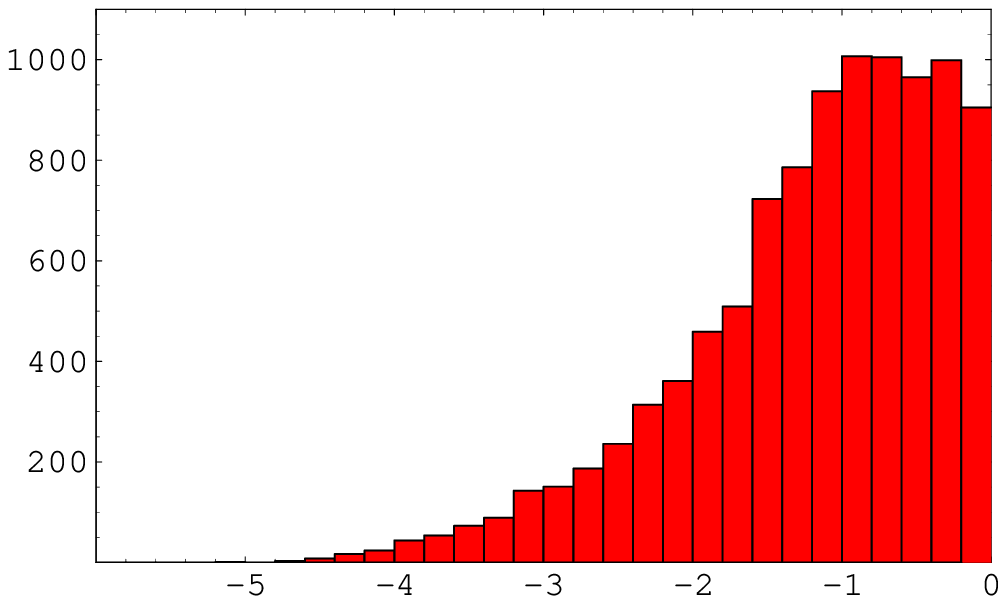} &
\includegraphics[width=0.3\linewidth]{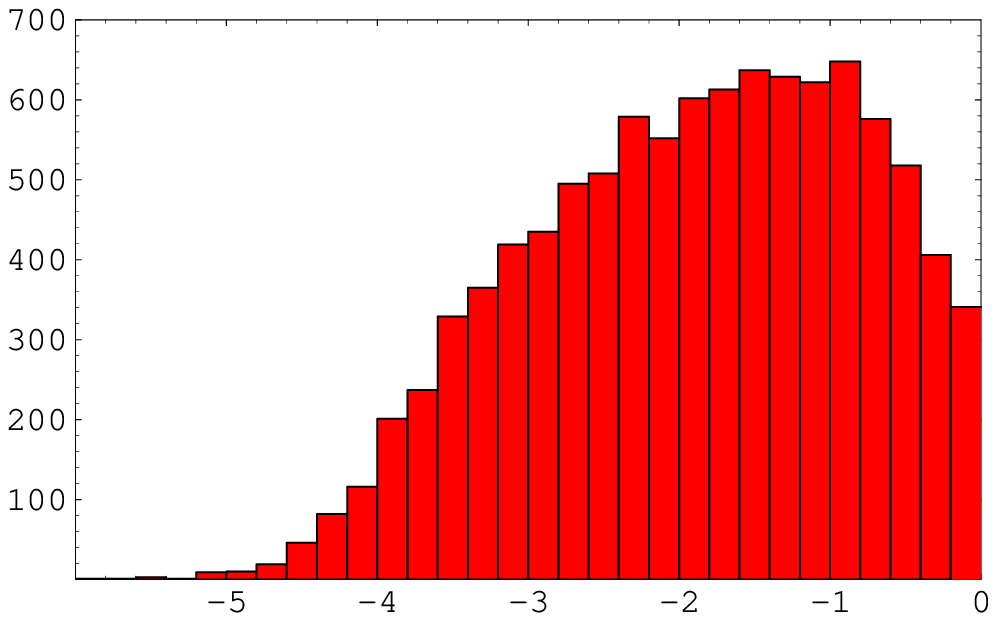} \\
$\log_{10} (2\theta_{12}/\pi)$ & $\log_{10} (2\theta_{23}/\pi)$ & 
$\log_{10} \sin \theta_{13}$ \\
\end{tabular}
\caption{\label{fig:cpx-quark-mix} Distribution of the quark mixing 
angles in the Gaussian landscape on $S^1$, with complex-valued 
wavefunctions with $r=3.0$, $d_{\bf 10}/L = d_{H}/L = 0.08$ and  
$d_{\bar{\bf 5}}/L = 0.3$.}
\end{center}
\end{figure}
We see that the mixing angles of solar and atmospheric neutrino 
oscillations, $\theta_{12}$ and $\theta_{23}$, are likely to be any value 
between $0$ and $\pi/2$. Distributions of the oscillation amplitudes 
$\sin (2\theta_{ij})$ are almost flat for both of these mixing angles.
Peaks at maximal mixing come from the Jacobian~\cite{HM}.  Comparing the 
distributions of $\theta_{23}$ between Figure~\ref{fig:cpx-lepton-mix} 
and Figure~\ref{fig:fail2get23}, we see that the introduction of complex 
phases (and hence CP violation) is an essential ingredient in obtaining 
large (and even maximal) mixing angles in neutrino oscillations. 

The mixing angle $\theta_{13}$ can also be of order unity, which is  
similar to the situation in neutrino anarchy~\cite{HMW, HM}. In the 
U(3)-invariant measure to be expected in neutrino anarchy, however, the 
distribution function of $\sin (2\theta_{13})$ is peaked at 
$\sin(2\theta_{13})=1$, and vanishes at $\sin(2\theta_{13})=0$. Thus all 
three mixing angles in the lepton sector are likely to be large. 
In particular, according to neutrino anarchy using the U(3)-invariant
measure, about 60--90\% of the statistics are already excluded by the 
current experimental limit on $\theta_{13}$~\cite{HMW, HM}.  In the 
Gaussian landscape on $S^1$, on the other hand, the $\theta_{13}$ 
distribution is weighted more toward $\sin(2\theta_{13})=0$.  As we see 
in section~\ref{ssec:prediction}, only about 30\% of the statistics have 
been excluded.  

For comparison, the distributions of the quark mixing angles are also 
simulated in the same Gaussian landscape and are presented in 
Figure~\ref{fig:cpx-quark-mix}.  We see that the complex-valued Gaussian 
profiles maintain small mixing angles when the overlap integrals involve 
narrow Gaussian widths for ${\bf 10}$'s and the Higgs.  Interestingly, 
simply choosing different widths for particles in the ${\bf\bar{5}}$ 
and ${\bf 10}$ representations, in the presence of large CP-violating
phases, allows for very different flavor structures between the quark and 
lepton sectors.  Thus pursuing a microscopic description behind
the statistical distributions of Yukawa couplings has enabled us to 
go beyond the results of~\cite{HMW,DDR}.      

\begin{figure}[t]
\begin{center}
\begin{tabular}{ccc}
\includegraphics[width=0.3\linewidth]{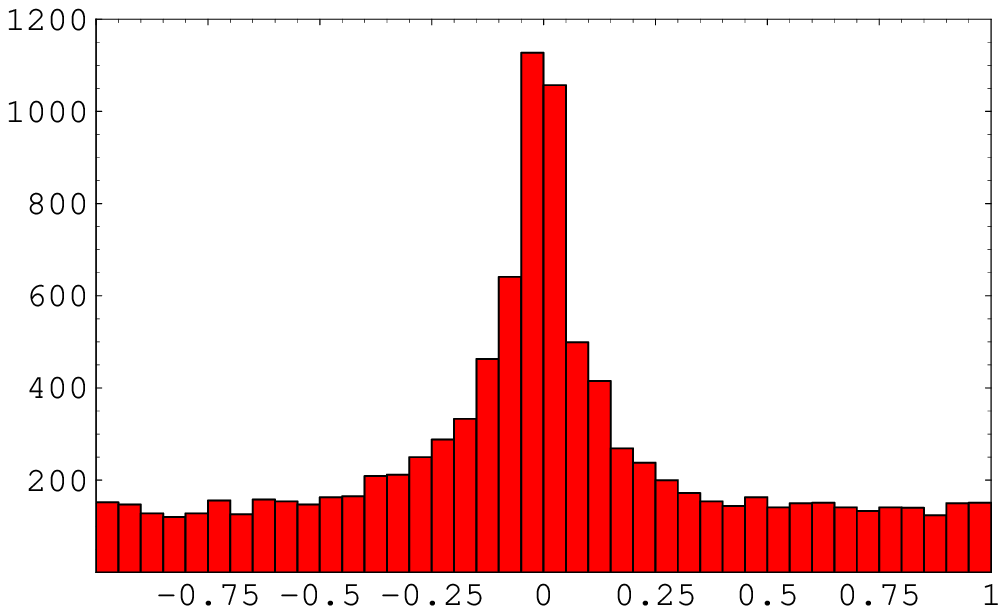} & 
\includegraphics[width=0.3\linewidth]{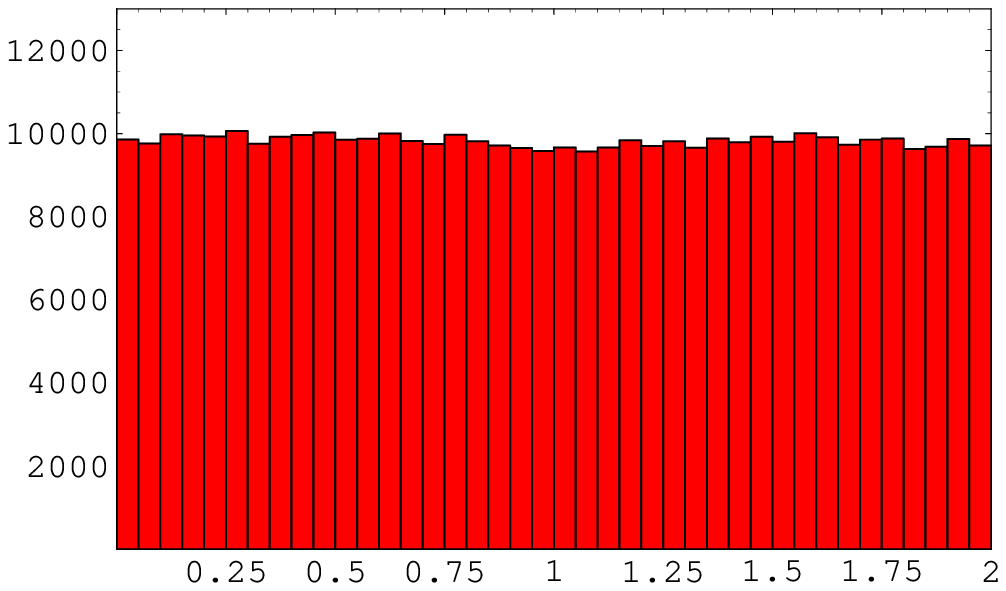} &
\includegraphics[width=0.3\linewidth]{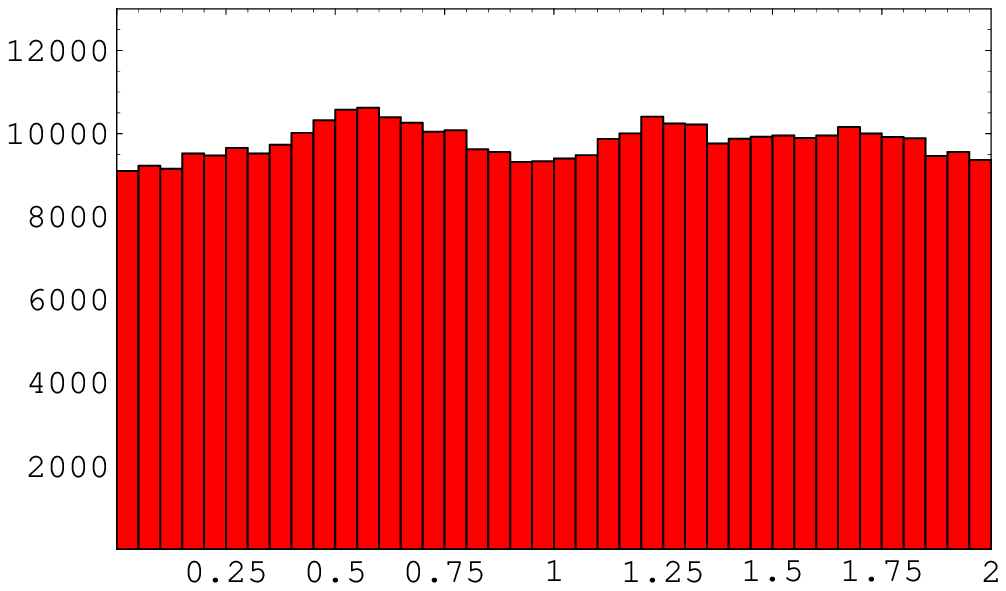} \\
$\delta_{CKM}/\pi$ & $\alpha_1/\pi$ & $\alpha_2/\pi$ \\
\includegraphics[width=0.3\linewidth]{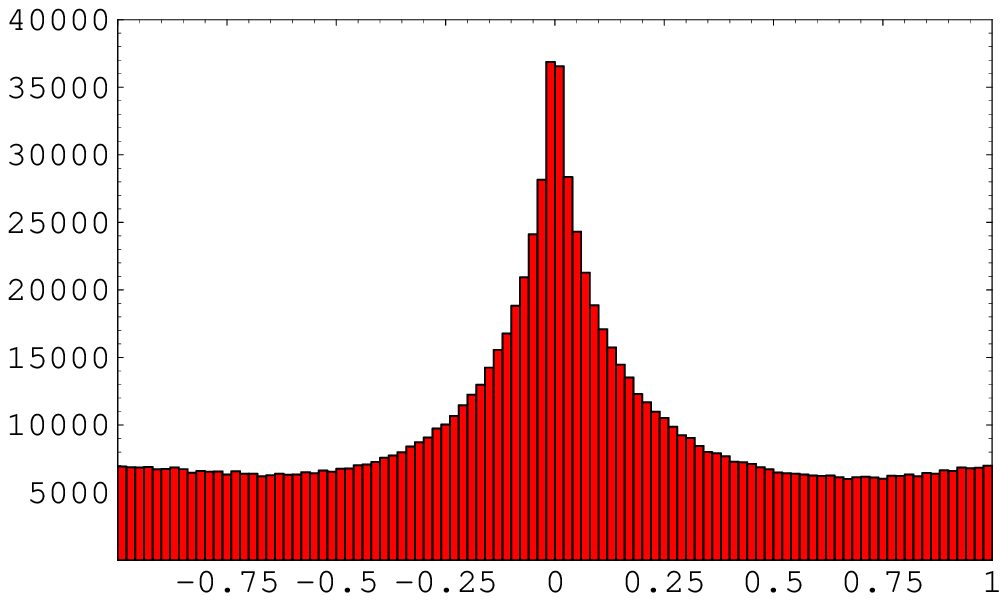} &
\includegraphics[width=0.3\linewidth]{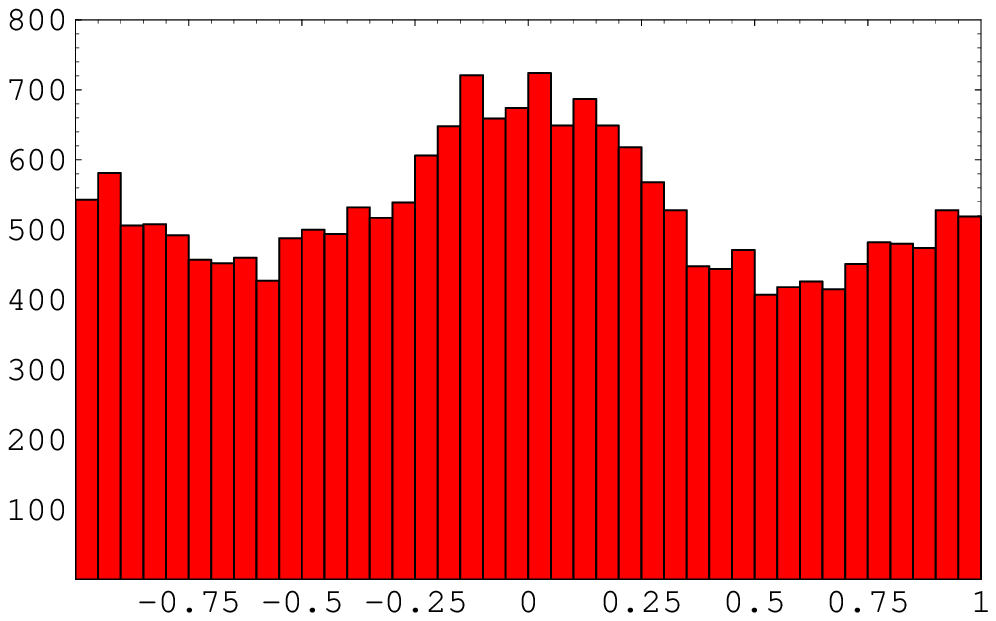} &
\includegraphics[width=0.3\linewidth]{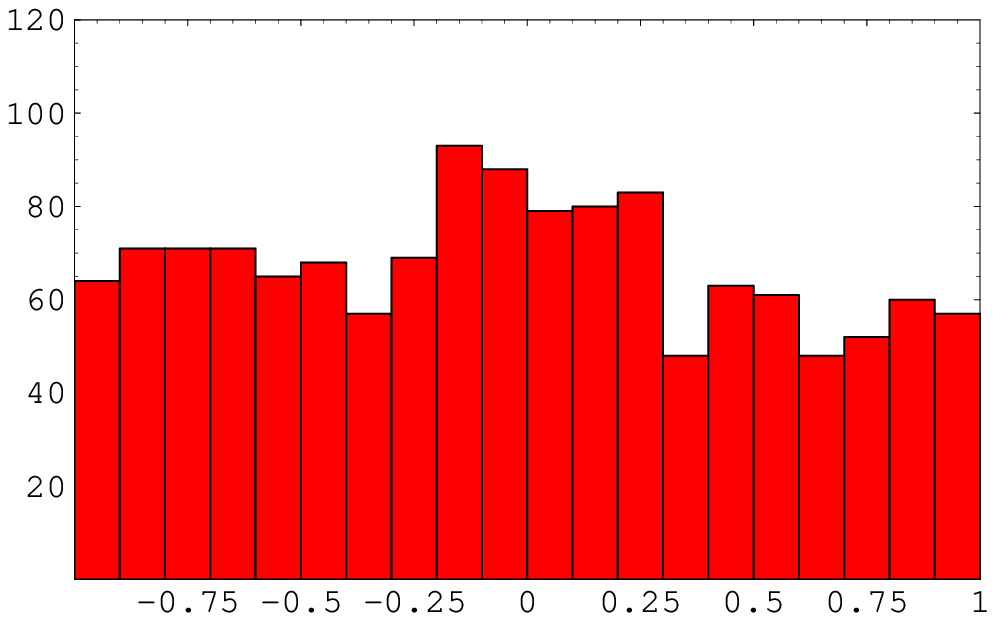} \\
$\delta_{\nu}/\pi$ (no cut) & $\delta_{\nu}/\pi$ ($B \cap C \cap D$) &  
$\delta_{\nu}/\pi$ ($A \cap B \cap C \cap D$) 
\end{tabular}
\caption{\label{fig:CP-phase} Distribution of the CP phases in the 
quark and lepton sectors, based on the Gaussian landscape on $S^1$ with 
$r=3.0$, $d/L = 0.08$ for all fields except     
$d_{\bar{\bf 5}}/L = 0.3$. }
\end{center}
\end{figure}

The distributions of the CP violating phases are also shown in 
Figure~\ref{fig:CP-phase}. 
Our parametrization of the CKM and leptonic mixing matrices $V$ and $U$ 
is quite standard:
\begin{eqnarray}
 V \!\!&=&\!\! \left(\!\!\begin{array}{ccc}
    c_{12} c_{13} & s_{12} c_{13} & s_{13} e^{-i \delta_{\rm CKM}} \\
    -s_{12} c_{23} - c_{12} s_{23} s_{13} e^{i \delta_{\rm CKM}} &
        c_{12} c_{23} - s_{12} s_{23} s_{13} e^{i\delta_{\rm CKM}} &
        s_{23} c_{13} \\
    s_{12} s_{23} - c_{12} c_{23} s_{13} e^{i \delta_{\rm CKM}} &
        - c_{12} s_{23} - s_{12} c_{23} s_{13} e^{i \delta_{\rm CKM}} &
        c_{23} c_{13} \\
		     \end{array}\!\!\right), \\
 U  \!\!&=&\!\!  \left(\!\!\begin{array}{ccc}
    c_{12} c_{13} & s_{12} c_{13} & s_{13} e^{-i \delta_{\nu}} \\
    -s_{12} c_{23} - c_{12} s_{23} s_{13} e^{i \delta_{\nu}} &
        c_{12} c_{23} - s_{12} s_{23} s_{13} e^{i\delta_{\nu}} &
        s_{23} c_{13} \\
    s_{12} s_{23} - c_{12} c_{23} s_{13} e^{i \delta_{\nu}} &
        - c_{12} s_{23} - s_{12} c_{23} s_{13} e^{i \delta_{\nu}} &
        c_{23} c_{13} \\
		     \end{array}\!\right)\!\! 
  \left(\!\begin{array}{ccc}
   e^{i\alpha_1/2}& & \\
	 & e^{i\alpha_2/2} & \\
	 & & 1\\
	\end{array}\!\right), \quad\,\,
\end{eqnarray}
where $s_{ij} \equiv \sin \theta_{ij}$ and $c_{ij}\equiv\cos\theta_{ij}$.
The Majorana mass phases of the neutrinos, 
$\alpha_{1,2} = {\rm Arg}(m_{1,2}/m_3)$, have almost flat distributions. 
On the other hand, the distributions of the two other CP phases, 
$\delta_{\rm CKM}$ and $\delta_\nu$, have peaks at $\delta\sim 0$ on 
top of otherwise flat distributions. We have not studied where this 
structure comes from.  The scatter plot of $V^e_{23}$ in 
Figure~\ref{fig:rotation-angle} reveals some structure---the scatter is 
dense in some places and thin in others---and the peaks at $\delta = 0$ in 
the CP phase distributions may have something to do with this structure,
which presumably originates from the specific form of the wavefunction 
(\ref{eq:cpxGaussian}).  If this guess is correct, then the peak plus flat 
structure in the CP phase distributions is likely to be an artifact of 
the particular way in which we have introduced complex phases into this 
landscape.  It would be interesting to see how the CP phase distribution 
changes when complex phases are introduced in a different ways, but this 
subject is beyond the scope of this paper.  Although we expect that the 
peak structure may deform, disappear, or become less significant, we 
consider it unlikely that the flat part of distribution would disappear 
and that CP phases would be predicted to be very small.  If this 
expectation is correct, then these landscapes are consistent with the 
measured value of the CP phase in the quark sector.

\subsection{Conditional Probabilities}
\label{ssec:prediction}

The neutrino sector has three observables yet to be measured, 
$\sin\theta_{13}$, $\delta_{\nu}$ and $m_{\beta\beta}$.  It would be 
interesting if landscape approaches to understanding flavor could make 
predictions for these observables.  We have developed our Gaussian 
landscapes so that all the observables measured in this universe are not 
too atypical. In this section, we use the Gaussian landscape on $S^1$, 
with the set of parameters $r=3$, 
$d_H/L=d_{\bf 10}/L=d_N/L=d_{SB}/L=0.08$, and $d_{\bar{\bf 5}}/L=0.3$, 
as an initial example to explore what kind of predictions can be obtained 
from this approach.\footnote{This Gaussian landscape has 19 scanning 
parameters, corresponding to the center coordinates of the wavefunctions 
of the quarks, leptons, Higgs boson, and symmetry-breaking field(s) 
$\varphi^{SB}$.  Meanwhile, this landscape now predicts distributions for 
all 22 flavor parameters.  Although this implies that the landscape makes 
three precise (zero-width) predictions; this is a result of our arbitrary 
choice to consider parameters such as $g$, $L/d_{\bar{\bf 5},{\bf 10},N,SB}$, 
and $r$ as fixed.  Specifically, these predictions disappear when these 
parameters also have probability distributions around the values we fixed.  
Therefore we do not pursue the possibility of a truly predictive Gaussian 
landscape.}    

The mixing angle $\theta_{13}$ is typically of order unity, and there is 
no reason to expect from this Gaussian landscape that it is very small.  
Indeed, the current experimental limit has excluded a significant 
fraction of the ensemble of vacua:
\begin{equation}
 P(\sin \theta_{13} > 0.18) = 33 \%. 
\end{equation}
Nevertheless, the experimental limit is not very strong and so a 
significant fraction of vacua---67\% of the total ensemble in this 
Gaussian landscape---sits within the experimental bound.  On the other 
hand, future experiments are expected to have a sensitivity down to about 
$\theta_{13}\sim 10^{-2}$.  The probability that $\theta_{13}$ is too 
small to be measured by such future experiments is rather low within
this Gaussian landscape; we find
\begin{equation}
 P(\sin \theta_{13} < 10^{-2})  =  12\%.
\end{equation}

\begin{figure}[t]
\begin{center}
\begin{tabular}{ccc}
$\log_{10} \sin \theta_{13}$ & 
$\log_{10} \sin^2 (2\theta_{13})$--$\delta_{\nu}/\pi$ & 
$\log_{10} |\Delta R|$ \\
\includegraphics[width=0.3\linewidth]{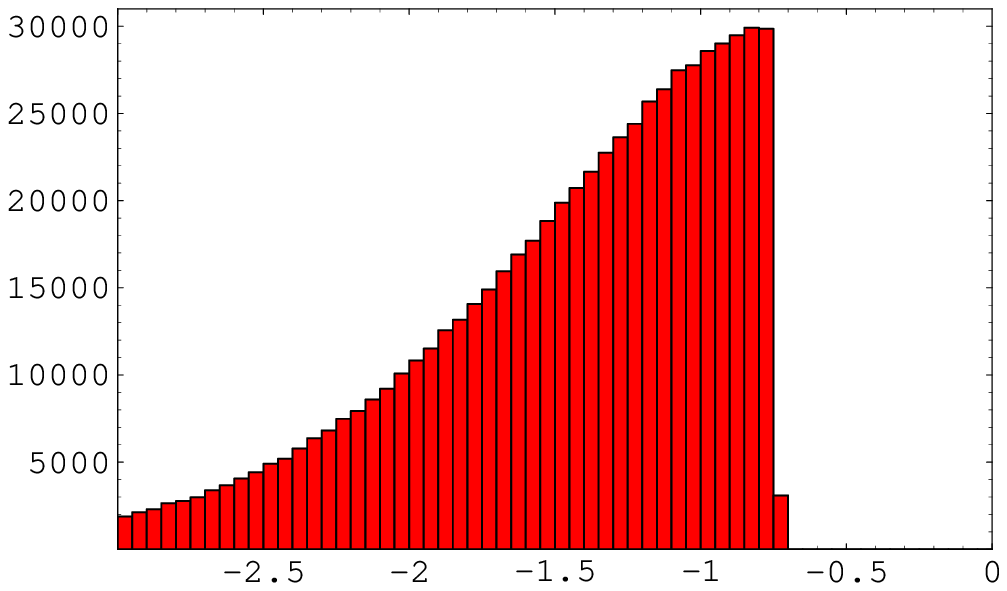} & 
\includegraphics[width=0.3\linewidth]{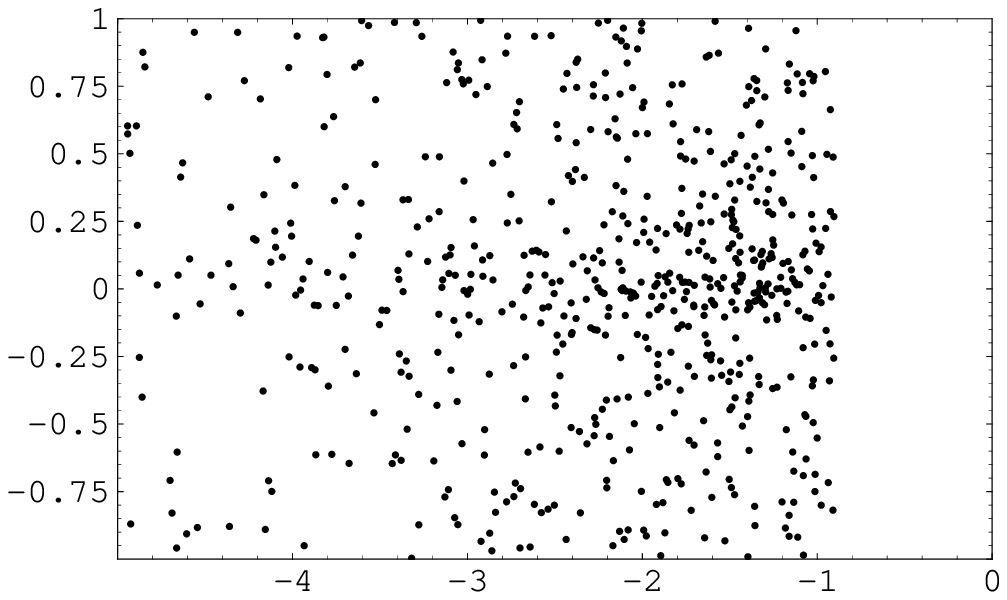} & 
\includegraphics[width=0.3\linewidth]{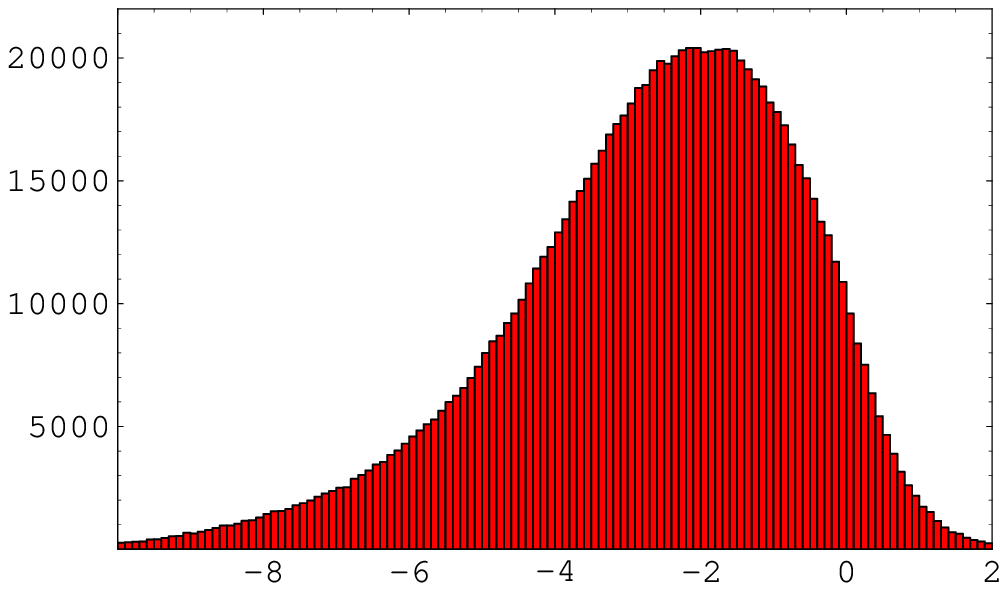} \\
($D$) & ($D$) & (no cut) \\
\includegraphics[width=0.3\linewidth]{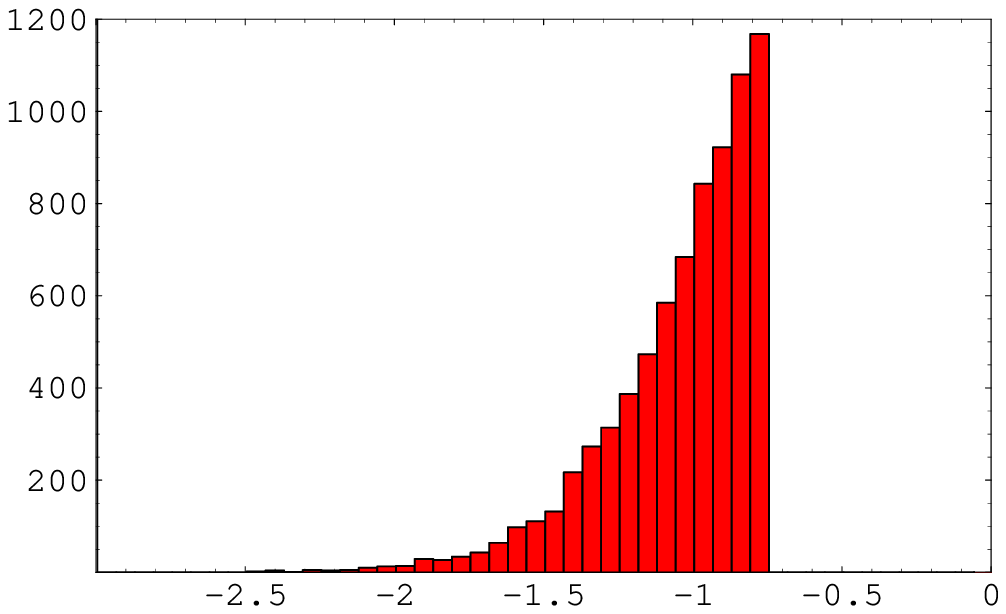} & 
\includegraphics[width=0.3\linewidth]{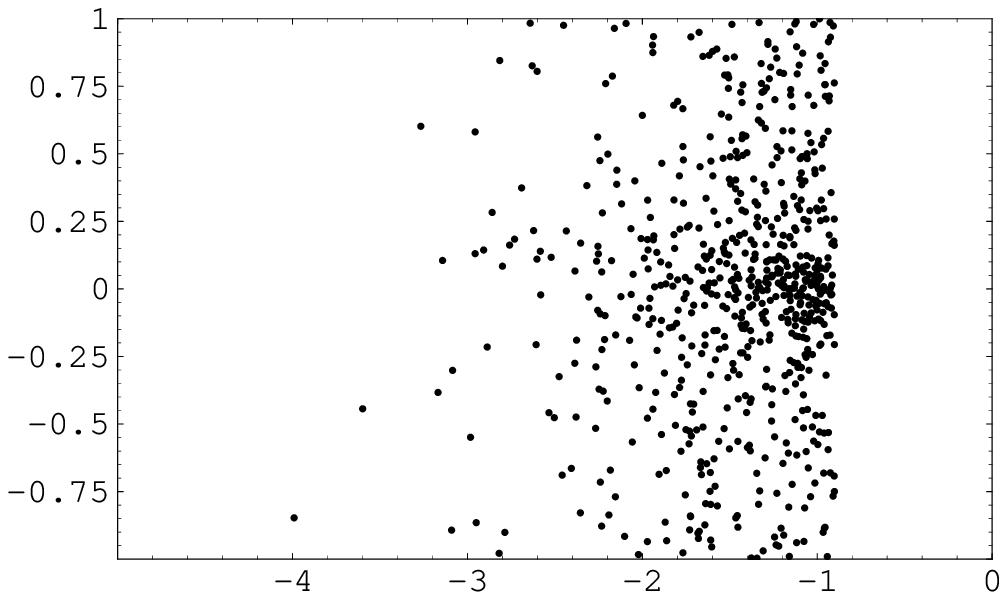} &
\includegraphics[width=0.3\linewidth]{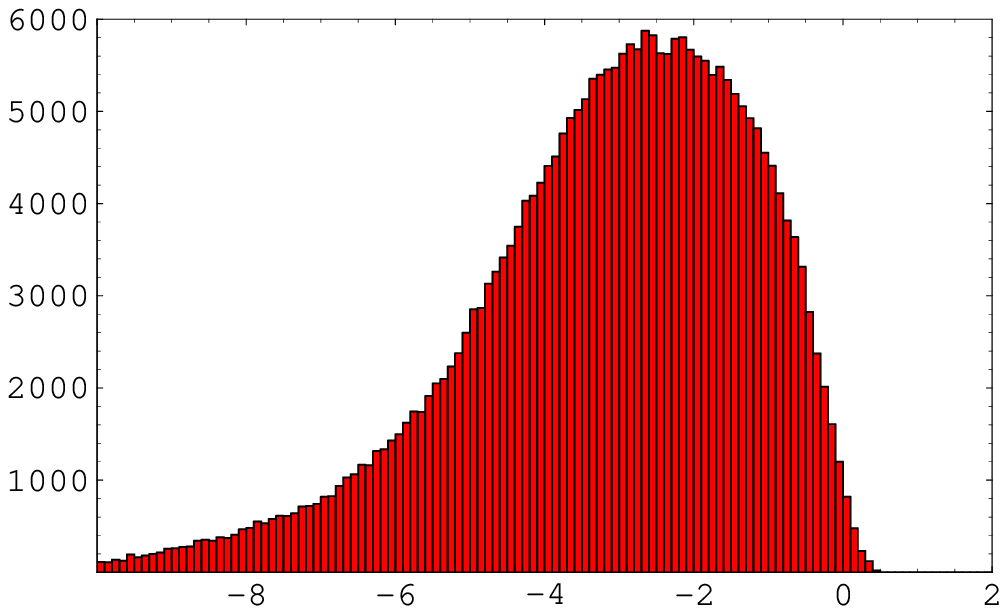} \\  
($A \cap B \cap D$) & ($A \cap B \cap D$) & ($B$) \\
\includegraphics[width=0.3\linewidth]{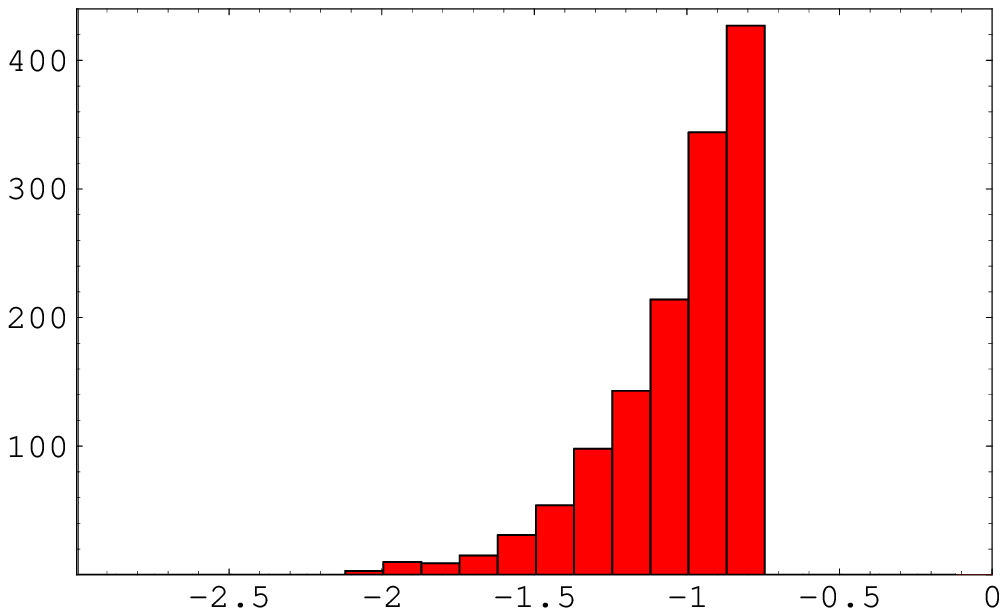} & 
\includegraphics[width=0.3\linewidth]{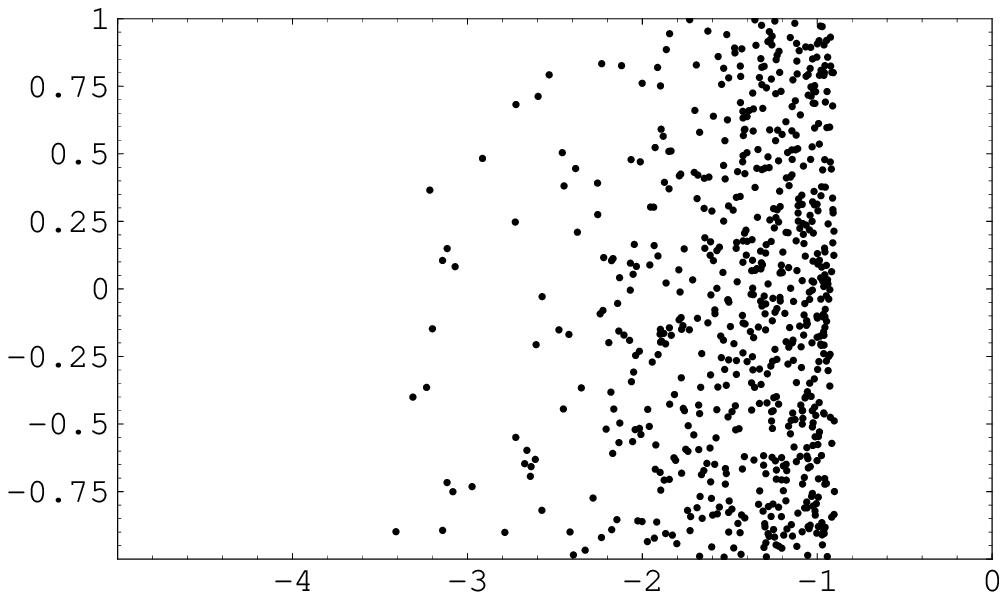} &
\includegraphics[width=0.3\linewidth]{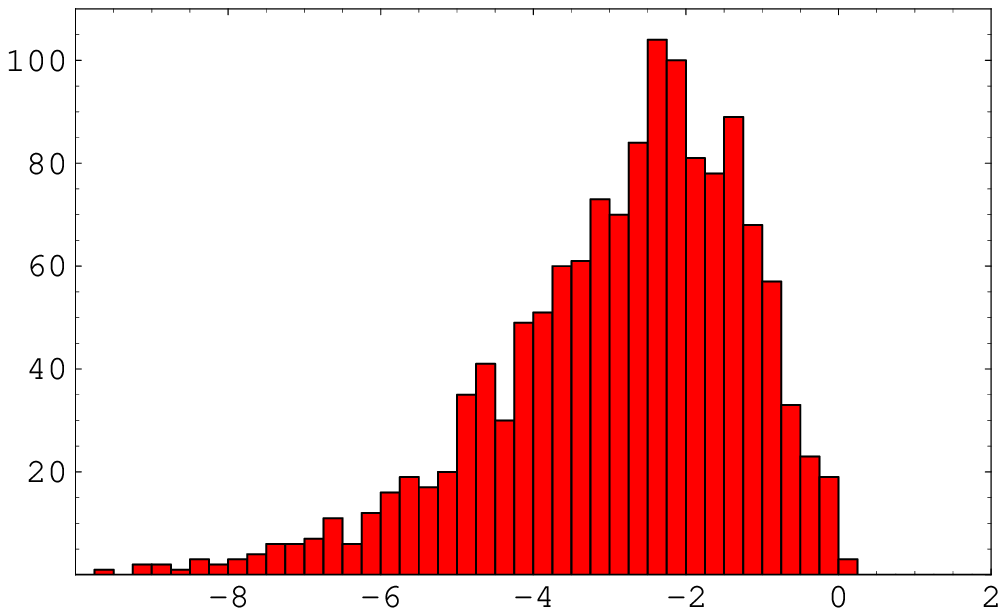} \\  
($A \cap B \cap C \cap D$) & ($A \cap B \cap C \cap D$) & 
($A \cap B \cap C \cap D$) 
\end{tabular}
\caption{\label{fig:cut-effect} Distributions of observables in the 
neutrino sector for different subsets of the total ensemble.  The 
Gaussian landscape on $S^1$ is used for this simulation, with 
$r=3$ and $d/L = 0.08$ for all fields except $d_{\bar{\bf 5}}/L = 0.3$.
In the $\log_{10}\sin^2 (2\theta_{13})$--$\delta_{\nu}$ scatter plots, the 
same number of points are displayed in all three figures.}
\end{center}
\end{figure}

The second yet-to-be measured parameter is the CP phase of neutrino 
oscillations.  The probability distribution of this is displayed in the 
bottom left panel of Figure~\ref{fig:CP-phase}.  Finally, neutrinoless 
double beta decay measures $|m_{\beta\beta}|$, where 
\begin{eqnarray}
m_{\beta\beta} &=& \sum_{i=1}^3 U_{ei}^2 m_i \equiv  
m_2 U_{e2}^2 \left[1 + \left(\frac{U_{e3}}{U_{e2}}\right)^2 
     \frac{m_3}{m_2} + \Delta R \right], \label{eq:mee} \\
\Delta R &=& \left(\frac{U_{e1}}{U_{e2}}\right)^2 \frac{m_1}{m_2}\,.
\end{eqnarray}
The second term of (\ref{eq:mee}) is roughly 
\begin{equation}
 \left| \frac{U_{e3}}{U_{e2}}\right|^2 \left|\frac{m_3}{m_2}\right|
 \sim 0.57 \times \left(\frac{\sin \theta_{13}}{0.18}\right)^2 
 \sim 0.18 \times \left(\frac{\sin \theta_{13}}{0.10}\right)^2 ,
\end{equation}
after using (\ref{eq:ignorem1}) and (\ref{eq:s2a}).  Once $\sin\theta_{13}$ 
is measured, the second term of (\ref{eq:mee}) can be estimated.  Yet we 
still need to know the last term, $\Delta R$, to make a prediction for 
$|m_{\beta\beta}|$.  The Gaussian landscape distribution of $|\Delta R|$ 
is found in the top right panel of Figure~\ref{fig:cut-effect}.

We can ask more specific questions to further constrain the probability 
distributions of these three observables {\it in universes like ours}. 
That is, when making predictions for future measurements, we can 
condition the probability distributions based on quantities that we have
already measured, as opposed to using the a priori probability 
distributions in the landscape.  However, if we were to impose the 
current experimental limit on all lepton flavor parameters, an enormous
numerical simulation would be required, since most of the simulated 
universes would not pass the cut.  This does not imply that our universe 
is very atypical, only that the present experimental error bars are small. 
Therefore, for practical reasons we employ much looser experimental 
cuts\footnote{One might consider these cuts as a tool to study the effects 
of possible cosmological or environmental selection in the multiverse, just 
like a large top-Yukawa cut (\ref{eq:cut-t}) was employed in section 
\ref{sec:toy1} in an attempt to study the impact of environmental 
selection in favor of a large top Yukawa coupling.  Here, however, we know
of no evidence that the weight factors are in favor of the range of 
parameters selected by the cut conditions $A$--$D$.} 
\begin{eqnarray}
 A: & &
 10^{-2} \,<\, \Delta m^2_{\odot}/\Delta m^2_{\rm atm} \,<\, 10^{-1}\,, \\
 B: & & \sin^2 (2\theta_{12}) > 0.7\,, \\
 C: & & \sin^2 (2\theta_{23}) > 0.8\,, \\
 D: & & \sin \theta_{13} < 0.18\,,  
\end{eqnarray}
and study whether these cuts influence the distributions of $\theta_{13}$, 
$\Delta R$ and $\delta_{\nu}$.

Table~\ref{tab:cutonchooz} shows how the probability to measure 
$\sin\theta_{13} > 3\times 10^{-2}$ changes depending on whether or not 
the cut conditions $A$--$C$ are imposed. 
\begin{table}[t]
\begin{center}
 \begin{tabular}{|c|ccc|ccc|c|}
 $D$   & $A \cap D$ & $B \cap D$ & $C \cap D$ & $A \cap B \cap D$ &
 $A \cap C \cap D$ & $B \cap C \cap D$ & $A \cap B \cap C \cap D$ \\
\hline
61\% & 75\% & 78\% & 77\% & 94--95\% & 79\% & 87\% & $96\pm 1$\%
 \end{tabular}
\caption{\label{tab:cutonchooz} Probability to measure  
$P(\sin \theta_{13}>3\times 10^{-2})$ under various cuts. 
Up to $\pm 1$\% uncertainties arise from the limited statistics 
gathered in the numerical simulation.  We use the Gaussian landscape on 
$S^1$ with $r=3$ and $d/L=0.08$ for all fields except 
$d_{\bar{\bf 5}}/L = 0.3$.} 
\end{center}
\end{table}
The loose cuts $A$--$C$ have a significant impact on the prediction. 
Among all the samples that are consistent with the current experimental 
limit on $\sin\theta_{13}$, 40\% of them are below $3\times 10^{-2}$. 
However, after imposing various other experimental constraints, $A$--$C$, 
only a few percent have $\sin \theta_{13} < 3 \times 10^{-2}$.  Therefore, 
for the wavefunction parameter choices used in this section, the Gaussian 
landscape on $S^1$ predicts a very high probability of measuring 
$\theta_{13}$ in future experiments in our universe. 

When $\sin \theta_{13}$ is large, there is a good chance to observe CP 
violation in neutrino oscillation experiments. The a priori vacuum
statistics for this parameter are peaked at $\delta_\nu = 0$ (bottom left 
panel of Figure~\ref{fig:CP-phase}).  However the distribution of 
$\delta_{\nu}$ changes as some of the cuts $A$--$D$ are imposed.  The 
impact of these changes is clearly visible when the mixing angles of both 
the solar and atmospheric neutrino oscillations are required to be large; 
then the peak of the CP phase distribution is significantly reduced 
(bottom middle panel of Figure~\ref{fig:CP-phase}).  This is reasonable, 
since complex phases play a crucial role in avoiding cancellation between 
large mixing angles in $V^e$ and $V^\nu$.  In scatter plots in the 
$\sin^2(2 \theta_{13})$--$\delta_{\nu}$ plane 
(Figure~\ref{fig:cut-effect}), we see that the distribution of the CP 
phase becomes uniform after cuts $A$--$D$ are imposed.  Therefore, after 
conditioning distributions based on our loose ``experimental cuts,'' both 
$\sin \theta_{13}$ and the CP phase become more likely to be larger, and 
hence more likely to be discovered in future experiments.

The loose cuts $A$--$D$ also have some impact on the distribution of 
$|\Delta R|$ (Figure~\ref{fig:cut-effect}).  The condition $B$ alone, 
requiring $|U_{e2}|$ be large, removes almost all the distribution 
$\log_{10} |\Delta R|>0$ as well.  Examining the samples that pass all 
cuts $A$--$D$, we find that 
\begin{eqnarray}
 P(|\Delta R| > 1) & = & (0.3 \pm 0.2)\%, \\ 
 P(|\Delta R| > 0.1) & = & (9 \pm 1)\%,  
\end{eqnarray}
where the uncertainty comes from limited statistics in the numerical 
simulation.  Therefore, it is very unlikely that $\Delta R$ is so large 
as to be significant next to the first term $m_2 U_{e2}^2$ in 
$m_{\beta\beta}$.  Hence this landscape predicts 
$m_{\beta\beta}\simeq m_2 U_{e2}^2 + m_3 U_{e3}^2$ and 
\begin{equation}
|m_2 U_{e2}^2| \times
 \left[1 - 0.18 \left(\frac{\sin \theta_{13}}{0.10}\right)^2 \right]
 \,\,\simlt\,\, |m_{\beta\beta}| \,\,\simlt\,\, 
 |m_2 U_{e2}^2| \times
 \left[1 + 0.18 \left(\frac{\sin \theta_{13}}{0.10}\right)^2 \right], 
\label{eq:mbetabeta}
\end{equation}
where $|m_2 U_{e2}^2| \simeq (2\mbox{--}3)\times 10^{-3} \, \EV$.

We have seen that the loose cuts $A$--$D$ have a significant impact 
on the distributions of $\sin\theta_{13}$, $\delta_{\nu}$ 
and $|\Delta R|$.  This study gives us a feeling for how much 
landscape predictions can change when various weight factors multiply 
the simple vacuum statistics, or when various experimental measurements 
are used to condition predictions based on what we know about our universe.
Details about these predictions will depend on the specific weight factors
or how severely the experimental cuts are imposed.  However, as we have 
seen, the impact of these cuts can be understood qualitatively, and
we expect similar results for similar weights or cuts.

Although the width parameters of ${\bf 10}$, $\bar{\bf 5}$, and
the Higgs have been optimized to fit the observed data, there is still 
much room left to change $d_N$ and $d_{SB}$ (or even $r$ or how complex
phases are introduced into the Gaussian landscape). If $d_N$ and/or $d_{SB}$
were chosen slightly larger, then the distribution of $m_2/m_3$ covers a
smaller range, and the cut $A$ would have less impact while the other three 
cuts on the mixing angles would remain as important. Since it was cuts 
$B \cap C$ and $B$, respectively, that had impacts on the distributions of 
$\delta_\nu$ and $|\Delta R|$, the ``predictions'' on $\delta_\nu$ and 
$|\Delta R|$ would not be affected very much. On the other hand, since the 
$A \cap B$ cut was important in dragging the distribution of 
$\sin \theta_{13}$ upward, the distribution of $\sin \theta_{13}$ may shift 
downward for a slightly larger $d_N$ or $d_{SB}$. Thus the ``predictions'' 
in this section should be interpreted carefully, with these and other 
caveats\footnote{For example, landscapes of Yukawa couplings originating 
from super Yang--Mills theory in higher dimensions have basis-independent
distributions, but Gaussian landscapes---simplified versions of the
former---do not (see section~\ref{sec:SYM}).  This situation is compared
to the basis independent implementation of neutrino anarchy in~\cite{HM}
and the original implementation in~\cite{HMW}.  The latter predicted
anti-correlations between mixing angles in the lepton sector, but these
angles are uncorrelated in the former.  Thus some of the correlations in
Gaussian landscapes may be spurious.  Nevertheless, the disappearance of
correlations in the basis-independent neutrino anarchy was related to
invariance under the U(3) group.  Unless there is a similar underlying
symmetry, we expect that the probability distributions of Gaussian
landscapes are not qualitatively different from those of higher
dimensional gauge theories.}
in mind.

\section{Yukawa Couplings from Super Yang--Mills Interactions}
\label{sec:SYM}

In the preceding sections, we introduced a number of toy models, 
collectively termed Gaussian landscapes, to generate statistical ensembles 
of Yukawa matrices.  The pairing and generation structures of the CKM 
matrix along with the large mixing angles of the lepton sector were 
successfully explained as random selections from these ensembles.  The 
most crucial ingredient to these Gaussian landscapes was that Yukawa 
couplings are given by overlap integration of localized zero-mode 
wavefunctions on extra dimensions.  The correlation necessary to account 
for generation structure arose because the wavefunctions of quark 
doublets and the Higgs are relevant for both the up-type and down-type 
Yukawa matrices.  We also introduced a number of additional assumptions 
concerning which parameters are scanned and which not.  In this section 
we give give greater theoretical consideration to these assumptions.

Consider a supersymmetric Yang--Mills theory on a higher dimensional 
manifold.  The internal space of $D$ extra dimensions is denoted $X_D$.
The gauge group $G$ is chosen so that it contains a subgroup $H$ such 
as $\SU(3)_C\times\SU(2)_L\times\U(1)_Y$ of the Standard Model or a 
unified gauge group such as $\SU(5)_{\rm GUT}$.  The gauge fields $A_m$ 
of the theory may have non-trivial configurations on $X_D$ as long as 
they are stable (and hence satisfy the equations of motion).  Such a 
gauge field configuration can break the underlying gauge symmetry $G$ 
down to $H$. 

In this context the fields $A_\mu$ ($\mu = 0,1,2,3$) in $\mathfrak{h}$ 
become the gauge fields of the Standard Model or the unified theory. 
Meanwhile, the Kaluza--Klein spectrum of $A_m$ ($m = 5,\cdots,4+D$) in 
$\mathfrak{g}$, under the gauge field background, may have zero modes 
(i.e., massless modes), and such quantum fluctuations become scalar 
fields in the four-dimensional low-energy effective theory.  Gauginos 
in $\mathfrak{g}$ may also have zero modes; since a spinor in a 
higher-dimensional space is bi-spinor of the Lorentz group $\SO(3,1)$ 
and $\SO(D)$, and the zero-modes of the Dirac equation on $X_D$ 
become massless fermions in the effective theory.
These low-energy scalar and spinor fields may be charged under the 
Standard Model or unified theory gauge group, or they may be singlets 
under these gauge symmetries. This is worked out by decomposing the 
adjoint representation of $\mathfrak{g}$ into pieces irreducible 
under $H$.  Charged fields arising in this way may well be identified 
with quarks, leptons and the Higgs.

Fermion and scalar fields arising from the super Yang--Mills fields in 
higher dimensions have Yukawa couplings in the low-energy effective theory.  
To see this more explicitly, consider the Lagrangian of a super 
Yang--Mills theory in higher dimensions: 
\begin{equation}
{\cal L}_{4+D} = 
- \frac{1}{4}\left[
 \frac{M_*^D}{g_*^2} \tr \left( F_{\mu\nu} F^{\mu \nu} \right)
+ 2\frac{M_*^D}{g_*^2} \tr \left( F_{\mu m} F^{\mu m} \right)
+ \frac{M_*^D}{g_*^2} 
  \tr \left( \overline{\Psi} \Gamma^\mu \partial_\mu \Psi \right)
+ \cdots \right] .
\end{equation}
These three terms become the kinetic terms for the vector, scalar, and 
spinor fields in the effective theory.  Meanwhile, the gauge coupling 
constant of the low-energy effective theory is 
\begin{equation}
 \frac{1}{g^2_{\rm eff.}} = \frac{M_*^D V}{g_*^2} \,,
\end{equation}
where $V$ is the volume of the internal space.  When the zero-mode wave 
functions $\varphi_m(y)$ and $\psi(y)$ are normalized such that 
\begin{equation}
 M_*^D \int_{X_D} d^D y \left| \varphi_m \right|^2 = 1\,, \qquad 
 M_*^D \int_{X_D} d^D y \left| \psi \right|^2 = 1\,, 
\end{equation}
then canonically normalized kinetic terms result when the low-energy 
fields are related to those in the higher dimensional theory by 
\begin{equation}
A_m(x,y) = g_* \varphi_m(y) \phi(x)\,, \qquad  
\Psi(x,y) = g_* \psi(y) \cdot \chi(x)\,.
\end{equation}
Then low-energy Yukawa couplings originate from the gauge interactions of 
gauginos
\begin{eqnarray}
\int_{X_D} d^D y \; {\cal L}_{4+D} \,=\, 
\int_{X_D} d^D y \; \frac{M_*^D}{4 g^2_*} 
 \tr\left(\overline{\Psi} i \Gamma^n \left[ A_n, \Psi \right] 
  \right) \,\longrightarrow\, \lambda \overline{\chi} \phi \chi\,,  
\end{eqnarray}
with the effective Yukawa coupling constants given by 
\begin{equation}
\lambda = \frac{M_*^D g_*}{4} \int_{X_D} d^D y 
 \tr \left( \overline{\psi} i \gamma^m 
  \left[ \varphi_m, \psi \right] \right) \,. 
\end{equation}
Thus the low-energy Yukawa couplings are given by overlap integrations, 
just as in (\ref{eq:overlap}), where now $g_*$ is interpreted as the 
gauge coupling constant of the super Yang--Mills theory.

Having specified the origin of the quarks, leptons, and their 
Yukawa couplings, we know what the underlying gauge group $G$ has 
to be. For $H=\SU(5)_{\rm GUT}$, super Yang--Mills interactions of $G=E_7$ 
or $E_8$ give rise to all of the Yukawa couplings of the quarks, charged 
leptons and neutrinos \cite{TW}.  For the Pati--Salam group 
$H=\SU(4)_C\times\SU(2)_L\times\SU(2)_R$, $G=\SU(8)$ is sufficient.
When the Heterotic $E_8 \times E_8'$ or $\SO(32)$ string theory is
compactified on a real six-dimensional compact manifold, we obtain 
Yukawa couplings in this way; for example, with $G=E_8$ and a
six-dimensional manifold $X_{D=6}$ for the $E_8 \times E_8'$ string
theory.  In some compactifications of the Type IIA 
string theory (or supergravity on eleven dimensions) and Type IIB string 
theory (or F-theory), the Yukawa couplings of quarks and leptons arise 
from open string interactions.  For example in the Type IIA string theory 
with an intersecting D6--D6 system, with quarks and leptons localized at 
the intersections of D6-branes, the Yukawa couplings are generated 
by spanning a world sheet to three intersection points with D6-branes 
connecting them. However, some of these compactifications can be 
understood as a certain limit of Heterotic compactification; this is the 
essence of string duality.  Thus the toy models that we use to 
calculate Yukawa couplings can capture certain aspects (and maybe all) of 
these compactifications as well.  Therefore, the numerical analyses of 
this paper can be used to understand the statistics of flavor structure
arising from a large region of the string theory landscape.

If quarks, leptons and the Higgs originate from a super Yang--Mills theory 
of $G$ containing $H$, and the Yukawa couplings from the super Yang--Mills 
interaction of $G$, then we cannot arbitrarily assume the following:  
\begin{itemize}
\item the shape of zero-mode wavefunctions---these must be solutions to the 
equations of motion.
\item the number of zero-modes---this is determined by the topology of the 
geometry of the internal manifold $X_D$ and of the gauge field configuration 
on $X_D$.
\item the number of independent parameters that freely scan---only moduli 
parameters, i.e. deformations of the manifold and gauge field 
configuration that cost no energy, are scanned.
\end{itemize} 
The possible restrictions above were ignored in the preceding sections, 
as we introduced assumptions instead. In this section we discuss when such 
assumptions are justified, when they are not, and when not what one should 
expect instead. 
In sections \ref{ssec:T2} and \ref{ssec:T4} we provide a pedagogical 
and bottom-up introduction to ideas that motivated us to assume Gaussian 
zero-mode wavefunctions with center coordinates that scan over the 
landscape. Most of the content of these sections is not particularly new,
and the intended audience is non-string theorists.  We use the remainder 
of this section to describe what the toy models of sections \ref{sec:toy1}, 
\ref{sec:Geometry} and \ref{sec:Lepton}---i.e. Gaussian landscapes---mean 
in the context of the string theory landscape.

Before we proceed, let us comment on the basis-independence of observables. 
The Standard Model has three independent fermions in a given representation, 
and the $3 \times 3$ Yukawa matrices transform under a group
$\U(3)_q \times \U(3)_{\bar{u}} \times \U(3)_{\bar{d}} \times \U(3)_l 
\times \U(3)_{\bar{e}}$ that preserves the canonical kinetic terms of 
the fermions. Observables such as mass eigenvalues and mixing angles
also do not change under this transformation. Thus, any probability 
distribution of Yukawa matrices in a landscape-based theory should be
either invariant under the U(3) transformations, or defined only for 
classes of Yukawa matrices equivalent under the U(3) transformations.
This basis independence follows in landscapes that derive from 
super Yang--Mills interactions.  Indeed, the zero modes---i.e. the 
solutions of massless Dirac equations in a given representation---form 
a vector space, the rank of which gives the number of generations.
One should find an orthonormal basis of the vector space 
$\left\{ \psi_I \right\}$, such that the metric of the kinetic term is 
given by 
\begin{equation}
 M_*^D \int_{X_D} d^D y \psi_I^* \psi_J = \delta_{IJ}\,,
\label{eq:orthog-condition}
\end{equation}
and use the zero-mode wavefunctions of the basis vectors to calculate 
the elements of Yukawa matrices through 
\begin{equation}
  \lambda_{IJ} = \frac{M_*^D g_*}{4} \int_{X_D} d^D y 
  \tr \left( \overline{\psi}_I i \gamma^m 
    \left[ \varphi_m, \psi_J \right] \right). 
\label{eq:Yukawa-overlap}
\end{equation}
The above U(3) transformations correspond to basis transformations of the 
vector space of zero modes.  Here the U(3) basis transformations correspond
to no more than choosing different descriptions of the same 
vacuum,\footnote{Only the vector space composed of solutions to the 
zero-mode equations of motion have these U(3) ``symmetries;''  the 
interaction terms---which are trilinear (not quadratic) in fields on a 
given background---do not respect these flavor symmetries at all.} 
and the actual statistical elements of the landscape correspond to vacua, 
not Yukawa matrices.  Thus, any landscape generating vacuum statistics 
through the overlap integrals of zero-mode wavefunctions is basis 
independent.\footnote{The U(3)-invariant Harr measure for mixing angles 
follows in basis independent landscapes~\cite{HM} when $\lambda^u$ and 
$\lambda^d$ ($\lambda^e$, $\lambda^{\nu}$, and $c_{\alpha\beta}$) are 
independent. In landscapes derived from overlap integration, however, 
there are correlations between the relevant matrices, and hence the 
mixing angles do not follow the Harr measure.}

The Gaussian landscapes introduced in sections~\ref{sec:toy1}, 
\ref{sec:Geometry} and \ref{sec:Lepton} are meant to be simplified versions 
of (subsets of) the landscape that derives from string compactification. 
Because of the way we have simplified the landscape---scanning the center 
coordinates of zero modes completely randomly and independently---the 
basis independence of the landscape from string compactification is lost 
in our Gaussian landscapes. However if the Gaussian landscapes are regarded 
as tools to acquire a better understanding of flavor physics in the 
landscape from string compactification---and this is certainly our 
attitude---then the lack of basis independence in Gaussian landscapes 
is itself not a major problem.

\subsection{Domain Wall Fermion and $T^2$ Compactification 
of Field Theories on Six Dimensions}
\label{ssec:T2}

\vspace{7pt}
\noindent
{\bf Domain Wall Fermion} 
\vspace{7pt}

\noindent
It is well known that a chiral fermion in a four-dimensional 
effective theory is localized at a domain wall in a five-dimensional 
theory~\cite{domain-wall}.  Consider a fermion in a five dimensions, 
\begin{equation}
d^4x \, dx_5 {\cal L}_{5} = d^4x \, dx_5 \, 
 \left( \chi, \overline{\psi} \right) 
 \left[
   i \left( \gamma^\mu \partial_\mu + \gamma^5 \partial_5 \right) 
   - \phi(x_5) \right]
     \left( \begin{array}{c} \!\psi\! \\ \!\overline{\chi}\! \end{array} 
\right),
\label{eq:5D-Dirac}
\end{equation}
where $\gamma^5 = i$ on $\psi$ and $-i$ on $\overline{\chi}$.  A chiral
zero mode is localized about the point $x_5 = x_{5,0}$ where a background 
scalar field configuration $\phi(x_5)$ equals zero, its wavefunction 
being  
\begin{equation}
 \psi(y) \propto e^{ - \int_{x_{5,0}} d x'_5 \, \phi(x'_5)}.
\end{equation}
When the background configuration is approximated by a linear function 
$\phi(x_5)\approx F (x_5 - x_{5,0})$, then the zero-mode 
wavefunction is approximately Gaussian:
\begin{equation}
 \psi(y) \approx e^{- F(x_5-x_{5,0})^2/2}.
\end{equation}
Thus the Gaussian profile follows from the zero-mode equations of motion 
rather easily.\footnote{On a domain-wall background, the zero-mode 
wavefunction is Gaussian in the middle of domain wall, where $\phi(x_5)$ 
is approximately linear, and decreases exponentially outside the wall, 
where $\phi(x_5)$ is nearly constant.}

However, when the $x_5$ direction is compactified into $S^1$, the scalar 
field $\phi(x_5)$ must be periodic.  If $\phi(x_5)$ hits zero with a 
positive slope $k$ times along $S^1$, it does with a negative slope $k$ 
times as well.  Thus, when there are $k$ chiral left-handed zero modes in 
the low-energy effective theory, there must be $k$ chiral right-handed 
zero modes as well, and the net chirality is zero.  Furthermore, the 
background field configuration $\phi(x_5)$ should satisfy its equation of 
motion, but kink--anti-kink configurations are in general not stable.

\vspace{7pt}
\noindent
{\bf Domain Wall Fermion from a $D=6$ U(1) Gauge Theory} 
\vspace{7pt}

\noindent
These problems are addressed if $\phi(x_5)$ is not a scalar field.
Consider for example a U(1) gauge theory on a six-dimensional spacetime 
with a charged fermion:
\begin{equation}
 {\cal L}_{6} = \overline{\Psi} i \Gamma^M D_M \Psi\,, \qquad 
 D_M = \partial_M - i A_M\,,
\end{equation}
where the gamma matrices are chosen to be 
\begin{equation}
 \Gamma^{M = 0,1,2,3} = \gamma^\mu \otimes {\bf 1}\,, \qquad 
 \Gamma^{M = 5,6} = \gamma^5 \otimes \tau^{1,2}\,. 
\end{equation}
When the gamma matrices are chosen as above, the SO(5,1) spinor $\Psi$ 
consists of four four-dimensional Weyl spinors, $\Psi = (\psi_{\uparrow}, 
\overline{\chi}_{\uparrow},\psi_{\downarrow},\overline{\chi}_{\downarrow})^T$. 
In a basis where $\Gamma^{012356} = 1 \otimes \tau^3$ and 
\begin{equation}
 \Gamma^\mu = \gamma^\mu \otimes \tau^1\,, \qquad 
 \Gamma^5 = \gamma^5 \otimes \tau^1\,, \qquad 
 \Gamma^6 = 1 \otimes i \tau^2\,,
\end{equation}
the six-dimensional Dirac equation is given by 
\begin{equation}
 \left( \begin{array}{cc}
  i \overline{D} & \sigma \cdot \partial \\ \bar{\sigma} \cdot \partial & -i D 
	\end{array}\right) 
 \left( \begin{array}{c}
	\psi_{\uparrow} \\ \overline{\chi}_{\downarrow}
	\end{array}\right) = 0\,, \qquad 
 \left( \begin{array}{cc}
  i D & \sigma \cdot \partial \\ \bar{\sigma} \cdot \partial & -i \overline{D} 
	\end{array}\right) 
 \left( \begin{array}{c}
	\psi_{\downarrow} \\ \overline{\chi}_{\uparrow}
	\end{array}\right) = 0\,,  
\end{equation}
where $D = (\partial_5 - i A_5) - i (\partial_6 - i A_6)$ and 
$\overline{D} = (\partial_5 - i A_5) + i (\partial_6 - i A_6)$.
Let us focus on the Weyl spinor 
$(\psi_\uparrow, \overline{\chi}_\downarrow)$.  When $\partial_6$ is 
ignored, this spinor's Dirac equation is the same as the five-dimensional
equation above---that is $A_6$ enters the equation just as $\phi(x_5)$ does 
in (\ref{eq:5D-Dirac}). Thus $A_6$ acts like a mass parameter for a five
dimensional fermion. Chiral fermions are localized where $A_6$ ``vanishes,'' 
just like with the domain wall fermion.  Below we make more precise what 
we mean by ``vanishing'' $A_6$.

Let us consider a U(1) gauge theory compactified on $T^2$, with 
coordinates $x_5$ and $x_6$ having periods $L_5$ and $L_6$.
Suppose that $A_6$ is linear in $x_5$ and $A_5$ is constant:
\begin{equation}
 A_5 = \xi_5\,, \qquad \qquad  A_6 = F x_5 + \xi_6\,. 
\label{eq:T2-gauge}
\end{equation} 
The field strength $F_{56} = F$ is constant, and satisfies an equation 
of motion $\partial_m F^{mn} = 0$, where $m,n = 5,6$.  There is no 
issue of kink--anti-kink instability (however see the discussion at the 
end of this section, where another instability is discussed).  Since 
$A_6$ is part of a gauge field, it must be periodic in 
$x_5 \rightarrow x_5 + L_5$ only up to a gauge transformation.  Since a 
gauge transformation $\varphi(x_6) = e^{2\pi i\,x_6/L_6}$ shifts 
$A_6$ by $2\pi/L_6$, $A_6(x_5)$ and $A_6(x_5 + L_5)$ can differ by an 
integral multiple of $2\pi/L_6$.  Therefore $F L_5$ (and the field strength 
$F$) have to be quantized:
\begin{equation}
F L_5 = \frac{2\pi k}{L_6}\,, \qquad 
F = \frac{2\pi \, k}{L_5 L_6}\,, \qquad k \in \Z\,.
\label{eq:quantize-T2}
\end{equation}
It follows from the index theorem that there are 
\begin{equation}
k = \int_{T^2} \frac{F}{2\pi} 
\end{equation}
chiral fermion zero modes. Because of the linear configuration of the 
effective mass term $A_6$, zero modes correspond only to left-handed 
spinors (if $k > 0$), and net chirality is obtained in the low-energy 
effective theory.

\vspace{7pt}
\noindent
{\bf Fermion Zero-Mode Wavefunctions:  I} 
\vspace{7pt}

\noindent
It is easy to find one of the $k$ zero-mode wavefunctions referred to
above: 
\begin{equation}
 \psi_\uparrow (x_5) \approx e^{i \xi_5 x_5} e^{- \frac{F}{2} 
    \left(x_5 + \frac{\xi_6}{F}\right)^2}\,, \qquad 
 \overline{\chi}_{\downarrow}(x_5) = 0\,,
\end{equation}
just like a domain wall fermion. This zero mode is localized where 
the effective mass parameter $A_6 = F x_5 + \xi_6$ is zero; 
$x_{5,0} = - \xi_6/F$.  Since a gauge transformation 
$\varphi(x_6)=e^{2\pi i k x_6/L_6}$ shifts $A_6$ by $2\pi k /L_6$ and 
shifts $-\xi_6/F$ by $-L_5$, the center coordinate $x_{5,0}$ of the 
Gaussian profile can always be chosen within the interval $[0, L_5]$.
To obtain the other $(k-1)$ zero modes, note that a Kaluza--Klein 
momentum in the $x_6$ direction effectively shifts the Wilson line $\xi_6$: 
\begin{equation}
(\partial_6 - i \xi_6)\, e^{2\pi i p \frac{x_6}{L_6}}\, \psi(x_5, x_6)
 = e^{2\pi i p \frac{x_6}{L_6}} 
  \left[\partial_6 - i \left(\xi_6 - \frac{2\pi p}{L_6}\right)\right]
  \psi(x_5,x_6)\,.
\end{equation}
Hence the remaining zero modes are 
\begin{equation}
 \psi_\uparrow (x_5,x_6) \approx e^{i \xi_5 x_5} 
 e^{2\pi i p \frac{x_6}{L_6}} e^{- \frac{F}{2} 
 \left( x_5 + \frac{\xi_6}{F} - \frac{p}{k} L_5
\right)^2}\,, \qquad 
 \overline{\chi}_{\downarrow} (x_5,x_6) = 0\,,
\label{eq:0mode-tentative}
\end{equation}
for $p \in \Z$ (mod $k$). The center coordinates of these $k$ Gaussian 
zero modes are at 
\begin{equation}
x_5 = - \frac{\xi_6}{F} + \frac{p}{k} L_5 \qquad {\rm ~for} 
\qquad p \in \Z/k \Z.
\end{equation}

The width of the Gaussian profile is 
\begin{equation}
 d = \frac{1}{\sqrt{F}} = \sqrt{\frac{L_5 L_6}{2\pi k}}\,.
\label{eq:width}
\end{equation}
It has been assumed that the width of Gaussian wavefunctions $d$ can 
be parametrically smaller than the size of the extra dimension in the 
$x_5$ direction. This is equivalent to 
\begin{equation}
\frac{d}{L_5} = \frac{1}{\sqrt{2\pi k u}} \ll 1\,, 
\label{eq:dLandu}
\end{equation}
where $u\equiv L_5/L_6$ parametrizes the shape of $T^2$.  In the analysis 
of section~\ref{sec:toy1} (the Gaussian landscape on $S^1$), we found
$d/L_5\approx 0.08$ was sufficient to reproduce the hierarchy of the quark 
sector. This translates to $u \approx 8$ when $k = 3$.

\vspace{7pt}
\noindent
{\bf Fermion Zero-Mode Wavefunctions: II} 
\vspace{7pt}

\noindent
In fact the fermion zero-mode wavefunctions on $T^2$ are known exactly; 
the approximate form (\ref{eq:0mode-tentative}) is valid only when 
$\partial_6$ can be ignored, i.e. when $L_6$ is so small that only one 
Kaluza--Klein momentum is relevant at a time.  Let us take a detour here 
to see the form of the zero-mode wavefunctions when $L_5\gg L_6$ is not 
satisfied.  First, we note that the fermion obeys a twisted 
periodicity condition, 
\begin{equation}
 \psi_\uparrow (x_5 + L_5,  x_6) = e^{ 2\pi i \, k \frac{x_6}{L_6}}
  \psi_\uparrow (x_5,x_6) \,, 
 \qquad \psi_\uparrow (x_5 , x_6 + L_6) = \psi_\uparrow (x_5 , x_6) \,, 
\label{eq:twist}
\end{equation}
just like, 
\begin{eqnarray}
\partial_m - i A_m(x_5+L_5,x_6) & = & e^{2\pi i k \frac{x_6}{L_6}} 
  \left[\partial_m - i A_m(x_5,x_6)\right] 
e^{ - 2\pi i k \frac{x_6}{L_6}}\,, \\
\partial_m - i A_m(x_5,x_6 + L_6) & = & \partial_m - i A_m(x_5,x_6)\,.  
\end{eqnarray}
The wavefunctions of the $k$ chiral zero modes, which satisfy 
$i \overline{D} \psi_{\uparrow} = 0$, are given by \cite{HK} 
\begin{eqnarray}
\psi_{\uparrow}(x_5,x_6)^p & = & C 
  \sum_{m \in \Z} e^{ - 2\pi \, i \frac{m}{k} p} e^{i \xi_5 x_5}
    e^{ i F  \left(x_5 + \frac{\xi_6}{F} \right)
             \left(x_6+\frac{m}{k} L_6 - \frac{\xi_5}{F}\right)}
    e^{-\frac{F}{2}\left(x_6 + \frac{m}{k} L_6 -
		    \frac{\xi_5}{F}\right)^2 } 
   \label{eq:0mode-exactA} \\
 & = & C \, e^{ i \xi_5 x_5} 
   e^{ - \frac{F}{2}\left( \tilde{x}_5\right)^2 }
   e^{ \frac{F}{2} (\tilde{x}_5 + i \tilde{x}_6)^2 }
   \vartheta_{0 ; 0}\left(\tilde{v} - \frac{p}{k}; \tau \right) ,
\label{eq:0mode-exact}
\end{eqnarray}
for $p \in \Z / k \Z = \{ 0,1,2,\cdots,(k-1)\}$ (mod $k$). Here we have
defined 
\begin{eqnarray}
\tilde{x}_5\equiv x_5+\frac{\xi_6}{F}\,, \qquad 
\tilde{x}_6\equiv x_6-\frac{\xi_5}{F}\,, \qquad
\tilde{v}\equiv\frac{\tilde{x}_5+i\tilde{x}_6}{L_5}\,, \qquad 
\tau\equiv\frac{\tau_0}{k}\,; \qquad  \tau_0\equiv i\frac{L_6}{L_5} 
\equiv\frac{i}{u}\,.
\label{eq:tau-def}
\end{eqnarray}
In addition, $\vartheta_{0;0}$ is a theta function defined by 
\begin{equation}
\vartheta_{0; 0}(v; \tau) = \sum_{m\,\in\,\Z} 
   e^{\pi i\, \tau \, m^2 + 2 \pi i \, m \, v}\,.
\label{eq:theta-def}
\end{equation}
The wavefunctions (\ref{eq:0mode-exact}) form an orthonormal
basis when the normalization constant is 
\begin{equation}
C = \left(\frac{2}{k u}\right)^{\frac{1}{4}} 
    \frac{1}{\sqrt{M_*^2 L_5 L_6}} \,.
\end{equation}
If $L_6 \gg L_5$ and $L_6 \gg 1/\sqrt{F}\approx\sqrt{L_5 L_6}$, then the 
zero-mode wavefunction (\ref{eq:0mode-exactA}) receives dominant 
contributions only from the $k$ consecutive values of $m$ for which 
$x_{6,0} = \xi_5/F - (m/k)L_6$ is between $0$ and $L_6$.  For other 
values of $m$ the last factor is suppressed by a power of 
$e^{- (\pi/2) (L_6/L_5)} \ll 1$ when $x_6 \in \left[ 0, L_6 \right]$.
Thus the wavefunctions are linear combinations of Gaussian profiles 
localized in the $x_6$ direction, 
centered at $x_{6,0} = \xi_5/F - (m/k) L_6$. 

If $L_5 \gg L_6$, then the expression given in (\ref{eq:0mode-exactA}) 
is not useful to obtain an intuitive picture. Using a modular 
transformation of the theta function, 
\begin{equation}
\vartheta_{0; 0}(v; \tau) = 
  i \sqrt{\frac{i}{\tau}} e^{- \frac{\pi \, i }{\tau} v^2}
    \vartheta_{0; 0} \left(\frac{v}{\tau};-\frac{1}{\tau}\right),
\label{eq:theta-modular}
\end{equation}
(\ref{eq:0mode-exact}) can be rewritten 
\begin{eqnarray}
\psi(x_5,x_6)^p & =  & C \, e^{i \xi_5 x_5}
   e^{- \frac{F}{2}\left(\tilde{x}_5 \right)^2 }
   e^{ \frac{F}{2} \left(\tilde{x}_5 + i \tilde{x}_6 \right)^2 } 
  \times i \sqrt{k u} \,
      e^{- \frac{F}{2} \left(\tilde{x}_5 + i \tilde{x}_6
          - \frac{p}{k} L_5 \right)^2 }
    \Theta_{0;0} \left( v' ; \tau' \right) \\
  & = & C \, i \sqrt{k u} e^{i \xi_5 x_5} 
     e^{2\pi \, i p \frac{\tilde{x}_6}{L_6} }
     e^{-\frac{F}{2}\left(\tilde{x}_5 - \frac{p}{k}L_5 \right)^2 }
    \vartheta_{0; 0}\left( v' ; \tau' \right)\,, 
\end{eqnarray}
where 
\begin{equation}
v' = \frac{k L_5}{i L_6} 
  \left(\frac{\tilde{x}_5 + i \tilde{x}_6}{L_5} - \frac{p}{k}\right)\,, 
\qquad 
\tau'\equiv - \frac{1}{\tau} = - \frac{k}{\tau_0} = i k u\,.
\label{eq:tauprime}
\end{equation}
When $u\gg 1$, all but one term in the series expansion of the theta 
function (\ref{eq:theta-def}) are suppressed by powers of 
$e^{- \pi k u} \ll 1$, while the leading order term is a Gaussian 
wavefunction localized at $x_{5,0} = -\xi_6/F + (p/k)L_5$.  This is the 
solution we obtained in (\ref{eq:0mode-tentative}). 

Although we have chosen a particular gauge (\ref{eq:T2-gauge}), the
localization pattern of fermion zero modes does not depend on the choice
of gauge: the U(1) gauge transformation on fermion fields is not more 
than a phase multiplication.  Although we chose $A_6$ to be $x_5$ 
dependent while $A_5$ is not $x_6$ dependent, the fermion zero modes 
exhibit a localization in $x_5$ when $L_6\ll L_5$, and a localization in 
$x_6$ when $L_5 \ll L_6$.  Thus the width of the Gaussian profiles is 
given by (\ref{eq:width}) in both cases, and is smaller than $L_5$ and 
$L_6$, respectively.  As expected, despite the $x_5$--$x_6$ asymmetric 
gauge choice, the localization pattern is not $x_5$--$x_6$ asymmetric. 
The modular transformation property of theta function guarantees this.

Note that when $L_5$ and $L_6$ are comparable, the zero-mode 
wavefunctions are neither Gaussian nor localized.  For more about 
zero-mode wavefunctions on tori, see for example~\cite{CIM}.

\vspace{7pt}
\noindent
{\bf Zero Modes of Various Fields in Different Representations} 
\vspace{7pt}

\noindent
At the beginning of this section, we advertised the idea that all of the
fermions, the Higgs boson and the gauge bosons are unified into a super 
Yang--Mills multiplet of a gauge group $G$ that contains the gauge group 
of the Standard Model or some unified theory.  Let us describe this idea 
a little more explicitly, using the language of $T^2$ compactification 
with a U(1) gauge field background.  Consider a $G = \U(6)$ super 
Yang--Mills theory (a parallel description applies to Pati--Salam unified 
theories with $G=\SU(8)$ and $H=\SU(4)_C\times\SU(2)_L\times\SU(2)_R$).  
Among 
the generators of the U(6) symmetry, U(1) field strengths on $T^2$ 
are introduced along 
\begin{equation}
 {\bf t}_3 = \diag \left( 1,1,1,0,0,0 \right)\,, \qquad 
 {\bf t}_2 = \diag ( 0,0,0,1,1,0)\,, \qquad 
 {\bf t}_1 = \diag ( 0,0,0,0,0,1)\,.
\end{equation}
$\SU(3) \times \SU(2)$ and a couple of U(1) factors commute with 
these field strengths, and these are identified with the Standard Model 
gauge group. Gauginos in the $3 \times 2$ block and in the $1 \times 3$ 
block are in the $({\bf 3},{\bf 2})$ and $(\bar{\bf 3},{\bf 1})$
representations of $\SU(3)_C \times \SU(2)_L$, just like the quark
doublets and the anti up-quarks. Thus, the corresponding zero modes of the
gaugino can be identified with these Standard Model fields.  Meanwhile, 
the Higgs boson may arise from $A_m$ in the $2 \times 1$ block. 
Then the gauge interaction of the gauginos in (\ref{eq:Yukawa-overlap}) 
gives rise to the up-type Yukawa coupling: following the 
commutation relations of U(6) (c.f. \cite{Madrid-bottomup}) we have,  
\begin{equation}
 \tr \left( 
\left(\begin{array}{c|c|c}
 & & \\ \hline
 & & \\ \hline
\bar{u} & & 
\end{array}
\right) 
\left[ 
\left(\begin{array}{c|c|c}
 & & \\ \hline
 & & H_u \\ \hline
 & & 
\end{array}
\right), 
\left(\begin{array}{c|c|c}
 & q & \\ \hline
 & & \\ \hline
 & & 
\end{array}
\right)
\right]
\right) \longrightarrow \bar{u} q H_u.
\end{equation}
The Yukawa couplings are then calculated by overlap integrals of the zero 
modes.

The U(1) flux in U(6) is given by ${\bf t}_3 F^{(3)} + {\bf t}_2 F^{(2)} 
+ {\bf t}_1 F^{(1)}$, with each of $F^{(a)} = \partial_5 A_6^{(a)}$ 
($a= 1,2,3$) given by $A_6^{(a)} = 
(2\pi k^{(a)} / L_5 L_6) x_5 + \xi_6^{(a)}$, as in (\ref{eq:T2-gauge}) 
and (\ref{eq:quantize-T2}). One has to set the field strengths $F^{(3)}$, 
$F^{(2)}$ and $F^{(1)}$ so that there are three quark doublets and three 
anti-up-quarks in the low-energy spectrum. This means that 
\begin{equation}
 \int_{T^2} \frac{F^{(3)}}{2\pi} - \frac{F^{(2)}}{2\pi} = 
k^{(3)} -  k^{(2)} = 3, \qquad 
 \int_{T^2} \frac{F^{(1)}}{2\pi} - \frac{F^{(3)}}{2\pi} = 
k^{(1)} - k^{(3)} = 3.
\end{equation}
The three Gaussian zero-mode wavefunctions of quark doublets 
and anti-up-type quarks are localized at 
\begin{eqnarray}
 x^{q_j}_5 & = & \left( - \frac{(\xi_6^{(3)}-\xi_6^{(2)}) L_6}
                           {2\pi (k^{(3)} - k^{(2)})} 
                    + \frac{j}{k^{(3)}-k^{(2)}} \right) L_5, \qquad 
   j \in \Z  \mod (k^{(3)} - k^{(2)})\,,  \label{eq:q-center} \\ 
 x^{u^c_i}_5 & = & \left( - \frac{(\xi_6^{(1)}-\xi_6^{(3)}) L_6}
                             {2\pi (k^{(1)}-k^{(3)})} 
                    + \frac{i}{k^{(1)}-k^{(3)}} \right) L_5,  \qquad 
   i \in \Z  \mod (k^{(1)} - k^{(3)})\,. \label{eq:ubar-center}
\end{eqnarray}
This is an illustration of the picture described in the introduction to 
this section.  We started from U(6) super Yang--Mills theory on $T^2$ and 
broke the U(6) symmetry by turning on a gauge field background, so that 
an $\SU(3)_C\times\SU(2)_L$ gauge symmetry is left unbroken along with some 
U(1) factors.  Furthermore, the gauge-field background generated chirality 
in the low-energy spectrum, and determined the shape of the zero-mode 
wavefunctions (which are Gaussian when $L_5 \gg L_6$ or $L_6 \gg L_5$). 
Wilson lines $\xi_6^{(a)}$ ($a=1,2,3$) are chosen arbitrarily, yet the 
field strength satisfies the equation of motion. Thus, these constants 
determine the zero-mode wavefunctions, and hence the observables that 
arise from the Yukawa matrices. If the Wilson lines are scanned, the peaks
of the Gaussian zero modes are also scanned, and hence the Yukawa couplings 
are scanned. Since the zero modes of quark doublets and
anti-up-type quarks depend differently on the Wilson lines---see 
(\ref{eq:q-center}) and (\ref{eq:ubar-center})---the peak positions 
of the wavefunctions of different representations are scanned 
independently from one another. 

\vspace{7pt}
\noindent
{\bf Problems} 
\vspace{7pt}

\noindent
The $T^2$ compactification of super Yang--Mills theory reproduces 
certain aspects of the Gaussian landscape on $S^1$, but it also has 
serious problems.  The easiest problem to identify involves the sum rule 
on the chirality.  Because of the simple relation 
$(F^{(3)}-F^{(2)})+(F^{(2)}-F^{(1)}) + (F^{(1)}-F^{(3)}) = 0$, 
we have 
\begin{equation}
 \# \tilde{h}_u = - (\# q + \# \bar{u} ) = k^{(2)} - k^{(1)} = - 6\,,
\end{equation}
implying that the low-energy spectrum contains $n\geq 0$ up-type Higgsinos 
and $6+n$ fermions in a representation conjugate to that of the up-type
Higgsinos.  One encounters a similar sum rule in any compactification 
on a smooth two-dimensional manifold with U(1) field strengths; thus this 
is too simple a framework to provide both a realistic spectrum 
and Yukawa interactions.

A more serious problem is that the constant $\U(1)\times\U(1)\times\U(1)$ 
gauge field configuration in U(6) is not stable.  Although this 
configuration satisfies the equations of motion, this only means that it
is a stationary point of the action.  In fact, the equations of motion of 
the gauge field in the $3\times 2$ and $1\times 3$ blocks are given by 
\begin{eqnarray}
 \left[ \Delta_{56} + (F^{(3)}-F^{(2)})_{56} \right] (A_5 + i A_6) & = &
  0\,, \\
 \left[ \Delta_{56} + (F^{(1)}-F^{(3)})_{56} \right] (A_5 + i A_6) & = &
  0\,.
\end{eqnarray}
Here $\Delta_{56}$ is the Laplacian on the $x_5$--$x_6$ plane.
The positive $(F^{(3)} - F^{(2)})_{56}$ and $(F^{(1)} - F^{(3)})_{56}$ 
required for the proper quark chirality implies that the low-energy 
complex scalar fields coming from $A_5 + i A_6$ have negative 
mass-squared. Therefore the constant U(1) gauge field configuration is 
not stable on $T^2$. It is stable only when 
$F^{(3)}_{56} = F^{(2)}_{56} = F^{(1)}_{56}$, which yields no net chirality.

\vspace{7pt}
\noindent
{\bf Complex Valued Wavefunctions} 
\vspace{7pt}

\noindent
In section~\ref{ssec:T4} we move on to discuss more complicated 
compactifications of supersymmetric Yang--Mills theory in search of  
frameworks for constructing Gaussian landscapes that avoid the above
problems.  For the remainder of this section, however, we digress to
explore how CP-violating phases can be introduced into these landscapes.
We look at the simplest geometry imaginable---compactification on 
$T^2$---considering that the essence of obtaining CP-violating phases
will not be lost in more complicated frameworks.

Real-valued Gaussian wavefunctions never yield complex phases in the 
Yukawa matrices or CP-violating phases in the quark and lepton mixing 
matrices.  Although the zero-mode wavefunctions have phase factors 
$e^{i\xi_5 x_5} e^{2\pi i p (x_6/L_6)}$, these do not end up as complex 
phases in the Yukawa matrices.  The first factor $e^{i\xi_5 x_5}$ cancels 
in the integrand of the overlap integration.  For example, consider the 
$\U(6)\rightarrow\SU(3)_C\times\SU(2)_L\times\U(1)_Y$ symmetry breaking 
model.  The up-type Yukawa operator is neutral under any one of 
$\U(1)^{(a)}$ ($a=1,2,3$) symmetries, and this ensures that the 
$e^{i \xi_5^{(3)} x_5}$ phase factor in the quark-doublet wavefunctions 
is canceled by the factor $e^{-i \xi^{(3)}_5 x_5}$ in the anti-up-quark 
wavefunctions.  As for the second phase factor $e^{2\pi i p (x_6/L_6)}$, 
recall that the overlap integration is carried out on both the $x_5$ 
and $x_6$ coordinates. This phase factor is a plane wave associated with 
Kaluza--Klein momenta, and these momenta are conserved in $T^2$ 
compactifications. Therefore in any matrix element given by overlap 
integration on $T^2$, these plane wave phase factors cannot yield complex 
phases.\footnote{Kaluza--Klein momenta conservation leads to an 
approximate texture in the Yukawa matrices. When only the leading term in 
the series expansion of the theta function is kept for each of $q$, 
$\bar{u}$, and $h$, a given quark doublet has a non-vanishing Yukawa 
matrix element with only one anti-up-type quark. All of the other matrix 
elements are suppressed by at least $e^{-\pi u}$, which arises after 
sub-leading terms in the series expansion of the theta function are 
included.  On the other hand, when compactifying on generic manifolds 
there is not necessarily any Kaluza--Klein massless U(1) gauge field and 
its corresponding conserved Kaluza--Klein momenta.  This is why we 
consider that the Yukawa texture from Kaluza--Klein momenta conservation 
is an artifact of torus compactification, and have ignored it in the 
Gaussian landscapes of earlier sections.}

However, there is a simple way to obtain complex phases in Yukawa matrices 
without considering more complicated geometries.  So far we have assumed 
$T^2$ is rectangular; that is the two independent periods,
\begin{equation}
(x_5,x_6) \sim (x_5+L_5,x_6)\,, \qquad (x_5,x_6) \sim (x_5,x_6+L_6)\,,
\end{equation}
are rectangular. In other words, using the complex coordinate 
$v\equiv (x_5 + i x_6)/L_5$ we have, 
\begin{equation}
v\sim v+1\,, \qquad v\sim v + \tau_0\,, 
\label{eq:2periods}
\end{equation}
with a pure imaginary $\tau_0$ given by (\ref{eq:tau-def}).  When $\tau_0$ 
is not pure imaginary (while keeping the metric $ds^2 = dx_5^2 + dx_6^2$), 
zero-mode wavefunctions have more complicated complex phases, as we see 
explicitly below.

We now use the gauge field configuration, 
\begin{eqnarray}
A_5 = \xi_5\,, \qquad 
A_6 = F \left( x_5 - \frac{\tau_1}{\tau_2} x_6 \right) 
 + \xi_6 + \frac{\tau_1}{\tau_2}\xi_5 = 
 F \left( \tilde{x}_5 - \frac{\tau_1}{\tau_2} \tilde{x}_6 \right),
\end{eqnarray}
where $\tau_1$ and $\tau_2$ are the real and imaginary parts of 
$\tau = \tau_0 / k$, and  
\begin{equation}
F = \frac{2\pi \, k}{L_5^2 {\rm Im}(\tau_0)} = 
     \frac{2\pi}{L_5^2 \tau_2}.
\end{equation}
The fermion fields satisfy the twisted periodicity conditions in 
(\ref{eq:twist}) for the periods in (\ref{eq:2periods}). 
The zero-mode wavefunctions (\ref{eq:0mode-exact}) are modified 
and become 
\begin{eqnarray}
 \psi(x_5,x_6)^p & \propto & \sum_{m \in \Z} 
    e^{-2\pi i \frac{m}{k} p} 
    e^{i\xi_5 x_5}
    e^{i F 
       \left( \tilde{x}_5 - \frac{\tau_1}{\tau_2} \tilde{x}_6 \right)
       \left( \tilde{x}_6 + m \tau_2 L_5 \right)
       }
    e^{\frac{\pi i \tau}{L_5^2 \tau_2^2} 
       \left(\tilde{x}_6 + m \tau_2 L_5 \right)^2
      } \\
 & = &     e^{i\xi_5 x_5} 
    e^{-\frac{\pi i}{\tau} 
       \left( \frac{ \tilde{v}\bar{\tau} - \bar{\tilde{v}} \tau }
                   {\tau -\bar{\tau}} 
       \right)^2 }
    e^{\frac{\pi i}{\tau}\tilde{v}^2}
    \vartheta_{0;0} \left(\tilde{v} - \frac{p}{k}; \tau \right),
\end{eqnarray}
where $\bar{\tilde{v}}$ and $\bar{\tau}$ are the complex conjugates of 
$\tilde{v}$ and $\tau$, respectively. To see the behavior of these 
wavefunctions when $\tau_2={\rm Im}(\tau_0)$ is small (roughly equivalent 
to $u \gg 1$), we use (\ref{eq:theta-modular}). Dropping the sub-leading 
terms in the series expansion of the theta function and 
maintaining only the ${\cal O}(\tau'_2)$ and ${\cal O}(1)$ terms 
in the exponent, we have 
\begin{eqnarray}
 \psi^p(x_5,x_6) \approx e^{i \xi_5 x_5}
   e^{\pi i \tau' \left(\frac{\tilde{x}_5}{L_5} - \frac{p}{k}\right)^2}
   e^{2 \pi \tau' \frac{p}{k} \frac{\tilde{x}_6}{L_5}}
   e^{-2\pi \tau'_1 \frac{\tilde{x}_5 \tilde{x}_6}{L_5^2}}\,. 
\label{eq:cpx-theta} 
\end{eqnarray}
The factors $e^{i\xi_5 x_5}$ and $e^{2\pi i \tau'_2 (p/k)(\tilde{x}_6/L_5)}$
may cancel in overlap integration. The second factor,\footnote{Here we 
have used $2\pi\tau'_2\sim (L_5/d)^2$; see (\ref{eq:dLandu}) and 
(\ref{eq:tauprime}).} 
\begin{equation}
e^{-\pi \tau'_2 (1 - i \tau'_1/\tau'_2) (\tilde{x}_5/L_5 - p/k)^2} 
\sim  e^{ - (1 + r i)\frac{(\tilde{x}_5 - (p/k)L_5)^2}{2d^2} }\,, 
\label{eq:complexGs7}
\end{equation}
has a complex phase when $r \equiv - \tau'_1 / \tau'_2 \neq 0$.  This
phase will not cancel with overlap integration on $T^2$, which is why
we used this second factor in the zero-mode wavefunctions in the Gaussian 
landscape of section~\ref{ssec:mixcp}.

\subsection{Doubling $T^2$---$T^2 \times T^2$ Compactification of Field
Theory on Eight Dimensions}
\label{ssec:T4}

We have seen that the zero-mode wavefunctions are localized and Gaussian 
in a $T^2$ compactification with a constant U(1) gauge field background, 
when $L_5\gg L_6$ or vice versa. Thus, when the $T^2$ is doubled and 
$T^4 = T^2 \times T^2$ compactification with a constant U(1) background 
is considered, the zero-mode wavefunctions are localized and Gaussian in 
two of the four directions, when $T^4$ is short in the other two 
directions. As we will see in this section, in $T^4$ compactification
the net chirality of various representations is no longer subject to a 
linear sum rule, and the instability problem in the $T^2$ compactification 
can be avoided when the gauge field configuration satisfies an 
anti-self-dual condition.  Note that sections \ref{ssec:T4} and 
\ref{ssec:instanton} discuss supersymmetric Yang--Mills theory on eight 
dimensions as a possible origin for Gaussian landscapes, but not because
we regard a theory on eight dimensions as a candidate for a fundamental 
theory.  Instead, these sections are intended to provide a step-by-step 
introduction to ideas that formal theorists are already familiar with, but
with minimum technicality.  Later, in sections \ref{ssec:M} and 
\ref{ssec:F}, we consider a realistic framework for building Gaussian 
landscapes.

\vspace{7pt}
\noindent
{\bf Fermion Zero Modes in Anti-self-dual Gauge-Field Background} 
\vspace{7pt}

\noindent
We label the four coordinates of $T^4$ $x_{5,6,7,8}$ and the circumferences 
in these directions are denoted $L_{5,6,7,8}$.  We assume that these 
directions are all orthogonal, i.e. the internal metric is 
\begin{equation}
 ds^2 = (d x_5)^2 + (dx_6)^2 + (dx_7)^2 + (dx_8)^2\,.
\end{equation}
On this $T^2\times T^2$, we introduce a U(1)$_2$ gauge field configuration 
embedded in SU(2):
\begin{eqnarray}
A^{(2)}_5 = \xi^{(2)}_5 \tau^3\,, & & 
A^{(2)}_6 = \left( \frac{2\pi \, k}{L_5 L_6} x_5 + \xi^{(2)}_6 \right) 
             \tau^3\,, \\
A^{(2)}_7 = - \xi^{(2)}_7 \tau^3\,, & &
A^{(2)}_8 = - \left( \frac{2\pi \, k'}{L_7 L_8} x_7 + \xi^{(2)}_8
	       \right) \tau^3\,.
\label{eq:T4-gauge}
\end{eqnarray}
The field strength is constant and satisfies the equation of motion. 
Moreover, this configuration is stable when the anti-self-dual 
condition $F= - *F$ is satisfied; that is when
\begin{equation}
\quad F_{56} = - F_{78}\,, \qquad 
\frac{2\pi \, k}{L_5 L_6} = \frac{2\pi \, k'}{L_7 L_8} \,.
\label{eq:ASD}
\end{equation}

A Weyl fermion in the fundamental representation of the SU(2) has $2k k'$ 
zero modes on $T^4$. To see this, we use the following gamma matrices:
\begin{equation}
 \Gamma^{\mu = 0,1,2,3} = \gamma^\mu \otimes 1 \otimes 1\,, \qquad 
 \Gamma^{5,6} = \gamma^5 \otimes \tau^{1,2} \otimes 1\,, \qquad 
 \Gamma^{7,8} = \gamma^5 \otimes \tau^3 \otimes \tau^{1,2}\,.
\end{equation} 
A Weyl fermion on eight dimensions consists of four-dimensional Weyl 
fermions $(\psi^a_{\uparrow \downarrow}, \psi^a_{\downarrow\uparrow},
\overline{\chi}^a_{\uparrow \uparrow},
\overline{\chi}^a_{\downarrow \downarrow})$, where $a$ is a ``color'' 
index running $a=1,2$ in the case of a fermion in the fundamental 
representation of SU(2).  Corresponding to the $kk'$ different 
choices of $p\in\Z/k\Z$ and $p'\in\Z/k'\Z$, the Dirac equation on $T^4$ 
with the gauge field background (\ref{eq:T4-gauge}) has $kk'$ zero modes 
in $\psi^{a=1}$:
\begin{equation}
\psi^{a = 1; p,p'}_{\uparrow \downarrow} = 
   \psi_\uparrow (x_5,x_6;k)^p \psi_\uparrow (x_7,x_8;k')^{p'}\,, 
\qquad \psi^{a = 1;p,p'}_{\downarrow \uparrow} = 0\,,
\qquad \psi^{a = 2; p,p'} = 0\,, 
\qquad \bar{\chi} = 0\,,  
\end{equation}
Likewise, there are also $kk'$ zero modes in $\psi^{a=2}$:
\begin{equation}
 \psi^{a = 2; p,p'}_{\downarrow \uparrow} = 
   \psi_\uparrow (x_5,x_6;k)^p \psi_\uparrow (x_7,x_8;k')^{p'}\,, 
\qquad \psi^{a = 2;p,p'}_{\uparrow \downarrow} = 0\,,
\qquad \psi^{a = 1; p,p'} = 0\,, 
\qquad \bar{\chi} = 0\,.
\end{equation}
When $L_5 \gg L_6$ and $L_7 \gg L_8$, all of these wavefunctions are 
approximately Gaussian in the $x_5$--$x_7$ plane, since the wavefunction 
is a product of Gaussians in the $x_5$ and $x_7$ directions.  Two zero
modes, one in $\psi^{a=1}_{\uparrow \downarrow}$ and the other in 
$\psi^{a=2}_{\downarrow \uparrow}$, are localized at each of 
\begin{equation}
(x_5,x_7) = \left( \left(- \frac{\xi_6 L_6}{2\pi\, k} +
		     \frac{p}{k}\right) L_5,\,
                    \left(- \frac{\xi_8 L_8}{2\pi \, k'} 
                    + \frac{p'}{k'} \right) L_7 
             \right) .
\end{equation}

The width of all of these zero modes is $d=1/\sqrt{|F_{56}|}$ in the $x_5$ 
direction and $d'= 1/\sqrt{|F_{78}|}$ in the $x_7$ direction.  The 
anti-self-dual condition (\ref{eq:ASD}) implies that the two widths $d$ 
and $d'$ are equal, as was assumed in the $D = 2$ Gaussian landscapes of
section~\ref{sec:Geometry}.

\vspace{7pt}
\noindent
{\bf Absence of a Chirality Sum Rule} 
\vspace{7pt}

\noindent
The two serious problems of $T^2$ compactification with a U(1) gauge field 
background derived from instability of the gauge field configuration and
the sum rule satisfied by the chirality of different representations.
The anti-self-dual condition (\ref{eq:ASD}) addresses the first issue in
$T^4$ (or in any four-dimensional manifold) compactification, while in 
what follows we see that there is hope to resolve the second issue in 
four-fold compactifications as well.

The chirality sum rule on $T^2$ derived from the Yukawa interactions being
consistent with the gauge symmetry of an underlying group $G$. 
Therefore, to discuss the existence of such a sum 
rule\footnote{To be precise, chirality on four dimensions is {\it not} 
obtained from a super Yang--Mills theory on eight dimensions compactified 
on $T^4$ with an anti-self-dual gauge field configuration.  This is 
because the super Yang--Mills multiplet contains a pair of Weyl fermions 
with opposite chirality.  When a zero mode is found in 
$\psi^{a=1}_{\uparrow \downarrow}$ ($\psi^{p=2}_{\downarrow \uparrow}$) in 
a Weyl fermion $(\psi_{\uparrow \downarrow}, \psi_{\downarrow \uparrow}, 
\bar{\chi}_{\uparrow \uparrow}, \bar{\chi}_{\downarrow \downarrow})$, 
another Weyl fermion with the opposite chirality, 
$(\bar{\psi}_{\uparrow \downarrow}, \bar{\psi}_{\downarrow \uparrow}, 
\chi_{\uparrow \uparrow}, \chi_{\downarrow \downarrow})$, also has 
a zero mode $\bar{\psi}^{p=1}_{\uparrow \downarrow}$ 
($\bar{\psi}^{a=2}_{\downarrow \uparrow}$). The chirality we refer to 
in this section concerns $T^4$ compactification of ten to six dimensions, 
which has nothing to do with the chirality on four dimensions.  As we 
explain later, however, the chirality on six dimensions in $T^4$ 
compactification has non-linearity, which is shared by chirality on four 
dimensions in any six-dimensional compactification. This ``chirality'' is 
discussed in the context of $T^4$ compactification because this is the 
simplest system with this non-linearity.} 
on $T^4$, we must consider constraints from the underlying gauge symmetry 
whose super Yang--Mills interactions give rise to the Yukawa couplings of 
quarks and leptons.  Here we consider $G = E_6$ and $H = \SU(5)_{\rm GUT}$.  
$E_6$ contains a subgroup 
\begin{equation}
 E_6 \supset \SU(2) \times \SU(6) \supset \SU(2) 
   \times \U(1)_6 \times \SU(5)_{\rm GUT}\,.
\end{equation}
The $H=\SU(5)_{\rm GUT}$ symmetry is unbroken when a background gauge 
field on $T^4$ is contained within $\SU(2)\times\U(1)_6$. We use the 
U(1)$_2$ gauge-field configuration $A_m^{(2)}$ in (\ref{eq:T4-gauge}), 
embedded in the SU(2) factor, and introduce a U(1) gauge field 
background in the U(1)$_6$ factor, given by 
\begin{equation}
 A^{(6)}_6 = \frac{2\pi k''}{L_5 L_6} x_5 + \xi_6^{(6)}\,, \qquad 
 A^{(6)}_8 = - \left(\frac{2\pi k^{'''}}{L_7 L_8} x_7 + \xi_8^{(6)}\right)\,,
\end{equation}
with another anti-self-dual condition $2\pi k''/L_5 L_6 = 
2\pi k^{'''}/L_7 L_8$.

The irreducible decomposition of the $E_6$ Lie algebra, 
\begin{eqnarray}
 \mathfrak{e}_6 \mbox{-}{\rm adj.} & \longrightarrow & 
 ({\rm adj.},1) + (1,{\rm adj.}) + (\wedge^3 {\bf 6},{\bf 2})\,, \\
 & \longrightarrow & 
 ({\rm adj.},1)^0 + ({\bf 5},1)^6 + {\rm h.c.} + (1,{\rm adj.}) + 
 (\wedge^2 {\bf 5},{\bf 2})^{-3} + {\rm h.c.}\,,
\end{eqnarray}
shows that $\mathfrak{g}/\mathfrak{h}$ contains both 
$(\wedge^2 {\bf 5},{\bf 2})^{-3}$ and $({\bf 5},{\bf 1})^6$, 
candidates for the ${\bf 10}=(q,\overline{u},\overline{e})$ and 
$H({\bf 5})$ multiplets, respectively.  The above decomposition refers to
the the $\SU(6)\times\SU(2)$ subgroup in the first line, and 
$\SU(5)_{\rm GUT}\times\U(1)_6\times\SU(2)$ in the second line.  Gauge 
interactions of $\mathfrak{e}_6$ gauginos may in principle give rise to the 
four-dimensional up-type Yukawa 
couplings\footnote{Let us first consider how gauge indices are contracted. 
The SU(5)$_{\rm GUT}$ indices of the two fermions $\psi_{\bar{\bf 10}}$
are contracted with the SU(5)$_{\rm GUT}$ totally anti-symmetric tensor, 
and the contraction is symmetric under the exchange of the two. On the 
other hand, the internal gauge indices of $\SU(2)$ are anti-symmetric 
as the above interaction is neutral under the underlying gauge symmetry 
$\SU(2)\times\U(1)_6$, when the contraction is in the first term 
of the right-hand side of 
${\bf 2}^{-3} \otimes {\bf 2}^{-3} \otimes {\bf 1}^{+6} \simeq 
\wedge^2 {\bf 2} \oplus \cdots$. 
Generally, gauge indices are contracted anti-symmetrically under the 
exchange of the two zero modes in the $(\wedge^2 {\bf 5},{\bf 2})^{-3}$ 
representation.  This should be the case because the gauge indices are 
contracted through the structure constant
of a Lie algebra in (\ref{eq:Yukawa-overlap}). 

Spinor indices are contracted symmetrically under the exchange of 
$\psi_{\bf 10}$'s;  both the SO(3,1) contraction of left-handed spinors 
in four dimensions and the contraction of spinor indices of SU(2)
$\subset$ SO(4) in the internal space are anti-symmetric.
Therefore the combined contraction of gauge (anti-symmetric) and spinor
(symmetric) indices is anti-symmetric.  This is expected since the 
gauginos in eight dimensions are Grassmann variables.
}\raisebox{1mm}{,}\footnote{\label{fn:HiggsinVect}
The Higgs scalar should, then, originate from the complex scalar of a 
super Yang--Mills multiplet on eight-dimensions. However, because of the 
${\cal N} = 2$ supersymmetry, preserved in $T^4$ compactification with an 
anti-self-dual gauge field configuration, the existence of massless Higgs 
boson implies that its ${\cal N}=2$ super partner, a vector field in the 
same representation, should also be massless.  This implies that the 
symmetry is not broken down to $\SU(5)_{\rm GUT}$.  As we have seen, 
$T^4$ compactification with an anti-self-dual configuration has so many 
problems that it cannot be regarded as a realistic framework for landscapes. 
Our goal in section~\ref{ssec:T4} 
is to introduce field-theory ideas that apply to many 
compactifications, using the language of $T^4$ compactification.
} 
\begin{equation}
{\cal L}_{4} = \psi_{\bf 10} \cdot \psi_{\bf 10} \cdot 
\phi_{H({\bf 5})}\,. 
\end{equation}
As a quick check, note that in the $\mathfrak{e}_6$ algebra 
$\tr \left((\wedge^2 {\bf 5},{\bf 2})^{-3} \; \left[ ({\bf 5},{\bf 1})^{+6}, 
(\wedge^2 {\bf 5},{\bf 2})^{-3}\right] \right)$ does not vanish.
The coefficients (Yukawa matrix elements) are calculated by the
overlap integration of the zero-mode wavefunctions, picking up a pair of 
$\psi^{p=1}_{\uparrow \downarrow}$ and $\psi^{p=2}_{\downarrow \uparrow}$.
If the compactification preserves ${\cal N} = 1$ supersymmetry, a complex 
scalar and a chiral fermion in a chiral multiplet have the same wavefunctions 
on the internal manifold. Therefore the wavefunction of the Higgs 
boson is Gaussian, whenever the Higgsino wavefunction is Gaussian.

The number of zero modes is calculated in terms of the flux quanta 
$k, k', k''$ and $k^{'''}$. Up-type Higgsinos in low-energy spectrum 
are zero modes of the Dirac equation involving a U(1) gauge field
$6A^{(6)}_m$. Therefore the number of up-type Higgsino zero modes is 
\begin{equation}
 \# H_u = (6 k'')(6k^{'''}) = 36 k'' k^{'''}\,.
\end{equation}
Light fermions in the {\bf 10} representation are zero modes of a Dirac 
equation with gauge field $A^{(2)} - 3A^{(6)}$. The number of zero modes
is 
\begin{equation}
 \#  {\bf 10} = (k - 3k'')(k' - 3k^{'''}) + (k + 3k'')(k'+3k^{'''}) 
  = 2kk' + 18 k''k^{'''} \,,   
\end{equation}
with the first term coming from those in $\psi^{a=1}$ and 
the second term from those in $\psi^{a=2}$. We have four (discrete) 
parameters to choose by hand to fit the desired multiplicities of the fields 
in the two different representations, $H_u$ and ${\bf 10}$.

The discussion so far has not been realistic. The above multiplicities 
count only the number of hypermultiplets of ${\cal N} = 2$
supersymmetry in four dimensions. We have also commented in footnote 
\ref{fn:HiggsinVect} that the Higgs scalar field in the up-type Yukawa 
couplings should belong to an ${\cal N} = 2$ vector multiplet.  However, 
it will now be easy to take one more step and triple $T^2$ to a $T^6$ 
compactification of a super Yang--Mills theory in ten dimensions.  
Chirality on four dimensions is obtained in $D=6$ compactifications, and 
the net chirality of a given representation is cubic in discrete
parameters specifying the fluxes on $T^6$.  No sum rule holds among the 
chirality of various representations, since they are not linear functions 
of the flux parameters.  As more flux parameters are involved, it is easier 
to fit the multiplicities of fields in various representations.  It is 
also known that one of the gauge background stability conditions is 
$F_{\alpha\bar{\beta}} g^{\alpha\bar{\beta}} =0$ (in compactifications
preserving ${\cal N} = 1$ supersymmetry), and the $T^2$ compactification 
has instability because only one term contributes to the left-hand side, 
with no chance of cancellation. Hence this instability is an artifact of 
the $D=2$ compactification.  $E_8$ contains $E_6$, and therefore the 
algebra that led to the up-type Yukawa couplings still works in $E_8$. 
Thus super Yang--Mills theories of ten dimensions compactified on 
six-dimensional internal manifolds can be a good theoretical framework 
for building Gaussian landscapes. 

\vspace{7pt}
\noindent
{\bf Scanning of the Center Coordinates}
\vspace{7pt}

\noindent
Before closing this section, let us see how the localized wavefunctions 
behave as the constant gauge field background on $T^4$ changes.  We 
already know that all the zero-mode wavefunctions are approximately 
Gaussian if $L_5\gg L_6$ and $L_7\gg L_8$. The center coordinates of these 
wavefunctions depend on the Wilson lines $\xi_6^{(2),(6)}$ 
and $\xi_8^{(2),(6)}$, according to 
\begin{eqnarray}
 (x_5^{H_u},x_7^{H_u}) \!\!& = &\!\! \left(\! 
   \left( \frac{p_{H_u}}{6k''}-\frac{\xi_6^{(6)}L_6}{2\pi k''}\right)L_5, 
   \left( \frac{p'_{H_u}}{6k^{'''}}-\frac{\xi_8^{(8)}L_8}{2\pi k^{'''}}
    \right)L_7 \right), \\
 \left(x_5^{\bf 10}, x_7^{\bf 10}\right)_{a=1} \!\!& = &\!\!\! 
   \left(\! 
   \left( \frac{p_{\bf 10}^{a=1}}{k-3k''}
   -\frac{(\xi^{(2)}_6 -3 \xi_6^{(6)})L_6}{2\pi (k - 3k'')}\right) L_5, 
   \left( \frac{{p'}_{\bf 10}^{a=1}}{k'-3k^{'''}}
   -\frac{(\xi_8^{(2)}-3 \xi_8^{(8)})L_8}{2\pi (k'-3k^{'''})}\right) L_7 
   \right), \\
 (x_5^{\bf 10},x_7^{\bf 10})_{a=2} \!\!& = &\!\!\! 
   \left(\! 
   \left( \frac{p_{\bf 10}^{a=2}}{k+3k''}
   -\frac{(\xi^{(2)}_6 + 3 \xi_6^{(6)})L_6}{2\pi (k + 3k'')}\right) L_5, 
   \left( \frac{{p'}_{\bf 10}^{a=2}}{k'+3k^{'''}}
   -\frac{(\xi_8^{(2)}+3 \xi_8^{(8)})L_8}{2\pi (k'+ 3k^{'''})}\right) L_7 
   \right), \,\,\,\,\,\,\,\,\,\,\, 
\end{eqnarray}
where the coordinates in the first line are those of up-type
Higgsino(s), while the second and third lines are those of the {\bf 10} 
fermions, with the second and third lines coming from $\psi^{a=1}$ and 
$\psi^{a=2}$, respectively.  Note that the coordinates of the second and 
third lines show different dependences on the Wilson lines. Thus as the 
Wilson lines are scanned, the localized fermions 
in the {\bf 10} representation of $\SU(5)_{\rm GUT}$
change their relative positions.  We see that the scanning
of Wilson lines allows for more than the center-of-mass scanning of 
Gaussian peaks to zero-mode wavefunctions in $T^4$ compactification.  In
the Gaussian landscapes we scanned the peak positions of Gaussian 
wavefunctions without considering any correlations or constraints among 
the various peak positions.  The above situation is still far from this 
treatment, but so far we have only scanned the Wilson lines.  In section
\ref{ssec:instanton} we see that there are other stable gauge field 
configurations, and as more gauge field configurations are scanned, 
correlation will be lost among the Gaussian peak positions. This is the
rationale behind the absence of correlations in the Gaussian landscapes.

\subsection{Instanton Moduli and Random Peak-Position Scanning}
\label{ssec:instanton}

\noindent
{\bf Instanton Moduli} 
\vspace{7pt}

\noindent
We have now seen an explicit example of a stable gauge field background 
on $T^4$ parameterized by Wilson lines. In general, stable gauge field 
configurations allow for continuous deformations, such as Wilson lines, 
and the parameters of such deformations are called moduli. As for 
gauge field configurations on a four-dimensional manifold, we know that 
those satisfying the anti-self-dual condition satisfy both the Yang--Mills
theory equations of motion and the stability condition. Such gauge field 
configurations, called instantons, are known to have numerous moduli 
parameters; the Wilson lines on $T^4$ are just a subset of the instanton 
moduli parameters.  Therefore we now consider what happens when the 
instanton moduli parameters are scanned.  Our primary interest is to 
explore the moduli space of stable gauge field configurations on a 
six-dimensional manifold, as this is relevant to the compactification of 
the Heterotic string theory. However, to warm up we first study instanton 
moduli on a four-dimensional manifold.

The $E_6\rightarrow\SU(5)_{\rm GUT}$ symmetry breaking model in 
section~\ref{ssec:T4} generates the up-type Yukawa couplings.  As long as 
a background configuration of $\mathfrak{e}_6$ gauge field is contained in 
its $\mathfrak{su}(2) + \mathfrak{u}(1)_6$ subalgebra, the   
$\SU(5)_{\rm GUT}$ symmetry remains unbroken.  Note that the gauge field
configuration does not have to be pure Abelian over the entire $T^4$ 
as is assumed in (\ref{eq:T4-gauge}). The gauge field configuration 
$A^{(2)}$ in (\ref{eq:T4-gauge}) can be replaced by any one of the SU(2) 
instanton configurations\footnote{
The number of fermion zero modes is given by the topology of the gauge 
field configuration and of the geometry, as in the case of $T^2$ 
compactification.  It follows from the index theorem that the number of 
fermion zero modes in a representation $R$ of a gauge-field background is 
\begin{equation}
- \int_X {\rm ch}_R(F) \hat{A}(TX)  = - \int_X {\rm ch}_{2; \, R} (F) 
+ \frac{\dim R}{24} \int_X p_1(TX) \,,  
\end{equation}
where $\hat{A}(TX)$ is the $\hat{A}$ classes of $X$ and 
$p_1(TX)$ is in the first Pontrjagin class.  This expression is valid for 
an arbitrary four-dimensional manifold $X$. The first term on the 
right-hand side is equal to $2T_R I$, that is it is proportional to the 
instanton number $I$.  As long as the instanton number remains the same, 
the net chirality does not change for any instanton configuration.
See the text for the definitions of ${\rm ch}_{2, \; R}(F)$ and $I$.
} 
on $T^4$.  The 't Hooft solution is an SU(2) instanton configuration on 
a flat Euclidean four-dimensional space~\cite{tHooft}; it has $5 I$ 
moduli parameters, where the instanton number $I$ is topological and 
defined by 
\begin{equation}
 I = - \int {\rm ch}_{2,{\bf fund.}}\left(F \right) 
   = - \int \tr {}_{\bf fund.} 
            \left[ \frac{1}{2}\left(\frac{F}{2\pi}\right)^2 \right]
   = - \frac{1}{2 T_R} \int {\rm ch}_{2, R}\left( F \right)\,;
\end{equation}
here {\small\bf fund.} stands for fundamental representation, $R$ is an
arbitrary representation, and ``{\rm ch}$_{2;R}$'' stands for the second 
Chern character in the representation $R$. The gauge-field configuration
$A^{(2)}$ in (\ref{eq:T4-gauge}) has $I=2kk'$ instantons. 
Meanwhile, the 't Hooft solution, 
\begin{equation}
 A^a_m = - \bar{\eta}^a_{mn} \partial_n \ln \left( 1 + 
   \sum_{j = 1}^I \frac{\rho_j^2}{(y - y_j)^2} \right),
\label{eq:tHooft}
\end{equation}
describes $I$ isolated instantons, centered at $(y_j)_m$ 
with a size $\rho_j$. Here, $a=1,2,3$ label the three generators 
of $\mathfrak{su}(2)$ and $\bar{\eta}^a_{mn}$ is the eta symbol 
of 't Hooft. Thus there are indeed $(4+1)I$ moduli parameters.
In addition to these five moduli parameters per instanton, 
there are three more moduli parameters describing how an instanton 
solution is embedded within the group $\SU(2)$. Thus, there are 
(roughly) $8 I$ moduli parameters for the SU(2) $I$-instanton configuration.

On a compact four-dimensional manifold $X$, the number of instanton moduli 
parameters of a gauge group $G$ is given by~\cite{AS}
\begin{equation}
-2 \int_X {\rm ch}_{2; {\rm adj.}} (F)\, td(TX) = 
4 T_G I - \frac{\dim G}{6}\int_X c_2(TX) + c_1(TX)^2\,,  
\end{equation}
where $T_G$ denotes the Dynkin index in the adjoint representation (also 
known as the dual Coxeter number), $td(TX)$ is the Todd classes, and 
$\dim G$ is the dimension of $G$.  
When $G$ is $\SU(N)$, $T_G=N$ and $\dim G = N^2-1$.  Apart from the second 
term, which is associated with compactness of the four-fold $X$, the 
first term reproduces the result of $G = \SU(2)$---$8I$ moduli parameters.
The second term corresponds to an obstruction for lifting isolated $I$ 
instantons into an anti-self-dual configuration over all of the $X$ 
(c.f.~\cite{Taubes}).  The Wilson lines $\xi_{5,6,7,8}^{(2)}$ on the 
$T^4$ compactification of section~\ref{ssec:T4} are part of the instanton 
moduli.  However unlike the Wilson lines of $T^4$, which are associated 
with a non-trivial $\pi_1(T^4)$, most of instanton moduli here are not 
strongly associated with the specific geometry of $T^4$. Thus the existence 
of instanton moduli is very robust.

\vspace{7pt}
\noindent
{\bf Wilson lines as Instanton Moduli Parameters} 
\vspace{7pt}

\noindent
So far it has not been clear how the Wilson lines on $X = T^4$ are related 
to the instanton moduli parameters, such as $(y_j)_m$ and $\rho_m$, in 
the 't Hooft solution (\ref{eq:tHooft}); in particular the constant-field 
configuration does not have a well-defined center. To see explicitly that 
the Wilson lines actually correspond to instanton center coordinates, ADHM 
data~\cite{ADHM} and the Nahm transformation~\cite{Nahm} are useful. 

The ADHM formalism~\cite{ADHM} allows one to parametrize the moduli space 
of an instanton configuration on a flat space $\R^4$.  A set of parameters 
called ADHM data describes all possible instanton gauge-field 
configurations and, in turn, the data can be extracted starting from the 
gauge field configuration.  For simplicity, we consider only the  
$I$-instanton configurations of an SU($N$) gauge group.  Among the data 
is an $I\times I$ matrix-valued $\hat{A}_m$. This part of the data is 
extracted by 
\begin{equation}
 \left(\hat{A}_m \right)^{p'p} = \bra{p'}x_m\ket{p} = 
\int_{\R^4} d^4x \, \psi^{p'\dagger} x_m \psi^p\,,
\end{equation}
where $\psi^p$ ($p,p' = 1,\cdots,I$) are the zero-mode wavefunctions 
of a fermion in the fundamental representation of the SU($N$) gauge 
group.  Since the zero-mode wavefunctions are determined by the gauge 
field configuration used in the Dirac equation on $\R^4$, the data 
$\hat{A}_m$ carry some of the information of the gauge field configuration. 
It is known that the data corresponding to the 't Hooft solution are 
\begin{equation}
 \left(\hat{A}_m \right)^{p'p} = (y_p)_m \; \delta_{p'p}\,.
\end{equation}
Thus the data $\hat{A}_m$ provide a way to extract ``center coordinates'' 
from the instanton gauge field configuration. 
When one considers a family of instanton gauge field configurations by 
modifying 
\begin{equation}
 A_m(x) \longrightarrow A'_m(x) = A_m(x) + \xi_m\,, 
\label{eq:family}
\end{equation}
the fermion zero-mode wavefunctions change according to  
\begin{equation}
 \psi(x) \longrightarrow \psi'(x) = e^{i \xi_m x_m} \psi(x) \,.
\end{equation}
Therefore the data $\bra{p'}x_m\ket{p}$ are also extracted by 
\begin{equation}
 \left(\hat{A}'_m \right)^{p'p} = \bra{p'}x_m\ket{p} = 
 \bra{p'}-i\partial_{\xi_m}\ket{p} = 
 -i \int_{\R^4} d^4x \, \psi^{' p' \dagger} \partial_{\xi_m} \psi^{'p}\,. 
\end{equation}

The gauge field configuration (\ref{eq:T4-gauge}) is already in the form 
(\ref{eq:family}), and the zero-mode wavefunctions of the fermions in 
the fundamental representation are also already provided. The corresponding
data, now calculated by integration on $T^4$, not $\R^4$, are 
(\cite{HK} and references therein)
\begin{equation}
 \left(\hat{A}'_{5,6,7,8} \right)^{p'p} = 
 \left( - \frac{\xi_6^{(2)}}{|F_{56}|},0,
        - \frac{\xi_8^{(2)}}{|F_{78}|},0\right)  \delta_{p'p} = 
\left( - \left(\frac{\xi_6^{(2)} L_6}{2\pi k}\right)L_5 , 0 , 
       - \left(\frac{\xi_8^{(2)} L_8}{2\pi k'}\right) L_7, 0 \right) 
\delta_{p'p}\,.
\end{equation}
Therefore the Wilson lines $\xi^{(2)}_6$ and $\xi^{(2)}_8$ can be 
regarded as center-of-mass modes of the instanton center coordinates in
the appropriate limit of the instanton moduli space. 

\vspace{7pt}
\noindent
{\bf Scanning over Instanton Moduli Space}
\vspace{7pt}

\noindent
If the moduli parameters of the instanton configurations are scanned 
randomly (according to some measure), the gauge field configuration is 
determined for each choice of moduli parameter, as are the zero-mode 
wavefunctions.  In the $E_6\rightarrow\SU(5)_{\rm GUT}$ symmetry breaking 
model of section~\ref{ssec:T4}, fermions in the {\bf 10} representation 
of $\SU(5)_{\rm GUT}$ in the effective theory are zero modes of fermions 
in the ${\bf 2}^{-3}$ representation of an $\SU(2)\times\U(1)_6$ instanton 
background.  As the 
moduli of an $\SU(2)$ instanton are scanned, zero-mode wavefunctions of 
the fields in ${\bf 10}=(q,\bar{u},\bar{e})$ vary.  The center coordinates 
of these wavefunctions will be scanned almost randomly, since the instanton 
center coordinates can be chosen arbitrarily in (\ref{eq:tHooft}).  The 
zero modes of a fermion in the ${\bf 2}^{-3}$ representation are not 
classified into the zero modes in $\psi^{a=1}$ and those in $\psi^{a=2}$.
The instanton configuration is contained in a Cartan $\U(1)_2$ subgroup 
for only limited points in the instanton moduli space; for generic 
points in the moduli space the $\SU(2)$ symmetry is completely broken 
and the distinction between $a=1$ and $a=2$ is lost. In general each zero 
mode has non-zero wavefunctions in both $a=1$ and $a=2$.  This means that 
there is no selection rule in the Yukawa couplings; if some zero modes 
were exclusively in $a=1$ and all others were in $a=2$, then the Yukawa 
couplings would have involved only a pair of zero modes, one from each 
group. 

The Gaussian landscapes of sections~\ref{sec:toy1},~\ref{sec:Geometry}, 
and~\ref{sec:Lepton} extracted these features.  In particular, we applied 
these features to all of the fields in the Standard Model.  Certainly the 
Higgs wavefunction is determined only by a gauge field background of an 
Abelian symmetry $\U(1)_6$ in the $E_6 \rightarrow \SU(5)_{\rm GUT}$ symmetry 
breaking model; therefore the complexity (and variety) of instanton field
configurations of non-Abelian symmetries has nothing to do with the scanning 
of the Higgs field's center coordinate.  But this is just an artifact of 
choosing $G$ to be minimal for the up-type Yukawa couplings.  For a larger 
underlying gauge symmetry $G$, such as $G=E_8$, the Higgs is regarded 
as a zero mode of a field in the $\wedge^2 {\bf 5}$ representation 
in $E_8\rightarrow \SU(5)_{\rm GUT}$ symmetry breaking due to an $\SU(5)$ 
instanton.  Gaussian landscapes are based on an expectation that the 
variety of stable gauge-field configurations is so rich that the random 
scanning of gauge field moduli results in (approximately) random and 
independent scanning of the center coordinates of the zero-mode 
wavefunctions.  

One will notice here that the three Gaussian zero-mode wavefunctions in a 
given representation do not necessarily satisfy the orthonormal condition 
(\ref{eq:orthog-condition}), when the center coordinates are chosen 
completely randomly.  Our approach has been to try to implement the rich 
scanning of gauge field moduli in Gaussian landscapes at the cost of giving 
up basis independence.  Since the orthonormal condition is violated 
especially when two center coordinates coincide, the probability 
distribution functions of the smaller eigenvalues of a Yukawa matrix 
in Gaussian landscapes may become unreliable as the approach the largest
possible value, $z_i\sim 0$ in the notation of this paper.  Although the
center coordinates of two instantons can coincide, the two zero-mode 
wavefunctions associated with them should be properly modified so that 
they remain orthogonal.  This modification is not taken into account in
the Gaussian landscape.  It is a yet-to-be tested question in string 
theory whether the gauge field moduli are such that the random scanning 
of center coordinates is a relatively good approximation or not, and if 
not how the correct distribution functions of observables would deviate 
from the predictions of Gaussian landscapes.

\vspace{7pt}
\noindent
{\bf Non-Gaussian Wavefunctions} 
\vspace{7pt}

\noindent
We here note that zero-mode wavefunctions are not always Gaussian for 
arbitrary choice of gauge field moduli parameters.  In fact, when the 
sizes of instantons $\rho_j$ in the 't Hooft solution (\ref{eq:tHooft}) 
are much smaller than the typical distance between the instanton centers 
$|y_k-y_l|$, the 't Hooft solution is a collection of isolated BPST 
instantons. Fermion zero modes are localized around the instanton centers 
$y \sim y_j$, and their wavefunctions decay in as a power of the distance 
$|y-y_j|$, not exponentially. Unless the size parameters $\rho_j$ are 
extremely small, however, overlap integration using these non-exponential 
wavefunctions tends to be larger than the $10^{-5}$--$10^{-6}$ required 
to match the quark and lepton Yukawa couplings of the first generation.

There are also situations where zero-mode wavefunctions decay
linear-exponentially, rather than as a power law.  In section~\ref{ssec:T2} 
(\ref{ssec:T4}), we discuss only $T^2$ ($T^4$) compactification, where 
no topological 1-cycles (2-cycles) can shrink while keeping the volume of 
$T^2$ ($T^4$) finite.  However in more complicated geometries there are 
topological cycles that can do this.  
When a U(1) flux is introduced on such a topological 2-cycle, and if for 
some reason the 2-cycle shrinks, then symmetry breaking by the U(1) flux 
can be localized in extra dimensions, c.f.~\cite{HN}.  
This contrasts with the 
situation in sections~\ref{ssec:T2} and \ref{ssec:T4}, where the 
symmetry-breaking U(1) field strength is spread out homogeneously over the 
extra dimensions. If the symmetry breaking U(1) flux is localized at a 
point in the extra dimensions, then fermion zero-mode wavefunctions decay 
linear-exponentially, with the exponent proportional to the distance from 
the symmetry-breaking source.  Depending on the choice of parameters, the 
linear-exponential wavefunctions may or may not lead to as 
large a hierarchy as that which results from Gaussian wavefunctions.  It 
would be interesting to study the effects on Yukawa-related observables of 
such a localized symmetry breaking source, but this is beyond the scope
of this paper.

To date, there has not been much investigation into what part of the 
moduli space of gauge field backgrounds is more statistically weighted in 
flux compactification.  It might be discovered that the 
statistic distribution is more weighted in regions that lead to Gaussian 
(or possibly linear-exponential) wavefunctions, so that large hierarchy 
among Yukawa couplings follows as a likely consequence. The other 
possibility is that environmental selection in favor of a light charged 
lepton (and/or quarks) enhances the statistical weight for such regions.  
We do not know which, if either, possibility is correct.  However, for 
the landscape to account for the observed hierarchical patterns in the mass 
eigenvalues and mixing angles, we must assume one of these possibilities.  
As long as either of these possibilities is correct, then Gaussian 
landscapes should not be too terrible an approximation (or an effective 
description) of the landscape formulated by a super Yang--Mills 
theory on higher-dimensional spacetime. 

\subsection{$T^3$-Fibered Compactification and M(IIA)-theory Dual}
\label{ssec:M}

In this paper we present numerical simulations of Gaussian landscapes
involving only $D\leq 2$ extra dimensions.  This is due to limited 
computational resources, not because there is any theoretical motivation
to study landscapes involving $D\leq 2$. Indeed, more interesting would
be to study Gaussian landscapes on $D=3$, since such a model would
directly simulate (some fraction of) the landscape of Yukawa couplings 
in string theory. A Gaussian landscape on a three-dimensional manifold 
$B$ corresponds to Heterotic string theory compactified on a 
six-dimensional manifold that is a $T^3$-fibration over $B$. To clarify,
when one says that a $m$-dimensional manifold $X$ is a $T^n$-fibration 
over a manifold $B$, this means that the manifold $X$ locally resembles 
$T^n \times \R^{m-n}$ almost everywhere in $X$. For example, 
$T^2 = T^1 \times T^1$ and 
$T^4(x_{5,6,7,8})= T^2(x_{6,8})\times T^2(x_{5,7})$ in 
sections~\ref{ssec:T2} and~\ref{ssec:T4} are trivial examples of 
$T^1=S^1$ and $T^2$-fibered geometry. Here $B$ is called a base manifold, 
and $T^n$ a fiber. 

Suppose that a six-dimensional manifold $X$ is a $T^3$-fibration over 
a three-dimensional manifold $B$.  We choose a coordinate system locally, 
so that the $T^3$ direction is parametrized by $(x_6,x_8,x_{10})$ and a 
local patch of $B$ by $(x_5, x_7, x_9)$. When the periods of $T^3$ in its
three directions, $L_6$, $L_8$, and $L_{10}$, are small compared with the 
size of $B$ (that is, the cubic root of the volume of $B$), then a stable 
gauge field configuration on $X$ is approximately described by fields
$A_{6,8,10}$ that vary slowly on the coordinates $x_{5,7,9}$ of $B$. 
Fermion zero modes are localized at points on $B$ where all of 
$A_6(x_5,x_7,x_9)$, $A_8(x_5,x_7,x_9)$ and $A_{10}(x_5,x_7,x_9)$ 
vanish. 
As long as $L_6$, $L_8$, and $L_{10}$ are all small, the 
zero-mode wavefunctions are approximately Gaussian~\cite{AW}. Although 
$X$ is not globally the same as $T^3\times\R^3$ or $T^3\times T^3$, the
local structure of $X$---its $T^3$-fibration---is sufficient to determine 
the approximately Gaussian zero-mode wavefunctions. Wavefunctions away 
from the Gaussian peak will depend on details of the global structure of 
$X$, but the wavefunctions are exponentially small in this region.  
Assuming a Gaussian profile with a fixed width for wavefunctions is surely
a very crude approximation, but it might suffice as a zeroth order 
approximation.  A Gaussian landscape on a $D=3$-dimensional manifold $B$ 
assumes that the moduli of stable gauge field configurations are so rich 
that the center coordinates of various fields are scanned (almost) 
randomly and independently.  The results of section~\ref{sec:Geometry}
provide a qualitative picture of how the distributions of observables
depend on the base manifold $B$.

The most important parameter in a Gaussian landscape is the ratio of the
Gaussian width to the size of the extra dimensions, $d/L$.  For example, 
the overall hierarchy of Yukawa couplings is proportional to $(L/d)^2$ on 
a logarithmic scale.  As we have seen in section~\ref{ssec:T2}, this 
ratio is proportional to $L_5/L_6 = {\rm vol}(B=S^1)/{\rm vol}(T^1)$. This 
is generalized to 
\begin{equation}
\Delta\ln\lambda \propto 
  \left[ \frac{{\rm vol}(B)}{{\rm vol}(T^3)} \right]^{\frac{1}{3}} 
  \sim
   \frac{1}{\alpha'}\left[{\rm vol}(B) \times {\rm
	  vol}(\hat{T}^3)\right]^{\frac{1}{3}} . 
\label{eq:BtoF-M}
\end{equation}
As in sections~\ref{ssec:T2} and~\ref{ssec:T4}, the volume of the
$T^3$ fiber has to be sufficiently smaller than that of the base
manifold $B$ in order for hierarchy to be generated. It is an 
interesting question whether this property can be understood within
string theory, not as a phenomenological requirement.

The Heterotic string theory compactified on a $T^3$-fibered geometry 
corresponds to an eleven-dimensional supergravity compactification 
on a $K3$-fibered geometry (this is an intersecting D6--D6 system 
of the Type IIA string theory) in the limit of small ${\rm vol}(T^3)$. 
Thus, Gaussian landscapes on three-folds for various values of 
$(d/L)$ are intended to simulate a class of vacua of string theory that
interpolates between the Heterotic theory and M-theory (ignoring 
stringy corrections). The expression in Type IIA language (the last
term) in (\ref{eq:BtoF-M}) is known.  Numerical simulation can tell us 
the value of $d/L$ that fits to the observed pattern of masses and 
mixings, which suggests the approximate value of (\ref{eq:BtoF-M}) for 
the vacuum of our universe.

\subsection{$T^2$-Fibered Compactification and F-theory Dual}
\label{ssec:F}

The analysis of $D=2$ Gaussian landscapes in section~\ref{sec:Geometry} 
may help one understand the landscape of Yukawa couplings of Heterotic 
compactification on a $T^2$-fibered geometry. Let us now consider a 
six-dimensional manifold $X$ that is a $T^2$-fibration over a four-fold 
$B$. Let $(x_9,x_{10})$ parametrize a local patch of the $T^2$-fibration 
and $(x_{5,6,7,8})$ a local patch of $B$. Fermion zero modes are localized 
on a two-dimensional sub-manifold (called the matter curves) on $B$ so 
that the gauge fields $A_9(x_{5,6,7,8})$ and $A_{10}(x_{5,6,7,8})$ 
both vanish~\cite{FMW}.  Furthermore, their wavefunctions around the matter 
curves are approximately Gaussian in the two transverse directions 
determined by $\nabla A_9$ and $\nabla A_{10}$. The existence of the two 
transverse directions in which zero-mode wavefunctions are Gaussian is 
quite similar to the $D=2$ Gaussian landscapes.  One can determine the 
value of $d/L$ that fits the observed hierarchy of masses and mixing 
angles, which might then be used to infer 
${\rm vol}(B)^{\frac{1}{4}}/{\rm vol}(T^2)^{\frac{1}{2}}$ 
for the $T^2$-fibered geometry. This class of vacua interpolates 
Heterotic string theory and F(type IIB string)-theory.

There are also some features that are not captured by the $D=2$ 
Gaussian landscapes.  Suppose that $A_9$ varies along $x_7$, and 
$A_{10}$ along $x_8$; at least it is possible to choose a coordinate system 
on $B$ locally so that this happens. Then a matter curve is along the 
($x_5,x_6$) directions.  However the net chirality of a fermion depends 
on $F_{56}$, the gauge field along the matter curve.  Therefore zero-mode 
wavefunctions should have some $(x_5,x_6)$ dependence, but no such 
dependence is taken into account in the $D=2$ Gaussian landscapes.  
Moreover, the global geometry of matter curves may be complicated in $B$,
and they can intersect one another, but the $D=2$ Gaussian landscapes 
ignore this as well. The behavior of zero-mode wavefunctions on a 
$T^2$-fibered geometry and the geometry of the intersection of matter curves 
must be studied further to determine whether the $D=2$ Gaussian landscapes 
can be useful in studying the landscapes of Yukawa couplings 
with $T^2$-fibered compactification.

\section{Conclusions}

This decade has seen the emergence of a major debate:  to what extent 
is nature fundamentally uniquely prescribed, for example by symmetries, 
vs.\! to what extent nature results from the statistics of a huge 
landscape of vacua, modified by cosmological and environmental selection.  
For the Standard Model symmetries play a key role, but for physics beyond 
the Standard Model the question remains largely open.  Unified gauge 
symmetries have striking achievements, for example a simple interpretation 
of the quantum numbers of a generation and a precise numerical prediction 
for the ratios of the measured gauge couplings.  For quark and lepton 
flavor, however, the picture offered by approximate flavor symmetries (AFS) 
is much less compelling, lacking both theoretical simplicity and
significant successful predictions.  Although the AFS description of flavor 
is apparently well suited to give an understanding of the hierarchical 
nature of charged fermion masses and the CKM mixing matrix, it comes with 
too much flexibility: with an appropriate choice of charges and symmetries 
any pattern of flavor can be generated. 

In this paper we introduce Gaussian landscapes as some of the simplest
landscapes in extra dimensions that can account for flavor.  Particles 
are assumed to possess localized (Gaussian) zero-mode wavefunctions over 
some geometry of extra dimensions, and the Gaussian landscape consists of
independently scanning the peak position of each of these wavefunctions.  
Small flavor-symmetry breaking parameters are replaced by small overlap 
integrals of these wavefunctions on the extra dimensions.  Localized 
zero-mode wavefunctions in extra dimensions are a natural expectation, and 
may have a more elegant realization than Higgs potentials for flavor 
symmetry breaking.  We claim neither precise predictions nor do we present 
a compelling top-down theoretical model, rather we present the patterns 
of flavor that emerge from these simple toy landscapes, the features of 
these extra-dimensional landscapes that are relevant to the flavor problem, 
and how these landscapes might be realized within the context of string 
theory.

In the simplest Gaussian landscape describing the quark sector, where all 
quarks and the Higgs have a universal Gaussian wavefunction (but with 
independently scanning center coordinates) over a single extra dimension 
with geometry $S^1$, we find the three major characteristics of quark 
flavor: a hierarchical distribution of quark masses, pairing structure 
(the $W$-current approximately connects distinct pairs of quarks), and 
generation structure (the electroweak pairing connects the heaviest up-type 
quark to the heaviest down-type quark, and similarly for the middle and 
lightest quarks).  The relevant probability distributions are shown in 
Figures~\ref{fig:entrydistr} and~\ref{fig:D=1-nocut} and result from 
inputting just two parameters: a universal constant $g_{\rm eff}$ of order 
unity setting the scale for overlap integrals and a universal constant 
$d/L$ setting the width of Gaussian wavefunctions relative the size of the 
extra dimension.  

The flavor structure of the quark sector can also be obtained from AFS 
using two free parameters. For example, by hypothesizing an approximate 
$U(1)$ symmetry with a leading Yukawa coupling of order unity and 
others suppressed by various powers of a small symmetry breaking parameter 
$\epsilon$.  However there is a crucial difference. In the AFS case one 
must carefully choose the $U(1)$ charges of each of the fifteen fermions 
of the Standard Model.  A huge variety of mass patterns could be 
accommodated by suitable charge choices.  In the Gaussian landscape no 
such choices are made. Each of the fifteen fermions is treated 
symmetrically. They differ only by the location of their Gaussian 
wavefunction, and these are scanned randomly over $S^1$.  Thus the 
hierarchies arise purely from statistics; they cannot be changed as they 
do not involve any free parameters beyond $g_{\rm eff}$ and $d/L$. 

While the above accomplishments of the Gaussian landscape on $S^1$ are 
striking, there are certain features that are less than ideal.  Although 
they are peaked, the probability distributions for flavor parameters are
quite broad, as can be seen in Figure~\ref{fig:D=1-nocut}.  At half 
maximum, the deviation from the peak value is typically an order of 
magnitude.  Thus the statistical nature of the landscape prevents us
from making precise predictions. We find that this order of magnitude 
width is also typical of Gaussian landscapes in more than one dimension.
The $S^1$ landscape makes no distinction between up and down sectors.  
This is a problem for explaining the observed large $t/b$ mass ratio, 
in particular because on $S^1$ the distribution for the top and bottom
masses is narrower than for the other generations and also because it is
peaked near the maximum value.  Thus even if there were a selection effect 
favoring a heavy top quark, the observed bottom mass would still be 
somewhat unlikely.  

One possibility to resolve this shortcoming is to replace the effective 
coupling $g_{\rm eff}$ with two parameters: one for the up sector that is 
about an order of magnitude larger than the one for the down sector.  It 
would be interesting to find the landscape origin for such a ``$\tan\beta$''
factor.  Another possibility is that the shape of the $t/b$ mass 
distribution is a special feature of the $S^1$ landscape.  We find that 
this is the case.  In the $T^2$ landscape, where Gaussian wavefunctions 
are distributed at random over the surface of a (square) torus, the 
distribution is wider, as shown in Figure \ref{fig:T2YukawaEgval}.  The
same result holds for Gaussian landscapes on $S^2$ and $S^3$.  A selection 
for a heavy quark, for example for electroweak symmetry breaking, could 
then more easily account for the $t/b$ mass ratio.  For any number of extra 
dimensions, we can analytically compute the Yukawa probability distribution 
near maximal values, and we find that it is suppressed as the dimension 
increases, strengthening this interpretation of the $t/b$ ratio.

Although flavor symmetries play no fundamental role in Gaussian landscapes, 
small overlap integrals may allow AFS to emerge in the low-energy theory.  
For example, if $q_1$, $q_2$ and $h$ have narrow Gaussian profiles with 
centers that are well-separated from each other, then AFS that act on $q_1$ 
and $q_2$ will emerge at low energy.  We have found that Yukawa matrices
have the form expected from Abelian AFS, namely $\lambda_{ij} \sim
\epsilon_i^{\bar{q}}\epsilon^q_j$, if the Higgs width is not too wide
and if the Yukawa coupling is not too small.  As the separation between the 
Higgs and quarks approaches the scale of the extra dimension, effects 
associated with the periodicity of the wavefunctions destroys the AFS form 
of the Yukawa coupling.  Nevertheless, the AFS approximation aids in 
understanding the numerical results, and we have used it to compute 
approximate analytic distributions for the AFS factors $\epsilon$, the 
quark mass eigenvalues, and the CKM mixing angles.  

This strategy is particularly useful for understanding Gaussian landscapes 
on higher dimensional geometries where numerical integration requires 
greater computational resources.  Analytic results for the geometries 
$S^1$, $S^2$, $T^2=S^1\times S^1$, and $S^3$ have been obtained and the 
distributions are compared in Figure~\ref{fig:FN-D23}.  We find that the 
qualitative predictions of the Gaussian landscape on $S^1$ remain intact.  
One important feature is that, independent of the dimension $D$, the AFS 
distribution functions and flavor observables are found to be polynomial 
or logarithmic functions of $\ln\lambda_i$ or $\ln(\sin\theta_{ij})$.  By 
comparing numerical distributions of flavor parameters in Gaussian 
landscapes on $T^2$ and $S^2$, we have investigated the importance of the 
shape and curvature of the extra dimensional manifold.  We find some 
differences in the details, but the more striking are the similarities, 
as seen for example in Figure~\ref{fig:compareT2S2}.

Large leptonic mixing angles suggest that in the low-energy theory
no AFS emerges for the lepton doublets $l_i$, implying that the
Gaussian widths for $l_i$ are comparable to the size of the extra
dimension(s).  Since the fermion mass hierarchies are smaller in the 
down and charged lepton sectors than in the up sector, we economize on
parameters and assume that the widths preserve an $SU(5)$ symmetry,
with $d_{\bar{5}} \gg d_{10}, d_H$.  The smallness of the neutrino 
masses suggests that there is no light right-handed neutrinos, such that 
the light neutrinos are Majorana and the origin of neutrino masses 
involves the breaking of $B-L$ symmetry. In the seesaw mechanism this 
implies Majorana masses for heavy right-handed neutrinos. In the Gaussian 
landscape we expect Gaussian profiles for both $\overline{\nu}_R$ and the 
$B-L$ breaking fields. The center of the $B-L$ breaking is not expected 
to be correlated with the center of $SU(2)\times U(1)$ breaking, leading 
to a very different statistical character for the Majorana and Dirac mass
matrices of neutrinos.  

We find that the physics of neutrino masses and mixings is quite
different depending on whether the profile for $\overline{\nu}_R$ is 
narrow or wide.  In the case that it is narrow, localization leads to 
significant hierarchies in the eigenvalues of both the Majorana and Dirac
matrices. In theories based on AFS these hierarchies cancel in the
light neutrino mass matrix, but in the Gaussian landscape these 
hierarchies add.  The result is that $m_2/m_3$ is typically too small to 
agree with observation, unless perhaps if strong selection effects are 
important.  For wide $\overline{\nu}_R$ profiles, the neutrino Yukawa and 
Majorana matrices take on a democratic form, and again have hierarchical 
eigenvalues. If the profiles are sufficiently wide, $m_2/m_3$ is always 
less than the observed value, as shown in Figure~\ref{fig:broad-broad}.  
Similar difficulties arise with the lepton mixing angle $\theta_{23}$.  
Although a large width for the lepton doublets leads to large 23 rotations 
to diagonalize the charged and neutral lepton mass matrices, these large 
angles cancel so that the physical mixing angle $\theta_{23}$ is typically 
small, as shown in the numerical simulation of Figure~\ref{fig:fail2get23}. 

These difficulties, that $m_2/m_3$ and $\theta_{23}$ are typically much 
smaller than the values observed, are both solved by a remarkably simple 
observation.  In order to account for CP violation in the quark sector a 
complex phase must be introduced to the Gaussian profile. Using the simple 
profile of (\ref{eq:cpxGaussian}), we find the following results for the 
Gaussian landscape on $S^1$: 
\begin{itemize}
\item Previous results for distributions of charged fermion masses are 
preserved, for example Figure~\ref{fig:5vs10} is essentially unchanged 
for CP-violating Gaussian landscapes.
\item The generation structure of the quark sector is preserved 
(Figure~\ref{fig:cpx-quark-mix}).
\item The CKM phase is of order unity (Figure~\ref{fig:CP-phase}).
\item Large leptonic mixing angles $\theta_{12}$ and $\theta_{23}$ are 
obtained (Figure~\ref{fig:cpx-lepton-mix}).
\item The large observed value for the neutrino mass ratio $m_2/m_3$ is
not atypical (Figure~\ref{fig:compare23}(h)).   
\end{itemize}

The distributions of flavor observables on Gaussian landscapes are 
typically broad. However, given the experimental measurement of some
subset of flavor observables, the conditional distributions for the 
remaining observables in our universe changes significantly.  We find 
that even quite loose cuts on the measured neutrino parameters are 
sufficient to considerably sharpen the predictions for the leptonic 
mixing angle, $\theta_{13}$, and the neutrinoless double beta parameter, 
$m_{\beta\beta}$, within a particular Gaussian landscape; see 
Figure~\ref{fig:cut-effect} and~(\ref{eq:mbetabeta}), respectively.  
Furthermore these cuts lead to a prediction of large CP violation in 
neutrino oscillation as shown in the last row of 
Figure~\ref{fig:CP-phase} and in the scatter plots of 
Figure~\ref{fig:cut-effect}.

When a supersymmetric gauge theory is compactified on an internal
manifold, quarks, leptons and the Higgs boson may originate from
gauginos and the components of gauge fields with polarization along the 
internal manifold.  Their Yukawa interactions originate from the gauge 
interactions of the gauginos, while the Yukawa coupling constants are 
calculated by overlap integration of zero-mode wavefunctions over the 
extra-dimensional space. Wavefunctions become Gaussian when the 
extra-dimensional manifold is a torus fibration, and the fiber is small 
relative to the base.  Thus, Gaussian landscapes fit very well into this
framework---compactification of a supersymmetric gauge theory on a
torus-fibered manifold. 

The ratio $d/L$, which controls the hierarchy, roughly reflects the 
ratio of the torus fiber size to the base size in Heterotic theory language. 
We now know the value of $d/L$ that is good for quark and lepton 
phenomenology, but it remains an open question why such a value has been 
selected from all the possibilities in the landscape (or if it represents
a typical subset of the landscape).  Indeed, there is no theoretical 
motivation to consider torus-fibered compactification over the myriad of
other geometries for compactification.  We also assumed in the Gaussian
landscape that the center coordinates of all Gaussian wavefunctions scan 
randomly and independently over the base space. This assumption is not 
totally without motivation, but it would be nice if the validity of this 
assumption were studied in string theory. 

Simple Gaussian landscapes, motivated by supersymmetric gauge theories
compactified on a torus-fibered manifold, can account for the broad
pattern of quark and lepton masses and mixings in terms of just five
parameters. While the probability distributions for the measured
parameters are broad, predictions from a given landscape for future
measurements in neutrino physics are more precise.

\section*{Acknowledgments}

This work was supported in part by the the NSF grant PHY-04-57315 (LJH),
the US DOE under contract No. DE-AC03-76SF00098 (LJH) and No. 
DE-FG03-92ER40701 (MPS, TW), and the Gordon and Betty Moore Foundation 
(TW).  We thank Aspen Center for Physics (LJH, TW), UC Berkeley CTP (TW) 
and MIT CTP (TW) for hospitality.

\appendix
\section{Approximate Probability Distribution Functions}

We here collect the probability distribution functions calculated using 
the AFS approximation, i.e. using (\ref{eq:mass-approx}) and 
(\ref{eq:Vus-approx}--\ref{eq:Vub-approx}), for the Gaussian landscape
on $S^1$.  First we list the distribution functions of the Yukawa 
eigenvalues that follow from (\ref{eq:FN-combined}) and the 
approximation (\ref{eq:mass-approx}):
\begin{eqnarray}
\frac{dP(z_3)}{dz_3} &=& \left\{\begin{array}{ll}
\displaystyle\frac{9}{4}\left(\pi - 8\sqrt{z_3}+(4+\pi)z_3 
   -\frac{8}{3} z_3^{3/2} + \frac{\pi}{8} z_3^2 \right) \qquad\quad & 
   {\rm for~} 0\leq z_3\leq 1\,, \phantom{\Bigg(\Bigg)} \label{eq:egval3-D1}\\
\displaystyle -\frac{3}{16}\bigg[ 32 - 76 \sqrt{z_3-1}+48z_3-26z_3 
   \sqrt{z_3-1} \bigg. \phantom{\Bigg(\Bigg)} & \\
   \qquad \,-\,3 \left(8 + 8z_3 + z_3^2 \right) 
          {\rm arccot}\left( \sqrt{z_3-1} \right)
          \phantom{\bigg(\bigg)} & \\ 
   \qquad \bigg. -3\left(8 + 8z_3 + z_3^2 \right)
                 \arctan\left( \sqrt{z_3-1}\right) \bigg] & 
    {\rm for~} 1<z_3\leq 2\,. \end{array} \right.
\end{eqnarray}
\vspace{-12pt}
\begin{eqnarray}
\frac{dP(z_2)}{dz_2} &=& \left\{ \begin{array}{ll}
\displaystyle \frac{3}{8} z_2 \left(24 - 32 \sqrt{z_2}+3\pi z_2\right) &
  {\rm for~} 0\leq z_2\leq 1\,, \phantom{\bigg(\bigg)} \label{eq:egval2-D1}\\
\displaystyle\frac{3}{2}\left(4-2 \sqrt{z_2-1} -6z_2 + 5z_2 
  \sqrt{z_2-1} \right) \phantom{\Bigg(\Bigg)} & \\
\displaystyle-\frac{9}{4} z_2^2 \Big[ \arctan\left(\sqrt{z_2-1}\right) 
   -{\rm arccot}\left(\sqrt{z_2-1}\right)\Big] \quad\,\,\,\, &  
  {\rm for~} 1<z_2\leq 2\,.   \end{array} \right.
\end{eqnarray}
\vspace{-12pt}
\begin{eqnarray}
\frac{dP(z_1)}{dz_1} &=& \left\{ \begin{array}{ll}
\displaystyle\frac{9\pi}{32} z_1^2 \phantom{\Bigg(\Bigg)} & 
    {\rm for~} 0\leq z_1\leq 1\,, \label{eq:egval1-D1} \\
\displaystyle\frac{9}{16}\bigg\{ 2(2 - z_1) \sqrt{z_1-1} \bigg. & \\
 \quad\, \bigg. -z_1^2 \Big[ \arctan\left(\sqrt{z_1-1}\right)- 
     {\rm arccot}\left(\sqrt{z_1-1}\right) \Big] \bigg\} & 
    {\rm for~} 1< z_1\leq 2\,. \end{array}  \right. \,\,\,\,\,\,\,\,\,
\end{eqnarray}
Here $z_i \equiv \ln(\lambda_i/\lambda_{\rm max})/\Delta\ln\epsilon$ 
where $\lambda_{\rm max}=(4L^2/9\pi d^2)^{1/4}g_{\rm eff}$ and  
$\Delta\ln\epsilon=-(L/d)^2/12$.  We also note the mean values of 
these distributions: 
\begin{equation}
\vev{z_3} = 0.2\,, \qquad \vev{z_2} = 0.6\,, \qquad \vev{z_1} = 1.2\,. 
\label{eq:yij}
\end{equation}

Meanwhile, the distribution functions of the CKM mixing angles follow 
from (\ref{eq:FN-combined}) and the set of approximations 
(\ref{eq:Vus-approx}--\ref{eq:Vub-approx}).  They are given by:
\begin{eqnarray}
\frac{dP(t_{12})}{dt_{12}} &=& 6\left(1-\sqrt{t_{12}}\right)^3 
   \left(1 + 2\sqrt{t_{12}}\right) dt_{12} \label{eq:th12-D1} \\
\frac{dP(t_{23})}{dt_{23}} &=& 
   \frac{3}{2} \left[ -2\sqrt{1-t_{23}} + {\rm arccosh}
   \left(\frac{2}{t_{23}}-1\right)\right] \nonumber \\
  && \quad \times \left[ 2\sqrt{1-t_{23}}(1+2t_{23}) 
     -3t_{23} {\rm arccosh}\left( \frac{2}{t_{23}}-1\right) \right], 
     \label{eq:th23-D1} \\
\frac{dP(t_{13})}{dt_{13}} &=& 6\left( \sqrt{1-t_{13}}+\sqrt{t_{13}}
     -1\right) \left( 1 - \sqrt{t_{13}}\right) \nonumber \\
  && \quad \times \left[2\sqrt{1-t_{13}}-1-\sqrt{t_{13}}+2t_{13} 
     +2\sqrt{t_{13}(1-t_{13})} \right], \label{eq:th13-D1}
\end{eqnarray}
where each applies over the full range $0\leq t_{ij}\leq 1$, with  
$t_{ij}\equiv \ln|\sin\theta_{ij}|/\Delta\ln\epsilon$.  The mean values
of these distributions are:  
\begin{equation}
\vev{t_{12}} = 0.17\,, \qquad \vev{t_{23}} = 0.10\,, \qquad 
\vev{t_{13}} = 0.35\,.
\label{eq:tij} 
\end{equation}
%


\end{document}